\documentclass[a4paper,11pt]{article}
\pdfoutput=1 

\usepackage{jheppub} 

\usepackage[T1]{fontenc} 

\usepackage{float}
\usepackage{color}
\usepackage{xcolor}
\usepackage{autobreak}
\usepackage{multirow}
\usepackage{diagbox}
\usepackage{amsmath}
\usepackage[vcentermath]{youngtab}
\usepackage{pifont}
\usepackage{slashed}
\usepackage{makecell}

\newcommand{\cmark}{\color{blue}\checkmark}

\newcommand{\beq}{\begin {equation}}  
\newcommand{\eeq}{\end   {equation}} 
\newcommand{\bea}{\begin {eqnarray}} 
\newcommand{\eea}{\end   {eqnarray}}  
\newcommand{\baa}{\begin {array}   } 
\newcommand{\eaa}{\end   {array}   }     
\newcommand{\bit}{\begin {itemize} }
\newcommand{\eit}{\end   {itemize} }
\newcommand{\be }{\begin {equation}} 
\newcommand{\ee }{\end   {equation}}

\newcommand{\vev}[1]{ \langle {#1}  \rangle }

\newcommand{\mc}[1]{\mathcal{#1}}

\allowdisplaybreaks

\title{\boldmath A complete tree-level dictionary between simplified BSM models and SMEFT (d $\leq$ 7) operators}

\author[a]{Xu-Xiang Li,}
\author[b]{Zhe Ren,}
\author[b, c, d, e, f]{Jiang-Hao Yu}

\affiliation[a]{Department of Physics and State Key Laboratory of Nuclear Physics and Technology, Peking University, Beijing 100871, China}
\affiliation[b]{CAS Key Laboratory of Theoretical Physics, Institute of Theoretical Physics, Chinese Academy of Sciences, Beijing 100190, China}
\affiliation[c]{School of Physical Sciences, University of Chinese Academy of Sciences, Beijing 100049, P.\ R.\ China}
\affiliation[d]{Center for High Energy Physics, Peking University, Beijing 100871, China}
\affiliation[e]{School of Fundamental Physics and Mathematical Sciences, Hangzhou Institute for Advanced
Study, UCAS, Hangzhou 310024, China}
\affiliation[f]{International Centre for Theoretical Physics Asia-Pacific, Beijing/Hangzhou, China}

\emailAdd{xuxiangli@pku.edu.cn}
\emailAdd{renzhe@itp.ac.cn}
\emailAdd{jhyu@itp.ac.cn}

\abstract{
Finding all possible UV resonances of effective operators is an important task in the bottom-up approach of effective field theory. We present all the tree-level UV resonances for the dimension-5, -6 and -7 operators in the Standard Model effective field theory (SMEFT), and then obtain the correspondence between the UV resonances and the effective operators from the relations among their Wilson coefficients, through the functional matching and operator reduction procedure. This provides a cross-dimension UV/IR dictionary for the SMEFT at tree-level, and the methods used here, especially the on-shell construction of general UV Lagrangian and the systematic reduction of operators, are extendable for UV resonances of $d \geq 8$ operators in SMEFT and other EFTs.
}

\begin{document} 
\maketitle
\flushbottom



\section{Introduction}
\label{sec:intro}

Being the most successful theory in particle physics, the Standard Model (SM) has been tested and verified by many experiments. However, it is not a complete theory of fundamental interactions because it fails to answer some important questions, such as the nature of the dark matter, the origin of the neutrino masses, matter-antimatter asymmetry, etc. Therefore, physicists are searching for physics beyond the SM to address these issues. Up to now, the direct searches on the Large Hadron Collider (LHC) has not found any signals beyond the Standard Model (BSM), which pushes the scale of new physics up to TeV or several TeV. Due to the considerable energy gap between the electroweak scale and the new physics scale, the effective field theory (EFT) approach provide a model independent way that parameterize the effects of the BSM physics into the Wilson coefficients of the higher dimensional operators in the EFT to probe BSM physics.

The standard model effective field theory (SMEFT) is an EFT at the electroweak scale and it is constructed based on the fields and symmetries of the SM. The Lagrangian of the SMEFT is formulated as the sum of effective operators including the SM Lagrangian and a series of possible higher dimensional operators according to the power counting. Among the higher dimensional operators in the SMEFT, the dimension-5 operator is first written by Weinberg~\cite{Weinberg:1979sa}, and since then the operator bases have been enumerated up to dimension 9~\cite{Buchmuller:1985jz,Grzadkowski:2010es,Lehman:2014jma,Henning:2015alf,Liao:2016hru,Li:2020gnx,Murphy:2020rsh,Li:2020xlh,Liao:2020jmn} and higher. After the complete set of the SMEFT operators is given, the Wilson coefficients of the effective operators that parameterize the deviation from the SM can be determined by analyzing the experimental data form the high energy colliders and low energy experiments, e.g. Ref.~\cite{Brivio:2017vri,Ellis:2020unq,Isidori:2023pyp}. If the experimental data exhibits significant difference from the SM prediction, physicists can find the corresponding effective operators that cause the difference. After that, the dictionary between the effective operators and their UV origins will greatly benefit the searching for BSM physics~\cite{Bechtle:2022tck}.


In the EFT framework, there are usually two ways to find the connection between the effective operators and possible UV origins. The first one is to start from an UV model and then perform the matching procedure by integrating out the heavy degrees of freedom. If the above matching procedure yields a non-zero Wilson coefficient of an effective operator, then the UV model should be considered as one of the UV origins of the effective operator. This method is called the top-down approach~\cite{Georgi:1993mps,Skiba:2010xn}. The advantage of this approach is that all effective operators that connect with an UV origin can be found at the same time, but one can not find all the UV origins of an effective operator without performing a exhaustive search of all possible UV origins, which is very time-consuming and error-prone due to the large variety of UV models.
The second approach is the bottom-up approach. With this approach, one can find all UV origins of an effective operator without writing down the explicit UV interactions, by simply
examining the Lorentz and gauge quantum numbers of the local on-shell amplitude generated by the operator. For a partition of the external particles, one organize the on-shell amplitudes into the eigenstates of angular momentum $J$ and gauge quantum number $\mathbf{R}$, and this amplitude basis, as well as the corresponding operator basis, is called the j-basis~\cite{Li:2022tec,Li:2022abx,Jiang:2020rwz,Li:2020zfq}.

In this work, we focus on the tree-level dictionary between the UV origins and the effective operators in the SMEFT. The reason that we consider the tree-level dictionary is that the tree-level UV origins are typically the leading contribution to the observable. What is more, the tree-level amplitude with heavy immediate
particle behave as a resonance and can be probed directly in the collider experiments such as the LHC. At mass dimension 5 in the SMEFT, the tree-level dictionary is straightforward since the only independent operator is the Weinberg operator and the UV origins are the three types of seasaw models~\cite{Yanagida:1979as,Gell-Mann:1979vob,Mohapatra:1979ia,Magg:1980ut,Schechter:1980gr,Foot:1988aq}. 
The dimension-6 tree-level dictionary is given by Ref.~\cite{deBlas:2017xtg}, and the complete tree-level UV resonances are also presented in Ref.~\cite{Li:2022abx} using the j-basis method. Here we extend our discussion to involve the dimension-7 effective operators and the UV resonances to give a complete dimension-5, -6 and -7 tree-level dictionary, where the cross-dimension contributions of effective operators induced by field re-definitions are included. The complete UV resonances that contribute to the SMEFT operators up to dimension 7 have been listed in Ref.~\cite{Li:2022abx}. We start by utilizing the spinor-helicity formalism for massless and massive amplitudes and the Young tableau method to generate the Lorentz structure and the gauge structure of the UV Lagrangian that involves all the UV resonances. Then we apply the functional matching method~\cite{Gaillard:1985uh,Cheyette:1987qz,Henning:2014wua,Cohen:2020qvb,Fuentes-Martin:2020udw} 
to integrate out the UV resonances at tree level and obtain a set of effective operators carrying all kinds of redundancies. After that, we propose a systematic reduction method inspired by the off-shell amplitude formalism to reduce the set of operators to the integrated dimension-5, -6 and -7 SMEFT operator basis listed in appendix~\ref{sec:basis567}. 
The above procedures, including constructing UV Lagrangian, matching and reduction, are fully systematic and can be applied to higher-dimensional operators in the SMEFT and other EFTs.

The paper is organized as follows. In section~\ref{sec:spinor}, we briefly introduce the spinor-helicity formalism for massless and massive amplitudes, and present the independent Lorentz structures of the UV Lagrangian using the spinor-helicity formalism. We then show the construction of the full UV Lagrangian involving the UV resonances that have tree-level contributions to the dimension-5, -6, and -7 operators in section~\ref{sec:uvlag}. In section~\ref{sec:procedure}, we will use some examples to illustrate the matching and reduction procedure. We translate the result into the correspondence between the UV resonances and the IR effective operators and present the result in section~\ref{sec:result}. Section~\ref{sec:summary} is our conclusion.







\section{Massive on-shell amplitude basis}
\label{sec:spinor}
In order to construct the UV Lagrangian, first we need to construct the complete and independent basis of the Lorentz structures in the UV Lagrangian. In this paper, the above basis of Lorentz structures is obtained by translating the massive amplitude basis into operator basis with the massive amplitude-operator correspondence. We will illustrate the procedure in this section.
\subsection{Massless and massive spinors}
\label{sec:component}

In this section, we will briefly introduce the spinor-helicity formalism for massless~\cite{DECAUSMAECKER198253,BERENDS198261,KLEISS1985235,Xu:1986xb} and massive spinors~\cite{Arkani-Hamed:2017jhn}. We start with the 4-momentum $p_{\mu}$, which can be expressed by a $2\times 2$ matrix after contracting with the $\sigma^{\mu}_{\alpha\dot{\alpha}}$ matrices
\begin{eqnarray}
	p_{\alpha\dot{\alpha}} = p_{\mu} \sigma^{\mu}_{\alpha\dot{\alpha}} = \left(\begin{array}{cc}
		p^0-p^3 & -p^1+i p^2 \\
		-p^1-i p^2 & p^0+p^3
	\end{array}\right).
\end{eqnarray}
It should be noted that $\det p_{\alpha\dot{\alpha}}=p^2=m^2$.

For massless particles, $\det p_{\alpha\dot{\alpha}}=0$. That means the matrix $p_{\alpha\dot{\alpha}}$ is rank-1, and thus can be written as the direct product of two 2-vectors $\lambda_{\alpha}$ and $\tilde{\lambda}_{\dot{\alpha}}$,
\begin{align}
	p_{\alpha\dot{\alpha}}=\lambda_{\alpha}\tilde{\lambda}_{\dot{\alpha}},
\end{align}
where the $\lambda_{\alpha}$ and $\tilde{\lambda}_{\dot{\alpha}}$ are independent complex dimension-2 vectors for general complex momentum while $\tilde{\lambda}_{\dot{\alpha}}= (\lambda_{\alpha})^*$ for real momentum. The $\lambda_{\alpha}$ and $\tilde{\lambda}_{\dot{\alpha}}$ are called spinor-helicity variables, and they transform under both the Lorentz group and the little group. For a fixed momentum $p_{\alpha\dot{\alpha}}$, the choice of $\lambda_{\alpha}$ and $\tilde{\lambda}_{\dot{\alpha}}$ are not unique since we can always perform the following little group rescaling
\begin{eqnarray}\label{eq:lgscaling}
	\lambda_{\alpha} \to \omega^{-1} \lambda_{\alpha}, \quad \tilde{\lambda}_{\dot{\alpha}} \to \omega \tilde{\lambda}_{\dot{\alpha}}
\end{eqnarray}
and keep $p_{\alpha\dot{\alpha}}$ invariant. Generally, $\omega$ is a complex number and the action is $GL(1)$. For real momentum, we have $\omega^{-1} = \omega^*$ and thus $\omega=e^{i\theta}$, so the little group is $U(1)$. 

The amplitudes for massless particles are functions of $\lambda_i$s and $\tilde{\lambda}_i$s, where $i$ label the $i$th external particle in an amplitude, and transform under the little group scaling of the $\lambda_i$ and $\tilde{\lambda}_i$ as
\begin{eqnarray}
	\mc{M}(\dots,\omega^{-1}\lambda_i,\omega\tilde{\lambda}_i,\dots)=\omega^{2h_i}\mc{M}(\dots,\lambda_i,\tilde{\lambda}_i,\dots),
\end{eqnarray}
where $h_i$ is the helicity of the $i$th particle in the amplitude.

For massive particles, the matrix $p_{\alpha\dot{\alpha}}$ satisfies $\det p_{\alpha\dot{\alpha}} =m^2 \neq 0$, so now $p_{\alpha\dot{\alpha}}$ is rank-2 instead of rank-1. Here we adopt the massive spinor notation introduced in Ref.~\cite{Arkani-Hamed:2017jhn} and express the matrix $p_{\alpha\dot{\alpha}}$ as the product of two rank-1 matrices $\lambda^I_{\alpha}$ and $\tilde{\lambda}_{\dot{\alpha}I}$
\begin{eqnarray}
	p_{\alpha\dot{\alpha}}=\lambda^I_{\alpha}\tilde{\lambda}_{\dot{\alpha}I}, \quad I=1,2.
\end{eqnarray}
Now that $\det p_{\alpha\dot{\alpha}} =\det \lambda^I_{\alpha} \times \det \tilde{\lambda}_{\dot{\alpha}I}=m^2$, we can choose to take $\det \lambda^I_{\alpha} = \det \tilde{\lambda}_{\dot{\alpha}I}=m$. Similarly, the $\lambda^I_{\alpha}$ and $\tilde{\lambda}_{\dot{\alpha}I}$ can not be uniquely determined for a fixed momentum $p_{\alpha\dot{\alpha}}$ since we can utilize a $SL(2)$ transformation
\begin{eqnarray}
	\lambda^I_{\alpha} \rightarrow W_J^I \lambda^J_{\alpha}, \quad \tilde{\lambda}_{I\dot\alpha} \rightarrow(W^{-1}){}_I^J \tilde{\lambda}_{J\dot\alpha}
\end{eqnarray}
to change $\lambda^I_{\alpha}$ and $\tilde{\lambda}_{\dot{\alpha}I}$ while keep $p_{\alpha\dot{\alpha}}$ invariant. For real momentum, $W^I_J=(W^{-1}){}_I^{J*}$, so the little group is $SU(2)$.

We can utilize the $SU(2)$ invariant tensor $\epsilon_{IJ}$ and $\epsilon^{IJ}$ to lower and raise the little group indices on $\lambda^I_{\alpha}$ and $\tilde{\lambda}_{\dot{\alpha}I}$, such that $p_{\alpha\dot{\alpha}}=\epsilon_{IJ}\lambda^I_{\alpha}\tilde{\lambda}^J_{\dot{\alpha}}$. Furthermore, taking account of $\det \lambda^I_{\alpha} = \det \tilde{\lambda}_{\dot{\alpha}I}=m$, we have the following relations
\begin{eqnarray}
	p_{\alpha \dot{\alpha}} \tilde{\lambda}^{\dot{\alpha} I}=m \lambda_\alpha^I, \quad p_{\alpha \dot{\alpha}} \lambda^{\alpha I}=-m \tilde{\lambda}_{\dot{\alpha}}^I,
\end{eqnarray}
which allow us to convert between $\lambda_\alpha^I$ and $\tilde{\lambda}_{\dot{\alpha}}^I$ with a coefficient $p_{\alpha \dot{\alpha}}/m$. With these relations, we can use only $\lambda$ or $\tilde{\lambda}$ to present a massive particle in an amplitude, and the amplitude for massive particles must be a symmetric rank-$2S_i$ tensor $\mc{M}^{\{I_1\dots I_{2S_i}\}}$ for the $i$th spin $S_i$ particle. Thus we have
\begin{eqnarray}
	\mc{M}^{\{I_1\dots I_{2S_i}\}}=\lambda^{I_1}_{\alpha_1} \cdots \lambda^{I_{2S}}_{\alpha_{2S}} \mc{M}^{\{\alpha_1\dots \alpha_{2S_i}\}}.
\end{eqnarray}

\subsection{Massive amplitude basis}
\label{sec:basis}

In this subsection we present a basis of the independent Lorentz structures of effective operators involving the SM particles and the UV states with spin $s \leq 1$. The correspondence between operators and massive amplitudes~\cite{Li:2022abx} indicates that constructing the Lorentz structures of an operator basis is equivalent to finding a basis of kinematically independent structures formed of spinor-helicity variables. In order to involve the massive UV state with $s=1$ in the operator basis, among the different methods of constructing amplitude basis~\cite{Durieux:2019eor,Durieux:2020gip,Dong:2021yak,DeAngelis:2022qco}, we adopt the method in Ref.~\cite{DeAngelis:2022qco}, and discuss how the massive amplitude basis corresponds to an operator basis involving massive UV states.

Ref.~\cite{DeAngelis:2022qco} provides an algorithm that can be used to find all kinematically independent massive amplitudes for certain external particles and dimension of the amplitudes. Here we simply use the results and refer the readers to the paper mentioned above for more details of the algorithm. The correspondence between massless amplitudes and operators has been elaborated in Ref.~\cite{Li:2020xlh,Li:2022tec}, and here we present the correspondence as
\begin{align}\label{eq:CorMassless}
    \begin{array}{ccc}
		F_{{\tiny\rm L/R}\,i}	&	\sim	&	\lambda^2_i/\tilde\lambda^2_i,			\\
		\psi_i/\psi^{\dagger}_i	& \sim	&	\lambda_i/\tilde\lambda_i,	\\
		\phi_i	&	\sim	&	1, \\
		D_i &	\sim	&	-i\lambda_i \tilde\lambda_i,
	\end{array}
\end{align}
where $i$ in the subscript of a field labels the $i$th field in an operator and $i$ in the subscript of a covariant derivative indicates that the covariant derivative acts on the $i$th field, similarly hereinafter. For the correspondence between massive amplitudes and operators, the massive scalar and fermion are similar to the massless ones since the degrees of freedom of the massive and massless fields are the same. However, the massive vector has 3 degrees of freedom instead of 2, so its correspondence to the massive spinors should include the 3 degrees of freedom, that is, the 3 transversities. The correspondence between massive amplitudes and operators reads
\begin{align}\label{eq:CorMassive}
    \begin{array}{ccc}
		(DV_i)_{\rm L}/V_i/(DV_i)_{\rm R}	&	\sim	&	\tilde{m}_i\lambda^I_i\lambda^J_i/\lambda^I_i\tilde\lambda^J_i/m_i\tilde\lambda^I_i\tilde\lambda^J_i,			\\
		\psi_i/\psi^{\dagger}_i	& \sim	&	\lambda_i^I/\tilde\lambda_i^I,	\\
		\phi_i	&	\sim	&	1, \\
		D_i &	\sim	&	-ip_i,
	\end{array}
\end{align}
where $(DV_i)_{\rm L} \equiv D_{\alpha\dot{\alpha}} V_{\beta}{}^{\dot{\alpha}}$ and $(DV_i)_{\rm R} \equiv D^{\alpha}{}_{\dot{\alpha}} V_{\alpha\dot{\beta}}$. As
\begin{align}
\begin{aligned}
    D_{\alpha\dot{\alpha}} V_{\beta}{}^{\dot{\alpha}} = D_{\mu} V^{\mu} \epsilon_{\beta\alpha} - i D_{\mu} V_{\nu} (\sigma^{\mu\nu})_{\alpha\beta}, \\
    D^{\alpha}{}_{\dot{\alpha}} V_{\alpha\dot{\beta}} = D_{\mu} V^{\mu} \epsilon_{\dot{\alpha}\dot{\beta}} - i D_{\mu} V_{\nu} (\bar\sigma^{\mu\nu})_{\dot{\alpha}\dot{\beta}},
\end{aligned}
\end{align}
and $D_{\mu} V^{\mu}$ is the EOM of $V$, it is equivalent to use $D_{\alpha\dot{\alpha}} V_{\beta}{}^{\dot{\alpha}}$ or $i D_{\mu} V_{\nu} (\sigma^{\mu\nu})_{\alpha\beta}$ to construct UV operators.

Now that we can generate the amplitude basis using the method in Ref.~\cite{DeAngelis:2022qco} and translate it to an UV operator basis using the amplitude-operator correspondence eq.~(\ref{eq:CorMassless}) and eq.~(\ref{eq:CorMassive}). However, we should be careful about the fact that different amplitude bases could contribute to one UV operator basis due to eq.~(\ref{eq:CorMassive}). For example, if we want to find the complete and independent UV operator basis in $\phi V^3 D$, where $\phi$ is considered massless and $V$s are massive, we need to consider the corresponding amplitude bases where the massive vectors can be of different transversities. The complete and independent set of amplitudes that correspond to the UV operators in $\phi V^3 D$ are
\begin{align}\label{eq:egampbasis}
    \begin{aligned}
        &\vev{1\mathbf{2}}\vev{\mathbf{3}\mathbf{4}}[1\mathbf{2}][\mathbf{3}\mathbf{4}], \quad \vev{1\mathbf{4}}\vev{\mathbf{2}\mathbf{3}}[1\mathbf{4}][\mathbf{2}\mathbf{3}], \quad
        -\vev{\mathbf{2}\mathbf{4}} \langle\mathbf{3}|p_2|\mathbf{3}][\mathbf{2}\mathbf{4}], \\
        &\tilde{m}_2\vev{\mathbf{2}\mathbf{3}}\vev{\mathbf{2}\mathbf{4}}[\mathbf{3}\mathbf{4}], \quad
        \tilde{m}_3\vev{\mathbf{2}\mathbf{3}}\vev{\mathbf{3}\mathbf{4}}[\mathbf{2}\mathbf{4}], \quad
        \tilde{m}_4\vev{\mathbf{2}\mathbf{4}}\vev{\mathbf{3}\mathbf{4}}[\mathbf{2}\mathbf{3}], \\
        &m_2\vev{\mathbf{3}\mathbf{4}}[\mathbf{2}\mathbf{3}][\mathbf{2}\mathbf{4}], \quad
        m_3\vev{\mathbf{2}\mathbf{4}}[\mathbf{2}\mathbf{3}][\mathbf{3}\mathbf{4}], \quad
        m_4\vev{\mathbf{2}\mathbf{3}}[\mathbf{2}\mathbf{4}][\mathbf{3}\mathbf{4}],
    \end{aligned}
\end{align}
where we adopt the "$\mathbf{BOLD}$" notation instead of writing the little group indices explicitly. For the three amplitudes in the first row of eq.~(\ref{eq:egampbasis}), the massive vectors are of transversity 0, while the other six amplitudes in eq.~(\ref{eq:egampbasis}) are not. The corresponding UV operator basis is given in eq.~(\ref{eq:Lordim5}).


We list the Lorentz structures of the UV operator basis involving massless fields with helicity $|h| \leq 1$ and massive fields with spin $s \leq 1$ for interacting Lagrangian in the following,
\begin{table}[h]
		\centering
		\begin{tabular}{|c|c|c|}
			\hline
			classes & spinor notation & Lorentz notation \\
        \hline
        $\phi^3$ & $\phi_1\phi_2\phi_3$ & $\phi_1\phi_2\phi_3$ \\
        \hline
        $\phi V^2$ & $\epsilon^{\alpha\beta}\epsilon_{\dot{\alpha}\dot{\beta}}\phi_1V_{2\alpha}{}^{\dot{\alpha}}V_{3\beta}{}^{\dot{\beta}}$ & $\phi_1 V_{2\mu} V_3^{\mu}$ \\
        \hline
        $\phi^4$ & $\phi_1\phi_2\phi_3\phi_4$ & $\phi_1\phi_2\phi_3\phi_4$ \\
        \hline
        $\phi\psi^2$ & $\epsilon^{\alpha\beta}\phi_1\psi_{2\alpha}\psi_{3\beta}$ & $\phi_1 (\psi_{2} \psi_{3})$ \\
        \hline
        $\phi^2 VD$ & $\epsilon^{\alpha\beta}\epsilon_{\dot{\alpha}\dot{\beta}}(D_{\alpha}{}^{\dot{\alpha}} \phi_1)\phi_2V_{3\beta}{}^{\dot{\beta}}$ & $(D_{\mu}\phi_1)\phi_2V_3^{\mu}$ \\
        \hline
        $\phi^2 V^2$ & $\epsilon^{\alpha\beta}\epsilon_{\dot{\alpha}\dot{\beta}}\phi_1\phi_2V_{3\alpha}{}^{\dot{\alpha}}V_{4\beta}{}^{\dot{\beta}}$ & $\phi_1\phi_2V_{3\mu}V_4^{\mu}$ \\
        \hline
        $\psi\psi^{\dagger}V$ & $\epsilon^{\alpha\beta}\epsilon_{\dot{\alpha}\dot{\beta}}\psi_{1\alpha}\psi^{\dagger\dot{\alpha}}_2 V_{3\beta}{}^{\dot{\beta}}$ & $(\psi_1 \sigma_{\mu} \psi_2^{\dagger})V^{\mu}$ \\
        \hline
        $F_{\rm L}V^2$ & $\epsilon^{\alpha\gamma}\epsilon^{\beta\delta}\epsilon_{\dot{\alpha}\dot{\beta}}F_{\rm L}{}_{1\alpha\beta} V_{2\gamma}{}^{\dot{\alpha}} V_{3\delta}{}^{\dot{\beta}}$ & $F_{\rm L}{}_{1\mu\nu} V_2^{\mu} V_3^{\nu}$ \\
        \hline
        $V^3D$ & $\epsilon^{\alpha\gamma}\epsilon^{\beta\delta}\epsilon_{\dot{\beta}\dot{\alpha}}\epsilon_{\dot{\gamma}\dot{\delta}}(D_{\alpha}{}^{\dot{\alpha}} V_{1\beta}{}^{\dot{\beta}}) V_{2\gamma}{}^{\dot{\gamma}} V_{3\delta}{}^{\dot{\delta}}$ & $(D_{\mu} V_{1}{}_{\nu}) V_{2}{}^{\nu} V_{3}{}^{\mu}$ \\
        & $\epsilon^{\beta\alpha}\epsilon^{\gamma\delta}\epsilon_{\dot{\alpha}\dot{\gamma}}\epsilon_{\dot{\beta}\dot{\delta}}(D_{\alpha}{}^{\dot{\alpha}} V_{1\beta}{}^{\dot{\beta}}) V_{2\gamma}{}^{\dot{\gamma}} V_{3\delta}{}^{\dot{\delta}}$ & $\epsilon^{\mu\nu\rho\lambda}(D_{\mu} V_{1}{}_{\nu}) V_{2}{}_{\rho} V_{3}{}_{\lambda}$ \\
        & $\epsilon^{\alpha\beta}\epsilon^{\gamma\delta}\epsilon_{\dot{\alpha}\dot{\delta}}\epsilon_{\dot{\gamma}\dot{\beta}}V_{1\alpha}{}^{\dot{\alpha}} (D_{\beta}{}^{\dot{\beta}} V_{2\gamma}{}^{\dot{\gamma}}) V_{3\delta}{}^{\dot{\delta}}$ & $V_{1}{}^{\nu} (D_{\mu} V_{2\nu}) V_{3}{}^{\mu}$ \\
        & $\epsilon^{\alpha\delta}\epsilon^{\gamma\beta}\epsilon_{\dot{\alpha}\dot{\beta}}\epsilon_{\dot{\gamma}\dot{\delta}}V_{1\alpha}{}^{\dot{\alpha}} (D_{\beta}{}^{\dot{\beta}} V_{2\gamma}{}^{\dot{\gamma}}) V_{3\delta}{}^{\dot{\delta}}$ & $\epsilon^{\mu\nu\rho\lambda}V_{1}{}_{\mu} (D_{\nu} V_{2\rho}) V_{3}{}_{\lambda}$ \\
        & $\epsilon^{\alpha\gamma}\epsilon^{\beta\delta}\epsilon_{\dot{\alpha}\dot{\beta}}\epsilon_{\dot{\delta}\dot{\gamma}}V_{1\alpha}{}^{\dot{\alpha}} V_{2\beta}{}^{\dot{\beta}} (D_{\gamma}{}^{\dot{\gamma}} V_{3\delta}{}^{\dot{\delta}})$ & $V_{1}{}^{\mu} V_{2}{}^{\nu} (D_{\mu} V_{3}{}_{\nu})$ \\
        & $\epsilon^{\alpha\beta}\epsilon^{\delta\gamma}\epsilon_{\dot{\alpha}\dot{\gamma}}\epsilon_{\dot{\beta}\dot{\delta}}V_{1\alpha}{}^{\dot{\alpha}} V_{2\beta}{}^{\dot{\beta}} (D_{\gamma}{}^{\dot{\gamma}} V_{3\delta}{}^{\dot{\delta}})$ & $V_{1}{}^{\nu} V_{2}{}^{\mu} (D_{\mu} V_{3}{}_{\nu})$ \\
        \hline
        $V^4$ & $\epsilon^{\alpha\beta}\epsilon^{\gamma\delta}\epsilon_{\dot{\alpha}\dot{\beta}}\epsilon_{\dot{\gamma}\dot{\delta}}V_{1\alpha}{}^{\dot{\alpha}} V_{2\beta}{}^{\dot{\beta}} V_{3\gamma}{}^{\dot{\gamma}} V_{4\delta}{}^{\dot{\delta}}$ & $V_{1}{}^{\mu} V_{2\mu} V_{3}{}^{\nu} V_{4\nu}$ \\
        & $\epsilon^{\alpha\beta}\epsilon^{\gamma\delta}\epsilon_{\dot{\alpha}\dot{\delta}}\epsilon_{\dot{\beta}\dot{\gamma}}V_{1\alpha}{}^{\dot{\alpha}} V_{2\beta}{}^{\dot{\beta}} V_{3\gamma}{}^{\dot{\gamma}} V_{4\delta}{}^{\dot{\delta}}$ & $V_{1}{}^{\mu} V_{2}{}^{\nu} V_{3\mu} V_{4\nu}$ \\
        & $\epsilon^{\alpha\delta}\epsilon^{\beta\gamma}\epsilon_{\dot{\alpha}\dot{\beta}}\epsilon_{\dot{\gamma}\dot{\delta}}V_{1\alpha}{}^{\dot{\alpha}} V_{2\beta}{}^{\dot{\beta}} V_{3\gamma}{}^{\dot{\gamma}} V_{4\delta}{}^{\dot{\delta}}$ & $V_{1}{}^{\mu} V_{2\nu} V_{3}{}^{\nu} V_{4\mu}$ \\
        & $\epsilon^{\alpha\delta}\epsilon^{\beta\gamma}\epsilon_{\dot{\alpha}\dot{\delta}}\epsilon_{\dot{\beta}\dot{\gamma}}V_{1\alpha}{}^{\dot{\alpha}} V_{2\beta}{}^{\dot{\beta}} V_{3\gamma}{}^{\dot{\gamma}} V_{4\delta}{}^{\dot{\delta}}$ & $\epsilon^{\mu\nu\rho\lambda} V_{1\mu} V_{2\nu} V_{3\rho} V_{4\lambda}$ \\
        \hline
		\end{tabular}
\caption{Dimension-3 and dimension-4 Lorentz structures in spinor indices and Lorentz indices.}
\label{tab:Dim8ClY}
\end{table}

\begin{align}
    \mathcal{B}_3&=\phi_1\phi_2\phi_3+\epsilon^{\alpha\beta}\epsilon_{\dot{\alpha}\dot{\beta}}\phi_1V_{2\alpha}{}^{\dot{\alpha}}V_{3\beta}{}^{\dot{\beta}}, \\
    \mathcal{B}_4&=\phi_1\phi_2\phi_3\phi_4+\epsilon^{\alpha\beta}\phi_1\psi_{2\alpha}\psi_{3\beta}+\epsilon^{\alpha\beta}\epsilon_{\dot{\alpha}\dot{\beta}}(D_{\alpha}{}^{\dot{\alpha}} \phi_1)\phi_2V_{3\beta}{}^{\dot{\beta}}+\epsilon^{\alpha\beta}\epsilon_{\dot{\alpha}\dot{\beta}}\phi_1\phi_2V_{3\alpha}{}^{\dot{\alpha}}V_{4\beta}{}^{\dot{\beta}} \nonumber \\
    &\quad +\epsilon^{\alpha\beta}\epsilon_{\dot{\alpha}\dot{\beta}}\psi_{1\alpha}\psi^{\dagger\dot{\alpha}}_2 V_{3\beta}{}^{\dot{\beta}}+\epsilon^{\alpha\gamma}\epsilon^{\beta\delta}\epsilon_{\dot{\alpha}\dot{\beta}}F_{\rm L}{}_{1\alpha\beta} V_{2\gamma}{}^{\dot{\alpha}} V_{3\delta}{}^{\dot{\beta}} \nonumber \\
    &\quad + \epsilon^{\alpha\gamma}\epsilon^{\beta\delta}\epsilon_{\dot{\beta}\dot{\alpha}}\epsilon_{\dot{\gamma}\dot{\delta}}(D_{\alpha}{}^{\dot{\alpha}} V_{1\beta}{}^{\dot{\beta}}) V_{2\gamma}{}^{\dot{\gamma}} V_{3\delta}{}^{\dot{\delta}} + \epsilon^{\beta\alpha}\epsilon^{\gamma\delta}\epsilon_{\dot{\alpha}\dot{\gamma}}\epsilon_{\dot{\beta}\dot{\delta}}(D_{\alpha}{}^{\dot{\alpha}} V_{1\beta}{}^{\dot{\beta}}) V_{2\gamma}{}^{\dot{\gamma}} V_{3\delta}{}^{\dot{\delta}} \nonumber \\
    &\quad + \epsilon^{\alpha\beta}\epsilon^{\gamma\delta}\epsilon_{\dot{\alpha}\dot{\delta}}\epsilon_{\dot{\gamma}\dot{\beta}}V_{1\alpha}{}^{\dot{\alpha}} (D_{\beta}{}^{\dot{\beta}} V_{2\gamma}{}^{\dot{\gamma}}) V_{3\delta}{}^{\dot{\delta}} + \epsilon^{\alpha\delta}\epsilon^{\gamma\beta}\epsilon_{\dot{\alpha}\dot{\beta}}\epsilon_{\dot{\gamma}\dot{\delta}}V_{1\alpha}{}^{\dot{\alpha}} (D_{\beta}{}^{\dot{\beta}} V_{2\gamma}{}^{\dot{\gamma}}) V_{3\delta}{}^{\dot{\delta}} \nonumber \\
    &\quad + \epsilon^{\alpha\gamma}\epsilon^{\beta\delta}\epsilon_{\dot{\alpha}\dot{\beta}}\epsilon_{\dot{\delta}\dot{\gamma}}V_{1\alpha}{}^{\dot{\alpha}} V_{2\beta}{}^{\dot{\beta}} (D_{\gamma}{}^{\dot{\gamma}} V_{3\delta}{}^{\dot{\delta}}) + \epsilon^{\alpha\beta}\epsilon^{\delta\gamma}\epsilon_{\dot{\alpha}\dot{\gamma}}\epsilon_{\dot{\beta}\dot{\delta}}V_{1\alpha}{}^{\dot{\alpha}} V_{2\beta}{}^{\dot{\beta}} (D_{\gamma}{}^{\dot{\gamma}} V_{3\delta}{}^{\dot{\delta}}) \nonumber \\
    &\quad + \epsilon^{\alpha\beta}\epsilon^{\gamma\delta}\epsilon_{\dot{\alpha}\dot{\beta}}\epsilon_{\dot{\gamma}\dot{\delta}}V_{1\alpha}{}^{\dot{\alpha}} V_{2\beta}{}^{\dot{\beta}} V_{3\gamma}{}^{\dot{\gamma}} V_{4\delta}{}^{\dot{\delta}} + \epsilon^{\alpha\beta}\epsilon^{\gamma\delta}\epsilon_{\dot{\alpha}\dot{\delta}}\epsilon_{\dot{\beta}\dot{\gamma}}V_{1\alpha}{}^{\dot{\alpha}} V_{2\beta}{}^{\dot{\beta}} V_{3\gamma}{}^{\dot{\gamma}} V_{4\delta}{}^{\dot{\delta}} \nonumber \\
    &\quad + \epsilon^{\alpha\delta}\epsilon^{\beta\gamma}\epsilon_{\dot{\alpha}\dot{\beta}}\epsilon_{\dot{\gamma}\dot{\delta}}V_{1\alpha}{}^{\dot{\alpha}} V_{2\beta}{}^{\dot{\beta}} V_{3\gamma}{}^{\dot{\gamma}} V_{4\delta}{}^{\dot{\delta}} + \epsilon^{\alpha\delta}\epsilon^{\beta\gamma}\epsilon_{\dot{\alpha}\dot{\delta}}\epsilon_{\dot{\beta}\dot{\gamma}}V_{1\alpha}{}^{\dot{\alpha}} V_{2\beta}{}^{\dot{\beta}} V_{3\gamma}{}^{\dot{\gamma}} V_{4\delta}{}^{\dot{\delta}}, \\
    \mathcal{B}_5&=\phi_1\phi_2\phi_3\phi_4\phi_5+\epsilon^{\alpha\beta}\phi_1\phi_2\psi_{3\alpha}\psi_{4\beta}+\epsilon^{\alpha\beta}\epsilon_{\dot{\alpha}\dot{\beta}}(D_{\alpha}{}^{\dot{\alpha}} \phi_1)\phi_2\phi_3 V_{4\beta}{}^{\dot{\beta}}+\epsilon^{\alpha\beta}\epsilon_{\dot{\alpha}\dot{\beta}}\phi_1(D_{\alpha}{}^{\dot{\alpha}} \phi_2)\phi_3 V_{4\beta}{}^{\dot{\beta}} \nonumber \\
    &\quad + \epsilon^{\alpha\beta}\epsilon_{\dot{\alpha}\dot{\beta}}\phi_1\phi_2\phi_3 V_{4\alpha}{}^{\dot{\alpha}} V_{5\beta}{}^{\dot{\beta}}+\epsilon^{\alpha\beta}\epsilon_{\dot{\alpha}\dot{\beta}}\phi_1 \psi_{2\alpha} \psi^{\dagger\dot{\alpha}}_3 V_{4\beta}{}^{\dot{\beta}}+ \epsilon^{\alpha\gamma}\epsilon^{\beta\delta}\phi_1 F_{\rm L2}{}_{\alpha\beta} F_{\rm L3}{}_{\gamma\delta} \nonumber \\
    &\quad +\epsilon^{\alpha\gamma}\epsilon^{\beta\delta}\psi_{1\alpha} \psi_{2\beta} F_{\rm L3}{}_{\gamma\delta}+ \epsilon^{\alpha\gamma}\epsilon^{\beta\delta}\epsilon_{\dot{\alpha}\dot{\beta}}\phi_1 F_{\rm L2}{}_{\alpha\beta} (D_{\gamma}{}^{\dot{\alpha}} V_{3\delta}{}^{\dot{\beta}})+\epsilon^{\alpha\gamma}\epsilon^{\beta\delta}\epsilon_{\dot{\alpha}\dot{\beta}}\phi_1 F_{\rm L2}{}_{\alpha\beta} V_{3\gamma}{}^{\dot{\alpha}} V_{4\delta}{}^{\dot{\beta}} \nonumber \\
    &\quad +\epsilon^{\alpha\gamma}\epsilon^{\beta\delta}\epsilon_{\dot{\alpha}\dot{\beta}}\psi_{1\alpha} \psi_{2\beta} (D_{\gamma}{}^{\dot{\alpha}} V_{3\delta}{}^{\dot{\beta}})+\epsilon^{\alpha\gamma}\epsilon^{\beta\delta}\epsilon_{\dot{\alpha}\dot{\beta}}\psi_{1\alpha} \psi_{2\beta} V_{3\gamma}{}^{\dot{\alpha}} V_{4\delta}{}^{\dot{\beta}}+\epsilon^{\alpha\beta}\epsilon^{\gamma\delta}\epsilon_{\dot{\alpha}\dot{\beta}}\psi_{1\alpha} \psi_{2\beta} V_{3\gamma}{}^{\dot{\alpha}} V_{4\delta}{}^{\dot{\beta}} \nonumber \\
    &\quad +\epsilon^{\alpha\gamma}\epsilon^{\beta\delta}\epsilon_{\dot{\beta}\dot{\alpha}}\epsilon_{\dot{\delta}\dot{\gamma}}\phi_1 (D_{\alpha}{}^{\dot{\alpha}} V_{2\beta}{}^{\dot{\beta}}) (D_{\gamma}{}^{\dot{\gamma}} V_{3\delta}{}^{\dot{\delta}})+\epsilon^{\beta\alpha}\epsilon^{\delta\gamma}\epsilon_{\dot{\alpha}\dot{\gamma}}\epsilon_{\dot{\beta}\dot{\delta}}\phi_1 (D_{\alpha}{}^{\dot{\alpha}} V_{2\beta}{}^{\dot{\beta}}) (D_{\gamma}{}^{\dot{\gamma}} V_{3\delta}{}^{\dot{\delta}}) \nonumber \\
    &\quad +\epsilon^{\alpha\gamma}\epsilon^{\beta\delta}\epsilon_{\dot{\beta}\dot{\alpha}}\epsilon_{\dot{\gamma}\dot{\delta}}\phi_1 (D_{\alpha}{}^{\dot{\alpha}} V_{2\beta}{}^{\dot{\beta}}) V_{3\gamma}{}^{\dot{\gamma}} V_{4\delta}{}^{\dot{\delta}}+\epsilon^{\alpha\beta}\epsilon^{\gamma\delta}\epsilon_{\dot{\alpha}\dot{\delta}}\epsilon_{\dot{\gamma}\dot{\beta}}\phi_1 V_{2\alpha}{}^{\dot{\alpha}} (D_{\beta}{}^{\dot{\beta}} V_{3\gamma}{}^{\dot{\gamma}}) V_{4\delta}{}^{\dot{\delta}} \nonumber \\
    &\quad +\epsilon^{\alpha\gamma}\epsilon^{\beta\delta}\epsilon_{\dot{\alpha}\dot{\beta}}\epsilon_{\dot{\delta}\dot{\gamma}}\phi_1 V_{2\alpha}{}^{\dot{\alpha}} V_{3\beta}{}^{\dot{\beta}} (D_{\gamma}{}^{\dot{\gamma}} V_{4\delta}{}^{\dot{\delta}})+\epsilon^{\alpha\beta}\epsilon^{\gamma\delta}\epsilon_{\dot{\alpha}\dot{\beta}}\epsilon_{\dot{\gamma}\dot{\delta}}(D_{\alpha}{}^{\dot{\alpha}}\phi_1)  V_{2\beta}{}^{\dot{\beta}} V_{3\gamma}{}^{\dot{\gamma}} V_{4\delta}{}^{\dot{\delta}} \nonumber \\
    &\quad+\epsilon^{\alpha\delta}\epsilon^{\beta\gamma}\epsilon_{\dot{\alpha}\dot{\delta}}\epsilon_{\dot{\beta}\dot{\gamma}}(D_{\alpha}{}^{\dot{\alpha}}\phi_1)  V_{2\beta}{}^{\dot{\beta}} V_{3\gamma}{}^{\dot{\gamma}} V_{4\delta}{}^{\dot{\delta}} +\epsilon^{\alpha\gamma}\epsilon^{\beta\delta}\epsilon_{\dot{\alpha}\dot{\gamma}}\epsilon_{\dot{\beta}\dot{\delta}}\phi_1 (D_{\alpha}{}^{\dot{\alpha}} V_{2\beta}{}^{\dot{\beta}}) V_{3\gamma}{}^{\dot{\gamma}} V_{4\delta}{}^{\dot{\delta}}\nonumber \\
    &\quad +\epsilon^{\beta\alpha}\epsilon^{\gamma\delta}\epsilon_{\dot{\alpha}\dot{\gamma}}\epsilon_{\dot{\beta}\dot{\delta}}\phi_1 (D_{\alpha}{}^{\dot{\alpha}} V_{2\beta}{}^{\dot{\beta}}) V_{3\gamma}{}^{\dot{\gamma}} V_{4\delta}{}^{\dot{\delta}} + \epsilon^{\alpha\delta}\epsilon^{\gamma\beta}\epsilon_{\dot{\alpha}\dot{\beta}}\epsilon_{\dot{\gamma}\dot{\delta}}\phi_1 V_{2\alpha}{}^{\dot{\alpha}} (D_{\beta}{}^{\dot{\beta}} V_{3\gamma}{}^{\dot{\gamma}}) V_{4\delta}{}^{\dot{\delta}}\nonumber \\
    &\quad +\epsilon^{\alpha\beta}\epsilon^{\delta\gamma}\epsilon_{\dot{\alpha}\dot{\gamma}}\epsilon_{\dot{\beta}\dot{\delta}}\phi_1 V_{2\alpha}{}^{\dot{\alpha}} V_{3\beta}{}^{\dot{\beta}} (D_{\gamma}{}^{\dot{\gamma}} V_{4\delta}{}^{\dot{\delta}})+ \epsilon^{\alpha\beta}\epsilon^{\gamma\delta}\epsilon_{\dot{\alpha}\dot{\beta}}\epsilon_{\dot{\gamma}\dot{\delta}}\phi_1 V_{2\alpha}{}^{\dot{\alpha}} V_{3\beta}{}^{\dot{\beta}} V_{4\gamma}{}^{\dot{\gamma}} V_{5\delta}{}^{\dot{\delta}} \nonumber \\
    &\quad +\epsilon^{\alpha\beta}\epsilon^{\gamma\delta}\epsilon_{\dot{\alpha}\dot{\delta}}\epsilon_{\dot{\beta}\dot{\gamma}}\phi_1 V_{2\alpha}{}^{\dot{\alpha}} V_{3\beta}{}^{\dot{\beta}} V_{4\gamma}{}^{\dot{\gamma}} V_{5\delta}{}^{\dot{\delta}} + \epsilon^{\alpha\delta}\epsilon^{\beta\gamma}\epsilon_{\dot{\alpha}\dot{\beta}}\epsilon_{\dot{\gamma}\dot{\delta}}\phi_1 V_{2\alpha}{}^{\dot{\alpha}} V_{3\beta}{}^{\dot{\beta}} V_{4\gamma}{}^{\dot{\gamma}} V_{5\delta}{}^{\dot{\delta}} \nonumber \\
    &\quad+\epsilon^{\alpha\delta}\epsilon^{\beta\gamma}\epsilon_{\dot{\alpha}\dot{\delta}}\epsilon_{\dot{\beta}\dot{\gamma}}\phi_1 V_{2\alpha}{}^{\dot{\alpha}} V_{3\beta}{}^{\dot{\beta}} V_{4\gamma}{}^{\dot{\gamma}} V_{5\delta}{}^{\dot{\delta}}.\label{eq:Lordim5}
\end{align}

\section{The general BSM model}
\label{sec:uvlag}

In this section, we will construct the UV Lagrangian with non-redundant operators, based on the massive basis in Sec.\,\ref{sec:spinor}. First, we should list all possible UV resonances labeled by their representations under the SM gauge symmetry, which could be obtained by drawing Feynman diagrams and enumerating the mediating particles. We will straightforwardly use the results in Ref.\,\cite{Li:2022abx}. After that we build up the general Lagrangian for all the states and pick up the non-redundant terms with the help of amplitude basis and Young diagrams. The Lagrangian is presented in App.\,\ref{sec:uvlagfull} with terms that could contribute to effective operators at classical level only. In this paper the Lagrangian for SM is written as
\begin{align}
    {\cal L}_{\rm SM} = & - \frac{1}{4}G_{\mu\nu}^A G^{A\mu\nu} - \frac{1}{4}W_{\mu\nu}^I W^{I\mu\nu} - \frac{1}{4}B_{\mu\nu}B^{\mu\nu} + (D_\mu H)^\dagger (D^\mu H)\nonumber \\
    & + \overline{q}_L i \slashed{D} q_L + \overline{\ell}_L i \slashed{D} \ell_L + \overline{u}_R i \slashed{D} u_R  + \overline{d}_R i \slashed{D} d_R  + \overline{e}_R i \slashed{D} e_R \nonumber \\
    & + \mu_H^2 H^\dagger H - \lambda_H (H^\dagger H)^2  - \left(\overline{q}_L Y_u u_R \tilde{H} + \overline{q}_L Y_d d_R H + \overline{\ell}_L Y_e e_R H + {\rm h.c.} \right).
    \label{eq:smlag}
\end{align}

Before digging into the details of the general model, we should make some assumptions on the UV theory as follows:
\begin{itemize}
    \item The UV theory follows the SM gauge symmetry $SU(3)_C \times SU(2)_L \times U(1)_Y$ which is linearly realized.
    \item The UV theory contains both the SM particles and new resonances.
    \item The interaction between the new resonances and the SM particles is perturbative. Same to the interaction among new resonances.
    \item The new fields decouple at the electroweak scale. That is to say, new particles do not generate non-zero vacuum expectation values (VEVs) which breaks the SM gauge symmetry, and new fermions are either vector-like or Majorana.\footnote{The fields except the SM Higgs doublet are supposed to be in the broken phase if any gauge symmetry except $SU(3)_C \times SU(2)_L \times U(1)_Y$ is spontaneously broken. Note that in the `broken' phase where we are working, the SM gauge symmetry is still conserved. Masses of the new particles, which are at the heavy scale, can receive contributions during the symmetry breaking. We assume that after the symmetry breaking of $SU(2)_L \times U(1)_Y$ the VEV of the SM doublet Higgs contributes only to a small portion.}
\end{itemize}
These assumptions allow us to describe the physics at the electroweak scale by the SMEFT instead of the more complicated HEFT \cite{Falkowski:2019tft,Agrawal:2019bpm,Cohen:2020xca}. 
A large coupling between the SM Higgs and new resonances may violate the last assumption so we restrict our focus only on weakly-coupled theories. Strong dynamics are out of the scope of this work.

\begin{table}[]
    \centering
    \begin{tabular}{lcccccccc}
        \hline \hline
        Notation & $S_1$ & $S_2$ & $S_3$ & $S_4$ & $S_5$ & $S_6$ & $S_7$ & $S_8$ \\
        Name & $\mathcal{S}$ & $\mathcal{S}_1$ & $\mathcal{S}_2$ & $\varphi$ & $\Xi$ & $\Xi_1$ & $\Theta_1$ & $\Theta_3$ \\
        Irrep & $({\bf 1,1})_0$ & $({\bf 1,1})_1$ & $({\bf 1,1})_2$ & $({\bf 1,2})_{\frac{1}{2}}$ & $({\bf 1,3})_0$ & $({\bf 1,3})_1$ & $({\bf 1,4})_{\frac{1}{2}}$ & $({\bf 1,4})_{\frac{3}{2}}$ \\
        \hline
        Notation & $S_9$ & $S_{10}$ & $S_{11}$ & $S_{12}$ & $S_{13}$ & $S_{14}$ & & \\
        Name & $\omega_4$ & $\omega_1$ & $\omega_2$ & $\Pi_1$ & $\Pi_7$ & $\zeta$ & & \\
        Irrep & $({\bf 3,1})_{-\frac{4}{3}}$ & $({\bf 3,1})_{-\frac{1}{3}}$ & $({\bf 3,1})_{\frac{2}{3}}$ & $({\bf 3,2})_{\frac{1}{6}}$ & $({\bf 3,2})_{\frac{7}{6}}$ & $({\bf 3,3})_{-\frac{1}{3}}$ & & \\
        \hline
        Notation & $S_{15}$ & $S_{16}$ & $S_{17}$ & $S_{18}$ & $S_{19}$ & & & \\
        Name & $\Omega_2$ & $\Omega_1$ & $\Omega_4$ & $\Upsilon_1$ & $\Phi$ & & & \\
        Irrep & $({\bf 6,1})_{-\frac{2}{3}}$ & $({\bf 6,1})_{\frac{1}{3}}$ & $({\bf 6,1})_{\frac{4}{3}}$ & $({\bf 6,3})_{\frac{1}{3}}$ & $({\bf 8,2})_{\frac{1}{2}}$ & & & \\
        \hline \hline
    \end{tabular}
    \caption{New scalars that can contribute to operators up to dimension 7 in SMEFT at classical level. The second row in each block represents their name in Ref.\,\cite{deBlas:2017xtg}.}
    \label{tab:uvstates/scalar}
\end{table}

\begin{table}[]
    \centering
    \begin{tabular}{lccccccc}
        \hline \hline
        Notation & $F_1$ & $F_2$ & $F_3$ & $F_4$ & $F_5$ & $F_6$ & $F_7$ \\
        Name & $N$ & $E^c$ & $\Delta_1^c$ & $\Delta_3^c$ & $\Sigma$ & $\Sigma_1^c$ & \\
        Irrep & $({\bf 1,1})_0$ & $({\bf 1,1})_{1}$ & $({\bf 1,2})_{\frac{1}{2}}$ & $({\bf 1,2})_{\frac{3}{2}}$ & $({\bf 1,3})_0$ & $({\bf 1,3})_{1}$ & $({\bf 1,4})_{\frac{1}{2}}$ \\
        \hline
        Notation & $F_8$ & $F_{9}$ & $F_{10}$ & $F_{11}$ & $F_{12}$ & $F_{13}$ & $F_{14}$ \\
        Name & $D$ & $U$ & $Q_5$ & $Q_1$ & $Q_7$ & $T_1$ & $T_2$ \\
        Irrep & $({\bf 3,1})_{-\frac{1}{3}}$ & $({\bf 3,1})_{\frac{2}{3}}$ & $({\bf 3,2})_{-\frac{5}{6}}$ & $({\bf 3,2})_{\frac{1}{6}}$ & $({\bf 3,2})_{\frac{7}{6}}$ & $({\bf 3,3})_{-\frac{1}{3}}$ & $({\bf 3,3})_{\frac{2}{3}}$ \\
        \hline \hline
    \end{tabular}
    \caption{New fermions that can contribute to operators up to dimension 7 in SMEFT at classical level. The second row in each block represents their name in Ref.\,\cite{deBlas:2017xtg}, with a superscript ${}^c$ if they are conjugated with each other. Majorana fermions $F_1$ and $F_5$ have parity left, i.e. $F = F_L = P_L F$.}
    \label{tab:uvstates/fermion}
\end{table}

\begin{table}[]
    \centering
    \begin{tabular}{lccccccc}
        \hline \hline
        Notation & $V_1$ & $V_2$ & $V_3$ & $V_4$ & $V_5$ & $V_6$ & $V_7$ \\
        Name & $\mathcal{B}$ & $\mathcal{B}_1$ & $\mathcal{L}_3^\dagger$ & $\mathcal{W}$ & $\mathcal{U}_2$ & $\mathcal{U}_5$ & $\mathcal{Q}_5$ \\
        Irrep & $({\bf 1,1})_0$ & $({\bf 1,1})_1$ & $({\bf 1,2})_{\frac{3}{2}}$ & $({\bf 1,3})_0$ & $({\bf 3,1})_{\frac{2}{3}}$ & $({\bf 3,1})_{\frac{5}{3}}$ &  $({\bf 3,2})_{-\frac{5}{6}}$ \\
        \hline
        Notation & $V_8$ & $V_{9}$ & $V_{10}$ & $V_{11}$ & $V_{12}$ & $V_{13}$ & $V_{14}$ \\
        Name & $\mathcal{Q}_1$ & $\mathcal{X}$ & $\mathcal{Y}_1^\dagger$ & $\mathcal{Y}_5^\dagger$ & $\mathcal{G}$ & $\mathcal{G}_1$ & $\mathcal{H}$ \\
        Irrep & $({\bf 3,2})_{\frac{1}{6}}$ &$({\bf 3,3})_{\frac{2}{3}}$ & $({\bf {6}, 2})_{-\frac{1}{6}}$ & $({\bf {6}, 2})_{\frac{5}{6}}$ & $({\bf 8,1})_0$ & $({\bf 8,1})_1$ & $({\bf 8,3})_0$ \\
        \hline \hline
    \end{tabular}
    \caption{New vectors that can contribute to operators up to dimension 7 in SMEFT at classical level. The second row in each block represents their name in Ref.\,\cite{deBlas:2017xtg}, with a superscript ${}^\dagger$ if they are conjugated with each other.}
    \label{tab:uvstates/vector}
\end{table}

Equally important, there is no need to enforce the theory to be UV complete, since the gravity will come into the theory at even higher energy. As a consequence, it is acceptable that the UV theory contains high spin fields like Rarita-Schwinger spinors and tensor fields. Also, the UV physics may be described by a theory with unrenormalizable operators which is called Resonance EFT~\cite{Ecker:1988te,Pich:2016lew,Pich:2020xzo} or BSMEFT in the Ref.\,\cite{deBlas:2017xtg}. We leave these possibilities to future works.


There could be unlimited new physics beyond the Standard Model, even for the extended resonances. One should arrange a priority for studying among possible resonances. Since the Poincaré and the SM gauge symmetry are followed by the UV theory, it is very natural to label the resonances by their spins and representations. Following Ref.\,\cite{deBlas:2017xtg}, this work focuses on those resonances that could contribute to effective operators in SMEFT at the classical level, with dimension up to 7. If the UV theory contains some of these particles, they probably lead to the leading contributions to the observables. In addition, once the deviation from the SM is doubted to be sourced by these particles, it could be verified by resonance search on future colliders. There is no reason to skip these resonances.

The quantum number of UV resonances can be fixed by the effective operators they contribute to. All possible partitions of the external particles of an effective operator correspond to tree Feynman diagrams whose internal lines indicate possible UV resonances. For each partition, the Poincaré and gauge Casimir eigen-basis\,\cite{Li:2022tec,Li:2022abx,Jiang:2020rwz}, i.e. the j-basis operators, classify the quantum number of heavy resonances by the proposed j-basis/UV correspondence\,\cite{Li:2022abx}. Some of the selected resonances should be excluded since they only contribute to high dimensional effective operators, which can be done through dimension selection.

Tab.\,\ref{tab:uvstates/scalar}-\ref{tab:uvstates/vector} present all the UV resonances with rows in each block representing its notation in our paper, its name in Ref.\,\cite{deBlas:2017xtg} and its representation under $(SU(3)_C, SU(2)_L)_{U(1)_Y}$. We only consider particles with spin $\leq 1$. The field content differing with that in Ref.\,\cite{deBlas:2017xtg} by conjugation is attached with a superscript ${}^\dagger$ for bosons or ${}^c$ for fermions. The $U(1)_Y$ hypercharge $Y$ is defined as
\begin{equation}
    Q = Y + T_3,
\end{equation}
where $Q$ is the electric charge after symmetry breaking and $T_3$ the weak isospin. 

Although most resonances and terms have been listed in Ref.\,\cite{deBlas:2017xtg}, it is necessary to point out the differences as follows:
\begin{itemize}
    \item A quartet fermion $F_7$ is not presented in Ref.\,\cite{deBlas:2017xtg} since it can only generate dimension-7 operator $O_{LH}$ as we could see later.
    \item Vector ${\cal L}_1$ $({\bf 1}, {\bf 2})_{\frac{1}{2}}$ and vector ${\cal W}_1$ $({\bf 1}, {\bf 3})_{1}$ listed in Ref.\,\cite{deBlas:2017xtg} are not presented in this work. Many terms that involve derivatives acting on vector fields are identified as redundant operators by the massive amplitude method introduced in Sec.\,\ref{sec:spinor}, e.g. 
    \begin{equation}
        \epsilon^{\alpha\beta}\epsilon_{\dot{\alpha}\dot{\beta}} \phi_1\phi_2 D_{\alpha}{}^{\dot{\alpha}} V_{\beta}{}^{\dot{\beta}} \sim \phi_1\phi_2 D_\mu V^\mu.
    \end{equation}
    In the same way, terms like $V^\mu D_\mu \phi$ are not included since they can also be eliminated by field redefinition. As a result, the doublet vector ${\cal L}_1$  and the triplet vector ${\cal W}_1$ do not contribute under our assumptions.
\end{itemize}







The Lagrangian of the UV theory can be constructed from these field contents by enumerating Lorentz and gauge invariant combinations. Sec.\,\ref{sec:spinor} provides a general method to list all Lorentz invariant terms in Tab.\,\ref{tab:Dim8ClY} with the help of massive amplitudes, while the gauge sector could be handled by the Littlewood-Richardson rule as in the case of SMEFT \cite{Li:2020gnx}. To clarify it more clearly, take the $S_4S_6^\dagger S_7$ term as an example. These three fields transform as $\bf 2$, $\bf 3$ adn $\bf 4$ representations under $SU(2)_L$ symmetry and the combination is written as\footnote{Subscripts and superscripts labeling the gauge components are indices of fundamental and anti-fundamental representation with $i_n, j_n, k_n = 1, 2$. The components are linked to the common ones by Clebsch-Gordan coefficients, e.g. $(S_6^\dagger)^I = \frac{1}{\sqrt{2}} (\tau^I \epsilon)_{ij} (S_6^\dagger)^{ij}$. 
}
\begin{equation}
    (S_4)_i \times (-1)\epsilon_{j_1j_3}\epsilon_{j_2j_4}(S_6^\dagger)^{j_3j_4} \times (S_7)_{k_1k_2k_3}.
    \label{eq:lagEx1}
\end{equation}
The factor that contracts this term into a gauge singlet can be found by building a $N$-block-height Young tableau for $SU(N)$ group as\footnote{One needs to further specify the symmetry of these (anti-)fundamental indices if calculations are performed under this notation, i.e. fields with (anti-)fundamental indices. For example, the factor is actually ${\cal Y}\left[\tiny{\young({{k_1}}{{k_2}}{{k_3}})}\right] \circ {\cal Y}\left[\tiny{\young({{j_1}}{{j_2}})}\right] \circ {\cal Y}\left[\tiny{\young({{i}})}\right] \circ \epsilon^{ik_1}\epsilon^{j_1k_2}\epsilon^{j_2k_3} = \left(\epsilon^{ik_1}\epsilon^{j_1k_2}\epsilon^{j_2k_3} + (j_1 \leftrightarrow j_2)\right) + ({\text{perm. of } k_1,k_2,k_3})$ in such a case.}
\begin{align}
    \young(i) \overset{\tiny{\young({{j_1}}{{j_2}})}}{\longrightarrow}  \young(i{{j_1}}{{j_2}})  \overset{\tiny{\young({{k_1}}{{k_2}}{{k_3}})}}{\longrightarrow}  \young({{i}}{{j_1}}{{j_2}},{{k_1}}{{k_2}}{{k_3}}) ~ \Rightarrow ~ \epsilon^{ik_1}\epsilon^{j_1k_2}\epsilon^{j_2k_3},
    \label{eq:dim7/youngtab}
\end{align}
where the Young tableaux of $i$ to the leftmost and $j,k$ above the arrows represent the representations of $S_4, S_6^\dagger$ and $S_7$ respectively. Other formats to build such a rectangle Young tableau will result in vanishing factors due to the symmetry between indices.
The flavor symmetry may lead to vanishing operators as well, which could be checked by imposing Young operators. 

After dealing with these three kinds of symmetries systematically, we present the results in App.\,\ref{sec:uvlagfull} which only include terms that are relevant to the effective operators in SMEFT with $d \leq 7$. In the appendix the notation for the indices of fields is a little different from what is used in Eq.\,\eqref{eq:lagEx1}. All fields under non-fundamental representations of gauge groups are attached with a single index for each group instead of several indices of fundamental representation. The transformation rules between the notation used in the appendix and in Eq.\,\eqref{eq:lagEx1} read as
\begin{itemize}
    \item $\mathbf{3}, \mathbf{\overline{3}}$ representations of $SU(3)_C$ remains unchanged as
        \begin{equation}
            \phi_a, ~ \phi^{\dagger a},
        \end{equation}
    \item $\mathbf{8}$ representations of $SU(3)_C$
        \begin{equation}
            \phi_a^b = \frac{1}{\sqrt{2}}\phi^A (\lambda^A)_a^b,
        \end{equation}
    \item $\mathbf{6}, \mathbf{\overline{6}}$ representations of $SU(3)_C$
        \begin{equation}
            \phi_{ab} = (C^{\frak{c}})_{ab} \phi_{\frak{c}},~ \phi^{\dagger ab} = (C_{\frak{c}})^{ab} \phi^{\dagger\frak{c}},
        \end{equation}
    \item $\mathbf{2}, \mathbf{\overline{2}}$ representations of $SU(2)_L$ remains unchanged as
        \begin{equation}
            \phi_i, ~ \phi^{\dagger i},
        \end{equation}
    \item $\mathbf{3}$ representations of $SU(2)_L$
        \begin{equation}
            \phi_{ij} = \frac{1}{\sqrt{2}}\phi^I (\tau^I\epsilon)_{ij},~ \phi^{\dagger ij} = (\phi_{ji})^\dagger = \frac{1}{\sqrt{2}}\phi^I (\tau^I)_k{}^j \epsilon^{ki} \text{~if $\phi$ is complex},
        \end{equation}
    \item $\mathbf{4}, \mathbf{\overline{4}}$ representations of $SU(2)_L$
        \begin{equation}
            \phi_{ijk} = (C^{\cal{I}})_{ijk} \phi_{\cal{I}},~ \phi^{\dagger ijk} = (C_{\cal{I}})^{ijk} \phi^{\dagger\cal{I}}.
        \end{equation}
\end{itemize}
The normal lowercase Latin letters stand for the indices of fundamental representations, $a,b,c,\cdots$ for $SU(3)_C$ and $i,j,k,\cdots$ for $SU(2)_L$. The corresponding capital letters are for the adjoint representations. $\bf 6$ representation of $SU(3)_C$ are denoted by Gothic lowercase Latin letters $\frak{a,b,c},\cdots$ in the subscripts, and $\bf 4$ representation of $SU(2)_L$ are denoted by calligraphic capital letters ${\cal{I, J, K}},\cdots$. Indices of conjugate representations are marked in the superscripts. Typically one can freely define the Clebsch-Gordan coefficients, but we suggest a normalized one as in Ref.\,\cite{Chao:2018xwz,Ramsey-Musolf:2021ldh,Carpenter:2021rkl}. In such case, $\lambda^A$ and $\tau^I$ are Gell-Mann and Pauli matrices. The Clebsch-Gordan coefficients for higher dimensional representations are $(C_{\frak{a}})^{ab}$ for ${\frak a} = 1,2,...,6$ with
\begin{align}
	(C_1)^{ab} = \left( \begin{array}{ccc}
		1 & 0 & 0 \\
		0 & 0 & 0 \\
		0 & 0 & 0
	\end{array}\right),
	(C_2)^{ab} = \left( \begin{array}{ccc}
		0 & 0 & 0 \\
		0 & 1 & 0 \\
		0 & 0 & 0
	\end{array}\right),
	(C_3)^{ab} = \left( \begin{array}{ccc}
		0 & 0 & 0 \\
		0 & 0 & 0 \\
		0 & 0 & 1
	\end{array}\right), \nonumber\\
	(C_4)^{ab} = \frac{1}{\sqrt 2}\left( \begin{array}{ccc}
		0 & 0 & 0 \\
		0 & 0 & 1 \\
		0 & 1 & 0
	\end{array}\right),
	(C_5)^{ab} = \frac{1}{\sqrt 2}\left( \begin{array}{ccc}
		0 & 0 & 1 \\
		0 & 0 & 0 \\
		1 & 0 & 0
	\end{array}\right),
	(C_6)^{ab} = \frac{1}{\sqrt 2}\left( \begin{array}{ccc}
		0 & 1 & 0 \\
		1 & 0 & 0 \\
		0 & 0 & 0
	\end{array}\right)
\end{align} 
and $(C_{\cal I})^{ijk} = \frac{1}{\sqrt{2}}(C_{\cal I})^{Jk}(\epsilon\tau^{J})^{ji}$ for ${\cal I} = 3/2, 1/2, -1/2, -3/2$ with
\begin{align}
	(C_{3/2})^{Ij} = \frac{1}{\sqrt{2}}\left( \begin{array}{ccc}
		1 & 0  \\
		-i & 0  \\
		0 & 0 
	\end{array}\right),
	(C_{1/2})^{Ij} = \frac{1}{\sqrt{6}}\left( \begin{array}{ccc}
		0 & 1  \\
		0 & -i  \\
		-2 & 0 
	\end{array}\right), \nonumber \\
	(C_{-1/2})^{Ij} = -\frac{1}{\sqrt{6}}\left( \begin{array}{ccc}
		1 & 0  \\
		i & 0  \\
		0 & 2 
	\end{array}\right),
	(C_{-3/2})^{Ij} = -\frac{1}{\sqrt{2}}\left( \begin{array}{ccc}
		0 & 1  \\
		0 & i  \\
		0 & 0 
	\end{array}\right).
\end{align}
The conjugated representation are given by
\begin{align}
    (C^I)_{ab\cdots c} = [(C_I)^{c\cdots ba}]^*,
\end{align}
i.e. $(C^{\frak{a}})_{cd} \equiv [(C_{\frak{a}})^{dc}]^*$ and $(C^{\cal I})_{jkl} = [(C_{\cal I})^{lkj}]^*$. Both of them satisfy the normalization condition
\begin{align}
    (C_{\frak{a}})^{cd}(C^{\frak{b}})_{dc} = \delta_{\frak{a}}^{\frak{b}}, ~ (C_{\cal I})^{jkl} (C^{\cal J})_{lkj} = \delta_{\cal I}^{\cal J}
\end{align}
and
\begin{align}
    (C_{\frak{a}})^{ab}(C^{\frak{a}})_{c_1 c_2} = \frac{1}{2}\sum_{{\cal P}\in S_2}\delta^a_{c_{{\cal P}(1)}}\delta^b_{c_{{\cal P}(2)}}, ~ (C_{\cal I})^{ijk} (C^{\cal I})_{l_1l_2l_3} = \frac{1}{3!}\sum_{{\cal P}\in S_3}\delta^i_{l_{{\cal P}(1)}}\delta^j_{l_{{\cal P}(2)}}\delta^k_{l_{{\cal P}(3)}}
\end{align}
where $S_n$ is the permutation group on $n$ letters.

\section{The matching procedure}
\label{sec:procedure}

The matching procedure aims to find the effective theory which contains only light degrees of freedom to precisely describe the full theory. We have proposed some assumptions on the UV theory in Sec.\,\ref{sec:uvlag} which validate SMEFT to describe the physics, and in the following we need to find the Wilson coefficients of the effective operators. The procedure consists of two parts: deriving the effective Lagrangian and reducing the operators to a basis.

\subsection{Tree-level matching}

Amplitude matching and functional matching are two alternative ways to derive the effective theory. For convenience, we choose the latter one which matches the effective actions of two theories. At classical level, the effective Lagrangian can be derived from replacing the heavy fields by their classical equations of motion\,\cite{Henning:2014wua,Cohen:2020fcu}. Specifically, the kinetic terms of heavy fields in the Lagrangian are
\begin{align}
    & \Delta{\cal L}_{\rm kin} = -\eta\Phi^{\dagger}(D^2 + M^2) \Phi, &  \text{for scalars}, & \\
    & \Delta{\cal L}_{\rm kin} = \eta\left(L^{\dagger}_{\dot{\alpha}}iD^{\alpha\dot{\alpha}}L_{\alpha} + R^{\dagger\alpha}iD_{\alpha\dot{\alpha}}R^{\dot{\alpha}} - M R^{\dagger\alpha}L_{\alpha} - M L^{\dagger}_{\dot{\alpha}}R^{\dot{\alpha}}\right), &  \text{for fermions}, & \\
    & \Delta{\cal L}_{\rm kin} = \eta V^{\dagger\mu}(g_{\mu\nu}D^2 - D_{\nu}D_{\mu} + g_{\mu\nu}M^2) V^{\nu}, &  \text{for vectors}, &
\end{align}
where $\eta = \frac{1}{2} (1)$ is the normalization factor for real bosonic and Majorana (complex bosonic and Dirac) fields and note that for Majorana fermions $R^{\dot\alpha} = L^{\dagger\dot\alpha} \equiv (L_{\alpha})^\dagger$. In the vector case, a term proportional to $V^{\dagger\mu}[D_\mu, D_\nu] V^\nu$ also appears as kinetic terms, which could be transformed into an interacting term since $[D_\mu, D_\nu] = -iF_{\mu\nu}$. Note that the factor of this term can be fixed by unitarity \cite{Henning:2014wua}, but in this work we treat it as a free parameter. The classical equations of motion, $\frac{\delta S}{\delta \Phi} = 0$, lead to the following equations (gauge indices are omitted):
\begin{align}
    \label{eq:eomofs}
    \Phi & = \frac{1}{M^2}\left(\frac{\delta {S}_{\rm int}}{\delta \Phi^\dagger} - \frac{1}{2}D_{\alpha\dot\alpha}D^{\alpha\dot\alpha} \Phi\right), \\
    \label{eq:eomofl}
    L_{\alpha} & = - \frac{1}{M}\epsilon_{\alpha\beta}\frac{\delta {S}_{\rm int}}{\delta R^\dagger_{\beta}} + \frac{1}{M}iD_{\alpha\dot{\alpha}}R^{\dot{\alpha}} ,  \\
    \label{eq:eomofr}
    R^{\dot{\alpha}} & = - \frac{1}{M}\epsilon^{\dot{\alpha}\dot{\beta}}\frac{\delta {S}_{\rm int}}{\delta L^{\dagger\dot{\beta}}} + \frac{1}{M}iD^{\alpha\dot{\alpha}}L_{\alpha} , \\
    \label{eq:eomofv}
    V_{\alpha}^{\phantom{\alpha}\dot{\alpha}} & = \frac{1}{2M^2} \left(4 \epsilon_{\alpha\beta}\epsilon^{\dot{\alpha}\dot{\beta}}\frac{\delta{S}_{\rm int}}{\delta V_{\beta}^{\dagger}{}^{\dot{\beta}}} - D_{\beta\dot{\beta}}D^{\beta\dot\beta} V_{\alpha}^{\phantom{\alpha}\dot{\alpha}} - D^{\beta}{}_{\dot{\beta}}D_{\alpha}{}^{\dot\alpha} V_{\beta}{}^{\dot\beta}\right).
\end{align}
To be consistent, all fields are written with 2-component spinor indices, and left(right) indices are in the subscripts(superscripts). One may raise and lower the indices by the anti-symmetric tensor $\epsilon$ with the index to be raised or lowered right after the $\epsilon$. For instance, $D^{\alpha\dot\alpha}\Phi \equiv \epsilon^{\alpha\beta}D_{\beta}{}^{\dot\alpha}\Phi$. We choose $\epsilon^{12} = \epsilon^{\dot 1 \dot 2} = \epsilon_{21} = \epsilon_{\dot 2 \dot 1} = 1$ for $\epsilon$ of the Lorentz group.\footnote{It should be noted that we use the convention $\epsilon^{12} = \epsilon_{12} = 1$ for $SU(2)_L$ group. Thus $\epsilon^{\alpha\beta}\epsilon_{\alpha\gamma} = \delta^{\beta}_{\gamma}$ for the Lorentz group while $\epsilon^{ij}\epsilon_{ik} = - \delta^{i}_{k}$ for $SU(2)_L$ group.} Note that Eq.\,\eqref{eq:eomofl} and Eq.\,\eqref{eq:eomofr} become conjugation of each other for Majorana fermions.

Eq.\,\eqref{eq:eomofs}-\eqref{eq:eomofv} can be solved iteratively. For every heavy field $\Phi$, the interaction terms in the Lagrangian are bound to contain terms with a single $\Phi$, which can be verified by Feynman diagrams. A heavy particle only has loop contributions to scattering of light particles if the heavy particle can only be produced in a pair. Therefore, after heavy fields replacement, the functional derivative terms to the right of these equations of motion always contain a term consisting of only light fields. Since we are considering scattering of light fields in low energy region, this term will be suppressed by the heavy scale $M$ in the pre-factor. The remaining terms, e.g. terms with derivatives, receive additional suppression from the $M^{-n}$ factor after heavy field replacement. We can stop the iteration of replacement by setting a cut-off order for $M^{-n}$. In the end, the heavy fields in the classical Lagrangian are replaced by the classical solutions to get the effective Lagrangian, like ${\cal L}_{\rm EFT}[\phi] = {\cal L}_{\rm UV}[\phi, \Phi_c[\phi]]$ where $\phi$ represents light fields and $\Phi_c$ represents the classical solution.

Generally, ${\cal L}_{\rm EFT}[\phi]$ contains operators in all kinds of forms which may not be in an on-shell operator basis. In order to compare the experimental results with the Wilson coefficients, it is necessary to eliminate redundant operators in the Lagrangian since combination of redundant operators has null contribution to the S-matrix and we are not able to fix the coefficient. Operator reduction is very complicated due to various kinds of redundancy. In the following subsection, we will provide a general method to reduce any redundant operators in the effective Lagrangian.

\subsection{Operator reduction}

In this subsection, we propose a systematic method to reduce any effective operator to a given operator basis. In the reduction, not only the equations of motion (EOMs) of the SM, but the EOM terms that come from the Weinberg operator~\cite{Weinberg:1979sa} are also involved, and the operator bases at mass dimension 5, 6 and 7 are integrated as one basis for a complete reduction result.

First of all, we adopt the off-shell amplitude formalism introduced in Ref.~\cite{Ren:2022tvi}, where a one-to-one mapping from operators to off-shell amplitudes is proposed as an extension of the amplitude-operator correspondence,
\begin{align}
    \label{eq:offshellmap}
	\begin{array}{ccc}
		F_{{\tiny\rm L/R}\,i}	&	\sim	&	\lambda_{i,0}\lambda_{i,0}/\tilde\lambda_{i,0}\tilde\lambda_{i,0},			\\
		\psi_i/\psi^{\dagger}_i	&	\sim	&	\lambda_{i,0}/\tilde\lambda_{i,0},	\\
		\phi_i &	\sim	& 1, \\
		D_{i,d_i}	&	\sim	&	-i\lambda_{i,d_i}\tilde\lambda_{i,d_i}.
	\end{array}
\end{align}
$i$ in the subscript of each field labels the $i$th field in an operator, and $i$ in the subscript of a covariant derivative denotes that the covariant derivative acts on the $i$th field. $0$ in the subscript of a spinor indicates the spinor index corresponds to a field, while $d_i$ indicates the order of covariant derivatives acting on the $i$th field, $d_i \in \{1,\dots,\hat{d}_i\}$, $\hat{d}_i \in \mathbb{Z}$. For example, $d_i=1$ labels the first covariant derivative acting on the $i$th field, and $d_i=2$ labels the second covariant derivative acting on that field, etc. We will also use $x_i$ to label a spinor, $x_i \in \{0,1,\dots,\hat{d}_i\}$. The Dirac brackets of off-shell spinors are defined as $\vev{i_{x_i}j_{x_j}}\equiv\lambda^{\alpha}_{i,x_i}\lambda_{j,x_j\alpha}$ and $[i_{x_i}j_{x_j}]\equiv\tilde\lambda_{i,x_i\dot{\alpha}}\tilde\lambda_{j,x_j}^{\dot{\alpha}}$. 

With the off-shell amplitude formalism, effective operators can be presented as off-shell amplitudes by the map eq.~(\ref{eq:offshellmap}). Furthermore, the redundancy relations among these operators can be formulated in the off-shell amplitude formalism. Specifically, the IBP relation for off-shell amplitudes reads 
\begin{align}
\label{eq:ruleIBP}
		 |i_{\hat{d}_i}\rangle[i_{\hat{d}_i}|=-\sum^N_{j=1,j \neq i}|j_{\hat{d}_j+1}\rangle[j_{\hat{d}_j+1}|,
\end{align}
where $N$ denotes the total number of particles in the off-shell amplitudes. Eq.~(\ref{eq:ruleIBP}) just means that the outermost derivative on the $i$th field is moved to the other $N-1$ fields by the IBP relation in operator perspective. The Schouten identity among off-shell amplitudes is written as
\begin{align}
\label{eq:Schouten}
		 \langle i_{x_i} l_{x_l} \rangle \langle j_{x_j} k_{x_k} \rangle + \langle i_{x_i} j_{x_j} \rangle \langle k_{x_k} l_{x_l} \rangle + \langle i_{x_i} k_{x_k} \rangle \langle l_{x_l} j_{x_j} \rangle=0.
\end{align}

Utilizing the IBP relation eq.~(\ref{eq:ruleIBP}) and the Schouten identity eq.~(\ref{eq:Schouten}), any off-shell amplitude of a certain operator type, where by "type" we mean that the fields and number of covariant derivatives in the operators are fixed, can be reduced into a set of independent off-shell amplitudes in that operator type. What is more, off-shell amplitudes that correspond to the EOM of the fields in the operator could appear during the reduction, and these off-shell amplitudes would change the type. For example, we list the EOM of scalar, spinor and gauge boson and the corresponding off-shell amplitudes in the following:
\begin{align}
\begin{aligned}
	D^{\alpha}{}_{\dot{\alpha}} D_{\alpha}{}^{\dot{\alpha}} \phi_i &\sim \langle i_2 i_1 \rangle [i_2 i_1], \\
	D^{\alpha\dot{\alpha}} \psi_{i}{}_{\alpha} &\sim \langle i_1 i_0 \rangle |i_1], \\
	D_{\alpha\dot{\alpha}} \psi_i^{\dagger}{}^{\dot{\alpha}} &\sim [ i_1 i_0 ] |i_1 \rangle, \\
	D^{\alpha\dot{\alpha}} F_{\rm L}{}_{i}{}_{\alpha\beta} &\sim \langle i_1 i_0 \rangle |i_1] |i_0 \rangle, \\
	D_{\alpha\dot{\alpha}} F_{\rm R}{}_{i}{}^{\dot{\alpha}\dot{\beta}} &\sim [ i_1 i_0 ] |i_1 \rangle |i_0].
\end{aligned}
\end{align}
For these off-shell amplitudes corresponding to the EOM, we can derive the specific expressions of the EOM of the fields for a model and substitute them into the off-shell amplitudes in other operator types with the EOM. In this work, the model is the SM and the EOMs include terms from the SM Lagrangian and the dimension-5 Weinberg operator. So the mass dimension of the off-shell amplitude, as well as the mass dimension of the corresponding operator, may change after the substitution of the EOM. Generally, the result would be the sum of some off-shell amplitudes in different types at several mass dimensions after the above reduction procedure is applied once, and the procedure should be applied repeatedly for all involved types until the result does not change any more in order to make sure the reduction is complete.

Here we take the operator $C_{pr}(H^{\dagger}iD^{\mu}H)(L_p^{\dagger}\bar{\sigma}_{\mu}L_r)$ as a simple example to illustrate the method. Labeling the fields in the operator as $L_1 H_2 H^{\dagger}_3 L^{\dagger}_4$, the corresponding off-shell amplitude reads1233
\begin{align}
    -C_{f_4f_1}\delta^{i_1}_{i_4}\delta^{i_2}_{i_3}\langle 1_0 2_1\rangle [2_1 4_0],
\end{align}
and this off-shell amplitude can be reduced with the IBP relation eq.~(\ref{eq:ruleIBP}) as
\begin{align}\label{eq:offshelleg1}
\begin{aligned}
    -C_{f_4f_1}\delta^{i_1}_{i_4}\delta^{i_2}_{i_3}\langle 1_0 2_1\rangle [2_1 4_0] =& \ C_{f_4f_1}\delta^{i_1}_{i_4}\delta^{i_2}_{i_3}\langle 1_0 1_1\rangle [1_1 4_0] + C_{f_4f_1}\delta^{i_1}_{i_4}\delta^{i_2}_{i_3}\langle 1_0 3_1\rangle [3_1 4_0] \\&+ C_{f_4f_1}\delta^{i_1}_{i_4}\delta^{i_2}_{i_3}\langle 1_0 4_1\rangle [4_1 4_0].
\end{aligned}
\end{align}
It is straightforward to see that the first term and the third term on the right-hand side of eq.~(\ref{eq:offshelleg1}) correspond to the EOM, and can be converted to other types by substituting the EOM. For example, the first term on the right-hand side of eq.~(\ref{eq:offshelleg1}) corresponds to the EOM of $L_1$, and becomes the following off-shell amplitudes:
\begin{align}\label{eq:offshelleg2}
    \begin{aligned}
    \underbrace{C_{f_4f_1}\delta^{i_1}_{i_4}\delta^{i_2}_{i_3}\langle 1_0 1_1\rangle [1_1 4_0]}_{L_1 H_2 H^{\dagger}_3 L^{\dagger}_4D} \Rightarrow&
    \underbrace{-C_{f_5 p} (y_E)_{pf_4}\delta^{i_1}_{i_3}\delta^{i_2}_{i_5} [4_0 5_0]}_{H_1H_2H^{\dagger}_3e_4L^{\dagger}_5} \\&\underbrace{+ \; C_{f_5 p} (C_5^{\dagger})_{pf_6}\delta^{i_1}_{i_2} \epsilon_{i_5i_4}\epsilon_{i_6i_3} [5_0 6_0] + C_{f_5 p}(C_5^{\dagger})_{f_6p}\delta^{i_1}_{i_2} \epsilon_{i_5i_3}\epsilon_{i_6i_4} [5_0 6_0]}_{H_1H^{\dagger}_2H^{\dagger}_3H^{\dagger}_4L^{\dagger}_5L^{\dagger}_6},
    \end{aligned}
\end{align}
after substituting the EOM of $L_1$. The first term on the right-hand side of eq.~(\ref{eq:offshelleg2}) still corresponds to a dimension-6 operator, while the second and third terms correspond to dimension-7 operators.

As demonstrated above, we obtain the sum of a set of off-shell amplitudes in different types at different mass dimensions after the reduction of one off-shell amplitude in a certain type at a certain mass dimension. It is straightforward to see that each of the off-shell amplitudes corresponds to an operator in the on-shell operator basis since the redundancies among off-shell amplitudes (operators), including the IBP, the EOM and the Schouten identity, are removed. In fact, the correspondence can be found by simply taking the off-shell amplitudes on-shell~\cite{Ren:2022tvi}, and the on-shell basis is chosen to be the y-basis~\cite{Li:2022tec}. However, as we are doing the reduction across types and dimensions, we should merge the y-bases in different types at different dimensions to a ”full" y-basis, such that any off-shell amplitude is reduced to this y-basis.

As illustrated in Ref.~\cite{Li:2022tec}, the f-basis, instead of the y-basis, is the independent and complete basis if the redundancy of the flavor structure is taken account of. After an operator is translated to an off-shell amplitude and reduced to the ”full" y-basis, we utilize the $K_{py}$ matrix that converts the ”full" y-basis to the ”full" p-basis to find the its coordinates on the p-basis, and it is straightforward to find the its coordinates on the ”full" f-basis since the additional basis vectors in the p-basis are related to the corresponding f-basis vectors by permutations of the flavor indices. Following the idea, we can obtain the coordinates of any operator on any on-shell operator basis if the basis is equivalent to the p(f)-basis. 
In this work, the selected SMEFT operator bases are listed in appendix.~\ref{sec:basis567}.

Here we comment on the above operator reduction procedure. Operators related by the EOM are redundant
because operators related by the field re-definitions are physically equivalent~\cite{CHISHOLM1961469,Kamefuchi:1961sb,Arzt:1993gz}. However, in the operator reduction, substituting the EOM of fields is equivalent to the leading-order contribution of field re-definitions~\cite{Criado:2018sdb}. For example, if one want to include the dimension-8 SMEFT operators in the cross-dimension reduction, the higher-order contribution of the field re-definitions of the dimension-6 operators should be considered.




\section{The UV-IR correspondence}
\label{sec:result}

\begin{table}
    \centering
    \resizebox{0.95\textwidth}{!}{
    \begin{tabular}{|c|c|c|c|c|c|c|c|c|c|c|c|c|c|c|c|c|c|c|c|}
    \hline
        ~ & $S_1$ & $S_2$ & $S_3$ & $S_4$ & $S_5$ & $S_6$ & $S_7$ & $S_8$ & $S_9$ & $S_{10}$ & $S_{11}$ & $S_{12}$ & $S_{13}$ & $S_{14}$ & $S_{15}$ & $S_{16}$ & $S_{17}$ & $S_{18}$ & $S_{19}$ \\ \hline
        ${\cal O}_{H}$ & \cmark & ~ & ~ & \cmark & \cmark & \cmark & \cmark & \cmark & ~ & ~ & ~ & ~ & ~ & ~ & ~ & ~ & ~ & ~ & ~ \\ \hline
        ${\cal O}_{H\square}$ & \cmark & ~ & ~ & ~ & \cmark & \cmark & ~ & ~ & ~ & ~ & ~ & ~ & ~ & ~ & ~ & ~ & ~ & ~ & ~ \\ \hline
        ${\cal O}_{HD}$ & ~ & ~ & ~ & ~ & \cmark & \cmark & ~ & ~ & ~ & ~ & ~ & ~ & ~ & ~ & ~ & ~ & ~ & ~ & ~ \\ \hline
        ${\cal O}_{eH}$ & * & ~ & ~ & \cmark & \cmark & \cmark & ~ & ~ & ~ & ~ & ~ & ~ & ~ & ~ & ~ & ~ & ~ & ~ & ~ \\ \hline
        ${\cal O}_{uH}$ & * & ~ & ~ & \cmark & \cmark & \cmark & ~ & ~ & ~ & ~ & ~ & ~ & ~ & ~ & ~ & ~ & ~ & ~ & ~ \\ \hline
        ${\cal O}_{dH}$ & * & ~ & ~ & \cmark & \cmark & \cmark & ~ & ~ & ~ & ~ & ~ & ~ & ~ & ~ & ~ & ~ & ~ & ~ & ~ \\ \hline
        ${\cal O}_{ll}$ & ~ & \cmark & ~ & ~ & ~ & \cmark & ~ & ~ & ~ & ~ & ~ & ~ & ~ & ~ & ~ & ~ & ~ & ~ & ~ \\ \hline
        ${\cal O}_{qq}^{(1)}$ & ~ & ~ & ~ & ~ & ~ & ~ & ~ & ~ & ~ & \cmark & ~ & ~ & ~ & \cmark & ~ & \cmark & ~ & \cmark & ~ \\ \hline
        ${\cal O}_{qq}^{(3)}$ & ~ & ~ & ~ & ~ & ~ & ~ & ~ & ~ & ~ & \cmark & ~ & ~ & ~ & \cmark & ~ & \cmark & ~ & \cmark & ~ \\ \hline
        ${\cal O}_{lq}^{(1)}$ & ~ & ~ & ~ & ~ & ~ & ~ & ~ & ~ & ~ & \cmark & ~ & ~ & ~ & \cmark & ~ & ~ & ~ & ~ & ~ \\ \hline
        ${\cal O}_{lq}^{(3)}$ & ~ & ~ & ~ & ~ & ~ & ~ & ~ & ~ & ~ & \cmark & ~ & ~ & ~ & \cmark & ~ & ~ & ~ & ~ & ~ \\ \hline
        ${\cal O}_{ee}$ & ~ & ~ & \cmark & ~ & ~ & ~ & ~ & ~ & ~ & ~ & ~ & ~ & ~ & ~ & ~ & ~ & ~ & ~ & ~ \\ \hline
        ${\cal O}_{uu}$ & ~ & ~ & ~ & ~ & ~ & ~ & ~ & ~ & \cmark & ~ & ~ & ~ & ~ & ~ & ~ & ~ & \cmark & ~ & ~ \\ \hline
        ${\cal O}_{dd}$ & ~ & ~ & ~ & ~ & ~ & ~ & ~ & ~ & ~ & ~ & \cmark & ~ & ~ & ~ & \cmark & ~ & ~ & ~ & ~ \\ \hline
        ${\cal O}_{eu}$ & ~ & ~ & ~ & ~ & ~ & ~ & ~ & ~ & ~ & \cmark & ~ & ~ & ~ & ~ & ~ & ~ & ~ & ~ & ~ \\ \hline
        ${\cal O}_{ed}$ & ~ & ~ & ~ & ~ & ~ & ~ & ~ & ~ & \cmark & ~ & ~ & ~ & ~ & ~ & ~ & ~ & ~ & ~ & ~ \\ \hline
        ${\cal O}_{ud}^{(1)}$ & ~ & ~ & ~ & ~ & ~ & ~ & ~ & ~ & ~ & \cmark & ~ & ~ & ~ & ~ & ~ & \cmark & ~ & ~ & ~ \\ \hline
        ${\cal O}_{ud}^{(8)}$ & ~ & ~ & ~ & ~ & ~ & ~ & ~ & ~ & ~ & \cmark & ~ & ~ & ~ & ~ & ~ & \cmark & ~ & ~ & ~ \\ \hline
        ${\cal O}_{le}$ & ~ & ~ & ~ & \cmark & ~ & ~ & ~ & ~ & ~ & ~ & ~ & ~ & ~ & ~ & ~ & ~ & ~ & ~ & ~ \\ \hline
        ${\cal O}_{lu}$ & ~ & ~ & ~ & ~ & ~ & ~ & ~ & ~ & ~ & ~ & ~ & ~ & \cmark & ~ & ~ & ~ & ~ & ~ & ~ \\ \hline
        ${\cal O}_{ld}$ & ~ & ~ & ~ & ~ & ~ & ~ & ~ & ~ & ~ & ~ & ~ & \cmark & ~ & ~ & ~ & ~ & ~ & ~ & ~ \\ \hline
        ${\cal O}_{qe}$ & ~ & ~ & ~ & ~ & ~ & ~ & ~ & ~ & ~ & ~ & ~ & ~ & \cmark & ~ & ~ & ~ & ~ & ~ & ~ \\ \hline
        ${\cal O}_{qu}^{(1)}$ & ~ & ~ & ~ & \cmark & ~ & ~ & ~ & ~ & ~ & ~ & ~ & ~ & ~ & ~ & ~ & ~ & ~ & ~ & \cmark \\ \hline
        ${\cal O}_{qu}^{(8)}$ & ~ & ~ & ~ & \cmark & ~ & ~ & ~ & ~ & ~ & ~ & ~ & ~ & ~ & ~ & ~ & ~ & ~ & ~ & \cmark \\ \hline
        ${\cal O}_{qd}^{(1)}$ & ~ & ~ & ~ & \cmark & ~ & ~ & ~ & ~ & ~ & ~ & ~ & ~ & ~ & ~ & ~ & ~ & ~ & ~ & \cmark \\ \hline
        ${\cal O}_{qd}^{(8)}$ & ~ & ~ & ~ & \cmark & ~ & ~ & ~ & ~ & ~ & ~ & ~ & ~ & ~ & ~ & ~ & ~ & ~ & ~ & \cmark \\ \hline
        ${\cal O}_{ledq}$ & ~ & ~ & ~ & \cmark & ~ & ~ & ~ & ~ & ~ & ~ & ~ & ~ & ~ & ~ & ~ & ~ & ~ & ~ & ~ \\ \hline
        ${\cal O}_{quqd}^{(1)}$ & ~ & ~ & ~ & \cmark & ~ & ~ & ~ & ~ & ~ & \cmark & ~ & ~ & ~ & ~ & ~ & \cmark & ~ & ~ & ~ \\ \hline
        ${\cal O}_{quqd}^{(8)}$ & ~ & ~ & ~ & ~ & ~ & ~ & ~ & ~ & ~ & \cmark & ~ & ~ & ~ & ~ & ~ & \cmark & ~ & ~ & \cmark \\ \hline
        ${\cal O}_{lequ}^{(1)}$ & ~ & ~ & ~ & \cmark & ~ & ~ & ~ & ~ & ~ & \cmark & ~ & ~ & \cmark & ~ & ~ & ~ & ~ & ~ & ~ \\ \hline
        ${\cal O}_{lequ}^{(3)}$ & ~ & ~ & ~ & ~ & ~ & ~ & ~ & ~ & ~ & \cmark & ~ & ~ & \cmark & ~ & ~ & ~ & ~ & ~ & ~ \\ \hline
        ${\cal O}_{duq}$ & ~ & ~ & ~ & ~ & ~ & ~ & ~ & ~ & ~ & \cmark & ~ & ~ & ~ & ~ & ~ & ~ & ~ & ~ & ~ \\ \hline
        ${\cal O}_{qqu}$ & ~ & ~ & ~ & ~ & ~ & ~ & ~ & ~ & ~ & \cmark & ~ & ~ & ~ & ~ & ~ & ~ & ~ & ~ & ~ \\ \hline
        ${\cal O}_{qqq}$ & ~ & ~ & ~ & ~ & ~ & ~ & ~ & ~ & ~ & \cmark & ~ & ~ & ~ & \cmark & ~ & ~ & ~ & ~ & ~ \\ \hline
        ${\cal O}_{duu}$ & ~ & ~ & ~ & ~ & ~ & ~ & ~ & ~ & \cmark & \cmark & ~ & ~ & ~ & ~ & ~ & ~ & ~ & ~ & ~ \\ \hline
    \end{tabular}
    }
    \caption{The correspondence between dimension-6 operators in SMEFT and single-scalar-extended UV models. Check mark inside each box means that the operator to the left can be generated by introducing the new resonance above. Note that the operators with a star mark in the $S_1$ column cannot generated by an $S_1$ extended model. ${\cal O}_{eH}$, ${\cal O}_{uH}$ and ${\cal O}_{uH}$ need $S_1$ with $S_4/F_2/F_3$, $S_4/F_9/F_{11}$ and $S_4/F_8/F_{11}$ respectively. The explicit forms of operators are listed in Tab.\,\ref{tab:basis6}.}
    \label{tab:uvop/scalar}
\end{table}
\begin{table}
    \centering
    \resizebox{0.75\textwidth}{!}{
    \begin{tabular}{|c|c|c|c|c|c|c|c|c|c|c|c|c|c|c|}
    \hline
        ~ & $F_1$ & $F_2$ & $F_3$ & $F_4$ & $F_5$ & $F_6$ & $F_7$ & $F_8$ & $F_9$ & $F_{10}$ & $F_{11}$ & $F_{12}$ & $F_{13}$ & $F_{14}$ \\ \hline
        ${\cal O}_{eH}$ & ~ & \cmark & \cmark & \cmark & \cmark & \cmark & ~ & ~ & ~ & ~ & ~ & ~ & ~ & ~ \\ \hline
        ${\cal O}_{uH}$ & ~ & ~ & ~ & ~ & ~ & ~ & ~ & ~ & \cmark & ~ & \cmark & \cmark & \cmark & \cmark \\ \hline
        ${\cal O}_{dH}$ & ~ & ~ & ~ & ~ & ~ & ~ & ~ & \cmark & ~ & \cmark & \cmark & ~ & \cmark & \cmark \\ \hline
        ${\cal O}_{Hl}^{(1)}$ & \cmark & \cmark & ~ & ~ & \cmark & \cmark & ~ & ~ & ~ & ~ & ~ & ~ & ~ & ~ \\ \hline
        ${\cal O}_{Hl}^{(3)}$ & \cmark & \cmark & ~ & ~ & \cmark & \cmark & ~ & ~ & ~ & ~ & ~ & ~ & ~ & ~ \\ \hline
        ${\cal O}_{He}$ & ~ & ~ & \cmark & \cmark & ~ & ~ & ~ & ~ & ~ & ~ & ~ & ~ & ~ & ~ \\ \hline
        ${\cal O}_{Hq}^{(1)}$ & ~ & ~ & ~ & ~ & ~ & ~ & ~ & \cmark & \cmark & ~ & ~ & ~ & \cmark & \cmark \\ \hline
        ${\cal O}_{Hq}^{(3)}$ & ~ & ~ & ~ & ~ & ~ & ~ & ~ & \cmark & \cmark & ~ & ~ & ~ & \cmark & \cmark \\ \hline
        ${\cal O}_{Hu}$ & ~ & ~ & ~ & ~ & ~ & ~ & ~ & ~ & ~ & ~ & \cmark & \cmark & ~ & ~ \\ \hline
        ${\cal O}_{Hd}$ & ~ & ~ & ~ & ~ & ~ & ~ & ~ & ~ & ~ & \cmark & \cmark & ~ & ~ & ~ \\ \hline
        ${\cal O}_{Hud}$ & ~ & ~ & ~ & ~ & ~ & ~ & ~ & ~ & ~ & ~ & \cmark & ~ & ~ & ~ \\ \hline
    \end{tabular}
    }
    \caption{The correspondence between dimension-6 operators in SMEFT and single-fermion-extended UV models. The notation is the same as Tab.\,\ref{tab:uvop/scalar}.}
    \label{tab:uvop/fermion}
\end{table}

After matching and operator reduction, we are able to project the effective Lagrangian onto a selected operator basis. The result can be translated as a correspondence between UV resonances and IR effective operators, which are listed in Tab.\,\ref{tab:uvop/scalar}-\ref{tab:uvop/bnv}. The complete expression for Wilson coefficients can be found in App.\,\ref{sec:resultwc}.

\paragraph{The dictionary tables}

The relationship between UV resonances and IR effective operators is a bit complicated, so we rearrange the correspondence relationship into several tables according to the dimension of operators:
\begin{itemize}
    \item The correspondence between single scalar/fermion/vector resonances and dimension-6 operators are shown in Tab.\,\ref{tab:uvop/scalar}-\ref{tab:uvop/vector} respectively. Check mark inside each box means that the operator to the left can be generated by introducing the new resonance above. For models with 2 or more kinds of UV resonances, one can just counts and combines the effective operators generated by each new resonance, except for $S_1$ and ${\cal O}_{fH}$ types of operators. ${\cal O}_{eH}$, ${\cal O}_{uH}$ and ${\cal O}_{dH}$ need $S_1$ with $S_4/F_2/F_3$, $S_4/F_9/F_{11}$ and $S_4/F_8/F_{11}$ respectively. Other than this, interaction between different types of new resonances does not result in new effective operators. Actually, the relation between dimension-6 operators and heavy field multiplets has been provided in Ref.\,\cite{deBlas:2017xtg}, and our result is consistent with theirs.
    \item The UV completions of operators with odd canonical dimension are listed in Tab.\,\ref{tab:uvop/lnv} and Tab.\,\ref{tab:uvop/bnv}, depending on whether the model preserves baryon number or not. Every box with check mark denotes that the operator above can be generated by introducing a single resonance to the left, while boxes with resonances means that the operator above needs both the resonance to the left and one of the resonance inside the box. Among single resonance extended models only 3 seesaw models could generate dimension-7 operators, which has been verified by Ref.\,\cite{Elgaard-Clausen:2017xkq} for type-I seesaw model. Same as above, introducing new resonance with new interactions will not change the type of effective operators shown in the tables but only the Wilson coefficients.
    \item Note that not all models are presented in the tables. Only least requirements of resonances are listed. For example, ${\cal O}_{LH}$ can be generated by the model with $F_5$ as well as the model with $F_5$ and $F_7$, but only $F_5$ are marked in the column of ${\cal O}_{LH}$ since it has covered the latter situation.
\end{itemize}

With these identification tables, one can check what kinds of operators can be generated by a specific UV model, and also what kinds of UV resonance is required if one Wilson coefficient is measured to be non-zero. Although in most case the correspondence relationship is one-to-one, a quantitative analysis requires analytical expression of Wilson coefficients.

\begin{table}[H]
    \centering
    \resizebox{0.92\textwidth}{!}{
    \begin{tabular}{|c|c|c|c|c|c|c|c|c|c|c|c|c|c|c|c|}
    \hline
        ~ & $V_1$ & $V_2$ & $V_3$ & $V_4$ & $V_5$ & $V_6$ & $V_7$ & $V_8$ & $V_9$ & $V_{10}$ & $V_{11}$ & $V_{12}$ & $V_{13}$ & $V_{14}$ \\ \hline
        ${\cal O}_{H}$ & ~ & \cmark & ~ & \cmark & ~ & ~ & ~ & ~ & ~ & ~ & ~ & ~ & ~ & ~ \\ \hline
        ${\cal O}_{H\square}$ & \cmark & \cmark & ~ & \cmark & ~ & ~ & ~ & ~ & ~ & ~ & ~ & ~ & ~ & ~ \\ \hline
        ${\cal O}_{HD}$ & \cmark & \cmark & ~ & \cmark & ~ & ~ & ~ & ~ & ~ & ~ & ~ & ~ & ~ & ~ \\ \hline
        ${\cal O}_{eH}$ & \cmark & \cmark & ~ & \cmark & ~ & ~ & ~ & ~ & ~ & ~ & ~ & ~ & ~ & ~ \\ \hline
        ${\cal O}_{uH}$ & \cmark & \cmark & ~ & \cmark & ~ & ~ & ~ & ~ & ~ & ~ & ~ & ~ & ~ & ~ \\ \hline
        ${\cal O}_{dH}$ & \cmark & \cmark & ~ & \cmark & ~ & ~ & ~ & ~ & ~ & ~ & ~ & ~ & ~ & ~ \\ \hline
        ${\cal O}_{Hl}^{(1)}$ & \cmark & ~ & ~ & ~ & ~ & ~ & ~ & ~ & ~ & ~ & ~ & ~ & ~ & ~ \\ \hline
        ${\cal O}_{Hl}^{(3)}$ & ~ & ~ & ~ & \cmark & ~ & ~ & ~ & ~ & ~ & ~ & ~ & ~ & ~ & ~ \\ \hline
        ${\cal O}_{He}$ & \cmark & ~ & ~ & ~ & ~ & ~ & ~ & ~ & ~ & ~ & ~ & ~ & ~ & ~ \\ \hline
        ${\cal O}_{Hq}^{(1)}$ & \cmark & ~ & ~ & ~ & ~ & ~ & ~ & ~ & ~ & ~ & ~ & ~ & ~ & ~ \\ \hline
        ${\cal O}_{Hq}^{(3)}$ & ~ & ~ & ~ & \cmark & ~ & ~ & ~ & ~ & ~ & ~ & ~ & ~ & ~ & ~ \\ \hline
        ${\cal O}_{Hu}$ & \cmark & ~ & ~ & ~ & ~ & ~ & ~ & ~ & ~ & ~ & ~ & ~ & ~ & ~ \\ \hline
        ${\cal O}_{Hd}$ & \cmark & ~ & ~ & ~ & ~ & ~ & ~ & ~ & ~ & ~ & ~ & ~ & ~ & ~ \\ \hline
        ${\cal O}_{Hud}$ & ~ & \cmark & ~ & ~ & ~ & ~ & ~ & ~ & ~ & ~ & ~ & ~ & ~ & ~ \\ \hline
        ${\cal O}_{ll}$ & \cmark & ~ & ~ & \cmark & ~ & ~ & ~ & ~ & ~ & ~ & ~ & ~ & ~ & ~ \\ \hline
        ${\cal O}_{qq}^{(1)}$ & \cmark & ~ & ~ & ~ & ~ & ~ & ~ & ~ & ~ & ~ & ~ & \cmark & ~ & \cmark \\ \hline
        ${\cal O}_{qq}^{(3)}$ & ~ & ~ & ~ & \cmark & ~ & ~ & ~ & ~ & ~ & ~ & ~ & \cmark & ~ & \cmark \\ \hline
        ${\cal O}_{lq}^{(1)}$ & \cmark & ~ & ~ & ~ & \cmark & ~ & ~ & ~ & \cmark & ~ & ~ & ~ & ~ & ~ \\ \hline
        ${\cal O}_{lq}^{(3)}$ & ~ & ~ & ~ & \cmark & \cmark & ~ & ~ & ~ & \cmark & ~ & ~ & ~ & ~ & ~ \\ \hline
        ${\cal O}_{ee}$ & \cmark & ~ & ~ & ~ & ~ & ~ & ~ & ~ & ~ & ~ & ~ & ~ & ~ & ~ \\ \hline
        ${\cal O}_{uu}$ & \cmark & ~ & ~ & ~ & ~ & ~ & ~ & ~ & ~ & ~ & ~ & \cmark & ~ & ~ \\ \hline
        ${\cal O}_{dd}$ & \cmark & ~ & ~ & ~ & ~ & ~ & ~ & ~ & ~ & ~ & ~ & \cmark & ~ & ~ \\ \hline
        ${\cal O}_{eu}$ & \cmark & ~ & ~ & ~ & ~ & \cmark & ~ & ~ & ~ & ~ & ~ & ~ & ~ & ~ \\ \hline
        ${\cal O}_{ed}$ & \cmark & ~ & ~ & ~ & \cmark & ~ & ~ & ~ & ~ & ~ & ~ & ~ & ~ & ~ \\ \hline
        ${\cal O}_{ud}^{(1)}$ & \cmark & \cmark & ~ & ~ & ~ & ~ & ~ & ~ & ~ & ~ & ~ & ~ & \cmark & ~ \\ \hline
        ${\cal O}_{ud}^{(8)}$ & ~ & \cmark & ~ & ~ & ~ & ~ & ~ & ~ & ~ & ~ & ~ & \cmark & \cmark & ~ \\ \hline
        ${\cal O}_{le}$ & \cmark & ~ & \cmark & ~ & ~ & ~ & ~ & ~ & ~ & ~ & ~ & ~ & ~ & ~ \\ \hline
        ${\cal O}_{lu}$ & \cmark & ~ & ~ & ~ & ~ & ~ & ~ & \cmark & ~ & ~ & ~ & ~ & ~ & ~ \\ \hline
        ${\cal O}_{ld}$ & \cmark & ~ & ~ & ~ & ~ & ~ & \cmark & ~ & ~ & ~ & ~ & ~ & ~ & ~ \\ \hline
        ${\cal O}_{qe}$ & \cmark & ~ & ~ & ~ & ~ & ~ & \cmark & ~ & ~ & ~ & ~ & ~ & ~ & ~ \\ \hline
        ${\cal O}_{qu}^{(1)}$ & \cmark & ~ & ~ & ~ & ~ & ~ & \cmark & ~ & ~ & ~ & \cmark & ~ & ~ & ~ \\ \hline
        ${\cal O}_{qu}^{(8)}$ & ~ & ~ & ~ & ~ & ~ & ~ & \cmark & ~ & ~ & ~ & \cmark & \cmark & ~ & ~ \\ \hline
        ${\cal O}_{qd}^{(1)}$ & \cmark & ~ & ~ & ~ & ~ & ~ & ~ & \cmark & ~ & \cmark & ~ & ~ & ~ & ~ \\ \hline
        ${\cal O}_{qd}^{(8)}$ & ~ & ~ & ~ & ~ & ~ & ~ & ~ & \cmark & ~ & \cmark & ~ & \cmark & ~ & ~ \\ \hline
        ${\cal O}_{ledq}$ & ~ & ~ & ~ & ~ & \cmark & ~ & \cmark & ~ & ~ & ~ & ~ & ~ & ~ & ~ \\ \hline
        ${\cal O}_{duq}$ & ~ & ~ & ~ & ~ & ~ & ~ & \cmark & \cmark & ~ & ~ & ~ & ~ & ~ & ~ \\ \hline
        ${\cal O}_{qqu}$ & ~ & ~ & ~ & ~ & ~ & ~ & \cmark & ~ & ~ & ~ & ~ & ~ & ~ & ~ \\ \hline
    \end{tabular}
    }
    \caption{The correspondence between dimension-6 operators in SMEFT and single-vector-extended UV models. The notation is the same as Tab.\,\ref{tab:uvop/scalar}.}
    \label{tab:uvop/vector}
\end{table}

\paragraph{Notation of the Wilson coefficients}

The operators in the appendix have been ordered by their dimension as well as the operator type. Dimension-6 operators are divided into bosonic, 4-fermion, 2-fermion and baryon-number-violating operators, while dimension-7 operators are divided by whether the operator violates baryon number or not. It is still necessary to note down the notation used in the full expression:
\begin{itemize}
    \item Mass terms in the denominator of each term denote the source of this contribution, i.e. the Feynman diagram. The mediating particles are just those appearing in the subscripts of the mass terms.
    \item The coupling parameters of the SM is the same as those in Eq.\,\eqref{eq:smlag}. New coupling parameters of UV models are denoted by ${\cal C}$ for mass-dimension-1 parameters or ${\cal D}$ for dimensionless parameters. The corresponding interacting particles are written in the subscript, whose flavor indices are listed in order in the superscript. For instance, ${\cal D}_{F_{7L}LS_5}^{prs}$ is the dimensionless coupling  of the interaction between $F_{7L}$,$L$ and $S_5$, $p,r,s$ are the flavor indices of $F_{7L}, L, S_5$ respectively.
    \item In each term of the Wilson coefficients, the flavor indices of the Wilson coefficients are denoted by $f_i$ while the one summed in a pair are denoted by $p_j$. The subscripts of $f_i$ are ordered by helicity and also the letter of the field appearing in the operator. For example, the fields appearing in ${\cal O}_{dLueH} = \epsilon^{ij} (\overline{d}^{a} \ell_{i}) (u^T_{a} C e) H_j$ are $d^c, L, u, e, H$, whose helicities are $-1/2, -1/2, 1/2, 1/2, 0$ respectively. Thus the flavor indices for the former 4 particles are $f_1, f_2, f_5, f_4$. 
    \item The flavor indices of the mass terms in the denominator are omitted. Every pair to be summed up in the numerator corresponds to a mediating particle with the flavor index, whose mass should appear in the denominator.
\end{itemize}




\begin{table}
    \centering
    \resizebox{0.95\textwidth}{!}{
    \begin{tabular}{|c|c|c|c|c|c|c|c|c|c|c|c|}
        \hline
        & ${\cal O}_5$ & ${\cal O}_{LH}$ & ${\cal O}_{LeHD}$ & ${\cal O}_{LHD1}$ & ${\cal O}_{LHD2}$ & ${\cal O}_{LHW}$ & ${\cal O}_{eLLLH}$ & ${\cal O}_{dLQLH1}$ & ${\cal O}_{dLQLH2}$ & ${\cal O}_{dLueH}$ & ${\cal O}_{QuLLH}$ \\
        \hline 
        $S_2$ & & & & & & & $S_4/F_4$ & & $S_4/F_9/F_{10}$ & & $S_4/F_8/F_{12}$ \\
        \hline 
        $S_4$ & & & & & & & $S_2/S_6$ & $S_6$ & $S_2/S_6$ & & $S_2/S_6$ \\
        \hline
        $S_6$ & \cmark & \cmark & $F_3$ & \cmark & \cmark & & $S_4/F_4$ & $S_4/F_{10}/F_{14}$ & $S_4/F_{10}/F_{14}$ & & $S_4/F_{12}/F_{13}$ \\
        \hline
        $S_8$ & & $F_6$ & & & & & & & & & \\
        \hline
        $S_{12}$ & & & & & & & & $F_{14}$ & $F_9/F_{14}$ & $F_3/F_{12}$ & \\
        \hline
        $F_1$ & \cmark & \cmark & \cmark & & \cmark & \cmark & \cmark & \cmark & \cmark & $V_2/V_5$ & \cmark \\
        \hline
        $F_3$ & & & $S_6/V_2$ & & & & & & & $S_{12}/V_2$ & \\
        \hline
        $F_4$ & & & & & & & $S_2/S_6$ & & & &\\
        \hline
        $F_5$ & \cmark & \cmark & \cmark & \cmark & \cmark & \cmark & \cmark & \cmark & \cmark & & \cmark \\
        \hline
        $F_6$ & & $S_8$ & & & & & & & & & \\
        \hline
        $F_8$ & & & & & & & & & & & $S_2$ \\
        \hline
        $F_9$ & & & & & & & & & $S_2/S_{12}$ & & \\
        \hline
        $F_{10}$ & & & & & & & & $S_6$ & $S_2/S_{6}$ & $V_3$ & \\
        \hline
        $F_{12}$ & & & & & & & & & & $S_{12}/V_3/V_5$ & $S_2/S_6/V_5/V_9$ \\
        \hline
        $F_{13}$ & & & & & & & & & & & $S_6$ \\
        \hline
        $F_{14}$ & & & & & & & & $S_6/S_{12}$ & $S_6/S_{12}$ & & \\
        \hline
        $V_2$ & & & $F_3/V_3$ & & & & & & & $F_1/F_3/V_3$ & \\
        \hline
        $V_3$ & & & $V_2$ & & & & & & & $F_{10}/F_{12}/V_2$ & \\
        \hline
        $V_5$ & & & & & & & & & & $F_1/F_{12}$ & $F_{12}$ \\
        \hline
        $V_9$ & & & & & & & & & & & $F_{12}$ \\
        \hline
    \end{tabular}
    }
    \caption{The correspondence between dimension-5 and -7 LNV operators in SMEFT and UV resonances. Only models conserving baryon number are listed here. Every box with check mark denotes that the operator above can be generated by introducing a single resonance to the left, while boxes with resonances means that the operator above needs both the resonance to the left and one of the resonance inside the box. Note that not all models are covered by the tables. Only least requirements of resonances are listed. The explicit forms of operators are listed in Tab.\,\ref{tab:basis5} and Tab.\,\ref{tab:basis7}.}
    \label{tab:uvop/lnv}
\end{table}

\begin{table}
    \centering
    \resizebox{0.95\textwidth}{!}{
    \begin{tabular}{|c|c|c|c|c|c|c|c|c|}
        \hline
        & ${\cal O}_{dLQLH1}$ & ${\cal O}_{dLQLH2}$ & ${\cal O}_{dLueH}$ & ${\cal O}_{QuLLH}$ & ${\cal O}_{LdudH}$ & ${\cal O}_{LdddH}$ & ${\cal O}_{eQddH}$ & ${\cal O}_{LdQQH}$ \\
        \hline 
        ${\color{red} S_{10}}$ & $S_{12}/F_1/F_{10}$ & $S_{12}/F_1/F_{10}$ & $S_{12}/F_1/F_{10}$ & & $S_{12}/F_1/F_{10}$ & & & $S_{12}/F_1/F_{10}$ \\
        \hline 
        $S_{11}$ & & & & & $S_{13}/F_{1}/F_{11}$ & $S_{12}/F_2/F_{11}$ & $S_{13}/F_3/F_8$ & \\
        \hline
        $S_{12}$ & ${\color{red} S_{10}}/{\color{red} S_{14}}$ & ${\color{red} S_{10}}/{\color{red} S_{14}}$ & ${\color{red} S_{10}}$ & & ${\color{red} S_{10}}/F_{10}/F_{11}$ & $S_{11}/F_{11}$ & & ${\color{red} S_{10}}/{\color{red} S_{14}}/F_8/F_{13}$ \\
        \hline
        $S_{13}$ & & & & & $S_{11}/F_{10}$ & & $S_{11}/F_{10}$ & \\
        \hline
        ${\color{red} S_{14}}$ & $S_{12}/F_5/F_{10}$ & $S_{12}/F_5/F_{10}$ & & & & & & $S_{12}/F_5/F_{10}$ \\
        \hline
        $F_1$ & ${\color{red} S_{10}}$ & ${\color{red} S_{10}}$ & ${\color{red} S_{10}}$ & ${\color{red} V_8}$ & ${\color{red} S_{10}}/S_{11}$ & & & ${\color{red} S_{10}}/{\color{red} V_8}$ \\
        \hline
        $F_2$ & & & & & & $S_{11}$ & & \\
        \hline
        $F_3$ & & & ${\color{red} V_8}$ & & & & $S_{11}/{\color{red} V_8}$ & \\
        \hline
        $F_5$ & ${\color{red} S_{14}}$ & ${\color{red} S_{14}}$ & & ${\color{red} V_8}$ & & & & ${\color{red} S_{14}}/{\color{red} V_8}$ \\
        \hline
        $F_8$ & & & & ${\color{red} V_8}$ & & & $S_{11}/V_5$ & $S_{12}/V_5/{\color{red} V_8}$ \\
        \hline
        $F_{10}$ & ${\color{red} S_{10}}/{\color{red} S_{14}}$ & ${\color{red} S_{10}}/{\color{red} S_{14}}$ & ${\color{red} S_{10}}/{\color{red} V_8}$ & & ${\color{red} S_{10}}/S_{12}/S_{13}$ & & $S_{13}/V_5/{\color{red} V_8}$ & ${\color{red} S_{10}}/{\color{red} S_{14}}/V_{5}/V_{9}$ \\
        \hline
        $F_{11}$ & & & & & $S_{11}/S_{12}$ & $S_{11}/S_{12}$ & & \\
        \hline
        $F_{13}$ & & & & ${\color{red} V_8}$ & & & & $S_{12}/{\color{red} V_8}/V_9$ \\
        \hline
        $V_5$ & & & ${\color{red} V_8}$ & ${\color{red} V_8}$ & & & $F_8/F_{10}/{\color{red} V_8}$ & $F_8/F_{10}/{\color{red} V_8}$ \\
        \hline
        ${\color{red} V_8}$ & & & $F_3/F_{10}/V_5$ & \makecell{$F_1/F_5/F_8/$\\$F_{13}/V_5/V_9$} & & & $F_3/F_{10}/V_5$ & \makecell{$F_1/F_5/F_8/$\\$F_{13}/V_5/V_9$} \\
        \hline
        $V_9$ & & & & ${\color{red} V_8}$ & & & & $F_{10}/F_{13}/{\color{red} V_8}$ \\
        \hline
    \end{tabular}
    }
    \caption{The correspondence between dimension-7 operators in SMEFT and UV resonances. Models listed here all violate baryon number. Models with resonances marked with red can generate dimension-6 baryon number violating operators. Other notation is the same as Tab.\,\ref{tab:uvop/lnv}.}
    \label{tab:uvop/bnv}
\end{table}

\paragraph{Example of usage}
To illustrate the usage of our dictionary more clearly, we will take a simple example by looking for one UV completion of neutrino-less double decay ($0\nu\beta\beta$). $0\nu\beta\beta$ receives 3 types of contributions at tree level: short range, long range and neutrino mass insertion \cite{Cirigliano:2017djv,Cirigliano:2018yza,Liao:2019tep}. Among dimension-5, -6 and -7 operators in the SMEFT, ${\cal O}_{LHD1}$ and ${\cal O}_{duLLD}$ have the short-range or contact contribution while ${\cal O}_{LeHD}$, ${\cal O}_{LHW}$, ${\cal O}_{dLQLH1/2}$, ${\cal O}_{dLueH}$ and ${\cal O}_{QuLLH}$ have the long-range contribution. Neutrino mass insertion can be induced by ${\cal O}_5$ and ${\cal O}_{LH}$. Each operator has several UV origins according to Tab.\,\ref{tab:uvop/lnv}-\ref{tab:uvop/bnv}. We just pick the model with $F_3$ and $V_2$ as an example. Terms in the full Lagrangian that involves $F_3$ and $V_2$ can be read  from App.\,\ref{sec:uvlagfull} as
\begin{align}
    \Delta{\cal L}_{\rm UV} = & \sum_p \left[(\overline{F}_{3p})^i i \slashed{D} (F_{3 p})_i - M_{F_3}^p (\overline{F}_{3p})^i (F_{3 p})_i  + V_{2p}^{\dagger\mu}(g_{\mu\nu}D^2 - D_{\nu}D_{\mu} + g_{\mu\nu}(M_{V_2}^p)^2) V_{2p}^{\nu} \right] \nonumber \\
    & 
    + \left[ - \mathcal{D}_{{e^\dagger F_{3R}^\dagger H^\dagger }}^{{rp}}\epsilon _{{ij}}[\bar{e}_{{r}}({F}_{{3p}})^{{i}}]{H}^{{\dagger j}} 
    - 2 \mathcal{D}_{{d^\dagger uV_{2}^\dagger }}^{{rsp}} [(\bar{d}_{{r}})^{{a}}\gamma _{\mu }({u}_{{s}})_{{a}}]{V}_{{2p}}^{\dagger \mu } \right. \nonumber \\
    & \left. - 2 \mathcal{D}_{{HHV_{2}^\dagger D}}^{{p}}\epsilon ^{{ij}}[D_{\mu }{H}_{{i}}]{H}_{{j}}{V}_{{2p}}^{\dagger \mu } 
     + \mathcal{D}_{{F_{3L}L^\dagger V_{2}^\dagger }}^{{psr}}\delta _{{j}}^{{i}}[(\bar{l}_{{s}})^{{j}}\gamma _{\mu }({F}_{{3p}})_{{i}}]{V}_{{2r}}^{\dagger \mu } + {\rm h.c.} \right]
     \label{eq:exuvlag}
\end{align}
From the $F_3$ row of Tab.\,\ref{tab:uvop/lnv} it can be verified that the model with $F_3$ and $V_2$ will generate both ${\cal O}_{LeHD}$ and ${\cal O}_{dLueH}$. Their Wilson coefficients are listed in App.\,\ref{sec:resultwc} as
\begin{align}
    C_{LeHD}^{f_1f_5} = & \sum_{p_1,p_2} \frac{2 i \mathcal{D}_{{e^\dagger F_{3R}^\dagger H^\dagger }}^{{f_5p_1*}} \mathcal{D}_{{F_{3L}L^\dagger V_{2}^\dagger }}^{{p_1f_1p_2*}} \mathcal{D}_{{HHV_{2}^\dagger D}}^{{p_2}}}{{M_{F_{3}}^{p_1}} {(M_{V_{2}}^{p_2})^2}}, \\
    C_{dLueH}^{f_1f_2f_5f_4} = & \sum_{p_1,p_2} \frac{4 \mathcal{D}_{{d^\dagger uV_{2}^\dagger }}^{{f_1f_5p_1}} \mathcal{D}_{{e^\dagger F_{3R}^\dagger H^\dagger }}^{{f_4p_2*}} \mathcal{D}_{{F_{3L}L^\dagger V_{2}^\dagger }}^{{p_2f_2p_1*}}}{{M_{F_{3}}^{p_1}} {(M_{V_{2}}^{p_2})^2}}.
\end{align}
We have attached the flavor indices to the Wilson coefficients. The indices are sorted by helicities and letters of the composing fields. For ${\cal O}_{LeHD}$, the helicities of the composing fields $\ell, e, H, H, H$ are $-1/2, 1/2, 0, 0, 0$, so $\ell, e$ are labeled by $f_1$ and $f_5$. The same rule applies for ${\cal O}_{dLueH}$, which has been mentioned above.
Thus the effective Lagrangian is
\begin{align}
    \Delta{\cal L}_{\rm EFT} = & C_{LeHD}^{f_1f_5} \times \epsilon^{ij}\epsilon^{kl} (\ell^T_{if_1} C \gamma^\mu e_{f_5}) H_j H_k (iD_\mu H_l) + C_{dLueH}^{f_1f_2f_5f_4} \times \epsilon^{ij} (\overline{d}^{a}_{f_1} \ell_{if_2}) (u^T_{af_5} C e_{f_4}) H_j.
\end{align}
Constraints on the couplings in Eq.\,\eqref{eq:exuvlag} can be deduced from current experimental constraints on the Wilson coefficients.

\begin{figure}
    \centering
    \includegraphics[width=0.4\textwidth]{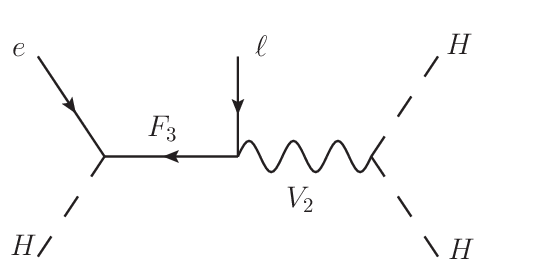}
    \includegraphics[width=0.4\textwidth]{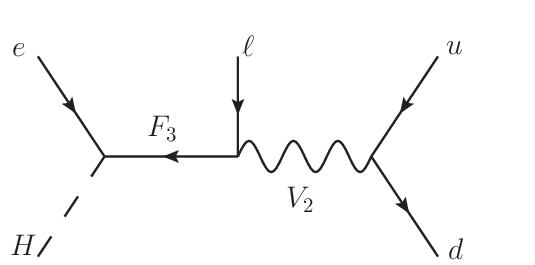}
    \caption{Contributions to dimension-7 operators ${\cal O}_{LeHD}$ and ${\cal O}_{dLueH}$ from the model with $F_3$ and $V_2$.}
    \label{fig:ex}
\end{figure}

One can also draw the corresponding Feynman diagrams from the Wilson coefficients\footnote{Some terms cannot be depicted by diagrams since they evolve from field redefinition. For example, the first term in $C_H$ \eqref{eq:wc/ch}, $2 \lambda_H (\mathcal{C}_{{HH^\dagger S_{5}}}^{{p_1}})^2/{M_{S_{5}}^4}$, originates from reduction of the redundant operators such as $H^\dagger H (D_\mu H)^\dagger (D^\mu H)$.}. The mediating propagators $F_3$ and $V_2$ are encoded in the denominator, while the vertices are just presented as the couplings in the numerator. The Feynman diagrams shows as Fig.\,\ref{fig:ex}.
\section{Summary}
\label{sec:summary}

The EFT approach provides a systematic way to parameterize the BSM physics in terms of a series of Wilson coefficients of effective operators. One could get knowledge of UV physics by measuring the related low-energy experimental observables and determining the Wilson coefficients. In order not to miss any possibilities, a systematic bottom-up approach to link the UV physics and Wilson coefficients is needed. In this work, we present a correspondence between different UV resonances and dimension-5, -6 and -7 SMEFT operators in Tab.\,\ref{tab:uvop/scalar}-\ref{tab:uvop/bnv}. 
Information about the UV resonances is encoded in the relation among Wilson coefficients of effective operators, including same dimension and cross-dimension relation.
With the help of the dictionary tables, pattern of non-zero Wilson coefficients measured by experiments can be utilized to determine the resonance that possibly exists.
The complete expression of the Wilson coefficients is presented in App.\,\ref{sec:resultwc}.

Following Ref.\,\cite{Li:2022abx}, the UV resonances that have tree-level contributions to the effective operator can be enumerated by finding the eigenstates of the Casimir operators. We use spinor helicity formalism with massive amplitude to generate the renormalizable Lagrangian for UV physics containing all possible resonances. The Lorentz structures are listed in Tab.\,\ref{tab:Dim8ClY}. The gauge structures as well as the Clebsch-Gordan coefficients can be found by Young tableau formalism. After writing down the complete Lagrangian, we use the functional matching method to integrate out the heavy fields at the classical level. In order to reduce numerous effective operators to one operator basis, we provide a systematic method by the off-shell amplitude formalism, which can be applied to any redundant operators. Our enumeration, matching and reduction procedure is also applicable for all kinds of EFT-guided physics search.

The UV-IR dictionary listed in Tab.\,\ref{tab:uvop/scalar}-\ref{tab:uvop/bnv} and App.\,\ref{sec:resultwc} can be used in two ways: one may check what kinds of effective operators can be generated for one UV model, and if several effective coefficients are measured to be non-zero he can also check which heavy resonance has the most possibilities to exist. The complete expression of the Wilson coefficients is also presented in the appendix for qualitative analysis. Our result could provide an EFT-guided UV resonance searches in the future collider experiments.

\acknowledgments


We would like to thank Yu-Han Ni for helpful discussions on bottom-up EFT method. This work is supported by the National Science Foundation of China under Grants No. 12022514, No. 11875003 and No. 12047503, and National Key Research and Development Program of China Grant No. 2020YFC2201501, No. 2021YFA0718304, and CAS Project for Young Scientists in Basic Research YSBR-006, the Key Research Program of the CAS Grant No. XDPB15.
\appendix
\section{The UV Lagrangian}
\label{sec:uvlagfull}

In this appendix we provide the relevant UV Lagrangian. Some descriptions are given as follows:
\begin{itemize}
    \item New couplings with mass dimension 1 and 0 are denoted by ${\cal C}$ and ${\cal D}$ respectively. The subscript marks the interacting fields and the flavor indices are written in the superscript. 
    \item Every field has its gauge indices ($a,A,{\frak a},...$ or $i,I,{\cal I},...$) and flavor indices ($p,r,s,t,...$). 
    \item The transpose marks ${}^T$ of fermion fields are omitted for convenience. $f_1^T C f_2 = \overline{f}_1^c f_2$ and $\overline{f}_1 C \overline{f}_2^T = \overline{f}_1 f_2^c$ are denoted by $f_1 C f_2$ and $\overline{f}_1 C \overline{f}_2$  respectively. 
\end{itemize}
Detailed description can be found in Sec.\,\ref{sec:uvlag}.

\subsection{Kinetic terms}

The kinetic terms have been presented in Sec.\,\ref{sec:procedure}. For completeness, we list the terms here:
\begin{align}
    & \Delta{\cal L}_{\rm kin} = -\eta\Phi^{\dagger}(D^2 + M^2) \Phi, &  \text{for scalars}, & \\
    & \Delta{\cal L}_{\rm kin} = \overline{F}i\slashed{D}F - \frac{1}{2}M \left(\overline{F}^cF + \overline{F}F^c\right), &  \text{for Majorana fermions}, & \\
    & \Delta{\cal L}_{\rm kin} = \overline{F}i\slashed{D}F - M \overline{F}F, &  \text{for Dirac fermions}, & \\
    & \Delta{\cal L}_{\rm kin} = \eta V^{\dagger\mu}(g_{\mu\nu}D^2 - D_{\nu}D_{\mu} + g_{\mu\nu}M^2) V^{\nu}, &  \text{for vectors}, &
\end{align}
where $\eta = \frac{1}{2}(1)$ for real(complex) bosonic fields. Note that Majorana fermions have parity left, i.e. $F = F_L = P_L F$, $P_L$ is the projection operator. Contraction of gauge indices are straightforwards. For example,
\begin{align}
    \Delta{\cal L}_{\rm kin} =  -S_{15}^{\dagger \frak{a}}(D^2 + M_{S_{15}}^2) S_{15 \frak{a}}
\end{align}
for $S_{15}$ and 
\begin{align}
    \Delta{\cal L}_{\rm kin} = \overline{F}_5^Ii\slashed{D}F_5^I - \frac{1}{2}M_{F_{5}} \left(\overline{F}_5^{cI}F_5^I + \overline{F}_5^IF_5^{cI}\right)
\end{align}
for $F_5$.

\subsection{Interacting terms}

As we construct the UV Lagrangian with the two-component spinors, initially the fermions in the UV Lagrangian are all two-component Weyl spinors, and then we translate them to four-component Dirac spinors for readers' convenience, through the following relations:
\begin{align}
	q=\begin{pmatrix}Q\\0\end{pmatrix},\quad u=\left(\begin{array}{c}0\\u_{(R)}\end{array}\right),\quad d=\left(\begin{array}{c}0\\d_{(R)}\end{array}\right),\quad l=\left(\begin{array}{c}L\\0\end{array}\right),\quad e=\left(\begin{array}{c}0\\e_{(R)}\end{array}\right),\\
	\bar{q}=\left(0\,,\,Q^{\dagger} \right),\quad \bar{u}=\left(u^{\dagger}_{(R)}\,,\,0 \right),\quad \bar{d}=\left(d^{\dagger}_{(R)}\,,\,0\right),\quad \bar{l}=\left(0\,,\,L^{\dagger}\right),\quad \bar{e}=\left(e^{\dagger}_{(R)}\,,\,0\right),
\end{align}
for SM fermions,
\begin{align}
    F = \left(\begin{array}{c} F_{\rm L} \\ 0 \end{array}\right) \text{ and } \left(\begin{array}{c} F_{\rm L} \\ F_{\rm R} \end{array}\right)
\end{align}
for UV Majorana and Dirac fermions.

\subsubsection{New scalars}

\begin{align}
    \begin{autobreak}
        \Delta{\cal L}_{\rm UV,S}^{(d = 3)} = 

        \mathcal{C}_{{HH^\dagger S_{1}}}^{{p}}{H}_{{i}}{H}^{{\dagger i}}{S}_{{1p}} 

 - \frac{\mathcal{C}_{{HH^\dagger S_{5}}}^{{p}}}{\sqrt{2}}(\tau ^{{I}})_{{j}}^{{i}}{H}_{{i}}{H}^{{\dagger j}}({S}_{{5p}})^{{I}} 

 + \frac{\mathcal{C}_{{HHS_{6}^\dagger }}^{{p}}}{\sqrt{2}}\epsilon ^{{kj}} (\tau ^{{I}})_{{k}}^{{i}}{H}_{{i}}{H}_{{j}}({S}_{{6p}}^{\dagger })^{{I}} 

 + \mathcal{C}_{{HS_{1}S_{4}^\dagger }}^{{pr}}{H}_{{i}}({S}_{{4r}}^{\dagger })^{{i}}{S}_{{1p}} 

 + \mathcal{C}_{{HS_{2}^\dagger S_{4}}}^{{rp}}\epsilon ^{{ji}}{H}_{{j}}({S}_{{4r}})_{{i}}{S}_{{2p}}^{\dagger } 

 - \frac{\mathcal{C}_{{H^\dagger S_{4}S_{5}}}^{{pr}}}{\sqrt{2}}(\tau ^{{I}})_{{j}}^{{i}}{H}^{{\dagger j}}({S}_{{4p}})_{{i}}({S}_{{5r}})^{{I}} 

 + \frac{\mathcal{C}_{{HS_{4}S_{6}^\dagger }}^{{pr}}}{\sqrt{2}}\epsilon ^{{ki}} (\tau ^{{I}})_{{k}}^{{j}}{H}_{{j}}({S}_{{4p}})_{{i}}({S}_{{6r}}^{\dagger })^{{I}} 

 + \frac{\mathcal{C}_{{HS_{5}S_{7}^\dagger }}^{{pr}}}{\sqrt{2}}\epsilon _{{jl}} {C}_{\mathfrak{i}}^{{ikl}} (\tau ^{{I}})_{{k}}^{{j}}{H}_{{i}}({S}_{{5p}})^{{I}}({S}_{{7r}}^{\dagger })^{\mathfrak{i}} 

 + \frac{\mathcal{C}_{{HS_{6}^\dagger S_{7}}}^{{rp}}}{\sqrt{2}}\epsilon ^{{im}} \epsilon ^{{jl}} {C}_{{klm}}^{\mathfrak{i}} (\tau ^{{I}})_{{j}}^{{k}}{H}_{{i}}({S}_{{6p}}^{\dagger })^{{I}}({S}_{{7r}})^{\mathfrak{i}} 

 + \frac{\mathcal{C}_{{HS_{6}S_{8}^\dagger }}^{{pr}}}{\sqrt{2}}\epsilon _{{jl}} {C}_{\mathfrak{i}}^{{ikl}} (\tau ^{{I}})_{{k}}^{{j}}{H}_{{i}}({S}_{{6p}})^{{I}}({S}_{{8r}}^{\dagger })^{\mathfrak{i}} 

 + 2 \mathcal{C}_{{HS_{10}S_{12}^\dagger }}^{{pr}} {H}_{{i}}({S}_{{10p}})_{{a}}({S}_{{12r}}^{\dagger })^{{ai}} 

 + 2 \mathcal{C}_{{HS_{11}^\dagger S_{12}}}^{{rp}}\epsilon ^{{ji}} {H}_{{j}}({S}_{{11p}}^{\dagger })^{{a}}({S}_{{12r}})_{{ai}} 

 + 2 \mathcal{C}_{{HS_{11}S_{13}^\dagger }}^{{pr}} {H}_{{i}}({S}_{{11p}})_{{a}}({S}_{{13r}}^{\dagger })^{{ai}} 

 - \sqrt{2} \mathcal{C}_{{HS_{12}^\dagger S_{14}}}^{{rp}} (\tau ^{{I}})_{{i}}^{{j}}{H}_{{j}}({S}_{{12p}}^{\dagger })^{{ai}}({S}_{{14r}})_{{a}}^{{I}} 

 + \mathcal{C}_{S_{1}S_{1}S_{1}}^{{prs}} {S}_{{1p}}{S}_{{1r}}{S}_{{1s}} 

 - \mathcal{C}_{{S_{1}S_{5}S_{5}}}^{{prs}} ({S}_{{5r}})^{{I}}({S}_{{5s}})^{{I}}{S}_{{1p}} 

 + \mathcal{C}_{{S_{1}S_{6}S_{6}^\dagger }}^{{prs}}({S}_{{6r}})^{{I}}({S}_{{6s}}^{\dagger })^{{I}}{S}_{{1p}} 

 - \mathcal{C}_{{S_{2}^\dagger S_{5}S_{6}}}^{{spr}}({S}_{{5r}})^{{I}}({S}_{{6s}})^{{I}}{S}_{{2p}}^{\dagger } 

 - \mathcal{C}_{{S_{3}S_{6}^\dagger S_{6}^\dagger }}^{{prs}}({S}_{{6r}}^{\dagger })^{{I}}({S}_{{6s}}^{\dagger })^{{I}}{S}_{{3p}} 

 + \frac{i \mathcal{C}_{{S_{5}S_{5}S_{5}}}^{{prs}}}{\sqrt{2}}\epsilon ^{{IJK}}({S}_{{5p}})^{{I}}({S}_{{5r}})^{{J}}({S}_{{5s}})^{{K}} 

 - \frac{i \mathcal{C}_{{S_{5}S_{6}S_{6}^\dagger }}^{{prs}}}{\sqrt{2}}\epsilon ^{{IJK}}({S}_{{5p}})^{{I}}({S}_{{6r}})^{{J}}({S}_{{6s}}^{\dagger })^{{K}}

    \end{autobreak}
\end{align}

\begin{align}
    \begin{autobreak}
        \Delta{\cal L}_{\rm UV,S}^{(d = 4)} = 
        
        - \mathcal{D}_{{LLS_{2}}}^{{rsp}}\epsilon ^{{ij}}[({l}_{{r}})_{{i}}{C}({l}_{{s}})_{{j}}]{S}_{{2p}} 

 - \mathcal{D}_{{e^\dagger e^\dagger S_{3}^\dagger }}^{{rsp}}[\bar{e}_{{r}}{C}\bar{e}_{{s}}]{S}_{{3p}}^{\dagger } 

 - 2 \mathcal{D}_{{d^\dagger QS_{4}^\dagger }}^{{rsp}} [(\bar{d}_{{r}})^{{a}}({q}_{{s}})_{{ai}}]({S}_{{4p}}^{\dagger })^{{i}} 

 - \mathcal{D}_{{e^\dagger LS_{4}^\dagger }}^{{rsp}} [\bar{e}_{{r}}({l}_{{s}})_{{i}}]({S}_{{4p}}^{\dagger })^{{i}} 

 - 2 \mathcal{D}_{{QS_{4}u^\dagger }}^{{rps}}\epsilon ^{{ji}} [(\bar{u}_{{s}})^{{a}}({q}_{{r}})_{{aj}}]({S}_{{4p}})_{{i}} 

 - \frac{\mathcal{D}_{{LLS_{6}}}^{{rsp}}}{\sqrt{2}}\epsilon ^{{jk}} (\tau ^{{I}})_{{k}}^{{i}}[({l}_{{r}})_{{i}}{C}({l}_{{s}})_{{j}}]({S}_{{6p}})^{{I}} 

 - 2 \mathcal{D}_{{d^\dagger e^\dagger S_{9}}}^{{rsp}} [(\bar{d}_{{r}})^{{a}}{C}\bar{e}_{{s}}]({S}_{{9p}})_{{a}} 

 - 4 \mathcal{D}_{{S_{9}^\dagger u^\dagger u^\dagger }}^{{prs}}\epsilon ^{{abc}}[(\bar{u}_{{r}})^{{b}}{C}(\bar{u}_{{s}})^{{c}}]({S}_{{9p}}^{\dagger })^{{a}} 

 - 4 \mathcal{D}_{{d^\dagger S_{10}^\dagger u^\dagger }}^{{rps}}\epsilon ^{{bac}}[(\bar{d}_{{r}})^{{b}}{C}(\bar{u}_{{s}})^{{c}}]({S}_{{10p}}^{\dagger })^{{a}} 

 - 2 \mathcal{D}_{{e^\dagger S_{10}u^\dagger }}^{{rps}} [\bar{e}_{{r}}{C}(\bar{u}_{{s}})^{{a}}]({S}_{{10p}})_{{a}} 

 - 2 \mathcal{D}_{{LQS_{10}^\dagger }}^{{rsp}}\epsilon ^{{ij}}  [({l}_{{r}})_{{i}}{C}({q}_{{s}})_{{aj}}]({S}_{{10p}}^{\dagger })^{{a}} 

 - \mathcal{D}_{{QQS_{10}}}^{{rsp}}\epsilon ^{{bca}} \epsilon ^{{ij}}[({q}_{{r}})_{{bi}}{C}({q}_{{s}})_{{cj}}]({S}_{{10p}})_{{a}} 

 - 4 \mathcal{D}_{{d^\dagger d^\dagger S_{11}^\dagger }}^{{rsp}}\epsilon ^{{bca}}[(\bar{d}_{{r}})^{{b}}{C}(\bar{d}_{{s}})^{{c}}]({S}_{{11p}}^{\dagger })^{{a}} 

 - 2 \mathcal{D}_{{d^\dagger LS_{12}}}^{{rsp}}\epsilon ^{{ji}} [(\bar{d}_{{r}})^{{a}}({l}_{{s}})_{{j}}]({S}_{{12p}})_{{ai}} 

 - 2 \mathcal{D}_{{e^\dagger QS_{13}^\dagger }}^{{rsp}} [\bar{e}_{{r}}({q}_{{s}})_{{ai}}]({S}_{{13p}}^{\dagger })^{{ai}} 

 - 2 \mathcal{D}_{{LS_{13}u^\dagger }}^{{rps}}\epsilon ^{{ji}} [(\bar{u}_{{s}})^{{a}}({l}_{{r}})_{{j}}]({S}_{{13p}})_{{ai}} 

 - \sqrt{2} \mathcal{D}_{{LQS_{14}^\dagger }}^{{rsp}}\epsilon ^{{kj}} (\tau ^{{I}})_{{k}}^{{i}}[({l}_{{r}})_{{i}}{C}({q}_{{s}})_{{aj}}]({S}_{{14p}}^{\dagger })^{{aI}} 

 - \frac{\mathcal{D}_{{QQS_{14}}}^{{rsp}}}{2 \sqrt{2}}\epsilon ^{{bca}} \epsilon ^{{jk}} (\tau ^{{I}})_{{k}}^{{i}}[({q}_{{r}})_{{bi}}{C}({q}_{{s}})_{{cj}}]({S}_{{14p}})_{{a}}^{{I}} 

 - \frac{\mathcal{D}_{{QQS_{14}}}^{{rsp}}}{2 \sqrt{2}}\epsilon ^{{bca}} \epsilon ^{{ik}} (\tau ^{{I}})_{{k}}^{{j}}[({q}_{{r}})_{{bi}}{C}({q}_{{s}})_{{cj}}]({S}_{{14p}})_{{a}}^{{I}} 

 - 4 \mathcal{D}_{{d^\dagger d^\dagger S_{15}}}^{{rsp}}{C}_{{ab}}^{\mathfrak{a}}[(\bar{d}_{{r}})^{{a}}{C}(\bar{d}_{{s}})^{{b}}]({S}_{{15p}})^{\mathfrak{a}} 

 - 4 \mathcal{D}_{{d^\dagger S_{16}u^\dagger }}^{{rps}}{C}_{{ab}}^{\mathfrak{a}}[(\bar{d}_{{r}})^{{a}}{C}(\bar{u}_{{s}})^{{b}}]({S}_{{16p}})^{\mathfrak{a}} 

 - 4 \mathcal{D}_{{QQS_{16}^\dagger }}^{{rsp}}\epsilon ^{{ij}} {C}_{\mathfrak{a}}^{{ab}}[({q}_{{r}})_{{ai}}{C}({q}_{{s}})_{{bj}}]({S}_{{16p}}^{\dagger })^{\mathfrak{a}} 

 - 2 \mathcal{D}_{{S_{17}u^\dagger u^\dagger }}^{{prs}}{C}_{{ab}}^{\mathfrak{a}}[(\bar{u}_{{r}})^{{a}}{C}(\bar{u}_{{s}})^{{b}}]({S}_{{17p}})^{\mathfrak{a}} 

 - 2 \mathcal{D}_{{S_{17}u^\dagger u^\dagger }}^{{prs}}{C}_{{ba}}^{\mathfrak{a}}[(\bar{u}_{{r}})^{{a}}{C}(\bar{u}_{{s}})^{{b}}]({S}_{{17p}})^{\mathfrak{a}} 

 - \sqrt{2} \mathcal{D}_{{QQS_{18}^\dagger }}^{{rsp}}\epsilon ^{{kj}} {C}_{\mathfrak{a}}^{{ab}} (\tau ^{{I}})_{{k}}^{{i}}[({q}_{{r}})_{{ai}}{C}({q}_{{s}})_{{bj}}]({S}_{{18p}}^{\dagger })^{{\mathfrak{a}I}} 

 - \sqrt{2} \mathcal{D}_{{QQS_{18}^\dagger }}^{{rsp}}\epsilon ^{{ki}} {C}_{\mathfrak{a}}^{{ab}} (\tau ^{{I}})_{{k}}^{{j}}[({q}_{{r}})_{{ai}}{C}({q}_{{s}})_{{bj}}]({S}_{{18p}}^{\dagger })^{{\mathfrak{a}I}} 

 - 2 \sqrt{2} \mathcal{D}_{{d^\dagger QS_{19}^\dagger }}^{{rsp}}  (\lambda ^{{A}})_{{a}}^{{b}}[(\bar{d}_{{r}})^{{a}}({q}_{{s}})_{{bi}}]({S}_{{19p}}^{\dagger })^{{Ai}} 

 - \sqrt{2} \mathcal{D}_{{QS_{19}u^\dagger }}^{{rps}}\epsilon ^{{ji}} (\lambda ^{{A}})_{{b}}^{{a}}[(\bar{u}_{{s}})^{{b}}({q}_{{r}})_{{aj}}]({S}_{{19p}})_{{i}}^{{A}} 

 - \mathcal{D}_{{HH^\dagger H^\dagger S_{4}(1)}}^{{p}}\delta _{{k}}^{{i}} \delta _{{l}}^{{j}}{H}_{{j}}{H}^{{\dagger k}}{H}^{{\dagger l}}({S}_{{4p}})_{{i}} 

 - \mathcal{D}_{{HH^\dagger H^\dagger S_{4}(2)}}^{{p}}\delta _{{l}}^{{i}} \delta _{{k}}^{{j}}{H}_{{j}}{H}^{{\dagger k}}{H}^{{\dagger l}}({S}_{{4p}})_{{i}} 

 + \mathcal{D}_{{HHH^\dagger S_{7}^\dagger }}^{{p}}\epsilon _{{kl}} {C}_{\mathfrak{i}}^{{ijl}}{H}_{{i}}{H}_{{j}}{H}^{{\dagger k}}({S}_{{7p}}^{\dagger })^{\mathfrak{i}} 

 + \mathcal{D}_{{HHHS_{8}^\dagger }}^{{p}}{C}_{\mathfrak{i}}^{{ijk}}{H}_{{i}}{H}_{{j}}{H}_{{k}}({S}_{{8p}}^{\dagger })^{\mathfrak{i}} 

 + \mathcal{D}_{{HH^\dagger S_{1}S_{1}}}^{{pr}}\delta _{{j}}^{{i}}{H}_{{i}}{H}^{{\dagger j}}{S}_{{1p}}{S}_{{1r}} 

 - \frac{\mathcal{D}_{{HH^\dagger S_{1}S_{5}}}^{{pr}}}{\sqrt{2}}(\tau ^{{I}})_{{j}}^{{i}}{H}_{{i}}{H}^{{\dagger j}}({S}_{{5r}})^{{I}}{S}_{{1p}} 

 + \frac{\mathcal{D}_{{HHS_{1}S_{6}^\dagger }}^{{pr}}}{\sqrt{2}}\epsilon ^{{kj}} (\tau ^{{I}})_{{k}}^{{i}}{H}_{{i}}{H}_{{j}}({S}_{{6r}}^{\dagger })^{{I}}{S}_{{1p}} 

 + \frac{\mathcal{D}_{{HHS_{2}^\dagger S_{5}}}^{{rp}}}{\sqrt{2}}\epsilon ^{{jk}} (\tau ^{{I}})_{{k}}^{{i}}{H}_{{i}}{H}_{{j}}({S}_{{5r}})^{{I}}{S}_{{2p}}^{\dagger } 

 + \frac{\mathcal{D}_{{HH^\dagger S_{2}S_{6}^\dagger }}^{{pr}}}{\sqrt{2}}(\tau ^{{I}})_{{j}}^{{i}}{H}_{{i}}{H}^{{\dagger j}}({S}_{{6r}}^{\dagger })^{{I}}{S}_{{2p}} 

 + \frac{\mathcal{D}_{{HHS_{3}^\dagger S_{6}}}^{{rp}}}{\sqrt{2}}\epsilon ^{{jk}} (\tau ^{{I}})_{{k}}^{{i}}{H}_{{i}}{H}_{{j}}({S}_{{6r}})^{{I}}{S}_{{3p}}^{\dagger } 

 - \frac{\mathcal{D}_{{HH^\dagger S_{5}S_{5}(1)}}^{{pr}}}{2}(\tau ^{{I}})_{{k}}^{{i}} (\tau ^{{J}})_{{j}}^{{k}}{H}_{{i}}{H}^{{\dagger j}}({S}_{{5p}})^{{I}}({S}_{{5r}})^{{J}} 

 - \mathcal{D}_{{HH^\dagger S_{5}S_{5}(2)}}^{{pr}}\delta ^{{IJ}} \delta _{{j}}^{{i}}{H}_{{i}}{H}^{{\dagger j}}({S}_{{5p}})^{{I}}({S}_{{5r}})^{{J}} 

 + \frac{\mathcal{D}_{{HHS_{5}S_{6}^\dagger }}^{{pr}}}{2}\epsilon ^{{kl}} (\tau ^{{I}})_{{l}}^{{i}} (\tau ^{{J}})_{{k}}^{{j}}{H}_{{i}}{H}_{{j}}({S}_{{5p}})^{{I}}({S}_{{6r}}^{\dagger })^{{J}} 

 + \frac{\mathcal{D}_{{HH^\dagger S_{6}S_{6}^\dagger (1)}}^{{pr}}}{2}(\tau ^{{I}})_{{k}}^{{i}} (\tau ^{{J}})_{{j}}^{{k}}{H}_{{i}}{H}^{{\dagger j}}({S}_{{6p}})^{{I}}({S}_{{6r}}^{\dagger })^{{J}} 

 + \mathcal{D}_{{HH^\dagger S_{6}S_{6}^\dagger (2)}}^{{pr}}\delta ^{{IJ}} \delta _{{j}}^{{i}}{H}_{{i}}{H}^{{\dagger j}}({S}_{{6p}})^{{I}}({S}_{{6r}}^{\dagger })^{{J}} 
 
    \end{autobreak}
\end{align}

\subsubsection{New fermions}

\begin{align}
    \begin{autobreak}
        \Delta{\cal L}_{\rm UV,F}^{(d = 4)} = 

        - \mathcal{D}_{{F_{1}HL}}^{{pr}}\epsilon ^{{ji}}[{F}_{{1p}}{C}({l}_{{r}})_{{i}}]{H}_{{j}} 

 + \mathcal{D}_{{F_{2L}H^\dagger L}}^{{pr}} [{F}_{{2p}}{C}({l}_{{r}})_{{i}}]{H}^{{\dagger i}} 

 - \mathcal{D}_{{e^\dagger F_{3R}^\dagger H^\dagger }}^{{rp}}\epsilon _{{ij}}[\bar{e}_{{r}}C(\bar{F}_{{3p}})^{{i}}]{H}^{{\dagger j}} 

 + \mathcal{D}_{{e^\dagger F_{4R}^\dagger H}}^{{rp}} [\bar{e}_{{r}}C(\bar{F}_{{4p}})^{{i}}]{H}_{{i}} 

 + \frac{\mathcal{D}_{{F_{5}HL}}^{{pr}}}{2 \sqrt{2}}\epsilon ^{{kj}} (\tau ^{{I}})_{{k}}^{{i}}[({F}_{{5p}})^{{I}}{C}({l}_{{r}})_{{i}}]{H}_{{j}} 

 + \frac{\mathcal{D}_{{F_{5}HL}}^{{pr}}}{2 \sqrt{2}}\epsilon ^{{ki}} (\tau ^{{I}})_{{k}}^{{j}}[({F}_{{5p}})^{{I}}{C}({l}_{{r}})_{{i}}]{H}_{{j}} 

 + \frac{\mathcal{D}_{{F_{6L}H^\dagger L}}^{{pr}}}{\sqrt{2}}(\tau ^{{I}})_{{j}}^{{i}}[({F}_{{6p}})^{{I}}{C}({l}_{{r}})_{{i}}]{H}^{{\dagger j}} 

 + 2 \mathcal{D}_{{F_{8R}^\dagger H^\dagger Q}}^{{pr}} [(\bar{F}_{{8p}})^{{a}}({q}_{{r}})_{{ai}}]{H}^{{\dagger i}} 

 - 2 \mathcal{D}_{{F_{9R}^\dagger HQ}}^{{pr}}\epsilon ^{{ji}} [(\bar{F}_{{9p}})^{{a}}({q}_{{r}})_{{ai}}]{H}_{{j}} 

 - 2 \mathcal{D}_{{d^\dagger F_{10L}H}}^{{rp}}\epsilon ^{{ij}}  [(\bar{d}_{{r}})^{{a}}({F}_{{10p}})_{{ai}}]{H}_{{j}} 

 - 2 \mathcal{D}_{{d^\dagger F_{11L}H^\dagger }}^{{rp}}  [(\bar{d}_{{r}})^{{a}}({F}_{{11p}})_{{ai}}]{H}^{{\dagger i}} 

 - 2 \mathcal{D}_{{F_{11L}Hu^\dagger }}^{{pr}}\epsilon ^{{ij}}  [(\bar{u}_{{r}})^{{a}}({F}_{{11p}})_{{ai}}]{H}_{{j}} 

 - 2 \mathcal{D}_{{F_{12L}H^\dagger u^\dagger }}^{{pr}} [(\bar{u}_{{r}})^{{a}}({F}_{{12p}})_{{ai}}]{H}^{{\dagger i}} 

 - \sqrt{2} \mathcal{D}_{{F_{13R}^\dagger H^\dagger Q}}^{{pr}}  (\tau ^{{I}})_{{j}}^{{i}}[(\bar{F}_{{13p}})^{{aI}}({q}_{{r}})_{{ai}}]{H}^{{\dagger j}} 

 - \frac{\mathcal{D}_{{F_{14R}^\dagger HQ}}^{{pr}}}{\sqrt{2}}\epsilon ^{{kj}}  (\tau ^{{I}})_{{k}}^{{i}}[(\bar{F}_{{14p}})^{{aI}}({q}_{{r}})_{{ai}}]{H}_{{j}} 

 - \frac{\mathcal{D}_{{F_{14R}^\dagger HQ}}^{{pr}}}{\sqrt{2}}\epsilon ^{{ki}}  (\tau ^{{I}})_{{k}}^{{j}}[(\bar{F}_{{14p}})^{{aI}}({q}_{{r}})_{{ai}}]{H}_{{j}} 

 - \mathcal{D}_{{F_{1}F_{3L}H^\dagger }}^{{pr}}\delta _{{j}}^{{i}}[{F}_{{1p}}{C}({F}_{{3r}})_{{i}}]{H}^{{\dagger j}} 

 - \mathcal{D}_{{F_{1}F_{3R}^\dagger H}}^{{pr}}\delta _{{i}}^{{j}}[(\bar{F}_{{3r}})^{{i}}{F}_{{1p}}]{H}_{{j}}

 - \mathcal{D}_{{F_{2R}^\dagger F_{3L}H}}^{{rp}}\epsilon ^{{ij}}[\bar{F}_{{2p}}({F}_{{3r}})_{{i}}]{H}_{{j}} 

 - \mathcal{D}_{{F_{2R}^\dagger F_{4L}H^\dagger }}^{{rp}}\delta _{{j}}^{{i}}[\bar{F}_{{2p}}({F}_{{4r}})_{{i}}]{H}^{{\dagger j}} 

 - \frac{\mathcal{D}_{{F_{3L}F_{5}H^\dagger }}^{{pr}}}{\sqrt{2}}(\tau ^{{I}})_{{j}}^{{i}}[({F}_{{3p}})_{{i}}{C}({F}_{{5r}})^{{I}}]{H}^{{\dagger j}} 

 - \frac{\mathcal{D}_{{F_{3R}^\dagger F_{5}H}}^{{rp}}}{\sqrt{2}}(\tau ^{{I}})_{{i}}^{{j}}[(\bar{F}_{{3p}})^{{i}}({F}_{{5r}})^{{I}}]{H}_{{j}} 

 - \frac{\mathcal{D}_{{F_{3L}F_{6R}^\dagger H}}^{{pr}}}{\sqrt{2}}\epsilon ^{{kj}} (\tau ^{{I}})_{{k}}^{{i}}[(\bar{F}_{{6r}})^{{I}}({F}_{{3p}})_{{i}}]{H}_{{j}}

 - \frac{\mathcal{D}_{{F_{4L}F_{6R}^\dagger H^\dagger }}^{{pr}}}{\sqrt{2}}(\tau ^{{I}})_{{j}}^{{i}}[(\bar{F}_{{6r}})^{{I}}({F}_{{4p}})_{{i}}]{H}^{{\dagger j}} 

 + \frac{\mathcal{D}_{{F_{5}F_{7L}H^\dagger }}^{{pr}}}{\sqrt{2}}\epsilon ^{{lk}} {C}_{{ijk}}^{\mathfrak{i}} (\tau ^{{I}})_{{l}}^{{j}}[({F}_{{5p}})^{{I}}{C}({F}_{{7r}})^{\mathfrak{i}}]{H}^{{\dagger i}} 

 - \frac{\mathcal{D}_{{F_{5}F_{7R}^\dagger H}}^{{pr}}}{\sqrt{2}}\epsilon _{{jl}} {C}_{\mathfrak{i}}^{{kli}} (\tau ^{{I}})_{{k}}^{{j}}[(\bar{F}_{{7r}})^{\mathfrak{i}}({F}_{{5p}})^{{I}}]{H}_{{i}}

 - \frac{\mathcal{D}_{{F_{6R}^\dagger F_{7L}H}}^{{rp}}}{\sqrt{2}}\epsilon ^{{jl}} \epsilon ^{{mi}} {C}_{{klm}}^{\mathfrak{i}} (\tau ^{{I}})_{{j}}^{{k}}[(\bar{F}_{{6p}})^{{I}}({F}_{{7r}})^{\mathfrak{i}}]{H}_{{i}} 

 - 2 \mathcal{D}_{{F_{10R}^\dagger F_{8L}H^\dagger }}^{{rp}}\epsilon _{{ij}} \delta _{{b}}^{{a}}[(\bar{F}_{{10p}})^{{ai}}({F}_{{8r}})_{{b}}]{H}^{{\dagger j}} 

 + 2 \mathcal{D}_{{F_{11R}^\dagger F_{8L}H}}^{{rp}}\delta _{{b}}^{{a}} \delta _{{i}}^{{j}}[(\bar{F}_{{11p}})^{{ai}}({F}_{{8r}})_{{b}}]{H}_{{j}}

 - 2 \mathcal{D}_{{F_{11R}^\dagger F_{9L}H^\dagger }}^{{rp}}\epsilon _{{ij}} \delta _{{b}}^{{a}}[(\bar{F}_{{11p}})^{{ai}}({F}_{{9r}})_{{b}}]{H}^{{\dagger j}} 

 + 2 \mathcal{D}_{{F_{12R}^\dagger F_{9L}H}}^{{rp}}\delta _{{b}}^{{a}} \delta _{{i}}^{{j}}[(\bar{F}_{{12p}})^{{ai}}({F}_{{9r}})_{{b}}]{H}_{{j}} 

 + \sqrt{2} \mathcal{D}_{{F_{10R}^\dagger F_{13L}H^\dagger }}^{{rp}}\epsilon _{{kj}} \delta _{{b}}^{{a}} (\tau ^{{I}})_{{i}}^{{k}}[(\bar{F}_{{10p}})^{{ai}}({F}_{{13r}})_{{b}}^{{I}}]{H}^{{\dagger j}}

 - \sqrt{2} \mathcal{D}_{{F_{11R}^\dagger F_{13L}H}}^{{rp}}\delta _{{b}}^{{a}} (\tau ^{{I}})_{{i}}^{{j}}[(\bar{F}_{{11p}})^{{ai}}({F}_{{13r}})_{{b}}^{{I}}]{H}_{{j}} 

 + \sqrt{2} \mathcal{D}_{{F_{11R}^\dagger F_{14L}H^\dagger }}^{{rp}}\epsilon _{{kj}} \delta _{{b}}^{{a}} (\tau ^{{I}})_{{i}}^{{k}}[(\bar{F}_{{11p}})^{{ai}}({F}_{{14r}})_{{b}}^{{I}}]{H}^{{\dagger j}} 

 - \sqrt{2} \mathcal{D}_{{F_{12R}^\dagger F_{14L}H}}^{{rp}}\delta _{{b}}^{{a}} (\tau ^{{I}})_{{i}}^{{j}}[(\bar{F}_{{12p}})^{{ai}}({F}_{{14r}})_{{b}}^{{I}}]{H}_{{j}}

    \end{autobreak}
\end{align}

\subsubsection{New vectors}

\begin{align}
    \begin{autobreak}
        \Delta{\cal L}_{\rm UV,V}^{(d = 3)} = 

 - 2 \mathcal{C}_{{HV_{2}V_{3}^\dagger }}^{{pr}} {H}_{{i}}({V}_{{3r}}^{\dagger })^{{i\mu }}{V}_{{2p\mu }} 

 - 4 \mathcal{C}_{{HV_{5}^\dagger V_{8}}}^{{rp}}\epsilon ^{{ji}}  {H}_{{j}}({V}_{{5p}}^{\dagger })_{\mu }^{{a}}({V}_{{8r}})_{{ai}}^{\mu } 

 - 2 \sqrt{2} \mathcal{C}_{{HV_{8}V_{9}^\dagger }}^{{pr}}\epsilon ^{{ki}}  (\tau ^{{I}})_{{k}}^{{j}}{H}_{{j}}({V}_{{8p}})_{{ai\mu }}({V}_{{9r}}^{\dagger })^{{aI\mu }}

    \end{autobreak}
\end{align}

\begin{align}
    \begin{autobreak}
        \Delta{\cal L}_{\rm UV,V}^{(d = 4)} = 

        - 2 \mathcal{D}_{{d^\dagger dV_{1}}}^{{rsp}} [(\bar{d}_{{r}})^{{a}}\gamma _{\mu }({d}_{{s}})_{{a}}]{V}_{{1p}}^{\mu } 

 - \mathcal{D}_{{e^\dagger eV_{1}}}^{{rsp}}1[\bar{e}_{{r}}\gamma _{\mu }{e}_{{s}}]{V}_{{1p}}^{\mu } 

 - 2 \mathcal{D}_{{HH^\dagger V_{1}D}}^{{p}} [D_{\mu }{H}_{{i}}]{H}^{{\dagger i}}{V}_{{1p}}^{\mu } 

 + \mathcal{D}_{{LL^\dagger V_{1}}}^{{rsp}} [(\bar{l}_{{s}})^{{i}}\gamma _{\mu }({l}_{{r}})_{{i}}]{V}_{{1p}}^{\mu } 

 + 2 \mathcal{D}_{{QQ^\dagger V_{1}}}^{{rsp}} [(\bar{q}_{{s}})^{{ai}}\gamma _{\mu }({q}_{{r}})_{{ai}}]{V}_{{1p}}^{\mu } 

 - 2 \mathcal{D}_{{u^\dagger uV_{1}}}^{{rsp}} [(\bar{u}_{{r}})^{{a}}\gamma _{\mu }({u}_{{s}})_{{a}}]{V}_{{1p}}^{\mu } 

 - 2 \mathcal{D}_{{d^\dagger uV_{2}^\dagger }}^{{rsp}} [(\bar{d}_{{r}})^{{a}}\gamma _{\mu }({u}_{{s}})_{{a}}]{V}_{{2p}}^{\dagger \mu } 

 - 2 \mathcal{D}_{{HHV_{2}^\dagger D}}^{{p}}\epsilon ^{{ij}}[D_{\mu }{H}_{{i}}]{H}_{{j}}{V}_{{2p}}^{\dagger \mu } 

 - \mathcal{D}_{{e^\dagger L^\dagger V_{3}^\dagger }}^{{rsp}}\epsilon _{{ji}}[\bar{e}_{{r}}\gamma _{\mu }{C}(\bar{l}_{{s}})^{{j}}]({V}_{{3p}}^{\dagger })^{{i\mu }} 

 + \sqrt{2} \mathcal{D}_{{HH^\dagger V_{4}D}}^{{p}}(\tau ^{{I}})_{{j}}^{{i}}[D_{\mu }{H}_{{i}}]{H}^{{\dagger j}}({V}_{{4p}})^{{I\mu }} 

 - \frac{\mathcal{D}_{{LL^\dagger V_{4}}}^{{rsp}}}{\sqrt{2}}(\tau ^{{I}})_{{j}}^{{i}}[(\bar{l}_{{s}})^{{j}}\gamma _{\mu }({l}_{{r}})_{{i}}]({V}_{{4p}})^{{I\mu }} 

 - \sqrt{2} \mathcal{D}_{{QQ^\dagger V_{4}}}^{{rsp}}  (\tau ^{{I}})_{{j}}^{{i}}[(\bar{q}_{{s}})^{{aj}}\gamma _{\mu }({q}_{{r}})_{{ai}}]({V}_{{4p}})^{{I\mu }} 

 - 2 \mathcal{D}_{{d^\dagger eV_{5}}}^{{rsp}} [(\bar{d}_{{r}})^{{a}}\gamma _{\mu }{e}_{{s}}]({V}_{{5p}})_{{a}}^{\mu } 

 + 2 \mathcal{D}_{{L^\dagger QV_{5}^\dagger }}^{{srp}}  [(\bar{l}_{{s}})^{{i}}\gamma _{\mu }({q}_{{r}})_{{ai}}]({V}_{{5p}}^{\dagger })^{{a\mu }} 

 + 2 \mathcal{D}_{{eu^\dagger V_{6}}}^{{srp}} [(\bar{u}_{{r}})^{{a}}\gamma _{\mu }{e}_{{s}}]({V}_{{6p}})_{{a}}^{\mu } 

 + 2 \mathcal{D}_{{d^\dagger L^\dagger V_{7}}}^{{rsp}} [(\bar{d}_{{r}})^{{a}}\gamma _{\mu }{C}(\bar{l}_{{s}})^{{i}}]({V}_{{7p}})_{{ai}}^{\mu } 

 + 2 \mathcal{D}_{{eQV_{7}^\dagger }}^{{srp}} [({q}_{{r}})_{{ai}}{C}\gamma _{\mu }{e}_{{s}}]({V}_{{7p}}^{\dagger })^{{ai\mu }} 

 + \mathcal{D}_{{QuV_{7}}}^{{rsp}}\epsilon ^{{bca}} \epsilon ^{{ij}}[({q}_{{r}})_{{bj}}{C}\gamma _{\mu }({u}_{{s}})_{{c}}]({V}_{{7p}})_{{ai}}^{\mu } 

 - \mathcal{D}_{{dQV_{8}}}^{{srp}}\epsilon ^{{cba}} \epsilon ^{{ij}}[({q}_{{r}})_{{bj}}{C}\gamma _{\mu }({d}_{{s}})_{{c}}]({V}_{{8p}})_{{ai}}^{\mu } 

 - 2 \mathcal{D}_{{L^\dagger u^\dagger V_{8}}}^{{srp}} [(\bar{u}_{{r}})^{{a}}\gamma _{\mu }{C}(\bar{l}_{{s}})^{{i}}]({V}_{{8p}})_{{ai}}^{\mu } 

 - \sqrt{2} \mathcal{D}_{{L^\dagger QV_{9}^\dagger }}^{{srp}}  (\tau ^{{I}})_{{j}}^{{i}}[(\bar{l}_{{s}})^{{j}}\gamma _{\mu }({q}_{{r}})_{{ai}}]({V}_{{9p}}^{\dagger })^{{aI\mu }} 

 + 4 \mathcal{D}_{{dQV_{10}^\dagger }}^{{srp}}{C}_{\mathfrak{a}}^{{ba}}  [({q}_{{r}})_{{ai}}{C}\gamma _{\mu }({d}_{{s}})_{{b}}]({V}_{{10p}}^{\dagger })^{{\mathfrak{a}i\mu }} 

 - 4 \mathcal{D}_{{QuV_{11}^\dagger }}^{{rsp}}{C}_{\mathfrak{a}}^{{ab}} [({q}_{{r}})_{{ai}}{C}\gamma _{\mu }({u}_{{s}})_{{b}}]({V}_{{11p}}^{\dagger })^{{\mathfrak{a}i\mu }} 

 - 2 \sqrt{2} \mathcal{D}_{{d^\dagger dV_{12}}}^{{rsp}}(\lambda ^{{A}})_{{a}}^{{b}}[(\bar{d}_{{r}})^{{a}}\gamma _{\mu }({d}_{{s}})_{{b}}]({V}_{{12p}})^{{A\mu }} 

 + \sqrt{2} \mathcal{D}_{{QQ^\dagger V_{12}}}^{{rsp}} (\lambda ^{{A}})_{{b}}^{{a}}[(\bar{q}_{{s}})^{{bi}}\gamma _{\mu }({q}_{{r}})_{{ai}}]({V}_{{12p}})^{{A\mu }} 

 - 2 \sqrt{2} \mathcal{D}_{{u^\dagger uV_{12}}}^{{rsp}}(\lambda ^{{A}})_{{a}}^{{b}}[(\bar{u}_{{r}})^{{a}}\gamma _{\mu }({u}_{{s}})_{{b}}]({V}_{{12p}})^{{A\mu }} 

 - 2 \sqrt{2} \mathcal{D}_{{d^\dagger uV_{13}^\dagger }}^{{rsp}}(\lambda ^{{A}})_{{a}}^{{b}}[(\bar{d}_{{r}})^{{a}}\gamma _{\mu }({u}_{{s}})_{{b}}]({V}_{{13p}}^{\dagger })^{{A\mu }} 

 - \mathcal{D}_{{QQ^\dagger V_{14}}}^{{rsp}}(\lambda ^{{A}})_{{b}}^{{a}} (\tau ^{{I}})_{{j}}^{{i}}[(\bar{q}_{{s}})^{{bj}}\gamma _{\mu }({q}_{{r}})_{{ai}}]({V}_{{14p}})^{{AI\mu }} 


    \end{autobreak}
\end{align}

\subsubsection{Mixed terms}

\begin{align}
    \begin{autobreak}
        \Delta{\cal L}_{\rm UV,SF}^{(d = 4)} = 

        - \mathcal{D}_{{e^\dagger F_{2R}^\dagger S_{1}}}^{{spr}}[\bar{e}_{{s}}{F}_{{2p}}]{S}_{{1r}} 

 - \mathcal{D}_{{F_{3L}LS_{1}}}^{{psr}}\epsilon ^{{ij}}[({F}_{{3p}})_{{i}}{C}({l}_{{s}})_{{j}}]{S}_{{1r}} 

 - 2 \mathcal{D}_{{d^\dagger F_{8L}S_{1}}}^{{spr}}\delta _{{b}}^{{a}}[(\bar{d}_{{s}})^{{b}}({F}_{{8p}})_{{a}}]{S}_{{1r}} 

 - 2 \mathcal{D}_{{F_{9L}S_{1}u^\dagger }}^{{prs}}\delta _{{b}}^{{a}}[(\bar{u}_{{s}})^{{b}}({F}_{{9p}})_{{a}}]{S}_{{1r}} 

 + 2 \mathcal{D}_{{F_{11R}^\dagger QS_{1}}}^{{psr}}\delta _{{a}}^{{b}} \delta _{{i}}^{{j}}[({F}_{{11p}})^{{ai}}{C}({q}_{{s}})_{{bj}}]{S}_{{1r}} 

 - \mathcal{D}_{{e^\dagger F_{1}S_{2}^\dagger }}^{{spr}}1[\bar{e}_{{s}}{F}_{{1p}}]{S}_{{2r}}^{\dagger } 

 - \mathcal{D}_{{F_{4L}LS_{2}^\dagger }}^{{psr}}\epsilon ^{{ij}}[({F}_{{4p}})_{{i}}{C}({l}_{{s}})_{{j}}]{S}_{{2r}}^{\dagger } 

 - 2 \mathcal{D}_{{F_{8L}S_{2}u^\dagger }}^{{prs}}\delta _{{b}}^{{a}}[(\bar{u}_{{s}})^{{b}}({F}_{{8p}})_{{a}}]{S}_{{2r}} 

 - 2 \mathcal{D}_{{d^\dagger F_{9L}S_{2}^\dagger }}^{{spr}}\delta _{{b}}^{{a}}[(\bar{d}_{{s}})^{{b}}({F}_{{9p}})_{{a}}]{S}_{{2r}}^{\dagger } 

 + 2 \mathcal{D}_{{F_{10R}^\dagger QS_{2}^\dagger }}^{{psr}}\delta _{{a}}^{{b}} \delta _{{i}}^{{j}}[({F}_{{10p}})^{{ai}}{C}({q}_{{s}})_{{bj}}]{S}_{{2r}}^{\dagger } 

 + 2 \mathcal{D}_{{F_{12R}^\dagger QS_{2}}}^{{psr}}\delta _{{a}}^{{b}} \delta _{{i}}^{{j}}[({F}_{{12p}})^{{ai}}{C}({q}_{{s}})_{{bj}}]{S}_{{2r}} 

 - \mathcal{D}_{{F_{1L}S_{4}}}^{{psr}}\epsilon ^{{ji}}[{F}_{{1p}}{C}({l}_{{s}})_{{j}}]({S}_{{4r}})_{{i}} 

 + \frac{\mathcal{D}_{{F_{5L}S_{4}}}^{{psr}}}{2 \sqrt{2}}\epsilon ^{{kj}} (\tau ^{{I}})_{{k}}^{{i}}[({F}_{{5p}})^{{I}}{C}({l}_{{s}})_{{j}}]({S}_{{4r}})_{{i}} 

 + \frac{\mathcal{D}_{{F_{5L}S_{4}}}^{{psr}}}{2 \sqrt{2}}\epsilon ^{{ki}} (\tau ^{{I}})_{{k}}^{{j}}[({F}_{{5p}})^{{I}}{C}({l}_{{s}})_{{j}}]({S}_{{4r}})_{{i}} 

 - \frac{\mathcal{D}_{{F_{3L}LS_{5}}}^{{psr}}}{\sqrt{2}}\epsilon ^{{jk}} (\tau ^{{I}})_{{k}}^{{i}}[({F}_{{3p}})_{{i}}{C}({l}_{{s}})_{{j}}]({S}_{{5r}})^{{I}} 

 - \mathcal{D}_{{e^\dagger F_{6R}^\dagger S_{5}}}^{{spr}}\delta ^{{IJ}}[\bar{e}_{{s}}({F}_{{6p}})^{{I}}]({S}_{{5r}})^{{J}} 

 + \frac{\mathcal{D}_{{F_{7L}LS_{5}}}^{{psr}}}{3 \sqrt{2}}\epsilon ^{{lj}} \epsilon ^{{mk}} {C}_{{klm}}^{\mathfrak{i}} (\tau ^{{I}})_{{j}}^{{i}}[({F}_{{7p}})^{\mathfrak{i}}{C}({l}_{{s}})_{{i}}]({S}_{{5r}})^{{I}} 

 - \frac{1}{3} \sqrt{2} \mathcal{D}_{{F_{7L}LS_{5}}}^{{psr}}\epsilon ^{{lj}} \epsilon ^{{mi}} {C}_{{klm}}^{\mathfrak{i}} (\tau ^{{I}})_{{j}}^{{k}}[({F}_{{7p}})^{\mathfrak{i}}{C}({l}_{{s}})_{{i}}]({S}_{{5r}})^{{I}} 

 - \frac{\mathcal{D}_{{F_{7L}LS_{5}}}^{{psr}}}{3 \sqrt{2}}\epsilon ^{{li}} \epsilon ^{{mj}} {C}_{{klm}}^{\mathfrak{i}} (\tau ^{{I}})_{{j}}^{{k}}[({F}_{{7p}})^{\mathfrak{i}}{C}({l}_{{s}})_{{i}}]({S}_{{5r}})^{{I}} 

 + \sqrt{2} \mathcal{D}_{{F_{11R}^\dagger QS_{5}}}^{{psr}}\delta _{{a}}^{{b}} (\tau ^{{I}})_{{i}}^{{j}}[({F}_{{11p}})^{{ai}}{C}({q}_{{s}})_{{bj}}]({S}_{{5r}})^{{I}} 

 + 2 \mathcal{D}_{{d^\dagger F_{13L}S_{5}}}^{{spr}}\delta ^{{IJ}} \delta _{{b}}^{{a}}[(\bar{d}_{{s}})^{{b}}({F}_{{13p}})_{{a}}^{{I}}]({S}_{{5r}})^{{J}} 

 + 2 \mathcal{D}_{{F_{14L}S_{5}u^\dagger }}^{{prs}}\delta ^{{IJ}} \delta _{{b}}^{{a}}[(\bar{u}_{{s}})^{{b}}({F}_{{14p}})_{{a}}^{{I}}]({S}_{{5r}})^{{J}} 

 + \frac{\mathcal{D}_{{F_{3R}^\dagger LS_{6}}}^{{psr}}}{\sqrt{2}}(\tau ^{{I}})_{{i}}^{{j}}[({F}_{{3p}})^{{i}}{C}({l}_{{s}})_{{j}}]({S}_{{6r}})^{{I}} 

 - \frac{\mathcal{D}_{{F_{4L}LS_{6}^\dagger }}^{{psr}}}{\sqrt{2}}\epsilon ^{{kj}} (\tau ^{{I}})_{{k}}^{{i}}[({F}_{{4p}})_{{i}}{C}({l}_{{s}})_{{j}}]({S}_{{6r}}^{\dagger })^{{I}} 

 - \mathcal{D}_{{e^\dagger F_{5}S_{6}^\dagger }}^{{spr}}\delta ^{{IJ}}[\bar{e}_{{s}}({F}_{{5p}})^{{I}}]({S}_{{6r}}^{\dagger })^{{J}} 

 + \frac{\mathcal{D}_{{F_{7R}^\dagger LS_{6}}}^{{psr}}}{3 \sqrt{2}}\epsilon _{{jl}} {C}_{\mathfrak{i}}^{{ikl}} (\tau ^{{I}})_{{k}}^{{j}}[({F}_{{7p}})^{\mathfrak{i}}{C}({l}_{{s}})_{{i}}]({S}_{{6r}})^{{I}} 

 + \frac{\mathcal{D}_{{F_{7R}^\dagger LS_{6}}}^{{psr}}}{3 \sqrt{2}}\epsilon _{{jl}} {C}_{\mathfrak{i}}^{{kil}} (\tau ^{{I}})_{{k}}^{{j}}[({F}_{{7p}})^{\mathfrak{i}}{C}({l}_{{s}})_{{i}}]({S}_{{6r}})^{{I}} 

 + \frac{\mathcal{D}_{{F_{7R}^\dagger LS_{6}}}^{{psr}}}{3 \sqrt{2}}\epsilon _{{jl}} {C}_{\mathfrak{i}}^{{kli}} (\tau ^{{I}})_{{k}}^{{j}}[({F}_{{7p}})^{\mathfrak{i}}{C}({l}_{{s}})_{{i}}]({S}_{{6r}})^{{I}} 

 - \sqrt{2} \mathcal{D}_{{F_{10R}^\dagger QS_{6}^\dagger }}^{{psr}}\delta _{{a}}^{{b}} (\tau ^{{I}})_{{i}}^{{j}}[({F}_{{10p}})^{{ai}}{C}({q}_{{s}})_{{bj}}]({S}_{{6r}}^{\dagger })^{{I}} 

 + \sqrt{2} \mathcal{D}_{{F_{12R}^\dagger QS_{6}}}^{{psr}}\delta _{{a}}^{{b}} (\tau ^{{I}})_{{i}}^{{j}}[({F}_{{12p}})^{{ai}}{C}({q}_{{s}})_{{bj}}]({S}_{{6r}})^{{I}} 

 + 2 \mathcal{D}_{{F_{13L}S_{6}u^\dagger }}^{{prs}}\delta ^{{IJ}} \delta _{{b}}^{{a}}[(\bar{u}_{{s}})^{{b}}({F}_{{13p}})_{{a}}^{{I}}]({S}_{{6r}})^{{J}} 

 - 2 \mathcal{D}_{{d^\dagger F_{14L}S_{6}^\dagger }}^{{spr}}\delta ^{{IJ}} \delta _{{b}}^{{a}}[(\bar{d}_{{s}})^{{b}}({F}_{{14p}})_{{a}}^{{I}}]({S}_{{6r}}^{\dagger })^{{J}} 

 - \sqrt{2} \mathcal{D}_{{HS_{6}V_{3}^\dagger D}}^{{pr}}(\tau ^{{I}})_{{i}}^{{j}}[D_{\mu }{H}_{{j}}]({S}_{{6p}})^{{I}}({V}_{{3r}}^{\dagger })^{{i\mu }} 

 + \frac{\mathcal{D}_{{F_{5L}S_{7}}}^{{psr}}}{\sqrt{2}}\epsilon ^{{il}} \epsilon ^{{mk}} {C}_{{jkl}}^{\mathfrak{i}} (\tau ^{{I}})_{{m}}^{{j}}[({F}_{{5p}})^{{I}}{C}({l}_{{s}})_{{i}}]({S}_{{7r}})^{\mathfrak{i}} 

 - \frac{\mathcal{D}_{{F_{6R}^\dagger LS_{8}}}^{{psr}}}{\sqrt{2}}\epsilon ^{{im}} \epsilon ^{{jl}} {C}_{{klm}}^{\mathfrak{i}} (\tau ^{{I}})_{{j}}^{{k}}[({F}_{{6p}})^{{I}}{C}({l}_{{s}})_{{i}}]({S}_{{8r}})^{\mathfrak{i}} 

 - 2 \mathcal{D}_{{d^\dagger F_{1}S_{10}}}^{{spr}}\delta _{{b}}^{{a}}[(\bar{d}_{{s}})^{{b}}{F}_{{1p}}]({S}_{{10r}})_{{a}} 

 + 2 \mathcal{D}_{{F_{10R}^\dagger LS_{10}}}^{{psr}}\delta _{{a}}^{{b}} \delta _{{i}}^{{j}}[({F}_{{10p}})^{{ai}}{C}({l}_{{s}})_{{j}}]({S}_{{10r}})_{{b}} 

 - 2 \mathcal{D}_{{F_{1}S_{11}u^\dagger }}^{{prs}}\delta _{{b}}^{{a}}[(\bar{u}_{{s}})^{{b}}{F}_{{1p}}]({S}_{{11r}})_{{a}} 

 - 2 \mathcal{D}_{{d^\dagger F_{2R}^\dagger S_{11}}}^{{spr}}\delta _{{b}}^{{a}}[(\bar{d}_{{s}})^{{b}}{F}_{{2p}}]({S}_{{11r}})_{{a}} 

 - 2 \mathcal{D}_{{F_{3L}QS_{11}^\dagger }}^{{psr}}\epsilon ^{{ij}} \delta _{{a}}^{{b}}[({F}_{{3p}})_{{i}}{C}({q}_{{s}})_{{bj}}]({S}_{{11r}}^{\dagger })^{{a}} 

 - 2 \mathcal{D}_{{e^\dagger F_{8L}S_{11}^\dagger }}^{{spr}}\delta _{{b}}^{{a}}[\bar{e}_{{s}}({F}_{{8p}})_{{a}}]({S}_{{11r}}^{\dagger })^{{b}} 

 + 2 \mathcal{D}_{{F_{11R}^\dagger LS_{11}}}^{{psr}}\delta _{{a}}^{{b}} \delta _{{i}}^{{j}}[({F}_{{11p}})^{{ai}}{C}({l}_{{s}})_{{j}}]({S}_{{11r}})_{{b}} 

 - 2 \mathcal{D}_{{F_{1}QS_{12}^\dagger }}^{{psr}}\delta _{{a}}^{{b}} \delta _{{i}}^{{j}}[{F}_{{1p}}{C}({q}_{{s}})_{{bj}}]({S}_{{12r}}^{\dagger })^{{ai}} 

 - 2 \mathcal{D}_{{F_{3L}S_{12}u^\dagger }}^{{prs}}\epsilon ^{{ij}} \delta _{{b}}^{{a}}[(\bar{u}_{{s}})^{{b}}({F}_{{3p}})_{{i}}]({S}_{{12r}})_{{aj}} 

 + \sqrt{2} \mathcal{D}_{{F_{5}QS_{12}^\dagger }}^{{psr}}\delta _{{a}}^{{b}} (\tau ^{{I}})_{{i}}^{{j}}[({F}_{{5p}})^{{I}}{C}({q}_{{s}})_{{bj}}]({S}_{{12r}}^{\dagger })^{{ai}} 

 - \mathcal{D}_{{F_{8L}QS_{12}}}^{{psr}}\epsilon ^{{acb}} \epsilon ^{{ji}}[({F}_{{8p}})_{{a}}{C}({q}_{{s}})_{{cj}}]({S}_{{12r}})_{{bi}} 

 - 2 \mathcal{D}_{{F_{9L}LS_{12}^\dagger }}^{{psr}}\delta _{{b}}^{{a}} \delta _{{i}}^{{j}}[({F}_{{9p}})_{{a}}{C}({l}_{{s}})_{{j}}]({S}_{{12r}}^{\dagger })^{{bi}} 

 - 4 \mathcal{D}_{{F_{10R}^\dagger S_{12}^\dagger u^\dagger }}^{{prs}}\epsilon _{{ij}} \epsilon ^{{abc}}[(\bar{u}_{{s}})^{{c}}({F}_{{10p}})^{{ai}}]({S}_{{12r}}^{\dagger })^{{bj}} 

 - 4 \mathcal{D}_{{d^\dagger F_{11R}^\dagger S_{12}^\dagger }}^{{spr}}\epsilon _{{ij}} \epsilon ^{{cab}}[(\bar{d}_{{s}})^{{c}}({F}_{{11p}})^{{ai}}]({S}_{{12r}}^{\dagger })^{{bj}} 

 + 2 \mathcal{D}_{{e^\dagger F_{12R}^\dagger S_{12}}}^{{spr}}\delta _{{a}}^{{b}} \delta _{{i}}^{{j}}[\bar{e}_{{s}}({F}_{{12p}})^{{ai}}]({S}_{{12r}})_{{bj}} 

 - \frac{\mathcal{D}_{{F_{13L}QS_{12}}}^{{psr}}}{\sqrt{2}}\epsilon ^{{acb}} \epsilon ^{{ik}} (\tau ^{{I}})_{{k}}^{{j}}[({F}_{{13p}})_{{a}}^{{I}}{C}({q}_{{s}})_{{cj}}]({S}_{{12r}})_{{bi}} 

 + \sqrt{2} \mathcal{D}_{{F_{14L}LS_{12}^\dagger }}^{{psr}}\delta _{{b}}^{{a}} (\tau ^{{I}})_{{i}}^{{j}}[({F}_{{14p}})_{{a}}^{{I}}{C}({l}_{{s}})_{{j}}]({S}_{{12r}}^{\dagger })^{{bi}} 

 - 4 \mathcal{D}_{{d^\dagger F_{10R}^\dagger S_{13}^\dagger }}^{{spr}}\epsilon _{{ij}} \epsilon ^{{cab}}[(\bar{d}_{{s}})^{{c}}({F}_{{10p}})^{{ai}}]({S}_{{13r}}^{\dagger })^{{bj}} 

 + 2 \mathcal{D}_{{d^\dagger F_{5}S_{14}}}^{{spr}}\delta ^{{IJ}} \delta _{{b}}^{{a}}[(\bar{d}_{{s}})^{{b}}({F}_{{5p}})^{{I}}]({S}_{{14r}})_{{a}}^{{J}} 

 + \sqrt{2} \mathcal{D}_{{F_{10R}^\dagger LS_{14}}}^{{psr}}\delta _{{a}}^{{b}} (\tau ^{{I}})_{{i}}^{{j}}[({F}_{{10p}})^{{ai}}{C}({l}_{{s}})_{{j}}]({S}_{{14r}})_{{b}}^{{I}} 

 - \mathcal{D}_{{F_{1}F_{1}S_{1}}}^{{prs}}[{F}_{{1p}}{C}{F}_{{1r}}]{S}_{{1s}} 

 + \mathcal{D}_{{F_{5}F_{5}S_{1}}}^{{prs}}\delta ^{{IJ}}[({F}_{{5p}})^{{I}}{C}({F}_{{5r}})^{{J}}]{S}_{{1s}} 

 + \mathcal{D}_{{F_{1}F_{5}S_{5}}}^{{prs}}\delta ^{{IJ}}[{F}_{{1p}}{C}({F}_{{5r}})^{{I}}]({S}_{{5s}})^{{J}} 

 - \frac{i \mathcal{D}_{{F_{5}F_{5}S_{5}}}^{{prs}}}{\sqrt{2}}\epsilon ^{{IJK}}[({F}_{{5p}})^{{I}}{C}({F}_{{5r}})^{{J}}]({S}_{{5s}})^{{K}} 

 - \mathcal{D}_{{F_{1}F_{6R}^\dagger S_{6}}}^{{prs}}\delta ^{{IJ}}[{F}_{{1p}}{C}({F}_{{6r}})^{{I}}]({S}_{{6s}})^{{J}} 

 + \mathcal{D}_{{F_{2R}^\dagger F_{5}S_{6}}}^{{rps}}\delta ^{{IJ}}[{F}_{{2p}}{C}({F}_{{5r}})^{{I}}]({S}_{{6s}})^{{J}} 

 - \frac{\mathcal{D}_{{F_{3L}F_{3L}S_{6}^\dagger }}^{{prs}}}{\sqrt{2}}\epsilon ^{{kj}} (\tau ^{{I}})_{{k}}^{{i}}[({F}_{{3p}})_{{i}}{C}({F}_{{3r}})_{{j}}]({S}_{{6s}}^{\dagger })^{{I}} 

 + \frac{i \mathcal{D}_{{F_{5}F_{6R}^\dagger S_{6}}}^{{prs}}}{\sqrt{2}}\epsilon ^{{IJK}}[({F}_{{5p}})^{{I}}{C}({F}_{{6r}})^{{J}}]({S}_{{6s}})^{{K}}

    \end{autobreak}
\end{align}

\begin{align}
    \begin{autobreak}
        \Delta{\cal L}_{\rm UV,FV}^{(d = 4)} = 

        + \mathcal{D}_{{eF_{1}V_{2}}}^{{spr}}[{F}_{{1p}}{C}\gamma _{\mu }{e}_{{s}}]{V}_{{2r}}^{\mu } 

 - 2 \mathcal{D}_{{F_{1}^\dagger u^\dagger V_{5}}}^{{psr}}\delta _{{b}}^{{a}}[(\bar{u}_{{s}})^{{b}}\gamma _{\mu }{C}\bar{F}_{{1p}}]({V}_{{5r}})_{{a}}^{\mu } 

 + 2 \mathcal{D}_{{F_{1}^\dagger QV_{8}^\dagger }}^{{psr}}\delta _{{a}}^{{b}} \delta _{{i}}^{{j}}[\bar{F}_{{1p}}\gamma _{\mu }({q}_{{s}})_{{bj}}]({V}_{{8r}}^{\dagger })^{{ai\mu }} 

 + \mathcal{D}_{{F_{3L}L^\dagger V_{2}^\dagger }}^{{psr}}\delta _{{j}}^{{i}}[(\bar{l}_{{s}})^{{j}}\gamma _{\mu }({F}_{{3p}})_{{i}}]{V}_{{2r}}^{\dagger \mu } 

 - 2 \mathcal{D}_{{d^\dagger F_{3L}^\dagger V_{8}}}^{{spr}}\delta _{{b}}^{{a}} \delta _{{i}}^{{j}}[(\bar{d}_{{s}})^{{b}}\gamma _{\mu }{C}(\bar{F}_{{3p}})^{{i}}]({V}_{{8r}})_{{aj}}^{\mu } 

 + \sqrt{2} \mathcal{D}_{{F_{5}^\dagger QV_{8}^\dagger }}^{{psr}}\delta _{{a}}^{{b}} (\tau ^{{I}})_{{i}}^{{j}}[(\bar{F}_{{5p}})^{{I}}\gamma _{\mu }({q}_{{s}})_{{bj}}]({V}_{{8r}}^{\dagger })^{{ai\mu }} 

 - 2 \mathcal{D}_{{F_{5}^\dagger u^\dagger V_{9}}}^{{psr}}\delta ^{{IJ}} \delta _{{b}}^{{a}}[(\bar{u}_{{s}})^{{b}}\gamma _{\mu }{C}(\bar{F}_{{5p}})^{{I}}]({V}_{{9r}})_{{a}}^{{J\mu }} 

 + 4 \mathcal{D}_{{d^\dagger F_{8L}^\dagger V_{5}^\dagger }}^{{spr}}\epsilon ^{{cab}}[(\bar{d}_{{s}})^{{c}}\gamma _{\mu }{C}(\bar{F}_{{8p}})^{{a}}]({V}_{{5r}}^{\dagger })^{{b\mu }} 

 + 2 \mathcal{D}_{{F_{8L}L^\dagger V_{8}^\dagger }}^{{psr}}\epsilon _{{ji}} \delta _{{b}}^{{a}}[(\bar{l}_{{s}})^{{j}}\gamma _{\mu }({F}_{{8p}})_{{a}}]({V}_{{8r}}^{\dagger })^{{bi\mu }} 

 - 2 \mathcal{D}_{{F_{10R}^\dagger uV_{3}^\dagger }}^{{psr}}\epsilon _{{ij}} \delta _{{a}}^{{b}}[({F}_{{10p}})^{{ai}}{C}\gamma _{\mu }({u}_{{s}})_{{b}}]({V}_{{3r}}^{\dagger })^{{j\mu }} 

 - 4 \mathcal{D}_{{F_{10R}^\dagger Q^\dagger V_{5}^\dagger }}^{{psr}}\epsilon _{{ij}} \epsilon ^{{abc}}[(\bar{q}_{{s}})^{{cj}}\gamma _{\mu }({F}_{{10p}})^{{ai}}]({V}_{{5r}}^{\dagger })^{{b\mu }} 

 - 2 \mathcal{D}_{{eF_{10R}^\dagger V_{8}}}^{{spr}}\delta _{{a}}^{{b}} \delta _{{i}}^{{j}}[({F}_{{10p}})^{{ai}}{C}\gamma _{\mu }{e}_{{s}}]({V}_{{8r}})_{{bj}}^{\mu } 

 + \sqrt{2} \mathcal{D}_{{F_{10R}^\dagger Q^\dagger V_{9}^\dagger }}^{{psr}}\epsilon _{{kj}} \epsilon ^{{abc}} (\tau ^{{I}})_{{i}}^{{k}}[(\bar{q}_{{s}})^{{cj}}\gamma _{\mu }({F}_{{10p}})^{{ai}}]({V}_{{9r}}^{\dagger })^{{bI\mu }} 

 + \sqrt{2} \mathcal{D}_{{F_{10R}^\dagger Q^\dagger V_{9}^\dagger }}^{{psr}}\epsilon _{{ki}} \epsilon ^{{abc}} (\tau ^{{I}})_{{j}}^{{k}}[(\bar{q}_{{s}})^{{cj}}\gamma _{\mu }({F}_{{10p}})^{{ai}}]({V}_{{9r}}^{\dagger })^{{bI\mu }} 

 - 2 \mathcal{D}_{{dF_{12R}^\dagger V_{3}}}^{{spr}}\delta _{{a}}^{{b}} \delta _{{i}}^{{j}}[({F}_{{12p}})^{{ai}}{C}\gamma _{\mu }({d}_{{s}})_{{b}}]({V}_{{3r}})_{{j}}^{\mu } 

 + 2 \mathcal{D}_{{F_{12R}^\dagger L^\dagger V_{5}}}^{{psr}}\epsilon _{{ij}} \delta _{{a}}^{{b}}[(\bar{l}_{{s}})^{{j}}\gamma _{\mu }({F}_{{12p}})^{{ai}}]({V}_{{5r}})_{{b}}^{\mu } 

 + \frac{\mathcal{D}_{{F_{12R}^\dagger L^\dagger V_{9}}}^{{psr}}}{\sqrt{2}}\epsilon _{{kj}} \delta _{{a}}^{{b}} (\tau ^{{I}})_{{i}}^{{k}}[(\bar{l}_{{s}})^{{j}}\gamma _{\mu }({F}_{{12p}})^{{ai}}]({V}_{{9r}})_{{b}}^{{I\mu }} 

 + \frac{\mathcal{D}_{{F_{12R}^\dagger L^\dagger V_{9}}}^{{psr}}}{\sqrt{2}}\epsilon _{{ki}} \delta _{{a}}^{{b}} (\tau ^{{I}})_{{j}}^{{k}}[(\bar{l}_{{s}})^{{j}}\gamma _{\mu }({F}_{{12p}})^{{ai}}]({V}_{{9r}})_{{b}}^{{I\mu }} 

 + \sqrt{2} \mathcal{D}_{{F_{13L}L^\dagger V_{8}^\dagger }}^{{psr}}\epsilon _{{kj}} \delta _{{b}}^{{a}} (\tau ^{{I}})_{{i}}^{{k}}[(\bar{l}_{{s}})^{{j}}\gamma _{\mu }({F}_{{13p}})_{{a}}^{{I}}]({V}_{{8r}}^{\dagger })^{{bi\mu }} 

 + \mathcal{D}_{{dF_{13L}V_{9}}}^{{spr}}\epsilon ^{{cba}} \delta ^{{IJ}}[({F}_{{13p}})_{{a}}^{{I}}{C}\gamma _{\mu }({d}_{{s}})_{{c}}]({V}_{{9r}})_{{b}}^{{J\mu }} 

    \end{autobreak}
\end{align}

\section{The complete Wilson coefficients of matching result}
\label{sec:resultwc}

In this appendix we provide the complete expression of Wilson coefficients.
Some descriptions are given as follows:
\begin{itemize}
    \item $y, \mu_H, \lambda_H$ are SM Yukawa, Higgs quadratic, quartic couplings as in Sec.\,\ref{sec:uvlag}. ${\cal C}$ and ${\cal D}$ are new couplings with mass dimension 1 and 0 as described in App.\,\ref{sec:uvlagfull}. 
    \item Flavor indices of every effective operator and Wilson coefficients are marked by $f_1, f_2, ...$ in the order of the helicities and the letters of the composing fields. For example, $C_{dLueH}$ is the Wilson coefficients of 
    ${\cal O}_{dLueH}^{f_1f_2f_5f_4} = \epsilon^{ij} (\overline{d}^{a}_{f_1} \ell_{if_2}) (u^T_{af_5} C e_{f_4}) H_j$, since the helicities of the fields $\overline{d},\ell,u,e,H$ are $-1/2, -1/2, 1/2, 1/2, 0$ respectively.
    \item $C_5$ denotes the coefficient of the Weinberg operator. Since it depends on the model, we just leave in the expressions.
\end{itemize}
Detailed description of usage can be found in Sec.\,\ref{sec:result}. 

\subsection{Dimension-5}
\begin{align}
\begin{autobreak}
 {C_{5}} =
 
-\frac{\mathcal{D}_{{F_{1}HL}}^{{p_1f_1}} \mathcal{D}_{{F_{1}HL}}^{{p_1f_2}} \mu_H^2 }{2 {M_{F_{1}}^3}}
-\frac{\mathcal{D}_{{F_{5}HL}}^{{p_1f_1}} \mathcal{D}_{{F_{5}HL}}^{{p_1f_2}} \mu_H^2 }{4 {M_{F_{5}}^3}}
-\frac{\mathcal{D}_{{LLS_{6}}}^{{f_1f_2p_1}} \mathcal{C}_{{HHS_{6}^\dagger }}^{{p_1}}}{{M_{S_{6}}^2}}
+\frac{\mathcal{D}_{{F_{1}HL}}^{{p_1f_1}} \mathcal{D}_{{F_{1}HL}}^{{p_1f_2}}}{2 {M_{F_{1}}}}
+\frac{\mathcal{D}_{{F_{5}HL}}^{{p_1f_1}} \mathcal{D}_{{F_{5}HL}}^{{p_1f_2}}}{4 {M_{F_{5}}}} 
\end{autobreak}
\end{align}

\subsection{Dimension-6}
\subsubsection{Bosonic operators}
\begin{align}\label{eq:wc/ch}
\begin{autobreak}
 {C_{H}} =
 \frac{2 \lambda_H (\mathcal{C}_{{HH^\dagger S_{5}}}^{{p_1}})^2}{{M_{S_{5}}^4}}
+\frac{\mathcal{D}_{{HHHS_{8}^\dagger }}^{{p_1*}} \mathcal{D}_{{HHHS_{8}^\dagger }}^{{p_1}}}{{M_{S_{8}}^2}}
+\frac{\mathcal{D}_{{HHH^\dagger S_{7}^\dagger }}^{{p_1*}} \mathcal{D}_{{HHH^\dagger S_{7}^\dagger }}^{{p_1}}}{3 {M_{S_{7}}^2}}
+\frac{4 \lambda_H \mathcal{C}_{{HHS_{6}^\dagger }}^{{p_1*}} \mathcal{C}_{{HHS_{6}^\dagger }}^{{p_1}}}{{M_{S_{6}}^4}}
-\frac{8 \lambda_H \mathcal{D}_{{HHV_{2}^\dagger D}}^{{p_1*}} \mathcal{D}_{{HHV_{2}^\dagger D}}^{{p_1}}}{{M_{V_{2}}^2}}
+\frac{\mathcal{D}_{{HH^\dagger H^\dagger S_{4}(1)}}^{{p_1*}} \mathcal{D}_{{HH^\dagger H^\dagger S_{4}(1)}}^{{p_1}}}{{M_{S_{4}}^2}}
+\frac{\mathcal{D}_{{HHS_{1}S_{6}^\dagger }}^{{p_1p_2}} \mathcal{C}_{{HHS_{6}^\dagger }}^{{p_2*}} \mathcal{C}_{{HH^\dagger S_{1}}}^{{p_1}}}{{M_{S_{1}}^2} {M_{S_{6}}^2}}
+\frac{\mathcal{D}_{{HHS_{1}S_{6}^\dagger }}^{{p_1p_2*}} \mathcal{C}_{{HHS_{6}^\dagger }}^{{p_2}} \mathcal{C}_{{HH^\dagger S_{1}}}^{{p_1}}}{{M_{S_{1}}^2} {M_{S_{6}}^2}}
+\frac{\mathcal{C}_{{HH^\dagger S_{1}}}^{{p_1}} \mathcal{C}_{{HH^\dagger S_{1}}}^{{p_2}} \mathcal{D}_{{HH^\dagger S_{1}S_{1}}}^{{p_1p_2}}}{{M_{S_{1}}^4}}
-\frac{\mathcal{D}_{{HHS_{5}S_{6}^\dagger }}^{{p_1p_2}} \mathcal{C}_{{HHS_{6}^\dagger }}^{{p_2*}} \mathcal{C}_{{HH^\dagger S_{5}}}^{{p_1}}}{2 {M_{S_{5}}^2} {M_{S_{6}}^2}}
+\frac{\mathcal{D}_{{HHS_{5}S_{6}^\dagger }}^{{p_1p_2*}} \mathcal{C}_{{HHS_{6}^\dagger }}^{{p_2}} \mathcal{C}_{{HH^\dagger S_{5}}}^{{p_1}}}{2 {M_{S_{5}}^2} {M_{S_{6}}^2}}
+\frac{\mathcal{C}_{{HH^\dagger S_{1}}}^{{p_1}} \mathcal{D}_{{HH^\dagger S_{1}S_{5}}}^{{p_1p_2}} \mathcal{C}_{{HH^\dagger S_{5}}}^{{p_2}}}{2 {M_{S_{1}}^2} {M_{S_{5}}^2}}
-\frac{\mathcal{C}_{{HH^\dagger S_{5}}}^{{p_1}} \mathcal{C}_{{HH^\dagger S_{5}}}^{{p_2}} \mathcal{D}_{{HH^\dagger S_{5}S_{5}(1)}}^{{p_2p_1}}}{4 {M_{S_{5}}^4}}
-\frac{\mathcal{C}_{{HH^\dagger S_{5}}}^{{p_1}} \mathcal{C}_{{HH^\dagger S_{5}}}^{{p_2}} \mathcal{D}_{{HH^\dagger S_{5}S_{5}(2)}}^{{p_2p_1}}}{2 {M_{S_{5}}^4}}
+\frac{\mathcal{C}_{{HHS_{6}^\dagger }}^{{p_1*}} \mathcal{C}_{{HHS_{6}^\dagger }}^{{p_2}} \mathcal{D}_{{HH^\dagger S_{6}S_{6}^\dagger (2)}}^{{p_2p_1}}}{{M_{S_{6}}^4}}
+\frac{8 \lambda_H \mathcal{D}_{{HH^\dagger V_{4}D}}^{{p_1*}} \mathcal{D}_{{HH^\dagger V_{4}D}}^{{p_1}}}{{M_{V_{4}}^2}}
-\frac{\mathcal{D}_{{HH^\dagger H^\dagger S_{4}(1)}}^{{p_1*}} \mathcal{C}_{{HH^\dagger S_{1}}}^{{p_2}} \mathcal{C}_{{HS_{1}S_{4}^\dagger }}^{{p_2p_1*}}}{{M_{S_{1}}^2} {M_{S_{4}}^2}}
+\frac{\mathcal{C}_{{HH^\dagger S_{1}}}^{{p_1}} \mathcal{C}_{{HH^\dagger S_{1}}}^{{p_2}} \mathcal{C}_{{HS_{1}S_{4}^\dagger }}^{{p_2p_3*}} \mathcal{C}_{{HS_{1}S_{4}^\dagger }}^{{p_1p_3}}}{{M_{S_{1}}^4} {M_{S_{4}}^2}}
-\frac{\mathcal{D}_{{HH^\dagger H^\dagger S_{4}(1)}}^{{p_1}} \mathcal{C}_{{HH^\dagger S_{1}}}^{{p_2}} \mathcal{C}_{{HS_{1}S_{4}^\dagger }}^{{p_2p_1}}}{{M_{S_{1}}^2} {M_{S_{4}}^2}}
-\frac{\mathcal{C}_{{HHS_{6}^\dagger }}^{{p_1}} \mathcal{D}_{{HH^\dagger H^\dagger S_{4}(1)}}^{{p_2}} \mathcal{C}_{{HS_{4}S_{6}^\dagger }}^{{p_2p_1*}}}{{M_{S_{4}}^2} {M_{S_{6}}^2}}
+\frac{\mathcal{C}_{{HHS_{6}^\dagger }}^{{p_1}} \mathcal{C}_{{HH^\dagger S_{1}}}^{{p_2}} \mathcal{C}_{{HS_{1}S_{4}^\dagger }}^{{p_2p_3*}} \mathcal{C}_{{HS_{4}S_{6}^\dagger }}^{{p_3p_1*}}}{{M_{S_{1}}^2} {M_{S_{4}}^2} {M_{S_{6}}^2}}
-\frac{\mathcal{C}_{{HHS_{6}^\dagger }}^{{p_1*}} \mathcal{D}_{{HH^\dagger H^\dagger S_{4}(1)}}^{{p_2*}} \mathcal{C}_{{HS_{4}S_{6}^\dagger }}^{{p_2p_1}}}{{M_{S_{4}}^2} {M_{S_{6}}^2}}
+\frac{\mathcal{C}_{{HHS_{6}^\dagger }}^{{p_1*}} \mathcal{C}_{{HH^\dagger S_{1}}}^{{p_2}} \mathcal{C}_{{HS_{1}S_{4}^\dagger }}^{{p_2p_3}} \mathcal{C}_{{HS_{4}S_{6}^\dagger }}^{{p_3p_1}}}{{M_{S_{1}}^2} {M_{S_{4}}^2} {M_{S_{6}}^2}}
+\frac{\mathcal{C}_{{HHS_{6}^\dagger }}^{{p_1*}} \mathcal{C}_{{HHS_{6}^\dagger }}^{{p_2}} \mathcal{C}_{{HS_{4}S_{6}^\dagger }}^{{p_3p_2*}} \mathcal{C}_{{HS_{4}S_{6}^\dagger }}^{{p_3p_1}}}{{M_{S_{4}}^2} {M_{S_{6}}^4}}
+\frac{\mathcal{D}_{{HHH^\dagger S_{7}^\dagger }}^{{p_1}} \mathcal{C}_{{HH^\dagger S_{5}}}^{{p_2}} \mathcal{C}_{{HS_{5}S_{7}^\dagger }}^{{p_2p_1*}}}{3 {M_{S_{5}}^2} {M_{S_{7}}^2}}
-\frac{2 \mathcal{C}_{{HH^\dagger S_{5}}}^{{p_1}} \mathcal{C}_{{HH^\dagger S_{5}}}^{{p_2}} \mathcal{C}_{{HS_{5}S_{7}^\dagger }}^{{p_2p_3*}} \mathcal{C}_{{HS_{5}S_{7}^\dagger }}^{{p_1p_3}}}{3 {M_{S_{5}}^4} {M_{S_{7}}^2}}
-\frac{\mathcal{D}_{{HHH^\dagger S_{7}^\dagger }}^{{p_1*}} \mathcal{C}_{{HH^\dagger S_{5}}}^{{p_2}} \mathcal{C}_{{HS_{5}S_{7}^\dagger }}^{{p_2p_1}}}{3 {M_{S_{5}}^2} {M_{S_{7}}^2}}
+\frac{\mathcal{C}_{{HH^\dagger S_{5}}}^{{p_1}} \mathcal{C}_{{HH^\dagger S_{5}}}^{{p_2}} \mathcal{C}_{{HS_{5}S_{7}^\dagger }}^{{p_1p_3*}} \mathcal{C}_{{HS_{5}S_{7}^\dagger }}^{{p_2p_3}}}{3 {M_{S_{5}}^4} {M_{S_{7}}^2}}
+\frac{\mathcal{D}_{{HHHS_{8}^\dagger }}^{{p_1}} \mathcal{C}_{{HHS_{6}^\dagger }}^{{p_2*}} \mathcal{C}_{{HS_{6}S_{8}^\dagger }}^{{p_2p_1*}}}{{M_{S_{6}}^2} {M_{S_{8}}^2}}
+\frac{\mathcal{D}_{{HHHS_{8}^\dagger }}^{{p_1*}} \mathcal{C}_{{HHS_{6}^\dagger }}^{{p_2}} \mathcal{C}_{{HS_{6}S_{8}^\dagger }}^{{p_2p_1}}}{{M_{S_{6}}^2} {M_{S_{8}}^2}}
+\frac{\mathcal{C}_{{HHS_{6}^\dagger }}^{{p_1*}} \mathcal{C}_{{HHS_{6}^\dagger }}^{{p_2}} \mathcal{C}_{{HS_{6}S_{8}^\dagger }}^{{p_1p_3*}} \mathcal{C}_{{HS_{6}S_{8}^\dagger }}^{{p_2p_3}}}{{M_{S_{6}}^4} {M_{S_{8}}^2}}
+\frac{\mathcal{C}_{{HHS_{6}^\dagger }}^{{p_1}} \mathcal{C}_{{HH^\dagger S_{5}}}^{{p_2}} \mathcal{C}_{{HS_{5}S_{7}^\dagger }}^{{p_2p_3*}} \mathcal{C}_{{HS_{6}^\dagger S_{7}}}^{{p_1p_3*}}}{3 {M_{S_{5}}^2} {M_{S_{6}}^2} {M_{S_{7}}^2}}
+\frac{\mathcal{D}_{{HHH^\dagger S_{7}^\dagger }}^{{p_1*}} \mathcal{C}_{{HHS_{6}^\dagger }}^{{p_2}} \mathcal{C}_{{HS_{6}^\dagger S_{7}}}^{{p_2p_1*}}}{3 {M_{S_{6}}^2} {M_{S_{7}}^2}}
-\frac{\mathcal{C}_{{HHS_{6}^\dagger }}^{{p_1*}} \mathcal{C}_{{HH^\dagger S_{5}}}^{{p_2}} \mathcal{C}_{{HS_{5}S_{7}^\dagger }}^{{p_2p_3}} \mathcal{C}_{{HS_{6}^\dagger S_{7}}}^{{p_1p_3}}}{3 {M_{S_{5}}^2} {M_{S_{6}}^2} {M_{S_{7}}^2}}
+\frac{\mathcal{C}_{{HHS_{6}^\dagger }}^{{p_1*}} \mathcal{C}_{{HHS_{6}^\dagger }}^{{p_2}} \mathcal{C}_{{HS_{6}^\dagger S_{7}}}^{{p_2p_3*}} \mathcal{C}_{{HS_{6}^\dagger S_{7}}}^{{p_1p_3}}}{3 {M_{S_{6}}^4} {M_{S_{7}}^2}}
+\frac{\mathcal{D}_{{HHH^\dagger S_{7}^\dagger }}^{{p_1}} \mathcal{C}_{{HHS_{6}^\dagger }}^{{p_2*}} \mathcal{C}_{{HS_{6}^\dagger S_{7}}}^{{p_2p_1}}}{3 {M_{S_{6}}^2} {M_{S_{7}}^2}}
+\frac{\mathcal{D}_{{HH^\dagger H^\dagger S_{4}(1)}}^{{p_1}} \mathcal{C}_{{HH^\dagger S_{5}}}^{{p_2}} \mathcal{C}_{{H^\dagger S_{4}S_{5}}}^{{p_1p_2*}}}{2 {M_{S_{4}}^2} {M_{S_{5}}^2}}
-\frac{\mathcal{C}_{{HH^\dagger S_{1}}}^{{p_1}} \mathcal{C}_{{HH^\dagger S_{5}}}^{{p_2}} \mathcal{C}_{{HS_{1}S_{4}^\dagger }}^{{p_1p_3*}} \mathcal{C}_{{H^\dagger S_{4}S_{5}}}^{{p_3p_2*}}}{2 {M_{S_{1}}^2} {M_{S_{4}}^2} {M_{S_{5}}^2}}
-\frac{\mathcal{C}_{{HHS_{6}^\dagger }}^{{p_1*}} \mathcal{C}_{{HH^\dagger S_{5}}}^{{p_2}} \mathcal{C}_{{HS_{4}S_{6}^\dagger }}^{{p_3p_1}} \mathcal{C}_{{H^\dagger S_{4}S_{5}}}^{{p_3p_2*}}}{2 {M_{S_{4}}^2} {M_{S_{5}}^2} {M_{S_{6}}^2}}
-\frac{\mathcal{D}_{{HH^\dagger H^\dagger S_{4}(1)}}^{{p_1*}} \mathcal{C}_{{HH^\dagger S_{5}}}^{{p_2}} \mathcal{C}_{{H^\dagger S_{4}S_{5}}}^{{p_1p_2}}}{2 {M_{S_{4}}^2} {M_{S_{5}}^2}}
-\frac{\mathcal{C}_{{HH^\dagger S_{5}}}^{{p_1}} \mathcal{C}_{{HH^\dagger S_{5}}}^{{p_2}} \mathcal{C}_{{H^\dagger S_{4}S_{5}}}^{{p_3p_2*}} \mathcal{C}_{{H^\dagger S_{4}S_{5}}}^{{p_3p_1}}}{4 {M_{S_{4}}^2} {M_{S_{5}}^4}}
+\frac{\mathcal{C}_{{HH^\dagger S_{1}}}^{{p_1}} \mathcal{C}_{{HH^\dagger S_{5}}}^{{p_2}} \mathcal{C}_{{HS_{1}S_{4}^\dagger }}^{{p_1p_3}} \mathcal{C}_{{H^\dagger S_{4}S_{5}}}^{{p_3p_2}}}{2 {M_{S_{1}}^2} {M_{S_{4}}^2} {M_{S_{5}}^2}}
+\frac{\mathcal{C}_{{HHS_{6}^\dagger }}^{{p_1}} \mathcal{C}_{{HH^\dagger S_{5}}}^{{p_2}} \mathcal{C}_{{HS_{4}S_{6}^\dagger }}^{{p_3p_1*}} \mathcal{C}_{{H^\dagger S_{4}S_{5}}}^{{p_3p_2}}}{2 {M_{S_{4}}^2} {M_{S_{5}}^2} {M_{S_{6}}^2}}
+\frac{\mathcal{C}_{{HH^\dagger S_{1}}}^{{p_1}} \mathcal{C}_{{HH^\dagger S_{1}}}^{{p_2}} \mathcal{C}_{{HH^\dagger S_{1}}}^{{p_3}} \mathcal{C}_{{S_{1}S_{1}S_{1}}}^{{p_1p_2p_3}}}{{M_{S_{1}}^6}}
-\frac{\mathcal{C}_{{HH^\dagger S_{1}}}^{{p_1}} \mathcal{C}_{{HH^\dagger S_{5}}}^{{p_2}} \mathcal{C}_{{HH^\dagger S_{5}}}^{{p_3}} \mathcal{C}_{{S_{1}S_{5}S_{5}}}^{{p_1p_3p_2}}}{2 {M_{S_{1}}^2} {M_{S_{5}}^4}}
+\frac{\mathcal{C}_{{HHS_{6}^\dagger }}^{{p_1*}} \mathcal{C}_{{HHS_{6}^\dagger }}^{{p_2}} \mathcal{C}_{{HH^\dagger S_{1}}}^{{p_3}} \mathcal{C}_{{S_{1}S_{6}S_{6}^\dagger }}^{{p_3p_2p_1}}}{{M_{S_{1}}^2} {M_{S_{6}}^4}}
+\frac{\mathcal{C}_{{HHS_{6}^\dagger }}^{{p_1*}} \mathcal{C}_{{HHS_{6}^\dagger }}^{{p_2}} \mathcal{C}_{{HH^\dagger S_{5}}}^{{p_3}} \mathcal{C}_{{S_{5}S_{6}S_{6}^\dagger }}^{{p_3p_2p_1}}}{2 {M_{S_{5}}^2} {M_{S_{6}}^4}} 
\end{autobreak}
\end{align}
\begin{align}
\begin{autobreak}
 {C_{H\square}} =
 
-\frac{(\mathcal{C}_{{HH^\dagger S_{1}}}^{{p_1}})^2}{2 {M_{S_{1}}^4}}
+\frac{(\mathcal{C}_{{HH^\dagger S_{5}}}^{{p_1}})^2}{4 {M_{S_{5}}^4}}
+\frac{(\mathcal{D}_{{HH^\dagger V_{1}D}}^{{p_1}})^2}{{M_{V_{1}}^2}}
+\frac{(\mathcal{D}_{{HH^\dagger V_{1}D}}^{{p_1*}})^2}{{M_{V_{1}}^2}}
+\frac{(\mathcal{D}_{{HH^\dagger V_{4}D}}^{{p_1}})^2}{2 {M_{V_{4}}^2}}
+\frac{(\mathcal{D}_{{HH^\dagger V_{4}D}}^{{p_1*}})^2}{2 {M_{V_{4}}^2}}
+\frac{\mathcal{C}_{{HHS_{6}^\dagger }}^{{p_1}} \mathcal{C}_{{HHS_{6}^\dagger }}^{{p_1*}}}{{M_{S_{6}}^4}}
-\frac{2 \mathcal{D}_{{HHV_{2}^\dagger D}}^{{p_1}} \mathcal{D}_{{HHV_{2}^\dagger D}}^{{p_1*}}}{{M_{V_{2}}^2}}
+\frac{2 \mathcal{D}_{{HH^\dagger V_{4}D}}^{{p_1}} \mathcal{D}_{{HH^\dagger V_{4}D}}^{{p_1*}}}{{M_{V_{4}}^2}} 
\end{autobreak}
\end{align}
\begin{align}
\begin{autobreak}
 {C_{HD}} =
 
-\frac{(\mathcal{C}_{{HH^\dagger S_{5}}}^{{p_1}})^2}{{M_{S_{5}}^4}}
+\frac{2 (\mathcal{D}_{{HH^\dagger V_{1}D}}^{{p_1}})^2}{{M_{V_{1}}^2}}
+\frac{2 (\mathcal{D}_{{HH^\dagger V_{1}D}}^{{p_1*}})^2}{{M_{V_{1}}^2}}
+\frac{(\mathcal{D}_{{HH^\dagger V_{4}D}}^{{p_1}})^2}{{M_{V_{4}}^2}}
+\frac{(\mathcal{D}_{{HH^\dagger V_{4}D}}^{{p_1*}})^2}{{M_{V_{4}}^2}}
+\frac{2 \mathcal{C}_{{HHS_{6}^\dagger }}^{{p_1}} \mathcal{C}_{{HHS_{6}^\dagger }}^{{p_1*}}}{{M_{S_{6}}^4}}
+\frac{4 \mathcal{D}_{{HHV_{2}^\dagger D}}^{{p_1}} \mathcal{D}_{{HHV_{2}^\dagger D}}^{{p_1*}}}{{M_{V_{2}}^2}}
-\frac{4 \mathcal{D}_{{HH^\dagger V_{1}D}}^{{p_1}} \mathcal{D}_{{HH^\dagger V_{1}D}}^{{p_1*}}}{{M_{V_{1}}^2}}
-\frac{2 \mathcal{D}_{{HH^\dagger V_{4}D}}^{{p_1}} \mathcal{D}_{{HH^\dagger V_{4}D}}^{{p_1*}}}{{M_{V_{4}}^2}} 
\end{autobreak}
\end{align}
\begin{align}
\begin{autobreak}
 {C_{eH}} =
 \frac{(\mathcal{C}_{{HH^\dagger S_{5}}}^{{p_1}})^2 {y}_{{e}}^{{f_5f_4}}}{2 {M_{S_{5}}^4}}
-\frac{(\mathcal{D}_{{HH^\dagger V_{1}D}}^{{p_1*}})^2 {y}_{{e}}^{{f_5f_4}}}{{M_{V_{1}}^2}}
+\frac{(\mathcal{D}_{{HH^\dagger V_{1}D}}^{{p_1}})^2 {y}_{{e}}^{{f_5f_4}}}{{M_{V_{1}}^2}}
-\frac{(\mathcal{D}_{{HH^\dagger V_{4}D}}^{{p_1*}})^2 {y}_{{e}}^{{f_5f_4}}}{2 {M_{V_{4}}^2}}
+\frac{(\mathcal{D}_{{HH^\dagger V_{4}D}}^{{p_1}})^2 {y}_{{e}}^{{f_5f_4}}}{2 {M_{V_{4}}^2}}
+\frac{\mathcal{C}_{{HHS_{6}^\dagger }}^{{p_1*}} \mathcal{C}_{{HHS_{6}^\dagger }}^{{p_1}} {y}_{{e}}^{{f_5f_4}}}{{M_{S_{6}}^4}}
-\frac{2 \mathcal{D}_{{HHV_{2}^\dagger D}}^{{p_1*}} \mathcal{D}_{{HHV_{2}^\dagger D}}^{{p_1}} {y}_{{e}}^{{f_5f_4}}}{{M_{V_{2}}^2}}
+\frac{2 \mathcal{D}_{{HH^\dagger V_{4}D}}^{{p_1*}} \mathcal{D}_{{HH^\dagger V_{4}D}}^{{p_1}} {y}_{{e}}^{{f_5f_4}}}{{M_{V_{4}}^2}}
+\frac{{y}_{{e}}^{{f_5p_1}} \mathcal{D}_{{e^\dagger F_{3R}^\dagger H^\dagger }}^{{f_4p_2*}} \mathcal{D}_{{e^\dagger F_{3R}^\dagger H^\dagger }}^{{p_1p_2}}}{2 {M_{F_{3}}^2}}
+\frac{{y}_{{e}}^{{f_5p_1}} \mathcal{D}_{{e^\dagger F_{4R}^\dagger H}}^{{f_4p_2*}} \mathcal{D}_{{e^\dagger F_{4R}^\dagger H}}^{{p_1p_2}}}{2 {M_{F_{4}}^2}}
+\frac{{y}_{{e}}^{{p_1f_4}} \mathcal{D}_{{F_{2L}H^\dagger L}}^{{p_2f_5*}} \mathcal{D}_{{F_{2L}H^\dagger L}}^{{p_2p_1}}}{2 {M_{F_{2}}^2}}
+\frac{\mathcal{D}_{{e^\dagger F_{3R}^\dagger H^\dagger }}^{{f_4p_1*}} \mathcal{D}_{{F_{2L}H^\dagger L}}^{{p_2f_5*}} \mathcal{D}_{{F_{2R}^\dagger F_{3L}H}}^{{p_2p_1*}}}{{M_{F_{2}}} {M_{F_{3}}}}
-\frac{\mathcal{D}_{{e^\dagger F_{4R}^\dagger H}}^{{f_4p_1*}} \mathcal{D}_{{F_{2L}H^\dagger L}}^{{p_2f_5*}} \mathcal{D}_{{F_{2R}^\dagger F_{4L}H^\dagger }}^{{p_2p_1*}}}{{M_{F_{2}}} {M_{F_{4}}}}
+\frac{\mathcal{D}_{{e^\dagger F_{3R}^\dagger H^\dagger }}^{{f_4p_1*}} \mathcal{D}_{{F_{3L}F_{5}H^\dagger }}^{{p_1p_2*}} \mathcal{D}_{{F_{5}HL}}^{{p_2f_5*}}}{{M_{F_{3}}} {M_{F_{5}}}}
+\frac{{y}_{{e}}^{{p_1f_4}} \mathcal{D}_{{F_{5}HL}}^{{p_2f_5*}} \mathcal{D}_{{F_{5}HL}}^{{p_2p_1}}}{2 {M_{F_{5}}^2}}
+\frac{\mathcal{D}_{{e^\dagger F_{3R}^\dagger H^\dagger }}^{{f_4p_1*}} \mathcal{D}_{{F_{3L}F_{6R}^\dagger H}}^{{p_1p_2*}} \mathcal{D}_{{F_{6L}H^\dagger L}}^{{p_2f_5*}}}{2 {M_{F_{3}}} {M_{F_{6}}}}
+\frac{\mathcal{D}_{{e^\dagger F_{4R}^\dagger H}}^{{f_4p_1*}} \mathcal{D}_{{F_{4L}F_{6R}^\dagger H^\dagger }}^{{p_1p_2*}} \mathcal{D}_{{F_{6L}H^\dagger L}}^{{p_2f_5*}}}{2 {M_{F_{4}}} {M_{F_{6}}}}
+\frac{{y}_{{e}}^{{p_1f_4}} \mathcal{D}_{{F_{6L}H^\dagger L}}^{{p_2f_5*}} \mathcal{D}_{{F_{6L}H^\dagger L}}^{{p_2p_1}}}{4 {M_{F_{6}}^2}}
-\frac{\mathcal{D}_{{e^\dagger F_{4R}^\dagger H}}^{{f_4p_1*}} \mathcal{D}_{{F_{4L}LS_{6}^\dagger }}^{{p_1f_5p_2*}} \mathcal{C}_{{HHS_{6}^\dagger }}^{{p_2}}}{{M_{F_{4}}} {M_{S_{6}}^2}}
-\frac{\mathcal{D}_{{e^\dagger F_{5}S_{6}^\dagger }}^{{f_4p_1p_2*}} \mathcal{D}_{{F_{5}HL}}^{{p_1f_5*}} \mathcal{C}_{{HHS_{6}^\dagger }}^{{p_2}}}{{M_{F_{5}}} {M_{S_{6}}^2}}
+\frac{\mathcal{D}_{{e^\dagger LS_{4}^\dagger }}^{{f_4f_5p_1*}} \mathcal{D}_{{HH^\dagger H^\dagger S_{4}(1)}}^{{p_1*}}}{{M_{S_{4}}^2}}
-\frac{\mathcal{D}_{{e^\dagger F_{2R}^\dagger S_{1}}}^{{f_4p_1p_2*}} \mathcal{D}_{{F_{2L}H^\dagger L}}^{{p_1f_5*}} \mathcal{C}_{{HH^\dagger S_{1}}}^{{p_2}}}{{M_{F_{2}}} {M_{S_{1}}^2}}
+\frac{\mathcal{D}_{{e^\dagger F_{3R}^\dagger H^\dagger }}^{{f_4p_1*}} \mathcal{D}_{{F_{3L}LS_{1}}}^{{p_1f_5p_2*}} \mathcal{C}_{{HH^\dagger S_{1}}}^{{p_2}}}{{M_{F_{3}}} {M_{S_{1}}^2}}
+\frac{\mathcal{D}_{{e^\dagger F_{3R}^\dagger H^\dagger }}^{{f_4p_1*}} \mathcal{D}_{{F_{3L}LS_{5}}}^{{p_1f_5p_2*}} \mathcal{C}_{{HH^\dagger S_{5}}}^{{p_2}}}{2 {M_{F_{3}}} {M_{S_{5}}^2}}
-\frac{\mathcal{D}_{{e^\dagger F_{6R}^\dagger S_{5}}}^{{f_4p_1p_2*}} \mathcal{D}_{{F_{6L}H^\dagger L}}^{{p_1f_5*}} \mathcal{C}_{{HH^\dagger S_{5}}}^{{p_2}}}{2 {M_{F_{6}}} {M_{S_{5}}^2}}
+\frac{i {y}_{{e}}^{{f_5p_1}} \mathcal{D}_{{e^\dagger eV_{1}}}^{{p_1f_4p_2}} \mathcal{D}_{{HH^\dagger V_{1}D}}^{{p_2*}}}{{M_{V_{1}}^2}}
+\frac{i {y}_{{e}}^{{f_5p_1}} \mathcal{D}_{{e^\dagger eV_{1}}}^{{p_1f_4p_2}} \mathcal{D}_{{HH^\dagger V_{1}D}}^{{p_2}}}{{M_{V_{1}}^2}}
-\frac{\mathcal{D}_{{e^\dagger LS_{4}^\dagger }}^{{f_4f_5p_1*}} \mathcal{C}_{{HH^\dagger S_{1}}}^{{p_2}} \mathcal{C}_{{HS_{1}S_{4}^\dagger }}^{{p_2p_1}}}{{M_{S_{1}}^2} {M_{S_{4}}^2}}
-\frac{\mathcal{D}_{{e^\dagger LS_{4}^\dagger }}^{{f_4f_5p_1*}} \mathcal{C}_{{HHS_{6}^\dagger }}^{{p_2}} \mathcal{C}_{{HS_{4}S_{6}^\dagger }}^{{p_1p_2*}}}{{M_{S_{4}}^2} {M_{S_{6}}^2}}
+\frac{\mathcal{D}_{{e^\dagger LS_{4}^\dagger }}^{{f_4f_5p_1*}} \mathcal{C}_{{HH^\dagger S_{5}}}^{{p_2}} \mathcal{C}_{{H^\dagger S_{4}S_{5}}}^{{p_1p_2*}}}{2 {M_{S_{4}}^2} {M_{S_{5}}^2}}
+\frac{i {y}_{{e}}^{{p_1f_4}} \mathcal{D}_{{HH^\dagger V_{1}D}}^{{p_2*}} \mathcal{D}_{{LL^\dagger V_{1}}}^{{p_1f_5p_2}}}{{M_{V_{1}}^2}}
+\frac{i {y}_{{e}}^{{p_1f_4}} \mathcal{D}_{{HH^\dagger V_{1}D}}^{{p_2}} \mathcal{D}_{{LL^\dagger V_{1}}}^{{p_1f_5p_2}}}{{M_{V_{1}}^2}}
-\frac{i {y}_{{e}}^{{p_1f_4}} \mathcal{D}_{{HH^\dagger V_{4}D}}^{{p_2*}} \mathcal{D}_{{LL^\dagger V_{4}}}^{{p_1f_5p_2}}}{2 {M_{V_{4}}^2}}
+\frac{i {y}_{{e}}^{{p_1f_4}} \mathcal{D}_{{HH^\dagger V_{4}D}}^{{p_2}} \mathcal{D}_{{LL^\dagger V_{4}}}^{{p_1f_5p_2}}}{2 {M_{V_{4}}^2}} 
\end{autobreak}
\end{align}
\begin{align}
\begin{autobreak}
 {C_{uH}} =
 \frac{(\mathcal{C}_{{HH^\dagger S_{5}}}^{{p_1}})^2 {y}_{{u}}^{{f_4f_5}}}{2 {M_{S_{5}}^4}}
+\frac{(\mathcal{D}_{{HH^\dagger V_{1}D}}^{{p_1*}})^2 {y}_{{u}}^{{f_4f_5}}}{{M_{V_{1}}^2}}
-\frac{(\mathcal{D}_{{HH^\dagger V_{1}D}}^{{p_1}})^2 {y}_{{u}}^{{f_4f_5}}}{{M_{V_{1}}^2}}
+\frac{(\mathcal{D}_{{HH^\dagger V_{4}D}}^{{p_1*}})^2 {y}_{{u}}^{{f_4f_5}}}{2 {M_{V_{4}}^2}}
-\frac{(\mathcal{D}_{{HH^\dagger V_{4}D}}^{{p_1}})^2 {y}_{{u}}^{{f_4f_5}}}{2 {M_{V_{4}}^2}}
+\frac{\mathcal{C}_{{HHS_{6}^\dagger }}^{{p_1*}} \mathcal{C}_{{HHS_{6}^\dagger }}^{{p_1}} {y}_{{u}}^{{f_4f_5}}}{{M_{S_{6}}^4}}
-\frac{2 \mathcal{D}_{{HHV_{2}^\dagger D}}^{{p_1*}} \mathcal{D}_{{HHV_{2}^\dagger D}}^{{p_1}} {y}_{{u}}^{{f_4f_5}}}{{M_{V_{2}}^2}}
+\frac{2 \mathcal{D}_{{HH^\dagger V_{4}D}}^{{p_1*}} \mathcal{D}_{{HH^\dagger V_{4}D}}^{{p_1}} {y}_{{u}}^{{f_4f_5}}}{{M_{V_{4}}^2}}
+\frac{2 {y}_{{u}}^{{f_4p_1}} \mathcal{D}_{{F_{11L}Hu^\dagger }}^{{p_2f_5*}} \mathcal{D}_{{F_{11L}Hu^\dagger }}^{{p_2p_1}}}{{M_{F_{11}}^2}}
+\frac{2 {y}_{{u}}^{{f_4p_1}} \mathcal{D}_{{F_{12L}H^\dagger u^\dagger }}^{{p_2f_5*}} \mathcal{D}_{{F_{12L}H^\dagger u^\dagger }}^{{p_2p_1}}}{{M_{F_{12}}^2}}
-\frac{8 \mathcal{D}_{{F_{11L}Hu^\dagger }}^{{p_1f_5*}} \mathcal{D}_{{F_{11R}^\dagger F_{13L}H}}^{{p_1p_2*}} \mathcal{D}_{{F_{13R}^\dagger H^\dagger Q}}^{{p_2f_4*}}}{{M_{F_{11}}} {M_{F_{13}}}}
+\frac{2 {y}_{{u}}^{{p_1f_5}} \mathcal{D}_{{F_{13R}^\dagger H^\dagger Q}}^{{p_2f_4*}} \mathcal{D}_{{F_{13R}^\dagger H^\dagger Q}}^{{p_2p_1}}}{{M_{F_{13}}^2}}
+\frac{4 \mathcal{D}_{{F_{11L}Hu^\dagger }}^{{p_1f_5*}} \mathcal{D}_{{F_{11R}^\dagger F_{14L}H^\dagger }}^{{p_1p_2*}} \mathcal{D}_{{F_{14R}^\dagger HQ}}^{{p_2f_4*}}}{{M_{F_{11}}} {M_{F_{14}}}}
+\frac{4 \mathcal{D}_{{F_{12L}H^\dagger u^\dagger }}^{{p_1f_5*}} \mathcal{D}_{{F_{12R}^\dagger F_{14L}H}}^{{p_1p_2*}} \mathcal{D}_{{F_{14R}^\dagger HQ}}^{{p_2f_4*}}}{{M_{F_{12}}} {M_{F_{14}}}}
+\frac{{y}_{{u}}^{{p_1f_5}} \mathcal{D}_{{F_{14R}^\dagger HQ}}^{{p_2f_4*}} \mathcal{D}_{{F_{14R}^\dagger HQ}}^{{p_2p_1}}}{{M_{F_{14}}^2}}
+\frac{8 \mathcal{D}_{{F_{11L}Hu^\dagger }}^{{p_1f_5*}} \mathcal{D}_{{F_{11R}^\dagger F_{9L}H^\dagger }}^{{p_1p_2*}} \mathcal{D}_{{F_{9R}^\dagger HQ}}^{{p_2f_4*}}}{{M_{F_{11}}} {M_{F_{9}}}}
-\frac{8 \mathcal{D}_{{F_{12L}H^\dagger u^\dagger }}^{{p_1f_5*}} \mathcal{D}_{{F_{12R}^\dagger F_{9L}H}}^{{p_1p_2*}} \mathcal{D}_{{F_{9R}^\dagger HQ}}^{{p_2f_4*}}}{{M_{F_{12}}} {M_{F_{9}}}}
+\frac{2 {y}_{{u}}^{{p_1f_5}} \mathcal{D}_{{F_{9R}^\dagger HQ}}^{{p_2f_4*}} \mathcal{D}_{{F_{9R}^\dagger HQ}}^{{p_2p_1}}}{{M_{F_{9}}^2}}
-\frac{4 \mathcal{D}_{{F_{12L}H^\dagger u^\dagger }}^{{p_1f_5*}} \mathcal{D}_{{F_{12R}^\dagger QS_{6}}}^{{p_1f_4p_2*}} \mathcal{C}_{{HHS_{6}^\dagger }}^{{p_2*}}}{{M_{F_{12}}} {M_{S_{6}}^2}}
-\frac{4 \mathcal{D}_{{F_{13L}S_{6}u^\dagger }}^{{p_1p_2f_5*}} \mathcal{D}_{{F_{13R}^\dagger H^\dagger Q}}^{{p_1f_4*}} \mathcal{C}_{{HHS_{6}^\dagger }}^{{p_2*}}}{{M_{F_{13}}} {M_{S_{6}}^2}}
-\frac{4 \mathcal{D}_{{F_{11L}Hu^\dagger }}^{{p_1f_5*}} \mathcal{D}_{{F_{11R}^\dagger QS_{1}}}^{{p_1f_4p_2*}} \mathcal{C}_{{HH^\dagger S_{1}}}^{{p_2}}}{{M_{F_{11}}} {M_{S_{1}}^2}}
-\frac{4 \mathcal{D}_{{F_{9L}S_{1}u^\dagger }}^{{p_1p_2f_5*}} \mathcal{D}_{{F_{9R}^\dagger HQ}}^{{p_1f_4*}} \mathcal{C}_{{HH^\dagger S_{1}}}^{{p_2}}}{{M_{F_{9}}} {M_{S_{1}}^2}}
+\frac{2 \mathcal{D}_{{F_{11L}Hu^\dagger }}^{{p_1f_5*}} \mathcal{D}_{{F_{11R}^\dagger QS_{5}}}^{{p_1f_4p_2*}} \mathcal{C}_{{HH^\dagger S_{5}}}^{{p_2}}}{{M_{F_{11}}} {M_{S_{5}}^2}}
+\frac{2 \mathcal{D}_{{F_{14L}S_{5}u^\dagger }}^{{p_1p_2f_5*}} \mathcal{D}_{{F_{14R}^\dagger HQ}}^{{p_1f_4*}} \mathcal{C}_{{HH^\dagger S_{5}}}^{{p_2}}}{{M_{F_{14}}} {M_{S_{5}}^2}}
+\frac{2 i {y}_{{u}}^{{p_1f_5}} \mathcal{D}_{{HH^\dagger V_{1}D}}^{{p_2*}} \mathcal{D}_{{QQ^\dagger V_{1}}}^{{p_1f_4p_2}}}{{M_{V_{1}}^2}}
+\frac{2 i {y}_{{u}}^{{p_1f_5}} \mathcal{D}_{{HH^\dagger V_{1}D}}^{{p_2}} \mathcal{D}_{{QQ^\dagger V_{1}}}^{{p_1f_4p_2}}}{{M_{V_{1}}^2}}
+\frac{i {y}_{{u}}^{{p_1f_5}} \mathcal{D}_{{HH^\dagger V_{4}D}}^{{p_2*}} \mathcal{D}_{{QQ^\dagger V_{4}}}^{{p_1f_4p_2}}}{{M_{V_{4}}^2}}
-\frac{i {y}_{{u}}^{{p_1f_5}} \mathcal{D}_{{HH^\dagger V_{4}D}}^{{p_2}} \mathcal{D}_{{QQ^\dagger V_{4}}}^{{p_1f_4p_2}}}{{M_{V_{4}}^2}}
+\frac{2 \mathcal{D}_{{HH^\dagger H^\dagger S_{4}(1)}}^{{p_1}} \mathcal{D}_{{QS_{4}u^\dagger }}^{{f_4p_1f_5*}}}{{M_{S_{4}}^2}}
-\frac{2 \mathcal{C}_{{HH^\dagger S_{1}}}^{{p_1}} \mathcal{C}_{{HS_{1}S_{4}^\dagger }}^{{p_1p_2*}} \mathcal{D}_{{QS_{4}u^\dagger }}^{{f_4p_2f_5*}}}{{M_{S_{1}}^2} {M_{S_{4}}^2}}
-\frac{2 \mathcal{C}_{{HHS_{6}^\dagger }}^{{p_1*}} \mathcal{C}_{{HS_{4}S_{6}^\dagger }}^{{p_2p_1}} \mathcal{D}_{{QS_{4}u^\dagger }}^{{f_4p_2f_5*}}}{{M_{S_{4}}^2} {M_{S_{6}}^2}}
-\frac{\mathcal{C}_{{HH^\dagger S_{5}}}^{{p_1}} \mathcal{C}_{{H^\dagger S_{4}S_{5}}}^{{p_2p_1}} \mathcal{D}_{{QS_{4}u^\dagger }}^{{f_4p_2f_5*}}}{{M_{S_{4}}^2} {M_{S_{5}}^2}}
+\frac{2 i {y}_{{u}}^{{f_4p_1}} \mathcal{D}_{{HH^\dagger V_{1}D}}^{{p_2*}} \mathcal{D}_{{u^\dagger uV_{1}}}^{{p_1f_5p_2}}}{{M_{V_{1}}^2}}
+\frac{2 i {y}_{{u}}^{{f_4p_1}} \mathcal{D}_{{HH^\dagger V_{1}D}}^{{p_2}} \mathcal{D}_{{u^\dagger uV_{1}}}^{{p_1f_5p_2}}}{{M_{V_{1}}^2}} 
\end{autobreak}
\end{align}
\begin{align}
\begin{autobreak}
 {C_{dH}} =
 \frac{(\mathcal{C}_{{HH^\dagger S_{5}}}^{{p_1}})^2 {y}_{{d}}^{{f_5f_4}}}{2 {M_{S_{5}}^4}}
-\frac{(\mathcal{D}_{{HH^\dagger V_{1}D}}^{{p_1*}})^2 {y}_{{d}}^{{f_5f_4}}}{{M_{V_{1}}^2}}
+\frac{(\mathcal{D}_{{HH^\dagger V_{1}D}}^{{p_1}})^2 {y}_{{d}}^{{f_5f_4}}}{{M_{V_{1}}^2}}
-\frac{(\mathcal{D}_{{HH^\dagger V_{4}D}}^{{p_1*}})^2 {y}_{{d}}^{{f_5f_4}}}{2 {M_{V_{4}}^2}}
+\frac{(\mathcal{D}_{{HH^\dagger V_{4}D}}^{{p_1}})^2 {y}_{{d}}^{{f_5f_4}}}{2 {M_{V_{4}}^2}}
+\frac{\mathcal{C}_{{HHS_{6}^\dagger }}^{{p_1*}} \mathcal{C}_{{HHS_{6}^\dagger }}^{{p_1}} {y}_{{d}}^{{f_5f_4}}}{{M_{S_{6}}^4}}
-\frac{2 \mathcal{D}_{{HHV_{2}^\dagger D}}^{{p_1*}} \mathcal{D}_{{HHV_{2}^\dagger D}}^{{p_1}} {y}_{{d}}^{{f_5f_4}}}{{M_{V_{2}}^2}}
+\frac{2 \mathcal{D}_{{HH^\dagger V_{4}D}}^{{p_1*}} \mathcal{D}_{{HH^\dagger V_{4}D}}^{{p_1}} {y}_{{d}}^{{f_5f_4}}}{{M_{V_{4}}^2}}
+\frac{2 {y}_{{d}}^{{f_5p_1}} \mathcal{D}_{{d^\dagger F_{10L}H}}^{{f_4p_2*}} \mathcal{D}_{{d^\dagger F_{10L}H}}^{{p_1p_2}}}{{M_{F_{10}}^2}}
+\frac{2 {y}_{{d}}^{{f_5p_1}} \mathcal{D}_{{d^\dagger F_{11L}H^\dagger }}^{{f_4p_2*}} \mathcal{D}_{{d^\dagger F_{11L}H^\dagger }}^{{p_1p_2}}}{{M_{F_{11}}^2}}
-\frac{4 \mathcal{D}_{{d^\dagger F_{10L}H}}^{{f_4p_1*}} \mathcal{D}_{{F_{10R}^\dagger F_{13L}H^\dagger }}^{{p_1p_2*}} \mathcal{D}_{{F_{13R}^\dagger H^\dagger Q}}^{{p_2f_5*}}}{{M_{F_{10}}} {M_{F_{13}}}}
-\frac{4 \mathcal{D}_{{d^\dagger F_{11L}H^\dagger }}^{{f_4p_1*}} \mathcal{D}_{{F_{11R}^\dagger F_{13L}H}}^{{p_1p_2*}} \mathcal{D}_{{F_{13R}^\dagger H^\dagger Q}}^{{p_2f_5*}}}{{M_{F_{11}}} {M_{F_{13}}}}
+\frac{{y}_{{d}}^{{p_1f_4}} \mathcal{D}_{{F_{13R}^\dagger H^\dagger Q}}^{{p_2f_5*}} \mathcal{D}_{{F_{13R}^\dagger H^\dagger Q}}^{{p_2p_1}}}{{M_{F_{13}}^2}}
+\frac{8 \mathcal{D}_{{d^\dagger F_{11L}H^\dagger }}^{{f_4p_1*}} \mathcal{D}_{{F_{11R}^\dagger F_{14L}H^\dagger }}^{{p_1p_2*}} \mathcal{D}_{{F_{14R}^\dagger HQ}}^{{p_2f_5*}}}{{M_{F_{11}}} {M_{F_{14}}}}
+\frac{2 {y}_{{d}}^{{p_1f_4}} \mathcal{D}_{{F_{14R}^\dagger HQ}}^{{p_2f_5*}} \mathcal{D}_{{F_{14R}^\dagger HQ}}^{{p_2p_1}}}{{M_{F_{14}}^2}}
+\frac{8 \mathcal{D}_{{d^\dagger F_{10L}H}}^{{f_4p_1*}} \mathcal{D}_{{F_{10R}^\dagger F_{8L}H^\dagger }}^{{p_1p_2*}} \mathcal{D}_{{F_{8R}^\dagger H^\dagger Q}}^{{p_2f_5*}}}{{M_{F_{10}}} {M_{F_{8}}}}
-\frac{8 \mathcal{D}_{{d^\dagger F_{11L}H^\dagger }}^{{f_4p_1*}} \mathcal{D}_{{F_{11R}^\dagger F_{8L}H}}^{{p_1p_2*}} \mathcal{D}_{{F_{8R}^\dagger H^\dagger Q}}^{{p_2f_5*}}}{{M_{F_{11}}} {M_{F_{8}}}}
+\frac{2 {y}_{{d}}^{{p_1f_4}} \mathcal{D}_{{F_{8R}^\dagger H^\dagger Q}}^{{p_2f_5*}} \mathcal{D}_{{F_{8R}^\dagger H^\dagger Q}}^{{p_2p_1}}}{{M_{F_{8}}^2}}
+\frac{4 \mathcal{D}_{{d^\dagger F_{10L}H}}^{{f_4p_1*}} \mathcal{D}_{{F_{10R}^\dagger QS_{6}^\dagger }}^{{p_1f_5p_2*}} \mathcal{C}_{{HHS_{6}^\dagger }}^{{p_2}}}{{M_{F_{10}}} {M_{S_{6}}^2}}
+\frac{4 \mathcal{D}_{{d^\dagger F_{14L}S_{6}^\dagger }}^{{f_4p_1p_2*}} \mathcal{D}_{{F_{14R}^\dagger HQ}}^{{p_1f_5*}} \mathcal{C}_{{HHS_{6}^\dagger }}^{{p_2}}}{{M_{F_{14}}} {M_{S_{6}}^2}}
+\frac{2 \mathcal{D}_{{d^\dagger QS_{4}^\dagger }}^{{f_4f_5p_1*}} \mathcal{D}_{{HH^\dagger H^\dagger S_{4}(1)}}^{{p_1*}}}{{M_{S_{4}}^2}}
-\frac{4 \mathcal{D}_{{d^\dagger F_{11L}H^\dagger }}^{{f_4p_1*}} \mathcal{D}_{{F_{11R}^\dagger QS_{1}}}^{{p_1f_5p_2*}} \mathcal{C}_{{HH^\dagger S_{1}}}^{{p_2}}}{{M_{F_{11}}} {M_{S_{1}}^2}}
-\frac{4 \mathcal{D}_{{d^\dagger F_{8L}S_{1}}}^{{f_4p_1p_2*}} \mathcal{D}_{{F_{8R}^\dagger H^\dagger Q}}^{{p_1f_5*}} \mathcal{C}_{{HH^\dagger S_{1}}}^{{p_2}}}{{M_{F_{8}}} {M_{S_{1}}^2}}
-\frac{2 \mathcal{D}_{{d^\dagger F_{11L}H^\dagger }}^{{f_4p_1*}} \mathcal{D}_{{F_{11R}^\dagger QS_{5}}}^{{p_1f_5p_2*}} \mathcal{C}_{{HH^\dagger S_{5}}}^{{p_2}}}{{M_{F_{11}}} {M_{S_{5}}^2}}
-\frac{2 \mathcal{D}_{{d^\dagger F_{13L}S_{5}}}^{{f_4p_1p_2*}} \mathcal{D}_{{F_{13R}^\dagger H^\dagger Q}}^{{p_1f_5*}} \mathcal{C}_{{HH^\dagger S_{5}}}^{{p_2}}}{{M_{F_{13}}} {M_{S_{5}}^2}}
+\frac{2 i {y}_{{d}}^{{f_5p_1}} \mathcal{D}_{{d^\dagger dV_{1}}}^{{p_1f_4p_2}} \mathcal{D}_{{HH^\dagger V_{1}D}}^{{p_2*}}}{{M_{V_{1}}^2}}
+\frac{2 i {y}_{{d}}^{{f_5p_1}} \mathcal{D}_{{d^\dagger dV_{1}}}^{{p_1f_4p_2}} \mathcal{D}_{{HH^\dagger V_{1}D}}^{{p_2}}}{{M_{V_{1}}^2}}
-\frac{2 \mathcal{D}_{{d^\dagger QS_{4}^\dagger }}^{{f_4f_5p_1*}} \mathcal{C}_{{HH^\dagger S_{1}}}^{{p_2}} \mathcal{C}_{{HS_{1}S_{4}^\dagger }}^{{p_2p_1}}}{{M_{S_{1}}^2} {M_{S_{4}}^2}}
-\frac{2 \mathcal{D}_{{d^\dagger QS_{4}^\dagger }}^{{f_4f_5p_1*}} \mathcal{C}_{{HHS_{6}^\dagger }}^{{p_2}} \mathcal{C}_{{HS_{4}S_{6}^\dagger }}^{{p_1p_2*}}}{{M_{S_{4}}^2} {M_{S_{6}}^2}}
+\frac{\mathcal{D}_{{d^\dagger QS_{4}^\dagger }}^{{f_4f_5p_1*}} \mathcal{C}_{{HH^\dagger S_{5}}}^{{p_2}} \mathcal{C}_{{H^\dagger S_{4}S_{5}}}^{{p_1p_2*}}}{{M_{S_{4}}^2} {M_{S_{5}}^2}}
+\frac{2 i {y}_{{d}}^{{p_1f_4}} \mathcal{D}_{{HH^\dagger V_{1}D}}^{{p_2*}} \mathcal{D}_{{QQ^\dagger V_{1}}}^{{p_1f_5p_2}}}{{M_{V_{1}}^2}}
+\frac{2 i {y}_{{d}}^{{p_1f_4}} \mathcal{D}_{{HH^\dagger V_{1}D}}^{{p_2}} \mathcal{D}_{{QQ^\dagger V_{1}}}^{{p_1f_5p_2}}}{{M_{V_{1}}^2}}
-\frac{i {y}_{{d}}^{{p_1f_4}} \mathcal{D}_{{HH^\dagger V_{4}D}}^{{p_2*}} \mathcal{D}_{{QQ^\dagger V_{4}}}^{{p_1f_5p_2}}}{{M_{V_{4}}^2}}
+\frac{i {y}_{{d}}^{{p_1f_4}} \mathcal{D}_{{HH^\dagger V_{4}D}}^{{p_2}} \mathcal{D}_{{QQ^\dagger V_{4}}}^{{p_1f_5p_2}}}{{M_{V_{4}}^2}} 
\end{autobreak}
\end{align}

\subsubsection{Mixed operators}

\begin{align}
\begin{autobreak}
 {C_{Hl}^{(1)}} =
 
-\frac{i \mathcal{D}_{{LL^\dagger V_{1}}}^{{f_1f_4p_1}} \mathcal{D}_{{HH^\dagger V_{1}D}}^{{p_1}}}{{M_{V_{1}}^2}}
+\frac{i \mathcal{D}_{{LL^\dagger V_{1}}}^{{f_1f_4p_1}} \mathcal{D}_{{HH^\dagger V_{1}D}}^{{p_1*}}}{{M_{V_{1}}^2}}
+\frac{\mathcal{D}_{{F_{1}HL}}^{{p_1f_1}} \mathcal{D}_{{F_{1}HL}}^{{p_1f_4*}}}{4 {M_{F_{1}}^2}}
-\frac{\mathcal{D}_{{F_{2L}H^\dagger L}}^{{p_1f_1}} \mathcal{D}_{{F_{2L}H^\dagger L}}^{{p_1f_4*}}}{4 {M_{F_{2}}^2}}
+\frac{3 \mathcal{D}_{{F_{5}HL}}^{{p_1f_1}} \mathcal{D}_{{F_{5}HL}}^{{p_1f_4*}}}{8 {M_{F_{5}}^2}}
-\frac{3 \mathcal{D}_{{F_{6L}H^\dagger L}}^{{p_1f_1}} \mathcal{D}_{{F_{6L}H^\dagger L}}^{{p_1f_4*}}}{8 {M_{F_{6}}^2}} 
\end{autobreak}
\end{align}
\begin{align}
\begin{autobreak}
 {C_{Hl}^{(3)}} =
 
-\frac{i \mathcal{D}_{{LL^\dagger V_{4}}}^{{f_1f_4p_1}} \mathcal{D}_{{HH^\dagger V_{4}D}}^{{p_1}}}{2 {M_{V_{4}}^2}}
-\frac{i \mathcal{D}_{{LL^\dagger V_{4}}}^{{f_1f_4p_1}} \mathcal{D}_{{HH^\dagger V_{4}D}}^{{p_1*}}}{2 {M_{V_{4}}^2}}
-\frac{\mathcal{D}_{{F_{1}HL}}^{{p_1f_1}} \mathcal{D}_{{F_{1}HL}}^{{p_1f_4*}}}{4 {M_{F_{1}}^2}}
-\frac{\mathcal{D}_{{F_{2L}H^\dagger L}}^{{p_1f_1}} \mathcal{D}_{{F_{2L}H^\dagger L}}^{{p_1f_4*}}}{4 {M_{F_{2}}^2}}
+\frac{\mathcal{D}_{{F_{5}HL}}^{{p_1f_1}} \mathcal{D}_{{F_{5}HL}}^{{p_1f_4*}}}{8 {M_{F_{5}}^2}}
+\frac{\mathcal{D}_{{F_{6L}H^\dagger L}}^{{p_1f_1}} \mathcal{D}_{{F_{6L}H^\dagger L}}^{{p_1f_4*}}}{8 {M_{F_{6}}^2}} 
\end{autobreak}
\end{align}
\begin{align}
\begin{autobreak}
 {C_{He}} =
 \frac{i \mathcal{D}_{{e^\dagger eV_{1}}}^{{f_1f_4p_1}} \mathcal{D}_{{HH^\dagger V_{1}D}}^{{p_1}}}{{M_{V_{1}}^2}}
-\frac{i \mathcal{D}_{{e^\dagger eV_{1}}}^{{f_1f_4p_1}} \mathcal{D}_{{HH^\dagger V_{1}D}}^{{p_1*}}}{{M_{V_{1}}^2}}
+\frac{\mathcal{D}_{{e^\dagger F_{3R}^\dagger H^\dagger }}^{{f_1p_1}} \mathcal{D}_{{e^\dagger F_{3R}^\dagger H^\dagger }}^{{f_4p_1*}}}{2 {M_{F_{3}}^2}}
-\frac{\mathcal{D}_{{e^\dagger F_{4R}^\dagger H}}^{{f_1p_1}} \mathcal{D}_{{e^\dagger F_{4R}^\dagger H}}^{{f_4p_1*}}}{2 {M_{F_{4}}^2}} 
\end{autobreak}
\end{align}
\begin{align}
\begin{autobreak}
 {C_{Hq}^{(1)}} =
 
-\frac{3 \mathcal{D}_{{F_{13R}^\dagger H^\dagger Q}}^{{p_1f_1}} \mathcal{D}_{{F_{13R}^\dagger H^\dagger Q}}^{{p_1f_4*}}}{2 {M_{F_{13}}^2}}
+\frac{3 \mathcal{D}_{{F_{14R}^\dagger HQ}}^{{p_1f_1}} \mathcal{D}_{{F_{14R}^\dagger HQ}}^{{p_1f_4*}}}{2 {M_{F_{14}}^2}}
-\frac{2 i \mathcal{D}_{{QQ^\dagger V_{1}}}^{{f_1f_4p_1}} \mathcal{D}_{{HH^\dagger V_{1}D}}^{{p_1}}}{{M_{V_{1}}^2}}
+\frac{2 i \mathcal{D}_{{QQ^\dagger V_{1}}}^{{f_1f_4p_1}} \mathcal{D}_{{HH^\dagger V_{1}D}}^{{p_1*}}}{{M_{V_{1}}^2}}
-\frac{\mathcal{D}_{{F_{8R}^\dagger H^\dagger Q}}^{{p_1f_1}} \mathcal{D}_{{F_{8R}^\dagger H^\dagger Q}}^{{p_1f_4*}}}{{M_{F_{8}}^2}}
+\frac{\mathcal{D}_{{F_{9R}^\dagger HQ}}^{{p_1f_1}} \mathcal{D}_{{F_{9R}^\dagger HQ}}^{{p_1f_4*}}}{{M_{F_{9}}^2}} 
\end{autobreak}
\end{align}
\begin{align}
\begin{autobreak}
 {C_{Hq}^{(3)}} =
 \frac{\mathcal{D}_{{F_{13R}^\dagger H^\dagger Q}}^{{p_1f_1}} \mathcal{D}_{{F_{13R}^\dagger H^\dagger Q}}^{{p_1f_4*}}}{2 {M_{F_{13}}^2}}
+\frac{\mathcal{D}_{{F_{14R}^\dagger HQ}}^{{p_1f_1}} \mathcal{D}_{{F_{14R}^\dagger HQ}}^{{p_1f_4*}}}{2 {M_{F_{14}}^2}}
-\frac{i \mathcal{D}_{{QQ^\dagger V_{4}}}^{{f_1f_4p_1}} \mathcal{D}_{{HH^\dagger V_{4}D}}^{{p_1}}}{{M_{V_{4}}^2}}
-\frac{i \mathcal{D}_{{QQ^\dagger V_{4}}}^{{f_1f_4p_1}} \mathcal{D}_{{HH^\dagger V_{4}D}}^{{p_1*}}}{{M_{V_{4}}^2}}
-\frac{\mathcal{D}_{{F_{8R}^\dagger H^\dagger Q}}^{{p_1f_1}} \mathcal{D}_{{F_{8R}^\dagger H^\dagger Q}}^{{p_1f_4*}}}{{M_{F_{8}}^2}}
-\frac{\mathcal{D}_{{F_{9R}^\dagger HQ}}^{{p_1f_1}} \mathcal{D}_{{F_{9R}^\dagger HQ}}^{{p_1f_4*}}}{{M_{F_{9}}^2}} 
\end{autobreak}
\end{align}
\begin{align}
\begin{autobreak}
 {C_{Hu}} =
 
-\frac{2 \mathcal{D}_{{F_{11L}Hu^\dagger }}^{{p_1f_1}} \mathcal{D}_{{F_{11L}Hu^\dagger }}^{{p_1f_4*}}}{{M_{F_{11}}^2}}
+\frac{2 \mathcal{D}_{{F_{12L}H^\dagger u^\dagger }}^{{p_1f_1}} \mathcal{D}_{{F_{12L}H^\dagger u^\dagger }}^{{p_1f_4*}}}{{M_{F_{12}}^2}}
+\frac{2 i \mathcal{D}_{{u^\dagger uV_{1}}}^{{f_1f_4p_1}} \mathcal{D}_{{HH^\dagger V_{1}D}}^{{p_1}}}{{M_{V_{1}}^2}}
-\frac{2 i \mathcal{D}_{{u^\dagger uV_{1}}}^{{f_1f_4p_1}} \mathcal{D}_{{HH^\dagger V_{1}D}}^{{p_1*}}}{{M_{V_{1}}^2}} 
\end{autobreak}
\end{align}
\begin{align}
\begin{autobreak}
 {C_{Hd}} =
 \frac{2 i \mathcal{D}_{{d^\dagger dV_{1}}}^{{f_1f_4p_1}} \mathcal{D}_{{HH^\dagger V_{1}D}}^{{p_1}}}{{M_{V_{1}}^2}}
-\frac{2 i \mathcal{D}_{{d^\dagger dV_{1}}}^{{f_1f_4p_1}} \mathcal{D}_{{HH^\dagger V_{1}D}}^{{p_1*}}}{{M_{V_{1}}^2}}
-\frac{2 \mathcal{D}_{{d^\dagger F_{10L}H}}^{{f_1p_1}} \mathcal{D}_{{d^\dagger F_{10L}H}}^{{f_4p_1*}}}{{M_{F_{10}}^2}}
+\frac{2 \mathcal{D}_{{d^\dagger F_{11L}H^\dagger }}^{{f_1p_1}} \mathcal{D}_{{d^\dagger F_{11L}H^\dagger }}^{{f_4p_1*}}}{{M_{F_{11}}^2}} 
\end{autobreak}
\end{align}
\begin{align}
\begin{autobreak}
 {C_{Hud}} =
 \frac{4 \mathcal{D}_{{d^\dagger F_{11L}H^\dagger }}^{{f_4p_1*}} \mathcal{D}_{{F_{11L}Hu^\dagger }}^{{p_1f_1}}}{{M_{F_{11}}^2}}
+\frac{4 i \mathcal{D}_{{d^\dagger uV_{2}^\dagger }}^{{f_4f_1p_1*}} \mathcal{D}_{{HHV_{2}^\dagger D}}^{{p_1}}}{{M_{V_{2}}^2}} 
\end{autobreak}
\end{align}

\subsubsection{4-fermion operators}

\begin{align}
\begin{autobreak}
 {C_{ll}} =
 
-\frac{\mathcal{D}_{{LLS_{2}}}^{{f_1f_2p_1}} \mathcal{D}_{{LLS_{2}}}^{{f_4f_3p_1*}}}{2 {M_{S_{2}}^2}}
+\frac{\mathcal{D}_{{LLS_{6}}}^{{f_1f_2p_1}} \mathcal{D}_{{LLS_{6}}}^{{f_4f_3p_1*}}}{4 {M_{S_{6}}^2}}
-\frac{\mathcal{D}_{{LL^\dagger V_{1}}}^{{f_1f_3p_1}} \mathcal{D}_{{LL^\dagger V_{1}}}^{{f_2f_4p_1}}}{2 {M_{V_{1}}^2}}
+\frac{\mathcal{D}_{{LL^\dagger V_{4}}}^{{f_1f_3p_1}} \mathcal{D}_{{LL^\dagger V_{4}}}^{{f_2f_4p_1}}}{4 {M_{V_{4}}^2}}
-\frac{\mathcal{D}_{{LL^\dagger V_{4}}}^{{f_1f_4p_1}} \mathcal{D}_{{LL^\dagger V_{4}}}^{{f_2f_3p_1}}}{2 {M_{V_{4}}^2}}
+\frac{\mathcal{D}_{{LLS_{2}}}^{{f_2f_1p_1}} \mathcal{D}_{{LLS_{2}}}^{{f_4f_3p_1*}}}{2 {M_{S_{2}}^2}}
+\frac{\mathcal{D}_{{LLS_{6}}}^{{f_2f_1p_1}} \mathcal{D}_{{LLS_{6}}}^{{f_4f_3p_1*}}}{4 {M_{S_{6}}^2}} 
\end{autobreak}
\end{align}
\begin{align}
\begin{autobreak}
 {C_{qq}^{(1)}} =
 \frac{\mathcal{D}_{{QQS_{10}}}^{{f_1f_2p_1}} \mathcal{D}_{{QQS_{10}}}^{{f_4f_3p_1*}}}{4 {M_{S_{10}}^2}}
-\frac{3 \mathcal{D}_{{QQS_{14}}}^{{f_1f_2p_1}} \mathcal{D}_{{QQS_{14}}}^{{f_4f_3p_1*}}}{8 {M_{S_{14}}^2}}
-\frac{2 \mathcal{D}_{{QQS_{16}^\dagger }}^{{f_1f_2p_1}} \mathcal{D}_{{QQS_{16}^\dagger }}^{{f_4f_3p_1*}}}{{M_{S_{16}}^2}}
+\frac{3 \mathcal{D}_{{QQS_{18}^\dagger }}^{{f_1f_2p_1}} \mathcal{D}_{{QQS_{18}^\dagger }}^{{f_4f_3p_1*}}}{{M_{S_{18}}^2}}
-\frac{2 \mathcal{D}_{{QQ^\dagger V_{1}}}^{{f_1f_3p_1}} \mathcal{D}_{{QQ^\dagger V_{1}}}^{{f_2f_4p_1}}}{{M_{V_{1}}^2}}
+\frac{2 \mathcal{D}_{{QQ^\dagger V_{12}}}^{{f_1f_3p_1}} \mathcal{D}_{{QQ^\dagger V_{12}}}^{{f_2f_4p_1}}}{3 {M_{V_{12}}^2}}
-\frac{\mathcal{D}_{{QQ^\dagger V_{12}}}^{{f_1f_4p_1}} \mathcal{D}_{{QQ^\dagger V_{12}}}^{{f_2f_3p_1}}}{{M_{V_{12}}^2}}
-\frac{3 \mathcal{D}_{{QQ^\dagger V_{14}}}^{{f_1f_4p_1}} \mathcal{D}_{{QQ^\dagger V_{14}}}^{{f_2f_3p_1}}}{2 {M_{V_{14}}^2}}
+\frac{\mathcal{D}_{{QQS_{10}}}^{{f_2f_1p_1}} \mathcal{D}_{{QQS_{10}}}^{{f_4f_3p_1*}}}{4 {M_{S_{10}}^2}}
+\frac{3 \mathcal{D}_{{QQS_{14}}}^{{f_2f_1p_1}} \mathcal{D}_{{QQS_{14}}}^{{f_4f_3p_1*}}}{8 {M_{S_{14}}^2}}
+\frac{2 \mathcal{D}_{{QQS_{16}^\dagger }}^{{f_2f_1p_1}} \mathcal{D}_{{QQS_{16}^\dagger }}^{{f_4f_3p_1*}}}{{M_{S_{16}}^2}}
+\frac{3 \mathcal{D}_{{QQS_{18}^\dagger }}^{{f_2f_1p_1}} \mathcal{D}_{{QQS_{18}^\dagger }}^{{f_4f_3p_1*}}}{{M_{S_{18}}^2}} 
\end{autobreak}
\end{align}
\begin{align}
\begin{autobreak}
 {C_{qq}^{(3)}} =
 
-\frac{\mathcal{D}_{{QQS_{10}}}^{{f_1f_2p_1}} \mathcal{D}_{{QQS_{10}}}^{{f_4f_3p_1*}}}{4 {M_{S_{10}}^2}}
-\frac{\mathcal{D}_{{QQS_{14}}}^{{f_1f_2p_1}} \mathcal{D}_{{QQS_{14}}}^{{f_4f_3p_1*}}}{8 {M_{S_{14}}^2}}
+\frac{2 \mathcal{D}_{{QQS_{16}^\dagger }}^{{f_1f_2p_1}} \mathcal{D}_{{QQS_{16}^\dagger }}^{{f_4f_3p_1*}}}{{M_{S_{16}}^2}}
+\frac{\mathcal{D}_{{QQS_{18}^\dagger }}^{{f_1f_2p_1}} \mathcal{D}_{{QQS_{18}^\dagger }}^{{f_4f_3p_1*}}}{{M_{S_{18}}^2}}
+\frac{\mathcal{D}_{{QQ^\dagger V_{14}}}^{{f_1f_3p_1}} \mathcal{D}_{{QQ^\dagger V_{14}}}^{{f_2f_4p_1}}}{3 {M_{V_{14}}^2}}
-\frac{\mathcal{D}_{{QQ^\dagger V_{4}}}^{{f_1f_3p_1}} \mathcal{D}_{{QQ^\dagger V_{4}}}^{{f_2f_4p_1}}}{{M_{V_{4}}^2}}
-\frac{\mathcal{D}_{{QQ^\dagger V_{12}}}^{{f_1f_4p_1}} \mathcal{D}_{{QQ^\dagger V_{12}}}^{{f_2f_3p_1}}}{{M_{V_{12}}^2}}
+\frac{\mathcal{D}_{{QQ^\dagger V_{14}}}^{{f_1f_4p_1}} \mathcal{D}_{{QQ^\dagger V_{14}}}^{{f_2f_3p_1}}}{2 {M_{V_{14}}^2}}
-\frac{\mathcal{D}_{{QQS_{10}}}^{{f_2f_1p_1}} \mathcal{D}_{{QQS_{10}}}^{{f_4f_3p_1*}}}{4 {M_{S_{10}}^2}}
+\frac{\mathcal{D}_{{QQS_{14}}}^{{f_2f_1p_1}} \mathcal{D}_{{QQS_{14}}}^{{f_4f_3p_1*}}}{8 {M_{S_{14}}^2}}
-\frac{2 \mathcal{D}_{{QQS_{16}^\dagger }}^{{f_2f_1p_1}} \mathcal{D}_{{QQS_{16}^\dagger }}^{{f_4f_3p_1*}}}{{M_{S_{16}}^2}}
+\frac{\mathcal{D}_{{QQS_{18}^\dagger }}^{{f_2f_1p_1}} \mathcal{D}_{{QQS_{18}^\dagger }}^{{f_4f_3p_1*}}}{{M_{S_{18}}^2}} 
\end{autobreak}
\end{align}
\begin{align}
\begin{autobreak}
 {C_{lq}^{(1)}} =
 \frac{\mathcal{D}_{{LQS_{10}^\dagger }}^{{f_1f_2p_1}} \mathcal{D}_{{LQS_{10}^\dagger }}^{{f_3f_4p_1*}}}{{M_{S_{10}}^2}}
+\frac{3 \mathcal{D}_{{LQS_{14}^\dagger }}^{{f_1f_2p_1}} \mathcal{D}_{{LQS_{14}^\dagger }}^{{f_3f_4p_1*}}}{2 {M_{S_{14}}^2}}
-\frac{2 \mathcal{D}_{{LL^\dagger V_{1}}}^{{f_1f_3p_1}} \mathcal{D}_{{QQ^\dagger V_{1}}}^{{f_2f_4p_1}}}{{M_{V_{1}}^2}}
-\frac{2 \mathcal{D}_{{L^\dagger QV_{5}^\dagger }}^{{f_1f_4p_1*}} \mathcal{D}_{{L^\dagger QV_{5}^\dagger }}^{{f_3f_2p_1}}}{{M_{V_{5}}^2}}
-\frac{3 \mathcal{D}_{{L^\dagger QV_{9}^\dagger }}^{{f_1f_4p_1*}} \mathcal{D}_{{L^\dagger QV_{9}^\dagger }}^{{f_3f_2p_1}}}{{M_{V_{9}}^2}} 
\end{autobreak}
\end{align}
\begin{align}
\begin{autobreak}
 {C_{lq}^{(3)}} =
 
-\frac{\mathcal{D}_{{LQS_{10}^\dagger }}^{{f_1f_2p_1}} \mathcal{D}_{{LQS_{10}^\dagger }}^{{f_3f_4p_1*}}}{{M_{S_{10}}^2}}
+\frac{\mathcal{D}_{{LQS_{14}^\dagger }}^{{f_1f_2p_1}} \mathcal{D}_{{LQS_{14}^\dagger }}^{{f_3f_4p_1*}}}{2 {M_{S_{14}}^2}}
-\frac{\mathcal{D}_{{LL^\dagger V_{4}}}^{{f_1f_3p_1}} \mathcal{D}_{{QQ^\dagger V_{4}}}^{{f_2f_4p_1}}}{{M_{V_{4}}^2}}
-\frac{2 \mathcal{D}_{{L^\dagger QV_{5}^\dagger }}^{{f_1f_4p_1*}} \mathcal{D}_{{L^\dagger QV_{5}^\dagger }}^{{f_3f_2p_1}}}{{M_{V_{5}}^2}}
+\frac{\mathcal{D}_{{L^\dagger QV_{9}^\dagger }}^{{f_1f_4p_1*}} \mathcal{D}_{{L^\dagger QV_{9}^\dagger }}^{{f_3f_2p_1}}}{{M_{V_{9}}^2}} 
\end{autobreak}
\end{align}
\begin{align}
\begin{autobreak}
 {C_{ee}} =
 \frac{\mathcal{D}_{{e^\dagger e^\dagger S_{3}^\dagger }}^{{f_1f_2p_1}} \mathcal{D}_{{e^\dagger e^\dagger S_{3}^\dagger }}^{{f_4f_3p_1*}}}{2 {M_{S_{3}}^2}}
-\frac{\mathcal{D}_{{e^\dagger eV_{1}}}^{{f_1f_3p_1}} \mathcal{D}_{{e^\dagger eV_{1}}}^{{f_2f_4p_1}}}{2 {M_{V_{1}}^2}} 
\end{autobreak}
\end{align}
\begin{align}
\begin{autobreak}
 {C_{uu}} =
 
-\frac{2 \mathcal{D}_{{u^\dagger uV_{1}}}^{{f_1f_3p_1}} \mathcal{D}_{{u^\dagger uV_{1}}}^{{f_2f_4p_1}}}{{M_{V_{1}}^2}}
+\frac{8 \mathcal{D}_{{u^\dagger uV_{12}}}^{{f_1f_3p_1}} \mathcal{D}_{{u^\dagger uV_{12}}}^{{f_2f_4p_1}}}{3 {M_{V_{12}}^2}}
-\frac{8 \mathcal{D}_{{u^\dagger uV_{12}}}^{{f_1f_4p_1}} \mathcal{D}_{{u^\dagger uV_{12}}}^{{f_2f_3p_1}}}{{M_{V_{12}}^2}}
+\frac{4 \mathcal{D}_{{S_{17}u^\dagger u^\dagger }}^{{p_1f_1f_2}} \mathcal{D}_{{S_{17}u^\dagger u^\dagger }}^{{p_1f_4f_3*}}}{{M_{S_{17}}^2}}
+\frac{4 \mathcal{D}_{{S_{17}u^\dagger u^\dagger }}^{{p_1f_2f_1}} \mathcal{D}_{{S_{17}u^\dagger u^\dagger }}^{{p_1f_4f_3*}}}{{M_{S_{17}}^2}}
-\frac{8 \mathcal{D}_{{S_{9}^\dagger u^\dagger u^\dagger }}^{{p_1f_1f_2}} \mathcal{D}_{{S_{9}^\dagger u^\dagger u^\dagger }}^{{p_1f_4f_3*}}}{{M_{S_{9}}^2}}
+\frac{8 \mathcal{D}_{{S_{9}^\dagger u^\dagger u^\dagger }}^{{p_1f_2f_1}} \mathcal{D}_{{S_{9}^\dagger u^\dagger u^\dagger }}^{{p_1f_4f_3*}}}{{M_{S_{9}}^2}} 
\end{autobreak}
\end{align}
\begin{align}
\begin{autobreak}
 {C_{dd}} =
 
-\frac{2 \mathcal{D}_{{d^\dagger dV_{1}}}^{{f_1f_3p_1}} \mathcal{D}_{{d^\dagger dV_{1}}}^{{f_2f_4p_1}}}{{M_{V_{1}}^2}}
+\frac{8 \mathcal{D}_{{d^\dagger dV_{12}}}^{{f_1f_3p_1}} \mathcal{D}_{{d^\dagger dV_{12}}}^{{f_2f_4p_1}}}{3 {M_{V_{12}}^2}}
-\frac{8 \mathcal{D}_{{d^\dagger dV_{12}}}^{{f_1f_4p_1}} \mathcal{D}_{{d^\dagger dV_{12}}}^{{f_2f_3p_1}}}{{M_{V_{12}}^2}}
-\frac{8 \mathcal{D}_{{d^\dagger d^\dagger S_{11}^\dagger }}^{{f_1f_2p_1}} \mathcal{D}_{{d^\dagger d^\dagger S_{11}^\dagger }}^{{f_4f_3p_1*}}}{{M_{S_{11}}^2}}
+\frac{8 \mathcal{D}_{{d^\dagger d^\dagger S_{11}^\dagger }}^{{f_2f_1p_1}} \mathcal{D}_{{d^\dagger d^\dagger S_{11}^\dagger }}^{{f_4f_3p_1*}}}{{M_{S_{11}}^2}}
+\frac{4 \mathcal{D}_{{d^\dagger d^\dagger S_{15}}}^{{f_1f_2p_1}} \mathcal{D}_{{d^\dagger d^\dagger S_{15}}}^{{f_4f_3p_1*}}}{{M_{S_{15}}^2}}
+\frac{4 \mathcal{D}_{{d^\dagger d^\dagger S_{15}}}^{{f_2f_1p_1}} \mathcal{D}_{{d^\dagger d^\dagger S_{15}}}^{{f_4f_3p_1*}}}{{M_{S_{15}}^2}} 
\end{autobreak}
\end{align}
\begin{align}
\begin{autobreak}
 {C_{eu}} =
 
-\frac{4 \mathcal{D}_{{eu^\dagger V_{6}}}^{{f_1f_4p_1*}} \mathcal{D}_{{eu^\dagger V_{6}}}^{{f_3f_2p_1}}}{{M_{V_{6}}^2}}
-\frac{2 \mathcal{D}_{{e^\dagger eV_{1}}}^{{f_1f_3p_1}} \mathcal{D}_{{u^\dagger uV_{1}}}^{{f_2f_4p_1}}}{{M_{V_{1}}^2}}
+\frac{2 \mathcal{D}_{{e^\dagger S_{10}u^\dagger }}^{{f_1p_1f_2}} \mathcal{D}_{{e^\dagger S_{10}u^\dagger }}^{{f_3p_1f_4*}}}{{M_{S_{10}}^2}} 
\end{autobreak}
\end{align}
\begin{align}
\begin{autobreak}
 {C_{ed}} =
 
-\frac{2 \mathcal{D}_{{d^\dagger dV_{1}}}^{{f_1f_3p_1}} \mathcal{D}_{{e^\dagger eV_{1}}}^{{f_2f_4p_1}}}{{M_{V_{1}}^2}}
-\frac{4 \mathcal{D}_{{d^\dagger eV_{5}}}^{{f_1f_4p_1}} \mathcal{D}_{{d^\dagger eV_{5}}}^{{f_3f_2p_1*}}}{{M_{V_{5}}^2}}
+\frac{2 \mathcal{D}_{{d^\dagger e^\dagger S_{9}}}^{{f_1f_2p_1}} \mathcal{D}_{{d^\dagger e^\dagger S_{9}}}^{{f_3f_4p_1*}}}{{M_{S_{9}}^2}} 
\end{autobreak}
\end{align}
\begin{align}
\begin{autobreak}
 {C_{ud}^{(1)}} =
 
-\frac{4 \mathcal{D}_{{d^\dagger dV_{1}}}^{{f_1f_3p_1}} \mathcal{D}_{{u^\dagger uV_{1}}}^{{f_2f_4p_1}}}{{M_{V_{1}}^2}}
+\frac{16 \mathcal{D}_{{d^\dagger S_{10}^\dagger u^\dagger }}^{{f_1p_1f_2}} \mathcal{D}_{{d^\dagger S_{10}^\dagger u^\dagger }}^{{f_3p_1f_4*}}}{3 {M_{S_{10}}^2}}
+\frac{16 \mathcal{D}_{{d^\dagger S_{16}u^\dagger }}^{{f_1p_1f_2}} \mathcal{D}_{{d^\dagger S_{16}u^\dagger }}^{{f_3p_1f_4*}}}{3 {M_{S_{16}}^2}}
-\frac{128 \mathcal{D}_{{d^\dagger uV_{13}^\dagger }}^{{f_1f_4p_1}} \mathcal{D}_{{d^\dagger uV_{13}^\dagger }}^{{f_3f_2p_1*}}}{9 {M_{V_{13}}^2}}
-\frac{4 \mathcal{D}_{{d^\dagger uV_{2}^\dagger }}^{{f_1f_4p_1}} \mathcal{D}_{{d^\dagger uV_{2}^\dagger }}^{{f_3f_2p_1*}}}{3 {M_{V_{2}}^2}} 
\end{autobreak}
\end{align}
\begin{align}
\begin{autobreak}
 {C_{ud}^{(8)}} =
 
-\frac{32 \mathcal{D}_{{d^\dagger dV_{12}}}^{{f_1f_3p_1}} \mathcal{D}_{{u^\dagger uV_{12}}}^{{f_2f_4p_1}}}{{M_{V_{12}}^2}}
-\frac{16 \mathcal{D}_{{d^\dagger S_{10}^\dagger u^\dagger }}^{{f_1p_1f_2}} \mathcal{D}_{{d^\dagger S_{10}^\dagger u^\dagger }}^{{f_3p_1f_4*}}}{{M_{S_{10}}^2}}
+\frac{8 \mathcal{D}_{{d^\dagger S_{16}u^\dagger }}^{{f_1p_1f_2}} \mathcal{D}_{{d^\dagger S_{16}u^\dagger }}^{{f_3p_1f_4*}}}{{M_{S_{16}}^2}}
+\frac{32 \mathcal{D}_{{d^\dagger uV_{13}^\dagger }}^{{f_1f_4p_1}} \mathcal{D}_{{d^\dagger uV_{13}^\dagger }}^{{f_3f_2p_1*}}}{3 {M_{V_{13}}^2}}
-\frac{8 \mathcal{D}_{{d^\dagger uV_{2}^\dagger }}^{{f_1f_4p_1}} \mathcal{D}_{{d^\dagger uV_{2}^\dagger }}^{{f_3f_2p_1*}}}{{M_{V_{2}}^2}} 
\end{autobreak}
\end{align}
\begin{align}
\begin{autobreak}
 {C_{le}} =
 \frac{\mathcal{D}_{{e^\dagger eV_{1}}}^{{f_1f_3p_1}} \mathcal{D}_{{LL^\dagger V_{1}}}^{{f_2f_4p_1}}}{{M_{V_{1}}^2}}
-\frac{\mathcal{D}_{{e^\dagger LS_{4}^\dagger }}^{{f_1f_2p_1}} \mathcal{D}_{{e^\dagger LS_{4}^\dagger }}^{{f_3f_4p_1*}}}{2 {M_{S_{4}}^2}}
+\frac{\mathcal{D}_{{e^\dagger L^\dagger V_{3}^\dagger }}^{{f_1f_4p_1}} \mathcal{D}_{{e^\dagger L^\dagger V_{3}^\dagger }}^{{f_3f_2p_1*}}}{{M_{V_{3}}^2}} 
\end{autobreak}
\end{align}
\begin{align}
\begin{autobreak}
 {C_{lu}} =
 \frac{2 \mathcal{D}_{{LL^\dagger V_{1}}}^{{f_1f_3p_1}} \mathcal{D}_{{u^\dagger uV_{1}}}^{{f_2f_4p_1}}}{{M_{V_{1}}^2}}
+\frac{4 \mathcal{D}_{{L^\dagger u^\dagger V_{8}}}^{{f_1f_4p_1*}} \mathcal{D}_{{L^\dagger u^\dagger V_{8}}}^{{f_3f_2p_1}}}{{M_{V_{8}}^2}}
-\frac{2 \mathcal{D}_{{LS_{13}u^\dagger }}^{{f_1p_1f_2}} \mathcal{D}_{{LS_{13}u^\dagger }}^{{f_3p_1f_4*}}}{{M_{S_{13}}^2}} 
\end{autobreak}
\end{align}
\begin{align}
\begin{autobreak}
 {C_{ld}} =
 \frac{2 \mathcal{D}_{{d^\dagger dV_{1}}}^{{f_1f_3p_1}} \mathcal{D}_{{LL^\dagger V_{1}}}^{{f_2f_4p_1}}}{{M_{V_{1}}^2}}
-\frac{2 \mathcal{D}_{{d^\dagger LS_{12}}}^{{f_1f_2p_1}} \mathcal{D}_{{d^\dagger LS_{12}}}^{{f_3f_4p_1*}}}{{M_{S_{12}}^2}}
+\frac{4 \mathcal{D}_{{d^\dagger L^\dagger V_{7}}}^{{f_1f_4p_1}} \mathcal{D}_{{d^\dagger L^\dagger V_{7}}}^{{f_3f_2p_1*}}}{{M_{V_{7}}^2}} 
\end{autobreak}
\end{align}
\begin{align}
\begin{autobreak}
 {C_{qe}} =
 \frac{4 \mathcal{D}_{{eQV_{7}^\dagger }}^{{f_1f_4p_1*}} \mathcal{D}_{{eQV_{7}^\dagger }}^{{f_3f_2p_1}}}{{M_{V_{7}}^2}}
+\frac{2 \mathcal{D}_{{e^\dagger eV_{1}}}^{{f_1f_3p_1}} \mathcal{D}_{{QQ^\dagger V_{1}}}^{{f_2f_4p_1}}}{{M_{V_{1}}^2}}
-\frac{2 \mathcal{D}_{{e^\dagger QS_{13}^\dagger }}^{{f_1f_2p_1}} \mathcal{D}_{{e^\dagger QS_{13}^\dagger }}^{{f_3f_4p_1*}}}{{M_{S_{13}}^2}} 
\end{autobreak}
\end{align}
\begin{align}
\begin{autobreak}
 {C_{qu}^{(1)}} =
 \frac{4 \mathcal{D}_{{QQ^\dagger V_{1}}}^{{f_1f_3p_1}} \mathcal{D}_{{u^\dagger uV_{1}}}^{{f_2f_4p_1}}}{{M_{V_{1}}^2}}
+\frac{32 \mathcal{D}_{{QuV_{11}^\dagger }}^{{f_1f_4p_1}} \mathcal{D}_{{QuV_{11}^\dagger }}^{{f_3f_2p_1*}}}{3 {M_{V_{11}}^2}}
+\frac{2 \mathcal{D}_{{QuV_{7}}}^{{f_1f_4p_1}} \mathcal{D}_{{QuV_{7}}}^{{f_3f_2p_1*}}}{3 {M_{V_{7}}^2}}
-\frac{16 \mathcal{D}_{{QS_{19}u^\dagger }}^{{f_1p_1f_2}} \mathcal{D}_{{QS_{19}u^\dagger }}^{{f_3p_1f_4*}}}{9 {M_{S_{19}}^2}}
-\frac{2 \mathcal{D}_{{QS_{4}u^\dagger }}^{{f_1p_1f_2}} \mathcal{D}_{{QS_{4}u^\dagger }}^{{f_3p_1f_4*}}}{3 {M_{S_{4}}^2}} 
\end{autobreak}
\end{align}
\begin{align}
\begin{autobreak}
 {C_{qu}^{(8)}} =
 \frac{16 \mathcal{D}_{{QQ^\dagger V_{12}}}^{{f_1f_3p_1}} \mathcal{D}_{{u^\dagger uV_{12}}}^{{f_2f_4p_1}}}{{M_{V_{12}}^2}}
+\frac{16 \mathcal{D}_{{QuV_{11}^\dagger }}^{{f_1f_4p_1}} \mathcal{D}_{{QuV_{11}^\dagger }}^{{f_3f_2p_1*}}}{{M_{V_{11}}^2}}
-\frac{2 \mathcal{D}_{{QuV_{7}}}^{{f_1f_4p_1}} \mathcal{D}_{{QuV_{7}}}^{{f_3f_2p_1*}}}{{M_{V_{7}}^2}}
+\frac{4 \mathcal{D}_{{QS_{19}u^\dagger }}^{{f_1p_1f_2}} \mathcal{D}_{{QS_{19}u^\dagger }}^{{f_3p_1f_4*}}}{3 {M_{S_{19}}^2}}
-\frac{4 \mathcal{D}_{{QS_{4}u^\dagger }}^{{f_1p_1f_2}} \mathcal{D}_{{QS_{4}u^\dagger }}^{{f_3p_1f_4*}}}{{M_{S_{4}}^2}} 
\end{autobreak}
\end{align}
\begin{align}
\begin{autobreak}
 {C_{qd}^{(1)}} =
 \frac{32 \mathcal{D}_{{dQV_{10}^\dagger }}^{{f_1f_4p_1*}} \mathcal{D}_{{dQV_{10}^\dagger }}^{{f_3f_2p_1}}}{3 {M_{V_{10}}^2}}
+\frac{2 \mathcal{D}_{{dQV_{8}}}^{{f_1f_4p_1*}} \mathcal{D}_{{dQV_{8}}}^{{f_3f_2p_1}}}{3 {M_{V_{8}}^2}}
+\frac{4 \mathcal{D}_{{d^\dagger dV_{1}}}^{{f_1f_3p_1}} \mathcal{D}_{{QQ^\dagger V_{1}}}^{{f_2f_4p_1}}}{{M_{V_{1}}^2}}
-\frac{64 \mathcal{D}_{{d^\dagger QS_{19}^\dagger }}^{{f_1f_2p_1}} \mathcal{D}_{{d^\dagger QS_{19}^\dagger }}^{{f_3f_4p_1*}}}{9 {M_{S_{19}}^2}}
-\frac{2 \mathcal{D}_{{d^\dagger QS_{4}^\dagger }}^{{f_1f_2p_1}} \mathcal{D}_{{d^\dagger QS_{4}^\dagger }}^{{f_3f_4p_1*}}}{3 {M_{S_{4}}^2}} 
\end{autobreak}
\end{align}
\begin{align}
\begin{autobreak}
 {C_{qd}^{(8)}} =
 \frac{16 \mathcal{D}_{{dQV_{10}^\dagger }}^{{f_1f_4p_1*}} \mathcal{D}_{{dQV_{10}^\dagger }}^{{f_3f_2p_1}}}{{M_{V_{10}}^2}}
-\frac{2 \mathcal{D}_{{dQV_{8}}}^{{f_1f_4p_1*}} \mathcal{D}_{{dQV_{8}}}^{{f_3f_2p_1}}}{{M_{V_{8}}^2}}
+\frac{16 \mathcal{D}_{{d^\dagger dV_{12}}}^{{f_1f_3p_1}} \mathcal{D}_{{QQ^\dagger V_{12}}}^{{f_2f_4p_1}}}{{M_{V_{12}}^2}}
+\frac{16 \mathcal{D}_{{d^\dagger QS_{19}^\dagger }}^{{f_1f_2p_1}} \mathcal{D}_{{d^\dagger QS_{19}^\dagger }}^{{f_3f_4p_1*}}}{3 {M_{S_{19}}^2}}
-\frac{4 \mathcal{D}_{{d^\dagger QS_{4}^\dagger }}^{{f_1f_2p_1}} \mathcal{D}_{{d^\dagger QS_{4}^\dagger }}^{{f_3f_4p_1*}}}{{M_{S_{4}}^2}} 
\end{autobreak}
\end{align}
\begin{align}
\begin{autobreak}
 {C_{ledq}} =
 
-\frac{8 \mathcal{D}_{{d^\dagger eV_{5}}}^{{f_1f_3p_1}} \mathcal{D}_{{L^\dagger QV_{5}^\dagger }}^{{f_4f_2p_1}}}{{M_{V_{5}}^2}}
-\frac{8 \mathcal{D}_{{d^\dagger L^\dagger V_{7}}}^{{f_1f_4p_1}} \mathcal{D}_{{eQV_{7}^\dagger }}^{{f_3f_2p_1}}}{{M_{V_{7}}^2}}
+\frac{2 \mathcal{D}_{{d^\dagger QS_{4}^\dagger }}^{{f_1f_2p_1}} \mathcal{D}_{{e^\dagger LS_{4}^\dagger }}^{{f_3f_4p_1*}}}{{M_{S_{4}}^2}} 
\end{autobreak}
\end{align}
\begin{align}
\begin{autobreak}
 {C_{quqd}^{(1)}} =
 \frac{4 \mathcal{D}_{{d^\dagger QS_{4}^\dagger }}^{{f_1f_3p_1*}} \mathcal{D}_{{QS_{4}u^\dagger }}^{{f_2p_1f_4*}}}{{M_{S_{4}}^2}}
-\frac{8 \mathcal{D}_{{d^\dagger S_{10}^\dagger u^\dagger }}^{{f_1p_1f_4*}} \mathcal{D}_{{QQS_{10}}}^{{f_2f_3p_1*}}}{3 {M_{S_{10}}^2}}
-\frac{8 \mathcal{D}_{{d^\dagger S_{10}^\dagger u^\dagger }}^{{f_1p_1f_4*}} \mathcal{D}_{{QQS_{10}}}^{{f_3f_2p_1*}}}{3 {M_{S_{10}}^2}}
-\frac{32 \mathcal{D}_{{d^\dagger S_{16}u^\dagger }}^{{f_1p_1f_4*}} \mathcal{D}_{{QQS_{16}^\dagger }}^{{f_2f_3p_1*}}}{3 {M_{S_{16}}^2}}
+\frac{32 \mathcal{D}_{{d^\dagger S_{16}u^\dagger }}^{{f_1p_1f_4*}} \mathcal{D}_{{QQS_{16}^\dagger }}^{{f_3f_2p_1*}}}{3 {M_{S_{16}}^2}} 
\end{autobreak}
\end{align}
\begin{align}
\begin{autobreak}
 {C_{quqd}^{(8)}} =
 \frac{16 \mathcal{D}_{{d^\dagger QS_{19}^\dagger }}^{{f_1f_3p_1*}} \mathcal{D}_{{QS_{19}u^\dagger }}^{{f_2p_1f_4*}}}{{M_{S_{19}}^2}}
+\frac{8 \mathcal{D}_{{d^\dagger S_{10}^\dagger u^\dagger }}^{{f_1p_1f_4*}} \mathcal{D}_{{QQS_{10}}}^{{f_2f_3p_1*}}}{{M_{S_{10}}^2}}
+\frac{8 \mathcal{D}_{{d^\dagger S_{10}^\dagger u^\dagger }}^{{f_1p_1f_4*}} \mathcal{D}_{{QQS_{10}}}^{{f_3f_2p_1*}}}{{M_{S_{10}}^2}}
-\frac{16 \mathcal{D}_{{d^\dagger S_{16}u^\dagger }}^{{f_1p_1f_4*}} \mathcal{D}_{{QQS_{16}^\dagger }}^{{f_2f_3p_1*}}}{{M_{S_{16}}^2}}
+\frac{16 \mathcal{D}_{{d^\dagger S_{16}u^\dagger }}^{{f_1p_1f_4*}} \mathcal{D}_{{QQS_{16}^\dagger }}^{{f_3f_2p_1*}}}{{M_{S_{16}}^2}} 
\end{autobreak}
\end{align}
\begin{align}
\begin{autobreak}
 {C_{lequ}^{(1)}} =
 
-\frac{2 \mathcal{D}_{{e^\dagger LS_{4}^\dagger }}^{{f_1f_2p_1*}} \mathcal{D}_{{QS_{4}u^\dagger }}^{{f_3p_1f_4*}}}{{M_{S_{4}}^2}}
-\frac{2 \mathcal{D}_{{e^\dagger QS_{13}^\dagger }}^{{f_1f_3p_1*}} \mathcal{D}_{{LS_{13}u^\dagger }}^{{f_2p_1f_4*}}}{{M_{S_{13}}^2}}
-\frac{2 \mathcal{D}_{{e^\dagger S_{10}u^\dagger }}^{{f_1p_1f_4*}} \mathcal{D}_{{LQS_{10}^\dagger }}^{{f_2f_3p_1*}}}{{M_{S_{10}}^2}} 
\end{autobreak}
\end{align}
\begin{align}
\begin{autobreak}
 {C_{lequ}^{(3)}} =
 \frac{\mathcal{D}_{{e^\dagger S_{10}u^\dagger }}^{{f_1p_1f_4*}} \mathcal{D}_{{LQS_{10}^\dagger }}^{{f_2f_3p_1*}}}{2 {M_{S_{10}}^2}}
-\frac{\mathcal{D}_{{e^\dagger QS_{13}^\dagger }}^{{f_1f_3p_1*}} \mathcal{D}_{{LS_{13}u^\dagger }}^{{f_2p_1f_4*}}}{2 {M_{S_{13}}^2}} 
\end{autobreak}
\end{align}

\subsubsection{B-violating operators}

\begin{align}
\begin{autobreak}
 {C_{duq}} =
 
-\frac{4 \mathcal{D}_{{dQV_{8}}}^{{f_3f_2p_1}} \mathcal{D}_{{L^\dagger u^\dagger V_{8}}}^{{f_1f_4p_1*}}}{{M_{V_{8}}^2}}
-\frac{4 \mathcal{D}_{{d^\dagger L^\dagger V_{7}}}^{{f_3f_1p_1*}} \mathcal{D}_{{QuV_{7}}}^{{f_2f_4p_1}}}{{M_{V_{7}}^2}}
+\frac{8 \mathcal{D}_{{d^\dagger S_{10}^\dagger u^\dagger }}^{{f_3p_1f_4*}} \mathcal{D}_{{LQS_{10}^\dagger }}^{{f_1f_2p_1}}}{{M_{S_{10}}^2}} 
\end{autobreak}
\end{align}
\begin{align}
\begin{autobreak}
 {C_{qqu}} =
 \frac{2 \mathcal{D}_{{e^\dagger S_{10}u^\dagger }}^{{f_3p_1f_4*}} \mathcal{D}_{{QQS_{10}}}^{{f_1f_2p_1}}}{{M_{S_{10}}^2}}
-\frac{4 \mathcal{D}_{{eQV_{7}^\dagger }}^{{f_3f_1p_1}} \mathcal{D}_{{QuV_{7}}}^{{f_2f_4p_1}}}{{M_{V_{7}}^2}} 
\end{autobreak}
\end{align}
\begin{align}
\begin{autobreak}
 {C_{qqq}} =
 \frac{2 \mathcal{D}_{{LQS_{10}^\dagger }}^{{f_1f_4p_1}} \mathcal{D}_{{QQS_{10}}}^{{f_2f_3p_1}}}{{M_{S_{10}}^2}}
+\frac{\mathcal{D}_{{LQS_{14}^\dagger }}^{{f_1f_4p_1}} \mathcal{D}_{{QQS_{14}}}^{{f_2f_3p_1}}}{{M_{S_{14}}^2}}
+\frac{2 \mathcal{D}_{{LQS_{10}^\dagger }}^{{f_1f_4p_1}} \mathcal{D}_{{QQS_{10}}}^{{f_3f_2p_1}}}{{M_{S_{10}}^2}}
-\frac{\mathcal{D}_{{LQS_{14}^\dagger }}^{{f_1f_4p_1}} \mathcal{D}_{{QQS_{14}}}^{{f_3f_2p_1}}}{{M_{S_{14}}^2}} 
\end{autobreak}
\end{align}
\begin{align}
\begin{autobreak}
 {C_{duu}} =
 
-\frac{8 \mathcal{D}_{{d^\dagger e^\dagger S_{9}}}^{{f_1f_2p_1*}} \mathcal{D}_{{S_{9}^\dagger u^\dagger u^\dagger }}^{{p_1f_3f_4*}}}{{M_{S_{9}}^2}}
+\frac{8 \mathcal{D}_{{d^\dagger e^\dagger S_{9}}}^{{f_1f_2p_1*}} \mathcal{D}_{{S_{9}^\dagger u^\dagger u^\dagger }}^{{p_1f_4f_3*}}}{{M_{S_{9}}^2}}
-\frac{8 \mathcal{D}_{{d^\dagger S_{10}^\dagger u^\dagger }}^{{f_1p_1f_3*}} \mathcal{D}_{{e^\dagger S_{10}u^\dagger }}^{{f_2p_1f_4*}}}{{M_{S_{10}}^2}} 
\end{autobreak}
\end{align}

\subsection{Dimension-7}

\subsubsection{B-conserving operators}

\begin{align}
\begin{autobreak}
 {C_{LH}} =
 
-\frac{{C}_5^{{f_1f_2}} (\mathcal{C}_{{HH^\dagger S_{5}}}^{{p_1}})^2}{{M_{S_{5}}^4}}
+\frac{2 {C}_5^{{f_1f_2}} (\mathcal{D}_{{HH^\dagger V_{1}D}}^{{p_1*}})^2}{{M_{V_{1}}^2}}
-\frac{2 {C}_5^{{f_1f_2}} (\mathcal{D}_{{HH^\dagger V_{1}D}}^{{p_1}})^2}{{M_{V_{1}}^2}}
+\frac{{C}_5^{{f_1f_2}} (\mathcal{D}_{{HH^\dagger V_{4}D}}^{{p_1*}})^2}{{M_{V_{4}}^2}}
-\frac{{C}_5^{{f_1f_2}} (\mathcal{D}_{{HH^\dagger V_{4}D}}^{{p_1}})^2}{{M_{V_{4}}^2}}
+\frac{\lambda_H \mathcal{D}_{{F_{1}HL}}^{{p_1f_1}} \mathcal{D}_{{F_{1}HL}}^{{p_1f_2}}}{{M_{F_{1}}^3}}
-\frac{{C}_5^{{f_2p_1}} \mathcal{D}_{{F_{1}HL}}^{{p_2p_1*}} \mathcal{D}_{{F_{1}HL}}^{{p_2f_1}}}{2 {M_{F_{1}}^2}}
-\frac{{C}_5^{{p_1f_2}} \mathcal{D}_{{F_{1}HL}}^{{p_2p_1*}} \mathcal{D}_{{F_{1}HL}}^{{p_2f_1}}}{2 {M_{F_{1}}^2}}
-\frac{\mathcal{D}_{{F_{1}F_{3L}H^\dagger }}^{{p_1p_2}} \mathcal{D}_{{F_{1}F_{3R}^\dagger H}}^{{p_3p_2}} \mathcal{D}_{{F_{1}HL}}^{{p_1f_2}} \mathcal{D}_{{F_{1}HL}}^{{p_3f_1}}}{{M_{F_{1}}^2} {M_{F_{3}}}}
+\frac{\lambda_H \mathcal{D}_{{F_{5}HL}}^{{p_1f_1}} \mathcal{D}_{{F_{5}HL}}^{{p_1f_2}}}{2 {M_{F_{5}}^3}}
-\frac{{C}_5^{{f_2p_1}} \mathcal{D}_{{F_{5}HL}}^{{p_2p_1*}} \mathcal{D}_{{F_{5}HL}}^{{p_2f_1}}}{4 {M_{F_{5}}^2}}
-\frac{{C}_5^{{p_1f_2}} \mathcal{D}_{{F_{5}HL}}^{{p_2p_1*}} \mathcal{D}_{{F_{5}HL}}^{{p_2f_1}}}{4 {M_{F_{5}}^2}}
-\frac{\mathcal{D}_{{F_{1}F_{3R}^\dagger H}}^{{p_1p_2}} \mathcal{D}_{{F_{1}HL}}^{{p_1f_2}} \mathcal{D}_{{F_{3L}F_{5}H^\dagger }}^{{p_2p_3}} \mathcal{D}_{{F_{5}HL}}^{{p_3f_1}}}{2 {M_{F_{1}}} {M_{F_{3}}} {M_{F_{5}}}}
+\frac{\mathcal{D}_{{F_{5}F_{7L}H^\dagger }}^{{p_1p_2}} \mathcal{D}_{{F_{5}F_{7R}^\dagger H}}^{{p_3p_2}} \mathcal{D}_{{F_{5}HL}}^{{p_1f_2}} \mathcal{D}_{{F_{5}HL}}^{{p_3f_1}}}{3 {M_{F_{5}}^2} {M_{F_{7}}}}
+\frac{\mathcal{D}_{{F_{3L}F_{5}H^\dagger }}^{{p_1p_2}} \mathcal{D}_{{F_{3R}^\dagger F_{5}H}}^{{p_1p_3}} \mathcal{D}_{{F_{5}HL}}^{{p_2f_2}} \mathcal{D}_{{F_{5}HL}}^{{p_3f_1}}}{4 {M_{F_{3}}} {M_{F_{5}}^2}}
+\frac{\mathcal{D}_{{F_{1}F_{3R}^\dagger H}}^{{p_1p_2}} \mathcal{D}_{{F_{1}HL}}^{{p_1f_1}} \mathcal{D}_{{F_{3L}F_{5}H^\dagger }}^{{p_2p_3}} \mathcal{D}_{{F_{5}HL}}^{{p_3f_2}}}{{M_{F_{1}}} {M_{F_{3}}} {M_{F_{5}}}}
-\frac{\mathcal{D}_{{F_{1}F_{3L}H^\dagger }}^{{p_1p_2}} \mathcal{D}_{{F_{1}HL}}^{{p_1f_1}} \mathcal{D}_{{F_{3R}^\dagger F_{5}H}}^{{p_2p_3}} \mathcal{D}_{{F_{5}HL}}^{{p_3f_2}}}{2 {M_{F_{1}}} {M_{F_{3}}} {M_{F_{5}}}}
-\frac{\mathcal{D}_{{F_{3L}F_{6R}^\dagger H}}^{{p_1p_2}} \mathcal{D}_{{F_{3R}^\dagger F_{5}H}}^{{p_1p_3}} \mathcal{D}_{{F_{5}HL}}^{{p_3f_2}} \mathcal{D}_{{F_{6L}H^\dagger L}}^{{p_2f_1}}}{4 {M_{F_{3}}} {M_{F_{5}}} {M_{F_{6}}}}
-\frac{{C}_5^{{f_2p_1}} \mathcal{D}_{{F_{6L}H^\dagger L}}^{{p_2p_1*}} \mathcal{D}_{{F_{6L}H^\dagger L}}^{{p_2f_1}}}{2 {M_{F_{6}}^2}}
-\frac{{C}_5^{{p_1f_2}} \mathcal{D}_{{F_{6L}H^\dagger L}}^{{p_2p_1*}} \mathcal{D}_{{F_{6L}H^\dagger L}}^{{p_2f_1}}}{2 {M_{F_{6}}^2}}
+\frac{3 \mathcal{D}_{{F_{3L}F_{6R}^\dagger H}}^{{p_1p_2}} \mathcal{D}_{{F_{3R}^\dagger F_{5}H}}^{{p_1p_3}} \mathcal{D}_{{F_{5}HL}}^{{p_3f_1}} \mathcal{D}_{{F_{6L}H^\dagger L}}^{{p_2f_2}}}{4 {M_{F_{3}}} {M_{F_{5}}} {M_{F_{6}}}}
-\frac{3 \mathcal{D}_{{F_{1}F_{3R}^\dagger H}}^{{p_1p_2}} \mathcal{D}_{{F_{1}HL}}^{{p_1f_2}} \mathcal{D}_{{F_{3L}F_{6R}^\dagger H}}^{{p_2p_3}} \mathcal{D}_{{F_{6L}H^\dagger L}}^{{p_3f_1}}}{2 {M_{F_{1}}} {M_{F_{3}}} {M_{F_{6}}}}
+\frac{5 \mathcal{D}_{{F_{1}F_{3R}^\dagger H}}^{{p_1p_2}} \mathcal{D}_{{F_{1}HL}}^{{p_1f_1}} \mathcal{D}_{{F_{3L}F_{6R}^\dagger H}}^{{p_2p_3}} \mathcal{D}_{{F_{6L}H^\dagger L}}^{{p_3f_2}}}{2 {M_{F_{1}}} {M_{F_{3}}} {M_{F_{6}}}}
+\frac{\mathcal{D}_{{F_{5}F_{7R}^\dagger H}}^{{p_1p_2}} \mathcal{D}_{{F_{5}HL}}^{{p_1f_2}} \mathcal{D}_{{F_{6L}H^\dagger L}}^{{p_3f_1}} \mathcal{D}_{{F_{6R}^\dagger F_{7L}H}}^{{p_3p_2}}}{6 {M_{F_{5}}} {M_{F_{6}}} {M_{F_{7}}}}
-\frac{\mathcal{D}_{{F_{5}F_{7R}^\dagger H}}^{{p_1p_2}} \mathcal{D}_{{F_{5}HL}}^{{p_1f_1}} \mathcal{D}_{{F_{6L}H^\dagger L}}^{{p_3f_2}} \mathcal{D}_{{F_{6R}^\dagger F_{7L}H}}^{{p_3p_2}}}{2 {M_{F_{5}}} {M_{F_{6}}} {M_{F_{7}}}}
-\frac{\mathcal{D}_{{F_{6L}H^\dagger L}}^{{p_1f_2}} \mathcal{D}_{{F_{6R}^\dagger LS_{8}}}^{{p_1f_1p_2}} \mathcal{D}_{{HHHS_{8}^\dagger }}^{{p_2}}}{{M_{F_{6}}} {M_{S_{8}}^2}}
-\frac{\mathcal{D}_{{F_{5}HL}}^{{p_1f_2}} \mathcal{D}_{{F_{5L}S_{7}}}^{{p_1f_1p_2}} \mathcal{D}_{{HHH^\dagger S_{7}^\dagger }}^{{p_2}}}{6 {M_{F_{5}}} {M_{S_{7}}^2}}
-\frac{\mathcal{D}_{{F_{5}HL}}^{{p_1f_1}} \mathcal{D}_{{F_{5L}S_{7}}}^{{p_1f_2p_2}} \mathcal{D}_{{HHH^\dagger S_{7}^\dagger }}^{{p_2}}}{6 {M_{F_{5}}} {M_{S_{7}}^2}}
-\frac{2 {C}_5^{{f_1f_2}} \mathcal{C}_{{HHS_{6}^\dagger }}^{{p_1*}} \mathcal{C}_{{HHS_{6}^\dagger }}^{{p_1}}}{{M_{S_{6}}^4}}
-\frac{\mathcal{D}_{{F_{1}F_{3L}H^\dagger }}^{{p_1p_2}} \mathcal{D}_{{F_{1}HL}}^{{p_1f_2}} \mathcal{D}_{{F_{3R}^\dagger LS_{6}}}^{{p_2f_1p_3}} \mathcal{C}_{{HHS_{6}^\dagger }}^{{p_3}}}{2 {M_{F_{1}}} {M_{F_{3}}} {M_{S_{6}}^2}}
-\frac{\mathcal{D}_{{F_{1}F_{3L}H^\dagger }}^{{p_1p_2}} \mathcal{D}_{{F_{1}HL}}^{{p_1f_1}} \mathcal{D}_{{F_{3R}^\dagger LS_{6}}}^{{p_2f_2p_3}} \mathcal{C}_{{HHS_{6}^\dagger }}^{{p_3}}}{2 {M_{F_{1}}} {M_{F_{3}}} {M_{S_{6}}^2}}
-\frac{\mathcal{D}_{{F_{3L}F_{5}H^\dagger }}^{{p_1p_2}} \mathcal{D}_{{F_{3R}^\dagger LS_{6}}}^{{p_1f_2p_3}} \mathcal{D}_{{F_{5}HL}}^{{p_2f_1}} \mathcal{C}_{{HHS_{6}^\dagger }}^{{p_3}}}{4 {M_{F_{3}}} {M_{F_{5}}} {M_{S_{6}}^2}}
+\frac{3 \mathcal{D}_{{F_{3L}F_{5}H^\dagger }}^{{p_1p_2}} \mathcal{D}_{{F_{3R}^\dagger LS_{6}}}^{{p_1f_1p_3}} \mathcal{D}_{{F_{5}HL}}^{{p_2f_2}} \mathcal{C}_{{HHS_{6}^\dagger }}^{{p_3}}}{4 {M_{F_{3}}} {M_{F_{5}}} {M_{S_{6}}^2}}
+\frac{\mathcal{D}_{{F_{1}F_{6R}^\dagger S_{6}}}^{{p_1p_2p_3}} \mathcal{D}_{{F_{1}HL}}^{{p_1f_2}} \mathcal{D}_{{F_{6L}H^\dagger L}}^{{p_2f_1}} \mathcal{C}_{{HHS_{6}^\dagger }}^{{p_3}}}{2 {M_{F_{1}}} {M_{F_{6}}} {M_{S_{6}}^2}}
-\frac{\mathcal{D}_{{F_{3L}F_{6R}^\dagger H}}^{{p_1p_2}} \mathcal{D}_{{F_{3R}^\dagger LS_{6}}}^{{p_1f_2p_3}} \mathcal{D}_{{F_{6L}H^\dagger L}}^{{p_2f_1}} \mathcal{C}_{{HHS_{6}^\dagger }}^{{p_3}}}{{M_{F_{3}}} {M_{F_{6}}} {M_{S_{6}}^2}}
-\frac{\mathcal{D}_{{F_{5}F_{6R}^\dagger S_{6}}}^{{p_1p_2p_3}} \mathcal{D}_{{F_{5}HL}}^{{p_1f_2}} \mathcal{D}_{{F_{6L}H^\dagger L}}^{{p_2f_1}} \mathcal{C}_{{HHS_{6}^\dagger }}^{{p_3}}}{4 {M_{F_{5}}} {M_{F_{6}}} {M_{S_{6}}^2}}
-\frac{3 \mathcal{D}_{{F_{1}F_{6R}^\dagger S_{6}}}^{{p_1p_2p_3}} \mathcal{D}_{{F_{1}HL}}^{{p_1f_1}} \mathcal{D}_{{F_{6L}H^\dagger L}}^{{p_2f_2}} \mathcal{C}_{{HHS_{6}^\dagger }}^{{p_3}}}{2 {M_{F_{1}}} {M_{F_{6}}} {M_{S_{6}}^2}}
+\frac{2 \mathcal{D}_{{F_{3L}F_{6R}^\dagger H}}^{{p_1p_2}} \mathcal{D}_{{F_{3R}^\dagger LS_{6}}}^{{p_1f_1p_3}} \mathcal{D}_{{F_{6L}H^\dagger L}}^{{p_2f_2}} \mathcal{C}_{{HHS_{6}^\dagger }}^{{p_3}}}{{M_{F_{3}}} {M_{F_{6}}} {M_{S_{6}}^2}}
+\frac{3 \mathcal{D}_{{F_{5}F_{6R}^\dagger S_{6}}}^{{p_1p_2p_3}} \mathcal{D}_{{F_{5}HL}}^{{p_1f_1}} \mathcal{D}_{{F_{6L}H^\dagger L}}^{{p_2f_2}} \mathcal{C}_{{HHS_{6}^\dagger }}^{{p_3}}}{4 {M_{F_{5}}} {M_{F_{6}}} {M_{S_{6}}^2}}
+\frac{\mathcal{D}_{{F_{5}F_{7L}H^\dagger }}^{{p_1p_2}} \mathcal{D}_{{F_{5}HL}}^{{p_1f_2}} \mathcal{D}_{{F_{7R}^\dagger LS_{6}}}^{{p_2f_1p_3}} \mathcal{C}_{{HHS_{6}^\dagger }}^{{p_3}}}{6 {M_{F_{5}}} {M_{F_{7}}} {M_{S_{6}}^2}}
-\frac{\mathcal{D}_{{F_{6L}H^\dagger L}}^{{p_1f_2}} \mathcal{D}_{{F_{6R}^\dagger F_{7L}H}}^{{p_1p_2}} \mathcal{D}_{{F_{7R}^\dagger LS_{6}}}^{{p_2f_1p_3}} \mathcal{C}_{{HHS_{6}^\dagger }}^{{p_3}}}{3 {M_{F_{6}}} {M_{F_{7}}} {M_{S_{6}}^2}}
+\frac{\mathcal{D}_{{F_{5}F_{7L}H^\dagger }}^{{p_1p_2}} \mathcal{D}_{{F_{5}HL}}^{{p_1f_1}} \mathcal{D}_{{F_{7R}^\dagger LS_{6}}}^{{p_2f_2p_3}} \mathcal{C}_{{HHS_{6}^\dagger }}^{{p_3}}}{6 {M_{F_{5}}} {M_{F_{7}}} {M_{S_{6}}^2}}
+\frac{4 {C}_5^{{f_1f_2}} \mathcal{D}_{{HHV_{2}^\dagger D}}^{{p_1*}} \mathcal{D}_{{HHV_{2}^\dagger D}}^{{p_1}}}{{M_{V_{2}}^2}}
+\frac{\mathcal{D}_{{F_{1}HL}}^{{p_1f_2}} \mathcal{D}_{{F_{1L}S_{4}}}^{{p_1f_1p_2}} \mathcal{D}_{{HH^\dagger H^\dagger S_{4}(1)}}^{{p_2*}}}{2 {M_{F_{1}}} {M_{S_{4}}^2}}
+\frac{\mathcal{D}_{{F_{1}HL}}^{{p_1f_1}} \mathcal{D}_{{F_{1L}S_{4}}}^{{p_1f_2p_2}} \mathcal{D}_{{HH^\dagger H^\dagger S_{4}(1)}}^{{p_2*}}}{2 {M_{F_{1}}} {M_{S_{4}}^2}}
-\frac{\mathcal{D}_{{F_{5}HL}}^{{p_1f_2}} \mathcal{D}_{{F_{5L}S_{4}}}^{{p_1f_1p_2}} \mathcal{D}_{{HH^\dagger H^\dagger S_{4}(1)}}^{{p_2*}}}{4 {M_{F_{5}}} {M_{S_{4}}^2}}
-\frac{\mathcal{D}_{{F_{5}HL}}^{{p_1f_1}} \mathcal{D}_{{F_{5L}S_{4}}}^{{p_1f_2p_2}} \mathcal{D}_{{HH^\dagger H^\dagger S_{4}(1)}}^{{p_2*}}}{4 {M_{F_{5}}} {M_{S_{4}}^2}}
+\frac{\mathcal{D}_{{F_{3L}LS_{1}}}^{{p_1f_2p_2}} \mathcal{D}_{{F_{3R}^\dagger F_{5}H}}^{{p_1p_3}} \mathcal{D}_{{F_{5}HL}}^{{p_3f_1}} \mathcal{C}_{{HH^\dagger S_{1}}}^{{p_2}}}{4 {M_{F_{3}}} {M_{F_{5}}} {M_{S_{1}}^2}}
+\frac{\mathcal{D}_{{F_{3L}LS_{1}}}^{{p_1f_1p_2}} \mathcal{D}_{{F_{3R}^\dagger F_{5}H}}^{{p_1p_3}} \mathcal{D}_{{F_{5}HL}}^{{p_3f_2}} \mathcal{C}_{{HH^\dagger S_{1}}}^{{p_2}}}{4 {M_{F_{3}}} {M_{F_{5}}} {M_{S_{1}}^2}}
+\frac{\mathcal{D}_{{F_{3L}LS_{1}}}^{{p_1f_2p_2}} \mathcal{D}_{{F_{3R}^\dagger LS_{6}}}^{{p_1f_1p_3}} \mathcal{C}_{{HHS_{6}^\dagger }}^{{p_3}} \mathcal{C}_{{HH^\dagger S_{1}}}^{{p_2}}}{{M_{F_{3}}} {M_{S_{1}}^2} {M_{S_{6}}^2}}
-\frac{\mathcal{D}_{{F_{1}F_{1}S_{1}}}^{{p_1p_2p_3}} \mathcal{D}_{{F_{1}HL}}^{{p_1f_1}} \mathcal{D}_{{F_{1}HL}}^{{p_2f_2}} \mathcal{C}_{{HH^\dagger S_{1}}}^{{p_3}}}{{M_{F_{1}}^2} {M_{S_{1}}^2}}
-\frac{\mathcal{D}_{{F_{1}F_{3R}^\dagger H}}^{{p_1p_2}} \mathcal{D}_{{F_{1}HL}}^{{p_1f_2}} \mathcal{D}_{{F_{3L}LS_{1}}}^{{p_2f_1p_3}} \mathcal{C}_{{HH^\dagger S_{1}}}^{{p_3}}}{2 {M_{F_{1}}} {M_{F_{3}}} {M_{S_{1}}^2}}
+\frac{3 \mathcal{D}_{{F_{1}F_{3R}^\dagger H}}^{{p_1p_2}} \mathcal{D}_{{F_{1}HL}}^{{p_1f_1}} \mathcal{D}_{{F_{3L}LS_{1}}}^{{p_2f_2p_3}} \mathcal{C}_{{HH^\dagger S_{1}}}^{{p_3}}}{2 {M_{F_{1}}} {M_{F_{3}}} {M_{S_{1}}^2}}
+\frac{\mathcal{D}_{{F_{5}F_{5}S_{1}}}^{{p_1p_2p_3}} \mathcal{D}_{{F_{5}HL}}^{{p_1f_1}} \mathcal{D}_{{F_{5}HL}}^{{p_2f_2}} \mathcal{C}_{{HH^\dagger S_{1}}}^{{p_3}}}{2 {M_{F_{5}}^2} {M_{S_{1}}^2}}
+\frac{\mathcal{D}_{{F_{3L}LS_{5}}}^{{p_1f_2p_2}} \mathcal{D}_{{F_{3R}^\dagger F_{5}H}}^{{p_1p_3}} \mathcal{D}_{{F_{5}HL}}^{{p_3f_1}} \mathcal{C}_{{HH^\dagger S_{5}}}^{{p_2}}}{8 {M_{F_{3}}} {M_{F_{5}}} {M_{S_{5}}^2}}
+\frac{\mathcal{D}_{{F_{3L}LS_{5}}}^{{p_1f_1p_2}} \mathcal{D}_{{F_{3R}^\dagger F_{5}H}}^{{p_1p_3}} \mathcal{D}_{{F_{5}HL}}^{{p_3f_2}} \mathcal{C}_{{HH^\dagger S_{5}}}^{{p_2}}}{8 {M_{F_{3}}} {M_{F_{5}}} {M_{S_{5}}^2}}
+\frac{\mathcal{D}_{{F_{3L}LS_{5}}}^{{p_1f_2p_2}} \mathcal{D}_{{F_{3R}^\dagger LS_{6}}}^{{p_1f_1p_3}} \mathcal{C}_{{HHS_{6}^\dagger }}^{{p_3}} \mathcal{C}_{{HH^\dagger S_{5}}}^{{p_2}}}{2 {M_{F_{3}}} {M_{S_{5}}^2} {M_{S_{6}}^2}}
-\frac{\mathcal{D}_{{F_{7L}LS_{5}}}^{{p_1f_2p_2}} \mathcal{D}_{{F_{7R}^\dagger LS_{6}}}^{{p_1f_1p_3}} \mathcal{C}_{{HHS_{6}^\dagger }}^{{p_3}} \mathcal{C}_{{HH^\dagger S_{5}}}^{{p_2}}}{3 {M_{F_{7}}} {M_{S_{5}}^2} {M_{S_{6}}^2}}
-\frac{\mathcal{D}_{{F_{1}F_{3R}^\dagger H}}^{{p_1p_2}} \mathcal{D}_{{F_{1}HL}}^{{p_1f_2}} \mathcal{D}_{{F_{3L}LS_{5}}}^{{p_2f_1p_3}} \mathcal{C}_{{HH^\dagger S_{5}}}^{{p_3}}}{4 {M_{F_{1}}} {M_{F_{3}}} {M_{S_{5}}^2}}
+\frac{3 \mathcal{D}_{{F_{1}F_{3R}^\dagger H}}^{{p_1p_2}} \mathcal{D}_{{F_{1}HL}}^{{p_1f_1}} \mathcal{D}_{{F_{3L}LS_{5}}}^{{p_2f_2p_3}} \mathcal{C}_{{HH^\dagger S_{5}}}^{{p_3}}}{4 {M_{F_{1}}} {M_{F_{3}}} {M_{S_{5}}^2}}
+\frac{\mathcal{D}_{{F_{1}F_{5}S_{5}}}^{{p_1p_2p_3}} \mathcal{D}_{{F_{1}HL}}^{{p_1f_1}} \mathcal{D}_{{F_{5}HL}}^{{p_2f_2}} \mathcal{C}_{{HH^\dagger S_{5}}}^{{p_3}}}{2 {M_{F_{1}}} {M_{F_{5}}} {M_{S_{5}}^2}}
+\frac{\mathcal{D}_{{F_{5}F_{7R}^\dagger H}}^{{p_1p_2}} \mathcal{D}_{{F_{5}HL}}^{{p_1f_2}} \mathcal{D}_{{F_{7L}LS_{5}}}^{{p_2f_1p_3}} \mathcal{C}_{{HH^\dagger S_{5}}}^{{p_3}}}{6 {M_{F_{5}}} {M_{F_{7}}} {M_{S_{5}}^2}}
-\frac{\mathcal{D}_{{F_{5}F_{7R}^\dagger H}}^{{p_1p_2}} \mathcal{D}_{{F_{5}HL}}^{{p_1f_1}} \mathcal{D}_{{F_{7L}LS_{5}}}^{{p_2f_2p_3}} \mathcal{C}_{{HH^\dagger S_{5}}}^{{p_3}}}{2 {M_{F_{5}}} {M_{F_{7}}} {M_{S_{5}}^2}}
-\frac{4 {C}_5^{{f_1f_2}} \mathcal{D}_{{HH^\dagger V_{4}D}}^{{p_1*}} \mathcal{D}_{{HH^\dagger V_{4}D}}^{{p_1}}}{{M_{V_{4}}^2}}
-\frac{\mathcal{D}_{{F_{1}HL}}^{{p_1f_2}} \mathcal{D}_{{F_{1L}S_{4}}}^{{p_1f_1p_2}} \mathcal{C}_{{HH^\dagger S_{1}}}^{{p_3}} \mathcal{C}_{{HS_{1}S_{4}^\dagger }}^{{p_3p_2}}}{2 {M_{F_{1}}} {M_{S_{1}}^2} {M_{S_{4}}^2}}
-\frac{\mathcal{D}_{{F_{1}HL}}^{{p_1f_1}} \mathcal{D}_{{F_{1L}S_{4}}}^{{p_1f_2p_2}} \mathcal{C}_{{HH^\dagger S_{1}}}^{{p_3}} \mathcal{C}_{{HS_{1}S_{4}^\dagger }}^{{p_3p_2}}}{2 {M_{F_{1}}} {M_{S_{1}}^2} {M_{S_{4}}^2}}
+\frac{\mathcal{D}_{{F_{5}HL}}^{{p_1f_2}} \mathcal{D}_{{F_{5L}S_{4}}}^{{p_1f_1p_2}} \mathcal{C}_{{HH^\dagger S_{1}}}^{{p_3}} \mathcal{C}_{{HS_{1}S_{4}^\dagger }}^{{p_3p_2}}}{4 {M_{F_{5}}} {M_{S_{1}}^2} {M_{S_{4}}^2}}
+\frac{\mathcal{D}_{{F_{5}HL}}^{{p_1f_1}} \mathcal{D}_{{F_{5L}S_{4}}}^{{p_1f_2p_2}} \mathcal{C}_{{HH^\dagger S_{1}}}^{{p_3}} \mathcal{C}_{{HS_{1}S_{4}^\dagger }}^{{p_3p_2}}}{4 {M_{F_{5}}} {M_{S_{1}}^2} {M_{S_{4}}^2}}
-\frac{\mathcal{D}_{{F_{1}HL}}^{{p_1f_2}} \mathcal{D}_{{F_{1L}S_{4}}}^{{p_1f_1p_2}} \mathcal{C}_{{HHS_{6}^\dagger }}^{{p_3}} \mathcal{C}_{{HS_{4}S_{6}^\dagger }}^{{p_2p_3*}}}{2 {M_{F_{1}}} {M_{S_{4}}^2} {M_{S_{6}}^2}}
-\frac{\mathcal{D}_{{F_{1}HL}}^{{p_1f_1}} \mathcal{D}_{{F_{1L}S_{4}}}^{{p_1f_2p_2}} \mathcal{C}_{{HHS_{6}^\dagger }}^{{p_3}} \mathcal{C}_{{HS_{4}S_{6}^\dagger }}^{{p_2p_3*}}}{2 {M_{F_{1}}} {M_{S_{4}}^2} {M_{S_{6}}^2}}
+\frac{\mathcal{D}_{{F_{5}HL}}^{{p_1f_2}} \mathcal{D}_{{F_{5L}S_{4}}}^{{p_1f_1p_2}} \mathcal{C}_{{HHS_{6}^\dagger }}^{{p_3}} \mathcal{C}_{{HS_{4}S_{6}^\dagger }}^{{p_2p_3*}}}{4 {M_{F_{5}}} {M_{S_{4}}^2} {M_{S_{6}}^2}}
+\frac{\mathcal{D}_{{F_{5}HL}}^{{p_1f_1}} \mathcal{D}_{{F_{5L}S_{4}}}^{{p_1f_2p_2}} \mathcal{C}_{{HHS_{6}^\dagger }}^{{p_3}} \mathcal{C}_{{HS_{4}S_{6}^\dagger }}^{{p_2p_3*}}}{4 {M_{F_{5}}} {M_{S_{4}}^2} {M_{S_{6}}^2}}
+\frac{\mathcal{D}_{{F_{5}HL}}^{{p_1f_2}} \mathcal{D}_{{F_{5L}S_{7}}}^{{p_1f_1p_2}} \mathcal{C}_{{HH^\dagger S_{5}}}^{{p_3}} \mathcal{C}_{{HS_{5}S_{7}^\dagger }}^{{p_3p_2}}}{6 {M_{F_{5}}} {M_{S_{5}}^2} {M_{S_{7}}^2}}
+\frac{\mathcal{D}_{{F_{5}HL}}^{{p_1f_1}} \mathcal{D}_{{F_{5L}S_{7}}}^{{p_1f_2p_2}} \mathcal{C}_{{HH^\dagger S_{5}}}^{{p_3}} \mathcal{C}_{{HS_{5}S_{7}^\dagger }}^{{p_3p_2}}}{6 {M_{F_{5}}} {M_{S_{5}}^2} {M_{S_{7}}^2}}
-\frac{\mathcal{D}_{{F_{6L}H^\dagger L}}^{{p_1f_2}} \mathcal{D}_{{F_{6R}^\dagger LS_{8}}}^{{p_1f_1p_2}} \mathcal{C}_{{HHS_{6}^\dagger }}^{{p_3}} \mathcal{C}_{{HS_{6}S_{8}^\dagger }}^{{p_3p_2}}}{{M_{F_{6}}} {M_{S_{6}}^2} {M_{S_{8}}^2}}
-\frac{\mathcal{D}_{{F_{5}HL}}^{{p_1f_2}} \mathcal{D}_{{F_{5L}S_{7}}}^{{p_1f_1p_2}} \mathcal{C}_{{HHS_{6}^\dagger }}^{{p_3}} \mathcal{C}_{{HS_{6}^\dagger S_{7}}}^{{p_3p_2*}}}{6 {M_{F_{5}}} {M_{S_{6}}^2} {M_{S_{7}}^2}}
-\frac{\mathcal{D}_{{F_{5}HL}}^{{p_1f_1}} \mathcal{D}_{{F_{5L}S_{7}}}^{{p_1f_2p_2}} \mathcal{C}_{{HHS_{6}^\dagger }}^{{p_3}} \mathcal{C}_{{HS_{6}^\dagger S_{7}}}^{{p_3p_2*}}}{6 {M_{F_{5}}} {M_{S_{6}}^2} {M_{S_{7}}^2}}
+\frac{\mathcal{D}_{{F_{1}HL}}^{{p_1f_2}} \mathcal{D}_{{F_{1L}S_{4}}}^{{p_1f_1p_2}} \mathcal{C}_{{HH^\dagger S_{5}}}^{{p_3}} \mathcal{C}_{{H^\dagger S_{4}S_{5}}}^{{p_2p_3*}}}{4 {M_{F_{1}}} {M_{S_{4}}^2} {M_{S_{5}}^2}}
+\frac{\mathcal{D}_{{F_{1}HL}}^{{p_1f_1}} \mathcal{D}_{{F_{1L}S_{4}}}^{{p_1f_2p_2}} \mathcal{C}_{{HH^\dagger S_{5}}}^{{p_3}} \mathcal{C}_{{H^\dagger S_{4}S_{5}}}^{{p_2p_3*}}}{4 {M_{F_{1}}} {M_{S_{4}}^2} {M_{S_{5}}^2}}
-\frac{\mathcal{D}_{{F_{5}HL}}^{{p_1f_2}} \mathcal{D}_{{F_{5L}S_{4}}}^{{p_1f_1p_2}} \mathcal{C}_{{HH^\dagger S_{5}}}^{{p_3}} \mathcal{C}_{{H^\dagger S_{4}S_{5}}}^{{p_2p_3*}}}{8 {M_{F_{5}}} {M_{S_{4}}^2} {M_{S_{5}}^2}}
-\frac{\mathcal{D}_{{F_{5}HL}}^{{p_1f_1}} \mathcal{D}_{{F_{5L}S_{4}}}^{{p_1f_2p_2}} \mathcal{C}_{{HH^\dagger S_{5}}}^{{p_3}} \mathcal{C}_{{H^\dagger S_{4}S_{5}}}^{{p_2p_3*}}}{8 {M_{F_{5}}} {M_{S_{4}}^2} {M_{S_{5}}^2}}
-\frac{\mathcal{D}_{{HHS_{1}S_{6}^\dagger }}^{{p_1p_2}} \mathcal{C}_{{HH^\dagger S_{1}}}^{{p_1}} \mathcal{D}_{{LLS_{6}}}^{{f_1f_2p_2}}}{{M_{S_{1}}^2} {M_{S_{6}}^2}}
+\frac{\mathcal{D}_{{HHS_{5}S_{6}^\dagger }}^{{p_1p_2}} \mathcal{C}_{{HH^\dagger S_{5}}}^{{p_1}} \mathcal{D}_{{LLS_{6}}}^{{f_1f_2p_2}}}{2 {M_{S_{5}}^2} {M_{S_{6}}^2}}
-\frac{\mathcal{C}_{{HHS_{6}^\dagger }}^{{p_1}} \mathcal{D}_{{HH^\dagger S_{6}S_{6}^\dagger (2)}}^{{p_1p_2}} \mathcal{D}_{{LLS_{6}}}^{{f_1f_2p_2}}}{{M_{S_{6}}^4}}
+\frac{\mathcal{D}_{{HH^\dagger H^\dagger S_{4}(1)}}^{{p_1*}} \mathcal{C}_{{HS_{4}S_{6}^\dagger }}^{{p_1p_2}} \mathcal{D}_{{LLS_{6}}}^{{f_1f_2p_2}}}{{M_{S_{4}}^2} {M_{S_{6}}^2}}
-\frac{\mathcal{D}_{{HHHS_{8}^\dagger }}^{{p_1}} \mathcal{C}_{{HS_{6}S_{8}^\dagger }}^{{p_2p_1*}} \mathcal{D}_{{LLS_{6}}}^{{f_1f_2p_2}}}{{M_{S_{6}}^2} {M_{S_{8}}^2}}
-\frac{\mathcal{C}_{{HHS_{6}^\dagger }}^{{p_1}} \mathcal{C}_{{HS_{6}S_{8}^\dagger }}^{{p_2p_3*}} \mathcal{C}_{{HS_{6}S_{8}^\dagger }}^{{p_1p_3}} \mathcal{D}_{{LLS_{6}}}^{{f_1f_2p_2}}}{{M_{S_{6}}^4} {M_{S_{8}}^2}}
-\frac{\mathcal{D}_{{HHH^\dagger S_{7}^\dagger }}^{{p_1}} \mathcal{C}_{{HS_{6}^\dagger S_{7}}}^{{p_2p_1}} \mathcal{D}_{{LLS_{6}}}^{{f_1f_2p_2}}}{3 {M_{S_{6}}^2} {M_{S_{7}}^2}}
-\frac{\mathcal{C}_{{HH^\dagger S_{1}}}^{{p_1}} \mathcal{C}_{{HS_{1}S_{4}^\dagger }}^{{p_1p_2}} \mathcal{C}_{{HS_{4}S_{6}^\dagger }}^{{p_2p_3}} \mathcal{D}_{{LLS_{6}}}^{{f_1f_2p_3}}}{{M_{S_{1}}^2} {M_{S_{4}}^2} {M_{S_{6}}^2}}
-\frac{\mathcal{C}_{{HHS_{6}^\dagger }}^{{p_1}} \mathcal{C}_{{HS_{4}S_{6}^\dagger }}^{{p_2p_1*}} \mathcal{C}_{{HS_{4}S_{6}^\dagger }}^{{p_2p_3}} \mathcal{D}_{{LLS_{6}}}^{{f_1f_2p_3}}}{{M_{S_{4}}^2} {M_{S_{6}}^4}}
+\frac{\mathcal{C}_{{HH^\dagger S_{5}}}^{{p_1}} \mathcal{C}_{{HS_{5}S_{7}^\dagger }}^{{p_1p_2}} \mathcal{C}_{{HS_{6}^\dagger S_{7}}}^{{p_3p_2}} \mathcal{D}_{{LLS_{6}}}^{{f_1f_2p_3}}}{3 {M_{S_{5}}^2} {M_{S_{6}}^2} {M_{S_{7}}^2}}
-\frac{\mathcal{C}_{{HHS_{6}^\dagger }}^{{p_1}} \mathcal{C}_{{HS_{6}^\dagger S_{7}}}^{{p_1p_2*}} \mathcal{C}_{{HS_{6}^\dagger S_{7}}}^{{p_3p_2}} \mathcal{D}_{{LLS_{6}}}^{{f_1f_2p_3}}}{3 {M_{S_{6}}^4} {M_{S_{7}}^2}}
+\frac{\mathcal{C}_{{HH^\dagger S_{5}}}^{{p_1}} \mathcal{C}_{{HS_{4}S_{6}^\dagger }}^{{p_2p_3}} \mathcal{C}_{{H^\dagger S_{4}S_{5}}}^{{p_2p_1*}} \mathcal{D}_{{LLS_{6}}}^{{f_1f_2p_3}}}{2 {M_{S_{4}}^2} {M_{S_{5}}^2} {M_{S_{6}}^2}}
+\frac{i {C}_5^{{f_2p_1}} \mathcal{D}_{{HH^\dagger V_{1}D}}^{{p_2*}} \mathcal{D}_{{LL^\dagger V_{1}}}^{{f_1p_1p_2}}}{{M_{V_{1}}^2}}
+\frac{i {C}_5^{{p_1f_2}} \mathcal{D}_{{HH^\dagger V_{1}D}}^{{p_2*}} \mathcal{D}_{{LL^\dagger V_{1}}}^{{f_1p_1p_2}}}{{M_{V_{1}}^2}}
+\frac{i {C}_5^{{f_2p_1}} \mathcal{D}_{{HH^\dagger V_{1}D}}^{{p_2}} \mathcal{D}_{{LL^\dagger V_{1}}}^{{f_1p_1p_2}}}{{M_{V_{1}}^2}}
+\frac{i {C}_5^{{p_1f_2}} \mathcal{D}_{{HH^\dagger V_{1}D}}^{{p_2}} \mathcal{D}_{{LL^\dagger V_{1}}}^{{f_1p_1p_2}}}{{M_{V_{1}}^2}}
+\frac{i {C}_5^{{f_2p_1}} \mathcal{D}_{{HH^\dagger V_{4}D}}^{{p_2*}} \mathcal{D}_{{LL^\dagger V_{4}}}^{{f_1p_1p_2}}}{2 {M_{V_{4}}^2}}
+\frac{i {C}_5^{{p_1f_2}} \mathcal{D}_{{HH^\dagger V_{4}D}}^{{p_2*}} \mathcal{D}_{{LL^\dagger V_{4}}}^{{f_1p_1p_2}}}{2 {M_{V_{4}}^2}}
-\frac{i {C}_5^{{f_2p_1}} \mathcal{D}_{{HH^\dagger V_{4}D}}^{{p_2}} \mathcal{D}_{{LL^\dagger V_{4}}}^{{f_1p_1p_2}}}{2 {M_{V_{4}}^2}}
-\frac{i {C}_5^{{p_1f_2}} \mathcal{D}_{{HH^\dagger V_{4}D}}^{{p_2}} \mathcal{D}_{{LL^\dagger V_{4}}}^{{f_1p_1p_2}}}{2 {M_{V_{4}}^2}}
-\frac{\mathcal{C}_{{HHS_{6}^\dagger }}^{{p_1}} \mathcal{C}_{{HH^\dagger S_{1}}}^{{p_2}} \mathcal{D}_{{LLS_{6}}}^{{f_1f_2p_3}} \mathcal{C}_{{S_{1}S_{6}S_{6}^\dagger }}^{{p_2p_1p_3}}}{{M_{S_{1}}^2} {M_{S_{6}}^4}}
-\frac{\mathcal{C}_{{HHS_{6}^\dagger }}^{{p_1}} \mathcal{C}_{{HH^\dagger S_{5}}}^{{p_2}} \mathcal{D}_{{LLS_{6}}}^{{f_1f_2p_3}} \mathcal{C}_{{S_{5}S_{6}S_{6}^\dagger }}^{{p_2p_1p_3}}}{2 {M_{S_{5}}^2} {M_{S_{6}}^4}} 
\end{autobreak}
\end{align}
\begin{align}
\begin{autobreak}
 {C_{LeHD}} =
 
-\frac{{y}_{{e}}^{{p_1f_5}} \mathcal{D}_{{F_{1}HL}}^{{p_2f_1}} \mathcal{D}_{{F_{1}HL}}^{{p_2p_1}}}{2 {M_{F_{1}}^3}}
-\frac{{y}_{{e}}^{{p_1f_5}} \mathcal{D}_{{LLS_{6}}}^{{f_1p_1p_2}} \mathcal{C}_{{HHS_{6}^\dagger }}^{{p_2}}}{{M_{S_{6}}^4}}
-\frac{{y}_{{e}}^{{p_1f_5}} \mathcal{D}_{{F_{5}HL}}^{{p_2f_1}} \mathcal{D}_{{F_{5}HL}}^{{p_2p_1}}}{4 {M_{F_{5}}^3}}
+\frac{{y}_{{e}}^{{p_1f_5}} \mathcal{C}_{{HHS_{6}^\dagger }}^{{p_2}} \mathcal{D}_{{LLS_{6}}}^{{p_1f_1p_2}}}{{M_{S_{6}}^4}}
+\frac{2 i \mathcal{D}_{{eF_{1}V_{2}}}^{{f_5p_1p_2}} \mathcal{D}_{{F_{1}HL}}^{{p_1f_1}} \mathcal{D}_{{HHV_{2}^\dagger D}}^{{p_2}}}{{M_{F_{1}}} {M_{V_{2}}^2}}
-\frac{\mathcal{D}_{{e^\dagger F_{3R}^\dagger H^\dagger }}^{{f_5p_1*}} \mathcal{D}_{{F_{1}F_{3R}^\dagger H}}^{{p_2p_1}} \mathcal{D}_{{F_{1}HL}}^{{p_2f_1}}}{{M_{F_{1}}} {M_{F_{3}}^2}}
-\frac{\mathcal{D}_{{e^\dagger F_{3R}^\dagger H^\dagger }}^{{f_5p_1*}} \mathcal{D}_{{F_{3L}F_{5}H^\dagger }}^{{p_1p_2*}} \mathcal{D}_{{F_{5}HL}}^{{p_2f_1}}}{{M_{F_{3}}} {M_{F_{5}}^2}}
+\frac{2 i \mathcal{D}_{{e^\dagger F_{3R}^\dagger H^\dagger }}^{{f_5p_1*}} \mathcal{D}_{{F_{3L}L^\dagger V_{2}^\dagger }}^{{p_1f_1p_2*}} \mathcal{D}_{{HHV_{2}^\dagger D}}^{{p_2}}}{{M_{F_{3}}} {M_{V_{2}}^2}}
-\frac{\mathcal{D}_{{e^\dagger F_{3R}^\dagger H^\dagger }}^{{f_5p_1*}} \mathcal{D}_{{F_{3R}^\dagger F_{5}H}}^{{p_1p_2}} \mathcal{D}_{{F_{5}HL}}^{{p_2f_1}}}{2 {M_{F_{3}}^2} {M_{F_{5}}}}
-\frac{\mathcal{D}_{{e^\dagger F_{3R}^\dagger H^\dagger }}^{{f_5p_1*}} \mathcal{D}_{{F_{3R}^\dagger LS_{6}}}^{{p_1f_1p_2}} \mathcal{C}_{{HHS_{6}^\dagger }}^{{p_2}}}{{M_{F_{3}}^2} {M_{S_{6}}^2}}
+\frac{\mathcal{D}_{{e^\dagger F_{5}S_{6}^\dagger }}^{{f_5p_1p_2*}} \mathcal{D}_{{F_{5}HL}}^{{p_1f_1}} \mathcal{C}_{{HHS_{6}^\dagger }}^{{p_2}}}{{M_{F_{5}}^2} {M_{S_{6}}^2}}
-\frac{2 i \mathcal{D}_{{e^\dagger L^\dagger V_{3}^\dagger }}^{{f_5f_1p_1*}} \mathcal{C}_{{HHS_{6}^\dagger }}^{{p_2}} \mathcal{D}_{{HS_{6}V_{3}^\dagger D}}^{{p_2p_1}}}{{M_{S_{6}}^2} {M_{V_{3}}^2}}
-\frac{4 i \mathcal{D}_{{e^\dagger L^\dagger V_{3}^\dagger }}^{{f_5f_1p_1*}} \mathcal{D}_{{HHV_{2}^\dagger D}}^{{p_2}} \mathcal{C}_{{HV_{2}V_{3}^\dagger }}^{{p_2p_1}}}{{M_{V_{2}}^2} {M_{V_{3}}^2}} 
\end{autobreak}
\end{align}
\begin{align}
\begin{autobreak}
 {C_{LHD1}} =
 \frac{2 \mathcal{D}_{{LLS_{6}}}^{{f_1f_2p_1}} \mathcal{C}_{{HHS_{6}^\dagger }}^{{p_1}}}{{M_{S_{6}}^4}}
+\frac{\mathcal{D}_{{F_{5}HL}}^{{p_1f_1}} \mathcal{D}_{{F_{5}HL}}^{{p_1f_2}}}{{M_{F_{5}}^3}} 
\end{autobreak}
\end{align}
\begin{align}
\begin{autobreak}
 {C_{LHD2}} =
 
-\frac{4 \mathcal{D}_{{LLS_{6}}}^{{f_1f_2p_1}} \mathcal{C}_{{HHS_{6}^\dagger }}^{{p_1}}}{{M_{S_{6}}^4}}
-\frac{\mathcal{D}_{{F_{1}HL}}^{{p_1f_1}} \mathcal{D}_{{F_{1}HL}}^{{p_1f_2}}}{{M_{F_{1}}^3}}
-\frac{\mathcal{D}_{{F_{5}HL}}^{{p_1f_1}} \mathcal{D}_{{F_{5}HL}}^{{p_1f_2}}}{2 {M_{F_{5}}^3}} 
\end{autobreak}
\end{align}
\begin{align}
\begin{autobreak}
 {C_{LHB}} =
 
-\frac{i {g'} \mathcal{D}_{{F_{1}HL}}^{{p_1f_2}} \mathcal{D}_{{F_{1}HL}}^{{p_1f_3}}}{8 {M_{F_{1}}^3}}
-\frac{i {g'} \mathcal{D}_{{LLS_{6}}}^{{f_2f_3p_1}} \mathcal{C}_{{HHS_{6}^\dagger }}^{{p_1}}}{2 {M_{S_{6}}^4}}
-\frac{i {g'} \mathcal{D}_{{F_{5}HL}}^{{p_1f_2}} \mathcal{D}_{{F_{5}HL}}^{{p_1f_3}}}{16 {M_{F_{5}}^3}} 
\end{autobreak}
\end{align}
\begin{align}
\begin{autobreak}
 {C_{LHW}} =
 
-\frac{i {g} \mathcal{D}_{{F_{1}HL}}^{{p_1f_2}} \mathcal{D}_{{F_{1}HL}}^{{p_1f_3}}}{8 {M_{F_{1}}^3}}
-\frac{i {g} \mathcal{D}_{{LLS_{6}}}^{{f_2f_3p_1}} \mathcal{C}_{{HHS_{6}^\dagger }}^{{p_1}}}{4 {M_{S_{6}}^4}}
+\frac{i {g} \mathcal{D}_{{LLS_{6}}}^{{f_3f_2p_1}} \mathcal{C}_{{HHS_{6}^\dagger }}^{{p_1}}}{4 {M_{S_{6}}^4}}
+\frac{3 i {g} \mathcal{D}_{{F_{5}HL}}^{{p_1f_2}} \mathcal{D}_{{F_{5}HL}}^{{p_1f_3}}}{16 {M_{F_{5}}^3}} 
\end{autobreak}
\end{align}
\begin{align}
\begin{autobreak}
 {C_{eLLLH}} =
 
-\frac{{y}_{{e}}^{{f_1f_2*}} \mathcal{D}_{{F_{1}HL}}^{{p_1f_3}} \mathcal{D}_{{F_{1}HL}}^{{p_1f_4}}}{2 {M_{F_{1}}^3}}
-\frac{3 \mathcal{D}_{{e^\dagger F_{1}S_{2}^\dagger }}^{{f_1p_1p_2}} \mathcal{D}_{{LLS_{2}}}^{{f_2f_3p_2}} \mathcal{D}_{{F_{1}HL}}^{{p_1f_4}}}{2 {M_{F_{1}}} {M_{S_{2}}^2}}
+\frac{3 \mathcal{D}_{{e^\dagger F_{1}S_{2}^\dagger }}^{{f_1p_1p_2}} \mathcal{D}_{{LLS_{2}}}^{{f_3f_2p_2}} \mathcal{D}_{{F_{1}HL}}^{{p_1f_4}}}{2 {M_{F_{1}}} {M_{S_{2}}^2}}
-\frac{\mathcal{D}_{{e^\dagger LS_{4}^\dagger }}^{{f_1f_2p_1}} \mathcal{D}_{{F_{1}HL}}^{{p_2f_4}} \mathcal{D}_{{F_{1L}S_{4}}}^{{p_2f_3p_1}}}{{M_{F_{1}}} {M_{S_{4}}^2}}
-\frac{{y}_{{e}}^{{f_1f_2*}} \mathcal{D}_{{F_{5}HL}}^{{p_1f_3}} \mathcal{D}_{{F_{5}HL}}^{{p_1f_4}}}{4 {M_{F_{5}}^3}}
-\frac{\mathcal{D}_{{e^\dagger LS_{4}^\dagger }}^{{f_1f_4p_1}} \mathcal{D}_{{F_{5}HL}}^{{p_2f_3}} \mathcal{D}_{{F_{5L}S_{4}}}^{{p_2f_2p_1}}}{{M_{F_{5}}} {M_{S_{4}}^2}}
+\frac{\mathcal{D}_{{e^\dagger LS_{4}^\dagger }}^{{f_1f_3p_1}} \mathcal{D}_{{F_{5}HL}}^{{p_2f_4}} \mathcal{D}_{{F_{5L}S_{4}}}^{{p_2f_2p_1}}}{2 {M_{F_{5}}} {M_{S_{4}}^2}}
+\frac{\mathcal{D}_{{e^\dagger LS_{4}^\dagger }}^{{f_1f_4p_1}} \mathcal{D}_{{F_{5}HL}}^{{p_2f_2}} \mathcal{D}_{{F_{5L}S_{4}}}^{{p_2f_3p_1}}}{2 {M_{F_{5}}} {M_{S_{4}}^2}}
+\frac{\mathcal{D}_{{e^\dagger LS_{4}^\dagger }}^{{f_1f_3p_1}} \mathcal{D}_{{F_{5}HL}}^{{p_2f_2}} \mathcal{D}_{{F_{5L}S_{4}}}^{{p_2f_4p_1}}}{2 {M_{F_{5}}} {M_{S_{4}}^2}}
+\frac{\mathcal{D}_{{e^\dagger F_{4R}^\dagger H}}^{{f_1p_1}} \mathcal{D}_{{F_{4L}LS_{2}^\dagger }}^{{p_1f_4p_2}} \mathcal{D}_{{LLS_{2}}}^{{f_2f_3p_2}}}{{M_{F_{4}}} {M_{S_{2}}^2}}
+\frac{2 \mathcal{D}_{{e^\dagger LS_{4}^\dagger }}^{{f_1f_4p_1}} \mathcal{C}_{{HS_{2}^\dagger S_{4}}}^{{p_2p_1}} \mathcal{D}_{{LLS_{2}}}^{{f_2f_3p_2}}}{{M_{S_{2}}^2} {M_{S_{4}}^2}}
-\frac{\mathcal{D}_{{e^\dagger F_{4R}^\dagger H}}^{{f_1p_1}} \mathcal{D}_{{F_{4L}LS_{2}^\dagger }}^{{p_1f_4p_2}} \mathcal{D}_{{LLS_{2}}}^{{f_3f_2p_2}}}{{M_{F_{4}}} {M_{S_{2}}^2}}
-\frac{2 \mathcal{D}_{{e^\dagger LS_{4}^\dagger }}^{{f_1f_4p_1}} \mathcal{C}_{{HS_{2}^\dagger S_{4}}}^{{p_2p_1}} \mathcal{D}_{{LLS_{2}}}^{{f_3f_2p_2}}}{{M_{S_{2}}^2} {M_{S_{4}}^2}}
+\frac{\mathcal{D}_{{e^\dagger F_{1}S_{2}^\dagger }}^{{f_1p_1p_2}} \mathcal{D}_{{F_{1}HL}}^{{p_1f_2}} \mathcal{D}_{{LLS_{2}}}^{{f_3f_4p_2}}}{2 {M_{F_{1}}} {M_{S_{2}}^2}}
-\frac{\mathcal{D}_{{e^\dagger LS_{4}^\dagger }}^{{f_1f_2p_1}} \mathcal{C}_{{HS_{2}^\dagger S_{4}}}^{{p_2p_1}} \mathcal{D}_{{LLS_{2}}}^{{f_3f_4p_2}}}{{M_{S_{2}}^2} {M_{S_{4}}^2}}
-\frac{\mathcal{D}_{{e^\dagger F_{1}S_{2}^\dagger }}^{{f_1p_1p_2}} \mathcal{D}_{{F_{1}HL}}^{{p_1f_2}} \mathcal{D}_{{LLS_{2}}}^{{f_4f_3p_2}}}{2 {M_{F_{1}}} {M_{S_{2}}^2}}
+\frac{\mathcal{D}_{{e^\dagger LS_{4}^\dagger }}^{{f_1f_2p_1}} \mathcal{C}_{{HS_{2}^\dagger S_{4}}}^{{p_2p_1}} \mathcal{D}_{{LLS_{2}}}^{{f_4f_3p_2}}}{{M_{S_{2}}^2} {M_{S_{4}}^2}}
-\frac{\mathcal{D}_{{e^\dagger F_{4R}^\dagger H}}^{{f_1p_1}} \mathcal{D}_{{F_{4L}LS_{6}^\dagger }}^{{p_1f_4p_2}} \mathcal{D}_{{LLS_{6}}}^{{f_2f_3p_2}}}{2 {M_{F_{4}}} {M_{S_{6}}^2}}
-\frac{3 \mathcal{D}_{{e^\dagger F_{5}S_{6}^\dagger }}^{{f_1p_1p_2}} \mathcal{D}_{{F_{5}HL}}^{{p_1f_4}} \mathcal{D}_{{LLS_{6}}}^{{f_2f_3p_2}}}{4 {M_{F_{5}}} {M_{S_{6}}^2}}
-\frac{\mathcal{D}_{{e^\dagger LS_{4}^\dagger }}^{{f_1f_4p_1}} \mathcal{C}_{{HS_{4}S_{6}^\dagger }}^{{p_1p_2}} \mathcal{D}_{{LLS_{6}}}^{{f_2f_3p_2}}}{2 {M_{S_{4}}^2} {M_{S_{6}}^2}}
+\frac{\mathcal{D}_{{e^\dagger F_{4R}^\dagger H}}^{{f_1p_1}} \mathcal{D}_{{F_{4L}LS_{6}^\dagger }}^{{p_1f_4p_2}} \mathcal{D}_{{LLS_{6}}}^{{f_3f_2p_2}}}{2 {M_{F_{4}}} {M_{S_{6}}^2}}
+\frac{3 \mathcal{D}_{{e^\dagger F_{5}S_{6}^\dagger }}^{{f_1p_1p_2}} \mathcal{D}_{{F_{5}HL}}^{{p_1f_4}} \mathcal{D}_{{LLS_{6}}}^{{f_3f_2p_2}}}{4 {M_{F_{5}}} {M_{S_{6}}^2}}
+\frac{\mathcal{D}_{{e^\dagger LS_{4}^\dagger }}^{{f_1f_4p_1}} \mathcal{C}_{{HS_{4}S_{6}^\dagger }}^{{p_1p_2}} \mathcal{D}_{{LLS_{6}}}^{{f_3f_2p_2}}}{2 {M_{S_{4}}^2} {M_{S_{6}}^2}}
+\frac{\mathcal{D}_{{e^\dagger F_{5}S_{6}^\dagger }}^{{f_1p_1p_2}} \mathcal{D}_{{F_{5}HL}}^{{p_1f_2}} \mathcal{D}_{{LLS_{6}}}^{{f_3f_4p_2}}}{4 {M_{F_{5}}} {M_{S_{6}}^2}}
-\frac{\mathcal{D}_{{e^\dagger F_{4R}^\dagger H}}^{{f_1p_1}} \mathcal{D}_{{F_{4L}LS_{6}^\dagger }}^{{p_1f_2p_2}} \mathcal{D}_{{LLS_{6}}}^{{f_4f_3p_2}}}{{M_{F_{4}}} {M_{S_{6}}^2}}
-\frac{5 \mathcal{D}_{{e^\dagger F_{5}S_{6}^\dagger }}^{{f_1p_1p_2}} \mathcal{D}_{{F_{5}HL}}^{{p_1f_2}} \mathcal{D}_{{LLS_{6}}}^{{f_4f_3p_2}}}{4 {M_{F_{5}}} {M_{S_{6}}^2}}
-\frac{\mathcal{D}_{{e^\dagger LS_{4}^\dagger }}^{{f_1f_2p_1}} \mathcal{C}_{{HS_{4}S_{6}^\dagger }}^{{p_1p_2}} \mathcal{D}_{{LLS_{6}}}^{{f_4f_3p_2}}}{{M_{S_{4}}^2} {M_{S_{6}}^2}} 
\end{autobreak}
\end{align}
\begin{align}
\begin{autobreak}
 {C_{dLQLH1}} =
 
-\frac{{y}_{{d}}^{{f_1f_4*}} \mathcal{D}_{{F_{1}HL}}^{{p_1f_2}} \mathcal{D}_{{F_{1}HL}}^{{p_1f_3}}}{{M_{F_{1}}^3}}
-\frac{{y}_{{d}}^{{f_1f_4*}} \mathcal{D}_{{F_{5}HL}}^{{p_1f_2}} \mathcal{D}_{{F_{5}HL}}^{{p_1f_3}}}{2 {M_{F_{5}}^3}}
+\frac{8 \mathcal{D}_{{d^\dagger F_{10L}H}}^{{f_1p_1}} \mathcal{D}_{{F_{10R}^\dagger LS_{10}}}^{{p_1f_2p_2}} \mathcal{D}_{{LQS_{10}^\dagger }}^{{f_3f_4p_2}}}{{M_{F_{10}}} {M_{S_{10}}^2}}
+\frac{4 \mathcal{D}_{{d^\dagger F_{10L}H}}^{{f_1p_1}} \mathcal{D}_{{F_{10R}^\dagger LS_{14}}}^{{p_1f_2p_2}} \mathcal{D}_{{LQS_{14}^\dagger }}^{{f_3f_4p_2}}}{{M_{F_{10}}} {M_{S_{14}}^2}}
+\frac{4 \mathcal{D}_{{d^\dagger F_{10L}H}}^{{f_1p_1}} \mathcal{D}_{{F_{10R}^\dagger QS_{6}^\dagger }}^{{p_1f_4p_2}} \mathcal{D}_{{LLS_{6}}}^{{f_2f_3p_2}}}{{M_{F_{10}}} {M_{S_{6}}^2}}
+\frac{4 \mathcal{D}_{{d^\dagger F_{10L}H}}^{{f_1p_1}} \mathcal{D}_{{F_{10R}^\dagger QS_{6}^\dagger }}^{{p_1f_4p_2}} \mathcal{D}_{{LLS_{6}}}^{{f_3f_2p_2}}}{{M_{F_{10}}} {M_{S_{6}}^2}}
+\frac{4 \mathcal{D}_{{d^\dagger F_{14L}S_{6}^\dagger }}^{{f_1p_1p_2}} \mathcal{D}_{{F_{14R}^\dagger HQ}}^{{p_1f_4}} \mathcal{D}_{{LLS_{6}}}^{{f_2f_3p_2}}}{{M_{F_{14}}} {M_{S_{6}}^2}}
+\frac{4 \mathcal{D}_{{d^\dagger F_{14L}S_{6}^\dagger }}^{{f_1p_1p_2}} \mathcal{D}_{{F_{14R}^\dagger HQ}}^{{p_1f_4}} \mathcal{D}_{{LLS_{6}}}^{{f_3f_2p_2}}}{{M_{F_{14}}} {M_{S_{6}}^2}}
+\frac{4 \mathcal{D}_{{d^\dagger F_{1}S_{10}}}^{{f_1p_1p_2}} \mathcal{D}_{{F_{1}HL}}^{{p_1f_2}} \mathcal{D}_{{LQS_{10}^\dagger }}^{{f_3f_4p_2}}}{{M_{F_{1}}} {M_{S_{10}}^2}}
-\frac{2 \mathcal{D}_{{d^\dagger F_{5}S_{14}}}^{{f_1p_1p_2}} \mathcal{D}_{{LQS_{14}^\dagger }}^{{f_3f_4p_2}} \mathcal{D}_{{F_{5}HL}}^{{p_1f_2}}}{{M_{F_{5}}} {M_{S_{14}}^2}}
+\frac{8 \mathcal{D}_{{d^\dagger LS_{12}}}^{{f_1f_2p_1}} \mathcal{D}_{{F_{14L}LS_{12}^\dagger }}^{{p_2f_3p_1}} \mathcal{D}_{{F_{14R}^\dagger HQ}}^{{p_2f_4}}}{{M_{F_{14}}} {M_{S_{12}}^2}}
+\frac{4 \mathcal{D}_{{d^\dagger LS_{12}}}^{{f_1f_2p_1}} \mathcal{D}_{{F_{1}HL}}^{{p_2f_3}} \mathcal{D}_{{F_{1}QS_{12}^\dagger }}^{{p_2f_4p_1}}}{{M_{F_{1}}} {M_{S_{12}}^2}}
+\frac{8 \mathcal{D}_{{d^\dagger LS_{12}}}^{{f_1f_2p_1}} \mathcal{D}_{{LQS_{10}^\dagger }}^{{f_3f_4p_2}} \mathcal{C}_{{HS_{10}S_{12}^\dagger }}^{{p_2p_1}}}{{M_{S_{10}}^2} {M_{S_{12}}^2}}
+\frac{4 \mathcal{D}_{{d^\dagger LS_{12}}}^{{f_1f_2p_1}} \mathcal{D}_{{LQS_{14}^\dagger }}^{{f_3f_4p_2}} \mathcal{C}_{{HS_{12}^\dagger S_{14}}}^{{p_1p_2}}}{{M_{S_{12}}^2} {M_{S_{14}}^2}}
+\frac{2 \mathcal{D}_{{d^\dagger LS_{12}}}^{{f_1f_2p_1}} \mathcal{D}_{{F_{5}HL}}^{{p_2f_3}} \mathcal{D}_{{F_{5}QS_{12}^\dagger }}^{{p_2f_4p_1}}}{{M_{F_{5}}} {M_{S_{12}}^2}}
-\frac{2 \mathcal{D}_{{d^\dagger QS_{4}^\dagger }}^{{f_1f_4p_1}} \mathcal{D}_{{F_{1}HL}}^{{p_2f_2}} \mathcal{D}_{{F_{1L}S_{4}}}^{{p_2f_3p_1}}}{{M_{F_{1}}} {M_{S_{4}}^2}}
-\frac{2 \mathcal{D}_{{d^\dagger QS_{4}^\dagger }}^{{f_1f_4p_1}} \mathcal{D}_{{F_{1}HL}}^{{p_2f_3}} \mathcal{D}_{{F_{1L}S_{4}}}^{{p_2f_2p_1}}}{{M_{F_{1}}} {M_{S_{4}}^2}}
-\frac{2 \mathcal{D}_{{d^\dagger QS_{4}^\dagger }}^{{f_1f_4p_1}} \mathcal{D}_{{LLS_{6}}}^{{f_2f_3p_2}} \mathcal{C}_{{HS_{4}S_{6}^\dagger }}^{{p_1p_2}}}{{M_{S_{4}}^2} {M_{S_{6}}^2}}
-\frac{2 \mathcal{D}_{{d^\dagger QS_{4}^\dagger }}^{{f_1f_4p_1}} \mathcal{D}_{{LLS_{6}}}^{{f_3f_2p_2}} \mathcal{C}_{{HS_{4}S_{6}^\dagger }}^{{p_1p_2}}}{{M_{S_{4}}^2} {M_{S_{6}}^2}}
+\frac{\mathcal{D}_{{d^\dagger QS_{4}^\dagger }}^{{f_1f_4p_1}} \mathcal{D}_{{F_{5}HL}}^{{p_2f_2}} \mathcal{D}_{{F_{5L}S_{4}}}^{{p_2f_3p_1}}}{{M_{F_{5}}} {M_{S_{4}}^2}}
+\frac{\mathcal{D}_{{d^\dagger QS_{4}^\dagger }}^{{f_1f_4p_1}} \mathcal{D}_{{F_{5}HL}}^{{p_2f_3}} \mathcal{D}_{{F_{5L}S_{4}}}^{{p_2f_2p_1}}}{{M_{F_{5}}} {M_{S_{4}}^2}} 
\end{autobreak}
\end{align}
\begin{align}
\begin{autobreak}
 {C_{dLQLH2}} =
 
-\frac{4 \mathcal{D}_{{d^\dagger LS_{12}}}^{{f_1f_2p_1}} \mathcal{D}_{{F_{14L}LS_{12}^\dagger }}^{{p_2f_3p_1}} \mathcal{D}_{{F_{14R}^\dagger HQ}}^{{p_2f_4}}}{{M_{F_{14}}} {M_{S_{12}}^2}}
+\frac{{y}_{{d}}^{{f_1f_4*}} \mathcal{D}_{{F_{1}HL}}^{{p_1f_2}} \mathcal{D}_{{F_{1}HL}}^{{p_1f_3}}}{2 {M_{F_{1}}^3}}
+\frac{2 \mathcal{D}_{{d^\dagger QS_{4}^\dagger }}^{{f_1f_4p_1}} \mathcal{D}_{{F_{1}HL}}^{{p_2f_2}} \mathcal{D}_{{F_{1L}S_{4}}}^{{p_2f_3p_1}}}{{M_{F_{1}}} {M_{S_{4}}^2}}
+\frac{{y}_{{d}}^{{f_1f_4*}} \mathcal{D}_{{F_{5}HL}}^{{p_1f_2}} \mathcal{D}_{{F_{5}HL}}^{{p_1f_3}}}{4 {M_{F_{5}}^3}}
-\frac{2 \mathcal{D}_{{d^\dagger QS_{4}^\dagger }}^{{f_1f_4p_1}} \mathcal{D}_{{F_{5}HL}}^{{p_2f_3}} \mathcal{D}_{{F_{5L}S_{4}}}^{{p_2f_2p_1}}}{{M_{F_{5}}} {M_{S_{4}}^2}}
+\frac{\mathcal{D}_{{d^\dagger QS_{4}^\dagger }}^{{f_1f_4p_1}} \mathcal{D}_{{F_{5}HL}}^{{p_2f_2}} \mathcal{D}_{{F_{5L}S_{4}}}^{{p_2f_3p_1}}}{{M_{F_{5}}} {M_{S_{4}}^2}}
-\frac{4 \mathcal{D}_{{d^\dagger LS_{12}}}^{{f_1f_2p_1}} \mathcal{D}_{{F_{5}HL}}^{{p_2f_3}} \mathcal{D}_{{F_{5}QS_{12}^\dagger }}^{{p_2f_4p_1}}}{{M_{F_{5}}} {M_{S_{12}}^2}}
+\frac{8 \mathcal{D}_{{d^\dagger LS_{12}}}^{{f_1f_2p_1}} \mathcal{D}_{{F_{9L}LS_{12}^\dagger }}^{{p_2f_3p_1}} \mathcal{D}_{{F_{9R}^\dagger HQ}}^{{p_2f_4}}}{{M_{F_{9}}} {M_{S_{12}}^2}}
-\frac{4 \mathcal{D}_{{d^\dagger F_{10L}H}}^{{f_1p_1}} \mathcal{D}_{{F_{10R}^\dagger QS_{2}^\dagger }}^{{p_1f_4p_2}} \mathcal{D}_{{LLS_{2}}}^{{f_2f_3p_2}}}{{M_{F_{10}}} {M_{S_{2}}^2}}
-\frac{4 \mathcal{D}_{{d^\dagger F_{9L}S_{2}^\dagger }}^{{f_1p_1p_2}} \mathcal{D}_{{F_{9R}^\dagger HQ}}^{{p_1f_4}} \mathcal{D}_{{LLS_{2}}}^{{f_2f_3p_2}}}{{M_{F_{9}}} {M_{S_{2}}^2}}
+\frac{2 \mathcal{D}_{{d^\dagger QS_{4}^\dagger }}^{{f_1f_4p_1}} \mathcal{C}_{{HS_{2}^\dagger S_{4}}}^{{p_2p_1}} \mathcal{D}_{{LLS_{2}}}^{{f_2f_3p_2}}}{{M_{S_{2}}^2} {M_{S_{4}}^2}}
+\frac{4 \mathcal{D}_{{d^\dagger F_{10L}H}}^{{f_1p_1}} \mathcal{D}_{{F_{10R}^\dagger QS_{2}^\dagger }}^{{p_1f_4p_2}} \mathcal{D}_{{LLS_{2}}}^{{f_3f_2p_2}}}{{M_{F_{10}}} {M_{S_{2}}^2}}
+\frac{4 \mathcal{D}_{{d^\dagger F_{9L}S_{2}^\dagger }}^{{f_1p_1p_2}} \mathcal{D}_{{F_{9R}^\dagger HQ}}^{{p_1f_4}} \mathcal{D}_{{LLS_{2}}}^{{f_3f_2p_2}}}{{M_{F_{9}}} {M_{S_{2}}^2}}
-\frac{2 \mathcal{D}_{{d^\dagger QS_{4}^\dagger }}^{{f_1f_4p_1}} \mathcal{C}_{{HS_{2}^\dagger S_{4}}}^{{p_2p_1}} \mathcal{D}_{{LLS_{2}}}^{{f_3f_2p_2}}}{{M_{S_{2}}^2} {M_{S_{4}}^2}}
-\frac{2 \mathcal{D}_{{d^\dagger F_{10L}H}}^{{f_1p_1}} \mathcal{D}_{{F_{10R}^\dagger QS_{6}^\dagger }}^{{p_1f_4p_2}} \mathcal{D}_{{LLS_{6}}}^{{f_2f_3p_2}}}{{M_{F_{10}}} {M_{S_{6}}^2}}
-\frac{2 \mathcal{D}_{{d^\dagger F_{14L}S_{6}^\dagger }}^{{f_1p_1p_2}} \mathcal{D}_{{F_{14R}^\dagger HQ}}^{{p_1f_4}} \mathcal{D}_{{LLS_{6}}}^{{f_2f_3p_2}}}{{M_{F_{14}}} {M_{S_{6}}^2}}
+\frac{\mathcal{D}_{{d^\dagger QS_{4}^\dagger }}^{{f_1f_4p_1}} \mathcal{C}_{{HS_{4}S_{6}^\dagger }}^{{p_1p_2}} \mathcal{D}_{{LLS_{6}}}^{{f_2f_3p_2}}}{{M_{S_{4}}^2} {M_{S_{6}}^2}}
-\frac{2 \mathcal{D}_{{d^\dagger F_{10L}H}}^{{f_1p_1}} \mathcal{D}_{{F_{10R}^\dagger QS_{6}^\dagger }}^{{p_1f_4p_2}} \mathcal{D}_{{LLS_{6}}}^{{f_3f_2p_2}}}{{M_{F_{10}}} {M_{S_{6}}^2}}
-\frac{2 \mathcal{D}_{{d^\dagger F_{14L}S_{6}^\dagger }}^{{f_1p_1p_2}} \mathcal{D}_{{F_{14R}^\dagger HQ}}^{{p_1f_4}} \mathcal{D}_{{LLS_{6}}}^{{f_3f_2p_2}}}{{M_{F_{14}}} {M_{S_{6}}^2}}
+\frac{\mathcal{D}_{{d^\dagger QS_{4}^\dagger }}^{{f_1f_4p_1}} \mathcal{C}_{{HS_{4}S_{6}^\dagger }}^{{p_1p_2}} \mathcal{D}_{{LLS_{6}}}^{{f_3f_2p_2}}}{{M_{S_{4}}^2} {M_{S_{6}}^2}}
-\frac{8 \mathcal{D}_{{d^\dagger F_{10L}H}}^{{f_1p_1}} \mathcal{D}_{{F_{10R}^\dagger LS_{10}}}^{{p_1f_2p_2}} \mathcal{D}_{{LQS_{10}^\dagger }}^{{f_3f_4p_2}}}{{M_{F_{10}}} {M_{S_{10}}^2}}
-\frac{4 \mathcal{D}_{{d^\dagger F_{1}S_{10}}}^{{f_1p_1p_2}} \mathcal{D}_{{F_{1}HL}}^{{p_1f_2}} \mathcal{D}_{{LQS_{10}^\dagger }}^{{f_3f_4p_2}}}{{M_{F_{1}}} {M_{S_{10}}^2}}
-\frac{8 \mathcal{D}_{{d^\dagger LS_{12}}}^{{f_1f_2p_1}} \mathcal{C}_{{HS_{10}S_{12}^\dagger }}^{{p_2p_1}} \mathcal{D}_{{LQS_{10}^\dagger }}^{{f_3f_4p_2}}}{{M_{S_{10}}^2} {M_{S_{12}}^2}}
+\frac{4 \mathcal{D}_{{d^\dagger F_{10L}H}}^{{f_1p_1}} \mathcal{D}_{{F_{10R}^\dagger LS_{14}}}^{{p_1f_2p_2}} \mathcal{D}_{{LQS_{14}^\dagger }}^{{f_3f_4p_2}}}{{M_{F_{10}}} {M_{S_{14}}^2}}
-\frac{2 \mathcal{D}_{{d^\dagger F_{5}S_{14}}}^{{f_1p_1p_2}} \mathcal{D}_{{F_{5}HL}}^{{p_1f_2}} \mathcal{D}_{{LQS_{14}^\dagger }}^{{f_3f_4p_2}}}{{M_{F_{5}}} {M_{S_{14}}^2}}
+\frac{4 \mathcal{D}_{{d^\dagger LS_{12}}}^{{f_1f_2p_1}} \mathcal{C}_{{HS_{12}^\dagger S_{14}}}^{{p_1p_2}} \mathcal{D}_{{LQS_{14}^\dagger }}^{{f_3f_4p_2}}}{{M_{S_{12}}^2} {M_{S_{14}}^2}} 
\end{autobreak}
\end{align}
\begin{align}
\begin{autobreak}
 {C_{dLueH}} =
 \frac{8 \mathcal{D}_{{dF_{12R}^\dagger V_{3}}}^{{f_1p_1p_2*}} \mathcal{D}_{{e^\dagger L^\dagger V_{3}^\dagger }}^{{f_4f_2p_2*}} \mathcal{D}_{{F_{12L}H^\dagger u^\dagger }}^{{p_1f_5*}}}{{M_{F_{12}}} {M_{V_{3}}^2}}
-\frac{16 \mathcal{D}_{{d^\dagger eV_{5}}}^{{f_1f_4p_1}} \mathcal{D}_{{F_{12L}H^\dagger u^\dagger }}^{{p_2f_5*}} \mathcal{D}_{{F_{12R}^\dagger L^\dagger V_{5}}}^{{p_2f_2p_1*}}}{{M_{F_{12}}} {M_{V_{5}}^2}}
+\frac{8 \mathcal{D}_{{d^\dagger eV_{5}}}^{{f_1f_4p_1}} \mathcal{D}_{{F_{1}HL}}^{{p_2f_2}} \mathcal{D}_{{F_{1}^\dagger u^\dagger V_{5}}}^{{p_2f_5p_1*}}}{{M_{F_{1}}} {M_{V_{5}}^2}}
+\frac{32 \mathcal{D}_{{d^\dagger eV_{5}}}^{{f_1f_4p_1}} \mathcal{D}_{{L^\dagger u^\dagger V_{8}}}^{{f_2f_5p_2*}} \mathcal{C}_{{HV_{5}^\dagger V_{8}}}^{{p_1p_2}}}{{M_{V_{5}}^2} {M_{V_{8}}^2}}
+\frac{16 \mathcal{D}_{{d^\dagger F_{10L}H}}^{{f_1p_1}} \mathcal{D}_{{eF_{10R}^\dagger V_{8}}}^{{f_4p_1p_2}} \mathcal{D}_{{L^\dagger u^\dagger V_{8}}}^{{f_2f_5p_2*}}}{{M_{F_{10}}} {M_{V_{8}}^2}}
+\frac{8 \mathcal{D}_{{d^\dagger F_{10L}H}}^{{f_1p_1}} \mathcal{D}_{{e^\dagger L^\dagger V_{3}^\dagger }}^{{f_4f_2p_2*}} \mathcal{D}_{{F_{10R}^\dagger uV_{3}^\dagger }}^{{p_1f_5p_2}}}{{M_{F_{10}}} {M_{V_{3}}^2}}
+\frac{8 \mathcal{D}_{{d^\dagger F_{10L}H}}^{{f_1p_1}} \mathcal{D}_{{e^\dagger S_{10}u^\dagger }}^{{f_4p_2f_5*}} \mathcal{D}_{{F_{10R}^\dagger LS_{10}}}^{{p_1f_2p_2}}}{{M_{F_{10}}} {M_{S_{10}}^2}}
+\frac{4 \mathcal{D}_{{d^\dagger F_{1}S_{10}}}^{{f_1p_1p_2}} \mathcal{D}_{{e^\dagger S_{10}u^\dagger }}^{{f_4p_2f_5*}} \mathcal{D}_{{F_{1}HL}}^{{p_1f_2}}}{{M_{F_{1}}} {M_{S_{10}}^2}}
-\frac{8 \mathcal{D}_{{d^\dagger F_{3L}^\dagger V_{8}}}^{{f_1p_1p_2}} \mathcal{D}_{{e^\dagger F_{3R}^\dagger H^\dagger }}^{{f_4p_1*}} \mathcal{D}_{{L^\dagger u^\dagger V_{8}}}^{{f_2f_5p_2*}}}{{M_{F_{3}}} {M_{V_{8}}^2}}
+\frac{8 \mathcal{D}_{{d^\dagger LS_{12}}}^{{f_1f_2p_1}} \mathcal{D}_{{e^\dagger F_{12R}^\dagger S_{12}}}^{{f_4p_2p_1*}} \mathcal{D}_{{F_{12L}H^\dagger u^\dagger }}^{{p_2f_5*}}}{{M_{F_{12}}} {M_{S_{12}}^2}}
-\frac{4 \mathcal{D}_{{d^\dagger LS_{12}}}^{{f_1f_2p_1}} \mathcal{D}_{{e^\dagger F_{3R}^\dagger H^\dagger }}^{{f_4p_2*}} \mathcal{D}_{{F_{3L}S_{12}u^\dagger }}^{{p_2p_1f_5*}}}{{M_{F_{3}}} {M_{S_{12}}^2}}
+\frac{8 \mathcal{D}_{{d^\dagger LS_{12}}}^{{f_1f_2p_1}} \mathcal{D}_{{e^\dagger S_{10}u^\dagger }}^{{f_4p_2f_5*}} \mathcal{C}_{{HS_{10}S_{12}^\dagger }}^{{p_2p_1}}}{{M_{S_{10}}^2} {M_{S_{12}}^2}}
+\frac{4 \mathcal{D}_{{d^\dagger uV_{2}^\dagger }}^{{f_1f_5p_1}} \mathcal{D}_{{eF_{1}V_{2}}}^{{f_4p_2p_1}} \mathcal{D}_{{F_{1}HL}}^{{p_2f_2}}}{{M_{F_{1}}} {M_{V_{2}}^2}}
+\frac{4 \mathcal{D}_{{d^\dagger uV_{2}^\dagger }}^{{f_1f_5p_1}} \mathcal{D}_{{e^\dagger F_{3R}^\dagger H^\dagger }}^{{f_4p_2*}} \mathcal{D}_{{F_{3L}L^\dagger V_{2}^\dagger }}^{{p_2f_2p_1*}}}{{M_{F_{3}}} {M_{V_{2}}^2}}
-\frac{8 \mathcal{D}_{{d^\dagger uV_{2}^\dagger }}^{{f_1f_5p_1}} \mathcal{D}_{{e^\dagger L^\dagger V_{3}^\dagger }}^{{f_4f_2p_2*}} \mathcal{C}_{{HV_{2}V_{3}^\dagger }}^{{p_1p_2}}}{{M_{V_{2}}^2} {M_{V_{3}}^2}} 
\end{autobreak}
\end{align}
\begin{align}
\begin{autobreak}
 {C_{QuLLH}} =
 \frac{{y}_{{u}}^{{f_4f_5}} \mathcal{D}_{{F_{1}HL}}^{{p_1f_1}} \mathcal{D}_{{F_{1}HL}}^{{p_1f_2}}}{2 {M_{F_{1}}^3}}
+\frac{8 \mathcal{D}_{{F_{1}^\dagger u^\dagger V_{5}}}^{{p_1f_5p_2*}} \mathcal{D}_{{L^\dagger QV_{5}^\dagger }}^{{f_1f_4p_2*}} \mathcal{D}_{{F_{1}HL}}^{{p_1f_2}}}{{M_{F_{1}}} {M_{V_{5}}^2}}
+\frac{8 \mathcal{D}_{{F_{1}^\dagger QV_{8}^\dagger }}^{{p_1f_4p_2*}} \mathcal{D}_{{L^\dagger u^\dagger V_{8}}}^{{f_1f_5p_2*}} \mathcal{D}_{{F_{1}HL}}^{{p_1f_2}}}{{M_{F_{1}}} {M_{V_{8}}^2}}
+\frac{2 \mathcal{D}_{{F_{1L}S_{4}}}^{{p_1f_1p_2}} \mathcal{D}_{{QS_{4}u^\dagger }}^{{f_4p_2f_5*}} \mathcal{D}_{{F_{1}HL}}^{{p_1f_2}}}{{M_{F_{1}}} {M_{S_{4}}^2}}
+\frac{{y}_{{u}}^{{f_4f_5}} \mathcal{D}_{{F_{5}HL}}^{{p_1f_1}} \mathcal{D}_{{F_{5}HL}}^{{p_1f_2}}}{4 {M_{F_{5}}^3}}
-\frac{4 \mathcal{D}_{{F_{12L}H^\dagger u^\dagger }}^{{p_1f_5*}} \mathcal{D}_{{F_{12R}^\dagger QS_{2}}}^{{p_1f_4p_2*}} \mathcal{D}_{{LLS_{2}}}^{{f_1f_2p_2}}}{{M_{F_{12}}} {M_{S_{2}}^2}}
-\frac{4 \mathcal{D}_{{F_{8L}S_{2}u^\dagger }}^{{p_1p_2f_5*}} \mathcal{D}_{{F_{8R}^\dagger H^\dagger Q}}^{{p_1f_4*}} \mathcal{D}_{{LLS_{2}}}^{{f_1f_2p_2}}}{{M_{F_{8}}} {M_{S_{2}}^2}}
+\frac{4 \mathcal{D}_{{F_{12L}H^\dagger u^\dagger }}^{{p_1f_5*}} \mathcal{D}_{{F_{12R}^\dagger QS_{2}}}^{{p_1f_4p_2*}} \mathcal{D}_{{LLS_{2}}}^{{f_2f_1p_2}}}{{M_{F_{12}}} {M_{S_{2}}^2}}
+\frac{4 \mathcal{D}_{{F_{8L}S_{2}u^\dagger }}^{{p_1p_2f_5*}} \mathcal{D}_{{F_{8R}^\dagger H^\dagger Q}}^{{p_1f_4*}} \mathcal{D}_{{LLS_{2}}}^{{f_2f_1p_2}}}{{M_{F_{8}}} {M_{S_{2}}^2}}
+\frac{2 \mathcal{D}_{{F_{12L}H^\dagger u^\dagger }}^{{p_1f_5*}} \mathcal{D}_{{F_{12R}^\dagger QS_{6}}}^{{p_1f_4p_2*}} \mathcal{D}_{{LLS_{6}}}^{{f_1f_2p_2}}}{{M_{F_{12}}} {M_{S_{6}}^2}}
+\frac{2 \mathcal{D}_{{F_{13L}S_{6}u^\dagger }}^{{p_1p_2f_5*}} \mathcal{D}_{{F_{13R}^\dagger H^\dagger Q}}^{{p_1f_4*}} \mathcal{D}_{{LLS_{6}}}^{{f_1f_2p_2}}}{{M_{F_{13}}} {M_{S_{6}}^2}}
+\frac{2 \mathcal{D}_{{F_{12L}H^\dagger u^\dagger }}^{{p_1f_5*}} \mathcal{D}_{{F_{12R}^\dagger QS_{6}}}^{{p_1f_4p_2*}} \mathcal{D}_{{LLS_{6}}}^{{f_2f_1p_2}}}{{M_{F_{12}}} {M_{S_{6}}^2}}
+\frac{2 \mathcal{D}_{{F_{13L}S_{6}u^\dagger }}^{{p_1p_2f_5*}} \mathcal{D}_{{F_{13R}^\dagger H^\dagger Q}}^{{p_1f_4*}} \mathcal{D}_{{LLS_{6}}}^{{f_2f_1p_2}}}{{M_{F_{13}}} {M_{S_{6}}^2}}
-\frac{16 \mathcal{D}_{{F_{12L}H^\dagger u^\dagger }}^{{p_1f_5*}} \mathcal{D}_{{F_{12R}^\dagger L^\dagger V_{5}}}^{{p_1f_2p_2*}} \mathcal{D}_{{L^\dagger QV_{5}^\dagger }}^{{f_1f_4p_2*}}}{{M_{F_{12}}} {M_{V_{5}}^2}}
+\frac{8 \mathcal{D}_{{F_{12L}H^\dagger u^\dagger }}^{{p_1f_5*}} \mathcal{D}_{{F_{12R}^\dagger L^\dagger V_{9}}}^{{p_1f_2p_2*}} \mathcal{D}_{{L^\dagger QV_{9}^\dagger }}^{{f_1f_4p_2*}}}{{M_{F_{12}}} {M_{V_{9}}^2}}
+\frac{4 \mathcal{D}_{{F_{5}HL}}^{{p_1f_2}} \mathcal{D}_{{F_{5}^\dagger u^\dagger V_{9}}}^{{p_1f_5p_2*}} \mathcal{D}_{{L^\dagger QV_{9}^\dagger }}^{{f_1f_4p_2*}}}{{M_{F_{5}}} {M_{V_{9}}^2}}
-\frac{16 \mathcal{D}_{{F_{12L}H^\dagger u^\dagger }}^{{p_1f_5*}} \mathcal{D}_{{F_{12R}^\dagger L^\dagger V_{9}}}^{{p_1f_1p_2*}} \mathcal{D}_{{L^\dagger QV_{9}^\dagger }}^{{f_2f_4p_2*}}}{{M_{F_{12}}} {M_{V_{9}}^2}}
-\frac{8 \mathcal{D}_{{F_{5}HL}}^{{p_1f_1}} \mathcal{D}_{{F_{5}^\dagger u^\dagger V_{9}}}^{{p_1f_5p_2*}} \mathcal{D}_{{L^\dagger QV_{9}^\dagger }}^{{f_2f_4p_2*}}}{{M_{F_{5}}} {M_{V_{9}}^2}}
+\frac{32 \mathcal{C}_{{HV_{8}V_{9}^\dagger }}^{{p_1p_2}} \mathcal{D}_{{L^\dagger QV_{9}^\dagger }}^{{f_2f_4p_2*}} \mathcal{D}_{{L^\dagger u^\dagger V_{8}}}^{{f_1f_5p_1*}}}{{M_{V_{8}}^2} {M_{V_{9}}^2}}
-\frac{8 \mathcal{D}_{{F_{13L}L^\dagger V_{8}^\dagger }}^{{p_1f_2p_2*}} \mathcal{D}_{{F_{13R}^\dagger H^\dagger Q}}^{{p_1f_4*}} \mathcal{D}_{{L^\dagger u^\dagger V_{8}}}^{{f_1f_5p_2*}}}{{M_{F_{13}}} {M_{V_{8}}^2}}
-\frac{4 \mathcal{D}_{{F_{5}HL}}^{{p_1f_2}} \mathcal{D}_{{F_{5}^\dagger QV_{8}^\dagger }}^{{p_1f_4p_2*}} \mathcal{D}_{{L^\dagger u^\dagger V_{8}}}^{{f_1f_5p_2*}}}{{M_{F_{5}}} {M_{V_{8}}^2}}
-\frac{16 \mathcal{D}_{{F_{8L}L^\dagger V_{8}^\dagger }}^{{p_1f_2p_2*}} \mathcal{D}_{{F_{8R}^\dagger H^\dagger Q}}^{{p_1f_4*}} \mathcal{D}_{{L^\dagger u^\dagger V_{8}}}^{{f_1f_5p_2*}}}{{M_{F_{8}}} {M_{V_{8}}^2}}
-\frac{16 \mathcal{C}_{{HV_{8}V_{9}^\dagger }}^{{p_1p_2}} \mathcal{D}_{{L^\dagger QV_{9}^\dagger }}^{{f_1f_4p_2*}} \mathcal{D}_{{L^\dagger u^\dagger V_{8}}}^{{f_2f_5p_1*}}}{{M_{V_{8}}^2} {M_{V_{9}}^2}}
-\frac{8 \mathcal{D}_{{F_{13L}L^\dagger V_{8}^\dagger }}^{{p_1f_1p_2*}} \mathcal{D}_{{F_{13R}^\dagger H^\dagger Q}}^{{p_1f_4*}} \mathcal{D}_{{L^\dagger u^\dagger V_{8}}}^{{f_2f_5p_2*}}}{{M_{F_{13}}} {M_{V_{8}}^2}}
+\frac{8 \mathcal{D}_{{F_{5}HL}}^{{p_1f_1}} \mathcal{D}_{{F_{5}^\dagger QV_{8}^\dagger }}^{{p_1f_4p_2*}} \mathcal{D}_{{L^\dagger u^\dagger V_{8}}}^{{f_2f_5p_2*}}}{{M_{F_{5}}} {M_{V_{8}}^2}}
+\frac{16 \mathcal{D}_{{F_{8L}L^\dagger V_{8}^\dagger }}^{{p_1f_1p_2*}} \mathcal{D}_{{F_{8R}^\dagger H^\dagger Q}}^{{p_1f_4*}} \mathcal{D}_{{L^\dagger u^\dagger V_{8}}}^{{f_2f_5p_2*}}}{{M_{F_{8}}} {M_{V_{8}}^2}}
+\frac{32 \mathcal{C}_{{HV_{5}^\dagger V_{8}}}^{{p_1p_2}} \mathcal{D}_{{L^\dagger QV_{5}^\dagger }}^{{f_1f_4p_1*}} \mathcal{D}_{{L^\dagger u^\dagger V_{8}}}^{{f_2f_5p_2*}}}{{M_{V_{5}}^2} {M_{V_{8}}^2}}
+\frac{\mathcal{C}_{{HS_{4}S_{6}^\dagger }}^{{p_1p_2}} \mathcal{D}_{{LLS_{6}}}^{{f_1f_2p_2}} \mathcal{D}_{{QS_{4}u^\dagger }}^{{f_4p_1f_5*}}}{{M_{S_{4}}^2} {M_{S_{6}}^2}}
+\frac{\mathcal{C}_{{HS_{4}S_{6}^\dagger }}^{{p_1p_2}} \mathcal{D}_{{LLS_{6}}}^{{f_2f_1p_2}} \mathcal{D}_{{QS_{4}u^\dagger }}^{{f_4p_1f_5*}}}{{M_{S_{4}}^2} {M_{S_{6}}^2}}
+\frac{\mathcal{D}_{{F_{5}HL}}^{{p_1f_2}} \mathcal{D}_{{F_{5L}S_{4}}}^{{p_1f_1p_2}} \mathcal{D}_{{QS_{4}u^\dagger }}^{{f_4p_2f_5*}}}{{M_{F_{5}}} {M_{S_{4}}^2}}
-\frac{2 \mathcal{D}_{{F_{5}HL}}^{{p_1f_1}} \mathcal{D}_{{F_{5L}S_{4}}}^{{p_1f_2p_2}} \mathcal{D}_{{QS_{4}u^\dagger }}^{{f_4p_2f_5*}}}{{M_{F_{5}}} {M_{S_{4}}^2}}
-\frac{2 \mathcal{C}_{{HS_{2}^\dagger S_{4}}}^{{p_1p_2}} \mathcal{D}_{{LLS_{2}}}^{{f_1f_2p_1}} \mathcal{D}_{{QS_{4}u^\dagger }}^{{f_4p_2f_5*}}}{{M_{S_{2}}^2} {M_{S_{4}}^2}}
+\frac{2 \mathcal{C}_{{HS_{2}^\dagger S_{4}}}^{{p_1p_2}} \mathcal{D}_{{LLS_{2}}}^{{f_2f_1p_1}} \mathcal{D}_{{QS_{4}u^\dagger }}^{{f_4p_2f_5*}}}{{M_{S_{2}}^2} {M_{S_{4}}^2}} 
\end{autobreak}
\end{align}

\subsubsection{B-violating operators}

\begin{align}
\begin{autobreak}
 {C_{LdudH}} =
 \frac{16 \mathcal{D}_{{d^\dagger d^\dagger S_{11}^\dagger }}^{{f_2f_3p_1*}} \mathcal{D}_{{F_{11L}Hu^\dagger }}^{{p_2f_5*}} \mathcal{D}_{{F_{11R}^\dagger LS_{11}}}^{{p_2f_4p_1*}}}{{M_{F_{11}}} {M_{S_{11}}^2}}
-\frac{16 \mathcal{D}_{{d^\dagger d^\dagger S_{11}^\dagger }}^{{f_3f_2p_1*}} \mathcal{D}_{{F_{11L}Hu^\dagger }}^{{p_2f_5*}} \mathcal{D}_{{F_{11R}^\dagger LS_{11}}}^{{p_2f_4p_1*}}}{{M_{F_{11}}} {M_{S_{11}}^2}}
+\frac{8 \mathcal{D}_{{d^\dagger d^\dagger S_{11}^\dagger }}^{{f_2f_3p_1*}} \mathcal{D}_{{F_{1}HL}}^{{p_2f_4*}} \mathcal{D}_{{F_{1}S_{11}u^\dagger }}^{{p_2p_1f_5*}}}{{M_{F_{1}}} {M_{S_{11}}^2}}
-\frac{8 \mathcal{D}_{{d^\dagger d^\dagger S_{11}^\dagger }}^{{f_3f_2p_1*}} \mathcal{D}_{{F_{1}HL}}^{{p_2f_4*}} \mathcal{D}_{{F_{1}S_{11}u^\dagger }}^{{p_2p_1f_5*}}}{{M_{F_{1}}} {M_{S_{11}}^2}}
+\frac{16 \mathcal{D}_{{d^\dagger d^\dagger S_{11}^\dagger }}^{{f_2f_3p_1*}} \mathcal{D}_{{LS_{13}u^\dagger }}^{{f_4p_2f_5*}} \mathcal{C}_{{HS_{11}S_{13}^\dagger }}^{{p_1p_2*}}}{{M_{S_{11}}^2} {M_{S_{13}}^2}}
-\frac{16 \mathcal{D}_{{d^\dagger d^\dagger S_{11}^\dagger }}^{{f_3f_2p_1*}} \mathcal{D}_{{LS_{13}u^\dagger }}^{{f_4p_2f_5*}} \mathcal{C}_{{HS_{11}S_{13}^\dagger }}^{{p_1p_2*}}}{{M_{S_{11}}^2} {M_{S_{13}}^2}}
+\frac{16 \mathcal{D}_{{d^\dagger F_{10L}H}}^{{f_2p_1*}} \mathcal{D}_{{d^\dagger F_{10R}^\dagger S_{13}^\dagger }}^{{f_3p_1p_2*}} \mathcal{D}_{{LS_{13}u^\dagger }}^{{f_4p_2f_5*}}}{{M_{F_{10}}} {M_{S_{13}}^2}}
-\frac{16 \mathcal{D}_{{d^\dagger F_{10L}H}}^{{f_3p_1*}} \mathcal{D}_{{d^\dagger F_{10R}^\dagger S_{13}^\dagger }}^{{f_2p_1p_2*}} \mathcal{D}_{{LS_{13}u^\dagger }}^{{f_4p_2f_5*}}}{{M_{F_{10}}} {M_{S_{13}}^2}}
-\frac{16 \mathcal{D}_{{d^\dagger F_{10L}H}}^{{f_3p_1*}} \mathcal{D}_{{d^\dagger LS_{12}}}^{{f_2f_4p_2*}} \mathcal{D}_{{F_{10R}^\dagger S_{12}^\dagger u^\dagger }}^{{p_1p_2f_5*}}}{{M_{F_{10}}} {M_{S_{12}}^2}}
+\frac{16 \mathcal{D}_{{d^\dagger F_{10L}H}}^{{f_2p_1*}} \mathcal{D}_{{d^\dagger S_{10}^\dagger u^\dagger }}^{{f_3p_2f_5*}} \mathcal{D}_{{F_{10R}^\dagger LS_{10}}}^{{p_1f_4p_2*}}}{{M_{F_{10}}} {M_{S_{10}}^2}}
+\frac{16 \mathcal{D}_{{d^\dagger F_{11R}^\dagger S_{12}^\dagger }}^{{f_3p_1p_2*}} \mathcal{D}_{{d^\dagger LS_{12}}}^{{f_2f_4p_2*}} \mathcal{D}_{{F_{11L}Hu^\dagger }}^{{p_1f_5*}}}{{M_{F_{11}}} {M_{S_{12}}^2}}
+\frac{8 \mathcal{D}_{{d^\dagger F_{1}S_{10}}}^{{f_2p_1p_2*}} \mathcal{D}_{{d^\dagger S_{10}^\dagger u^\dagger }}^{{f_3p_2f_5*}} \mathcal{D}_{{F_{1}HL}}^{{p_1f_4*}}}{{M_{F_{1}}} {M_{S_{10}}^2}}
+\frac{16 \mathcal{D}_{{d^\dagger LS_{12}}}^{{f_2f_4p_1*}} \mathcal{D}_{{d^\dagger S_{10}^\dagger u^\dagger }}^{{f_3p_2f_5*}} \mathcal{C}_{{HS_{10}S_{12}^\dagger }}^{{p_2p_1*}}}{{M_{S_{10}}^2} {M_{S_{12}}^2}} 
\end{autobreak}
\end{align}
\begin{align}
\begin{autobreak}
 {C_{LdddH}} =
 
-\frac{16 \mathcal{D}_{{d^\dagger d^\dagger S_{11}^\dagger }}^{{f_2f_3p_1*}} \mathcal{D}_{{d^\dagger F_{11L}H^\dagger }}^{{f_4p_2*}} \mathcal{D}_{{F_{11R}^\dagger LS_{11}}}^{{p_2f_5p_1*}}}{{M_{F_{11}}} {M_{S_{11}}^2}}
+\frac{16 \mathcal{D}_{{d^\dagger d^\dagger S_{11}^\dagger }}^{{f_3f_2p_1*}} \mathcal{D}_{{d^\dagger F_{11L}H^\dagger }}^{{f_4p_2*}} \mathcal{D}_{{F_{11R}^\dagger LS_{11}}}^{{p_2f_5p_1*}}}{{M_{F_{11}}} {M_{S_{11}}^2}}
-\frac{8 \mathcal{D}_{{d^\dagger d^\dagger S_{11}^\dagger }}^{{f_2f_3p_1*}} \mathcal{D}_{{d^\dagger F_{2R}^\dagger S_{11}}}^{{f_4p_2p_1*}} \mathcal{D}_{{F_{2L}H^\dagger L}}^{{p_2f_5*}}}{{M_{F_{2}}} {M_{S_{11}}^2}}
+\frac{8 \mathcal{D}_{{d^\dagger d^\dagger S_{11}^\dagger }}^{{f_3f_2p_1*}} \mathcal{D}_{{d^\dagger F_{2R}^\dagger S_{11}}}^{{f_4p_2p_1*}} \mathcal{D}_{{F_{2L}H^\dagger L}}^{{p_2f_5*}}}{{M_{F_{2}}} {M_{S_{11}}^2}}
-\frac{32 \mathcal{D}_{{d^\dagger d^\dagger S_{11}^\dagger }}^{{f_2f_3p_1*}} \mathcal{D}_{{d^\dagger LS_{12}}}^{{f_4f_5p_2*}} \mathcal{C}_{{HS_{11}^\dagger S_{12}}}^{{p_1p_2}}}{{M_{S_{11}}^2} {M_{S_{12}}^2}}
+\frac{16 \mathcal{D}_{{d^\dagger d^\dagger S_{11}^\dagger }}^{{f_4f_3p_1*}} \mathcal{D}_{{d^\dagger LS_{12}}}^{{f_2f_5p_2*}} \mathcal{C}_{{HS_{11}^\dagger S_{12}}}^{{p_1p_2}}}{{M_{S_{11}}^2} {M_{S_{12}}^2}}
+\frac{32 \mathcal{D}_{{d^\dagger d^\dagger S_{11}^\dagger }}^{{f_3f_2p_1*}} \mathcal{D}_{{d^\dagger LS_{12}}}^{{f_4f_5p_2*}} \mathcal{C}_{{HS_{11}^\dagger S_{12}}}^{{p_1p_2}}}{{M_{S_{11}}^2} {M_{S_{12}}^2}}
-\frac{16 \mathcal{D}_{{d^\dagger F_{11L}H^\dagger }}^{{f_2p_1*}} \mathcal{D}_{{d^\dagger F_{11R}^\dagger S_{12}^\dagger }}^{{f_3p_1p_2*}} \mathcal{D}_{{d^\dagger LS_{12}}}^{{f_4f_5p_2*}}}{{M_{F_{11}}} {M_{S_{12}}^2}}
+\frac{16 \mathcal{D}_{{d^\dagger F_{11L}H^\dagger }}^{{f_3p_1*}} \mathcal{D}_{{d^\dagger F_{11R}^\dagger S_{12}^\dagger }}^{{f_2p_1p_2*}} \mathcal{D}_{{d^\dagger LS_{12}}}^{{f_4f_5p_2*}}}{{M_{F_{11}}} {M_{S_{12}}^2}} 
\end{autobreak}
\end{align}
\begin{align}
\begin{autobreak}
 {C_{eQddH}} =
 \frac{16 \mathcal{D}_{{dQV_{8}}}^{{f_4f_2p_2}} \mathcal{D}_{{d^\dagger eV_{5}}}^{{f_5f_1p_1*}} \mathcal{C}_{{HV_{5}^\dagger V_{8}}}^{{p_1p_2*}}}{{M_{V_{5}}^2} {M_{V_{8}}^2}}
+\frac{8 \mathcal{D}_{{dQV_{8}}}^{{f_4f_2p_2}} \mathcal{D}_{{d^\dagger F_{10L}H}}^{{f_5p_1*}} \mathcal{D}_{{eF_{10R}^\dagger V_{8}}}^{{f_1p_1p_2*}}}{{M_{F_{10}}} {M_{V_{8}}^2}}
-\frac{4 \mathcal{D}_{{dQV_{8}}}^{{f_4f_2p_2}} \mathcal{D}_{{d^\dagger F_{3L}^\dagger V_{8}}}^{{f_5p_1p_2*}} \mathcal{D}_{{e^\dagger F_{3R}^\dagger H^\dagger }}^{{f_1p_1}}}{{M_{F_{3}}} {M_{V_{8}}^2}}
-\frac{8 \mathcal{D}_{{d^\dagger d^\dagger S_{11}^\dagger }}^{{f_5f_4p_1*}} \mathcal{D}_{{e^\dagger F_{3R}^\dagger H^\dagger }}^{{f_1p_2}} \mathcal{D}_{{F_{3L}QS_{11}^\dagger }}^{{p_2f_2p_1}}}{{M_{F_{3}}} {M_{S_{11}}^2}}
+\frac{16 \mathcal{D}_{{d^\dagger d^\dagger S_{11}^\dagger }}^{{f_5f_4p_1*}} \mathcal{D}_{{e^\dagger F_{8L}S_{11}^\dagger }}^{{f_1p_2p_1}} \mathcal{D}_{{F_{8R}^\dagger H^\dagger Q}}^{{p_2f_2}}}{{M_{F_{8}}} {M_{S_{11}}^2}}
+\frac{16 \mathcal{D}_{{d^\dagger d^\dagger S_{11}^\dagger }}^{{f_5f_4p_1*}} \mathcal{D}_{{e^\dagger QS_{13}^\dagger }}^{{f_1f_2p_2}} \mathcal{C}_{{HS_{11}S_{13}^\dagger }}^{{p_1p_2*}}}{{M_{S_{11}}^2} {M_{S_{13}}^2}}
-\frac{32 \mathcal{D}_{{d^\dagger eV_{5}}}^{{f_4f_1p_1*}} \mathcal{D}_{{d^\dagger F_{10L}H}}^{{f_5p_2*}} \mathcal{D}_{{F_{10R}^\dagger Q^\dagger V_{5}^\dagger }}^{{p_2f_2p_1*}}}{{M_{F_{10}}} {M_{V_{5}}^2}}
+\frac{32 \mathcal{D}_{{d^\dagger eV_{5}}}^{{f_4f_1p_1*}} \mathcal{D}_{{d^\dagger F_{8L}^\dagger V_{5}^\dagger }}^{{f_5p_2p_1*}} \mathcal{D}_{{F_{8R}^\dagger H^\dagger Q}}^{{p_2f_2}}}{{M_{F_{8}}} {M_{V_{5}}^2}}
-\frac{16 \mathcal{D}_{{d^\dagger F_{10L}H}}^{{f_4p_1*}} \mathcal{D}_{{d^\dagger F_{10R}^\dagger S_{13}^\dagger }}^{{f_5p_1p_2*}} \mathcal{D}_{{e^\dagger QS_{13}^\dagger }}^{{f_1f_2p_2}}}{{M_{F_{10}}} {M_{S_{13}}^2}} 
\end{autobreak}
\end{align}
\begin{align}
\begin{autobreak}
 {C_{LdQQH}} =
 
-\frac{8 \mathcal{D}_{{dQV_{8}}}^{{f_4f_2p_1}} \mathcal{D}_{{F_{13L}L^\dagger V_{8}^\dagger }}^{{p_2f_5p_1}} \mathcal{D}_{{F_{13R}^\dagger H^\dagger Q}}^{{p_2f_1}}}{{M_{F_{13}}} {M_{V_{8}}^2}}
-\frac{4 \mathcal{D}_{{d^\dagger LS_{12}}}^{{f_4f_5p_1*}} \mathcal{D}_{{F_{13L}QS_{12}}}^{{p_2f_2p_1}} \mathcal{D}_{{F_{13R}^\dagger H^\dagger Q}}^{{p_2f_1}}}{{M_{F_{13}}} {M_{S_{12}}^2}}
-\frac{4 \mathcal{D}_{{dQV_{8}}}^{{f_4f_1p_1}} \mathcal{D}_{{F_{13L}L^\dagger V_{8}^\dagger }}^{{p_2f_5p_1}} \mathcal{D}_{{F_{13R}^\dagger H^\dagger Q}}^{{p_2f_2}}}{{M_{F_{13}}} {M_{V_{8}}^2}}
-\frac{2 \mathcal{D}_{{d^\dagger LS_{12}}}^{{f_4f_5p_1*}} \mathcal{D}_{{F_{13L}QS_{12}}}^{{p_2f_1p_1}} \mathcal{D}_{{F_{13R}^\dagger H^\dagger Q}}^{{p_2f_2}}}{{M_{F_{13}}} {M_{S_{12}}^2}}
+\frac{4 \mathcal{D}_{{dQV_{8}}}^{{f_4f_2p_1}} \mathcal{D}_{{F_{1}HL}}^{{p_2f_5*}} \mathcal{D}_{{F_{1}^\dagger QV_{8}^\dagger }}^{{p_2f_1p_1}}}{{M_{F_{1}}} {M_{V_{8}}^2}}
+\frac{4 \mathcal{D}_{{dQV_{8}}}^{{f_4f_1p_1}} \mathcal{D}_{{F_{1}HL}}^{{p_2f_5*}} \mathcal{D}_{{F_{1}^\dagger QV_{8}^\dagger }}^{{p_2f_2p_1}}}{{M_{F_{1}}} {M_{V_{8}}^2}}
+\frac{2 \mathcal{D}_{{dQV_{8}}}^{{f_4f_2p_1}} \mathcal{D}_{{F_{5}HL}}^{{p_2f_5*}} \mathcal{D}_{{F_{5}^\dagger QV_{8}^\dagger }}^{{p_2f_1p_1}}}{{M_{F_{5}}} {M_{V_{8}}^2}}
-\frac{2 \mathcal{D}_{{dQV_{8}}}^{{f_4f_1p_1}} \mathcal{D}_{{F_{5}HL}}^{{p_2f_5*}} \mathcal{D}_{{F_{5}^\dagger QV_{8}^\dagger }}^{{p_2f_2p_1}}}{{M_{F_{5}}} {M_{V_{8}}^2}}
-\frac{8 \mathcal{D}_{{dQV_{8}}}^{{f_4f_1p_1}} \mathcal{D}_{{F_{8L}L^\dagger V_{8}^\dagger }}^{{p_2f_5p_1}} \mathcal{D}_{{F_{8R}^\dagger H^\dagger Q}}^{{p_2f_2}}}{{M_{F_{8}}} {M_{V_{8}}^2}}
+\frac{4 \mathcal{D}_{{d^\dagger LS_{12}}}^{{f_4f_5p_1*}} \mathcal{D}_{{F_{8L}QS_{12}}}^{{p_2f_1p_1}} \mathcal{D}_{{F_{8R}^\dagger H^\dagger Q}}^{{p_2f_2}}}{{M_{F_{8}}} {M_{S_{12}}^2}}
+\frac{32 \mathcal{D}_{{d^\dagger F_{10L}H}}^{{f_4p_1*}} \mathcal{D}_{{F_{10R}^\dagger Q^\dagger V_{5}^\dagger }}^{{p_1f_2p_2*}} \mathcal{D}_{{L^\dagger QV_{5}^\dagger }}^{{f_5f_1p_2}}}{{M_{F_{10}}} {M_{V_{5}}^2}}
-\frac{32 \mathcal{D}_{{d^\dagger F_{8L}^\dagger V_{5}^\dagger }}^{{f_4p_1p_2*}} \mathcal{D}_{{F_{8R}^\dagger H^\dagger Q}}^{{p_1f_2}} \mathcal{D}_{{L^\dagger QV_{5}^\dagger }}^{{f_5f_1p_2}}}{{M_{F_{8}}} {M_{V_{5}}^2}}
+\frac{16 \mathcal{D}_{{dQV_{8}}}^{{f_4f_2p_1}} \mathcal{C}_{{HV_{5}^\dagger V_{8}}}^{{p_2p_1*}} \mathcal{D}_{{L^\dagger QV_{5}^\dagger }}^{{f_5f_1p_2}}}{{M_{V_{5}}^2} {M_{V_{8}}^2}}
+\frac{16 \mathcal{D}_{{d^\dagger F_{10L}H}}^{{f_4p_1*}} \mathcal{D}_{{F_{10R}^\dagger Q^\dagger V_{9}^\dagger }}^{{p_1f_2p_2*}} \mathcal{D}_{{L^\dagger QV_{9}^\dagger }}^{{f_5f_1p_2}}}{{M_{F_{10}}} {M_{V_{9}}^2}}
+\frac{4 \mathcal{D}_{{dF_{13L}V_{9}}}^{{f_4p_1p_2}} \mathcal{D}_{{F_{13R}^\dagger H^\dagger Q}}^{{p_1f_2}} \mathcal{D}_{{L^\dagger QV_{9}^\dagger }}^{{f_5f_1p_2}}}{{M_{F_{13}}} {M_{V_{9}}^2}}
+\frac{8 \mathcal{D}_{{dQV_{8}}}^{{f_4f_2p_1}} \mathcal{C}_{{HV_{8}V_{9}^\dagger }}^{{p_1p_2*}} \mathcal{D}_{{L^\dagger QV_{9}^\dagger }}^{{f_5f_1p_2}}}{{M_{V_{8}}^2} {M_{V_{9}}^2}}
+\frac{32 \mathcal{D}_{{d^\dagger F_{10L}H}}^{{f_4p_1*}} \mathcal{D}_{{F_{10R}^\dagger Q^\dagger V_{9}^\dagger }}^{{p_1f_1p_2*}} \mathcal{D}_{{L^\dagger QV_{9}^\dagger }}^{{f_5f_2p_2}}}{{M_{F_{10}}} {M_{V_{9}}^2}}
+\frac{8 \mathcal{D}_{{dF_{13L}V_{9}}}^{{f_4p_1p_2}} \mathcal{D}_{{F_{13R}^\dagger H^\dagger Q}}^{{p_1f_1}} \mathcal{D}_{{L^\dagger QV_{9}^\dagger }}^{{f_5f_2p_2}}}{{M_{F_{13}}} {M_{V_{9}}^2}}
+\frac{16 \mathcal{D}_{{dQV_{8}}}^{{f_4f_1p_1}} \mathcal{C}_{{HV_{8}V_{9}^\dagger }}^{{p_1p_2*}} \mathcal{D}_{{L^\dagger QV_{9}^\dagger }}^{{f_5f_2p_2}}}{{M_{V_{8}}^2} {M_{V_{9}}^2}}
-\frac{4 \mathcal{D}_{{d^\dagger F_{10L}H}}^{{f_4p_1*}} \mathcal{D}_{{F_{10R}^\dagger LS_{10}}}^{{p_1f_5p_2*}} \mathcal{D}_{{QQS_{10}}}^{{f_1f_2p_2}}}{{M_{F_{10}}} {M_{S_{10}}^2}}
-\frac{2 \mathcal{D}_{{d^\dagger F_{1}S_{10}}}^{{f_4p_1p_2*}} \mathcal{D}_{{F_{1}HL}}^{{p_1f_5*}} \mathcal{D}_{{QQS_{10}}}^{{f_1f_2p_2}}}{{M_{F_{1}}} {M_{S_{10}}^2}}
-\frac{4 \mathcal{D}_{{d^\dagger LS_{12}}}^{{f_4f_5p_1*}} \mathcal{C}_{{HS_{10}S_{12}^\dagger }}^{{p_2p_1*}} \mathcal{D}_{{QQS_{10}}}^{{f_1f_2p_2}}}{{M_{S_{10}}^2} {M_{S_{12}}^2}}
-\frac{4 \mathcal{D}_{{d^\dagger F_{10L}H}}^{{f_4p_1*}} \mathcal{D}_{{F_{10R}^\dagger LS_{10}}}^{{p_1f_5p_2*}} \mathcal{D}_{{QQS_{10}}}^{{f_2f_1p_2}}}{{M_{F_{10}}} {M_{S_{10}}^2}}
-\frac{2 \mathcal{D}_{{d^\dagger F_{1}S_{10}}}^{{f_4p_1p_2*}} \mathcal{D}_{{F_{1}HL}}^{{p_1f_5*}} \mathcal{D}_{{QQS_{10}}}^{{f_2f_1p_2}}}{{M_{F_{1}}} {M_{S_{10}}^2}}
-\frac{4 \mathcal{D}_{{d^\dagger LS_{12}}}^{{f_4f_5p_1*}} \mathcal{C}_{{HS_{10}S_{12}^\dagger }}^{{p_2p_1*}} \mathcal{D}_{{QQS_{10}}}^{{f_2f_1p_2}}}{{M_{S_{10}}^2} {M_{S_{12}}^2}}
+\frac{2 \mathcal{D}_{{d^\dagger F_{10L}H}}^{{f_4p_1*}} \mathcal{D}_{{F_{10R}^\dagger LS_{14}}}^{{p_1f_5p_2*}} \mathcal{D}_{{QQS_{14}}}^{{f_1f_2p_2}}}{{M_{F_{10}}} {M_{S_{14}}^2}}
-\frac{\mathcal{D}_{{d^\dagger F_{5}S_{14}}}^{{f_4p_1p_2*}} \mathcal{D}_{{F_{5}HL}}^{{p_1f_5*}} \mathcal{D}_{{QQS_{14}}}^{{f_1f_2p_2}}}{{M_{F_{5}}} {M_{S_{14}}^2}}
+\frac{2 \mathcal{D}_{{d^\dagger LS_{12}}}^{{f_4f_5p_1*}} \mathcal{C}_{{HS_{12}^\dagger S_{14}}}^{{p_1p_2*}} \mathcal{D}_{{QQS_{14}}}^{{f_1f_2p_2}}}{{M_{S_{12}}^2} {M_{S_{14}}^2}}
-\frac{2 \mathcal{D}_{{d^\dagger F_{10L}H}}^{{f_4p_1*}} \mathcal{D}_{{F_{10R}^\dagger LS_{14}}}^{{p_1f_5p_2*}} \mathcal{D}_{{QQS_{14}}}^{{f_2f_1p_2}}}{{M_{F_{10}}} {M_{S_{14}}^2}}
+\frac{\mathcal{D}_{{d^\dagger F_{5}S_{14}}}^{{f_4p_1p_2*}} \mathcal{D}_{{F_{5}HL}}^{{p_1f_5*}} \mathcal{D}_{{QQS_{14}}}^{{f_2f_1p_2}}}{{M_{F_{5}}} {M_{S_{14}}^2}}
-\frac{2 \mathcal{D}_{{d^\dagger LS_{12}}}^{{f_4f_5p_1*}} \mathcal{C}_{{HS_{12}^\dagger S_{14}}}^{{p_1p_2*}} \mathcal{D}_{{QQS_{14}}}^{{f_2f_1p_2}}}{{M_{S_{12}}^2} {M_{S_{14}}^2}} 
\end{autobreak}
\end{align}

\section{Dim-5, 6 Warsaw and Dim-7 Green basis}
\label{sec:basis567}

\begin{table}[h]
  \centering
  {  
\begin{tabular}{|c|c|}
\hline
\multicolumn{2}{|c|}{\color{blue} Type: $\psi^2 H^2$} \\

\hline 
${\cal O}_{5}$ & $\epsilon^{ik}\epsilon^{jl}(\ell^T_{i} C \ell_{j}) H_k H_l$ \\

\hline
\end{tabular}
}\caption{The Weinberg operator.}\label{tab:basis5}
\end{table}

\begin{table}[h]
  \centering
  \resizebox{1.0\textwidth}{!}{  
\begin{tabular}{|c|c|c|c|c|c|}
\hline 
\multicolumn{2}{|c|}{\color{blue} Type: $X^3$} & \multicolumn{2}{c|}{\color{blue} Type: $H^4D^2$} & \multicolumn{2}{c|}{\color{blue} Type: $\psi^2H^3$}  \\

\hline 
${\cal O}_{G}$ & $f^{ABC} G^{A \nu}_{\mu} G^{B\rho}_{\nu} G^{C \mu}_{\rho}$ & ${\cal O}_{H\square}$ & $\left(H^\dagger H\right)\square \left(H^\dagger H\right) $ & ${\cal O}_{eH}$ & $ (H^\dagger H) (\bar{\ell} e {H})$ \\

${\cal O}_{\tilde{G}}$ & $f^{ABC} \tilde{G}^{A \nu}_{\mu} G^{B\rho}_{\nu} G^{C \mu}_{\rho}$ & ${\cal O}_{HD}$ & $\left(H^\dagger D^\mu H\right)^* \left(H^\dagger D^\mu H\right) $ & ${\cal O}_{uH}$ & $ (H^\dagger H) (\bar{q} u \tilde{H})$ \\

\cline{3-4} 
${\cal O}_{W}$ & $\epsilon^{IJK} W^{I \nu}_{\mu} W^{J\rho}_{\nu} W^{K \mu}_{\rho}$ & \multicolumn{2}{c|}{\color{blue} Type: $H^6$} & ${\cal O}_{dH}$ & $ (H^\dagger H) (\bar{q} d {H})$ \\

\cline{3-4} 
${\cal O}_{\tilde{W}}$ & $\epsilon^{ABC} \tilde{W}^{I \nu}_{\mu} W^{J\rho}_{\nu} W^{K \mu}_{\rho}$ & ${\cal O}_H$ & $\left(H^\dagger H\right)^3$ &  &  \\

\hline 
\multicolumn{2}{|c|}{\color{blue} Type: $X^2H^2$} & \multicolumn{2}{c|}{\color{blue} Type: $\psi^2 X H$} & \multicolumn{2}{c|}{\color{blue} Type: $\psi^2 H^2 D$} \\

\hline 
${\cal O}_{HG}$ & $H^\dagger H G^A_{\mu\nu} G^{A\mu\nu}$ & ${\cal O}_{eW}$ & $(\bar{\ell}\sigma^{\mu\nu} e )\tau^I H W^I_{\mu\nu}$ & ${\cal O}_{H\ell}^{(1)}$ & $(H^\dagger i \overset{\leftrightarrow}{D}{}_\mu H)(\bar{\ell}\gamma^\mu \ell)$ \\

${\cal O}_{H\tilde{G}}$ & $H^\dagger H \tilde{G}^A_{\mu\nu} G^{A\mu\nu}$ & ${\cal O}_{eB}$ & $(\bar{\ell}_L\sigma^{\mu\nu} e_R ) H B_{\mu\nu}$ & ${\cal O}_{H\ell}^{(3)}$ & $(H^\dagger i \overset{\leftrightarrow}{D}{}_\mu^I H)(\bar{\ell}\tau^I\gamma^\mu \ell)$ \\

${\cal O}_{HW}$ & $H^\dagger H W^I_{\mu\nu} W^{I\mu\nu}$ & ${\cal O}_{uG}$ & $(\bar{q}\sigma^{\mu\nu}T^A u)\tilde{H} G^A_{\mu\nu}$ & ${\cal O}_{He}$ & $(H^\dagger i \overset{\leftrightarrow}{D}{}_\mu H)(\bar{e}\gamma^\mu e)$ \\

${\cal O}_{H\tilde{W}}$ & $H^\dagger H \tilde{W}^I_{\mu\nu} W^{I\mu\nu}$ & ${\cal O}_{uW}$ & $(\bar{q}\sigma^{\mu\nu} u)\tau^I \tilde{H} W^I_{\mu\nu}$ & ${\cal O}_{Hq}^{(1)}$ & $(H^\dagger i \overset{\leftrightarrow}{D}{}_\mu H)(\bar{q}\gamma^\mu q)$ \\

${\cal O}_{HB}$ & $H^\dagger H B_{\mu\nu} B^{\mu\nu}$ & ${\cal O}_{uB}$ & $(\bar{q}\sigma^{\mu\nu} u) \tilde{H} B_{\mu\nu}$ & ${\cal O}_{Hq}^{(3)}$ & $(H^\dagger i \overset{\leftrightarrow}{D}{}_\mu^I H)(\bar{q}\tau^I\gamma^\mu q)$ \\

${\cal O}_{H\tilde{B}}$ & $H^\dagger H \tilde{B}_{\mu\nu} B^{\mu\nu}$ & ${\cal O}_{dG}$ & $(\bar{q}\sigma^{\mu\nu}T^A d)H G^A_{\mu\nu}$ & ${\cal O}_{Hu}$ & $(H^\dagger i \overset{\leftrightarrow}{D}{}_\mu H)(\bar{u}\gamma^\mu u)$ \\

${\cal O}_{HWB}$ & $H^\dagger \tau^I H W^I_{\mu\nu} B^{\mu\nu}$ & ${\cal O}_{dW}$ & $(\bar{q}\sigma^{\mu\nu} d)\tau^I H W^I_{\mu\nu}$ & ${\cal O}_{Hd}$ & $(H^\dagger i \overset{\leftrightarrow}{D}{}_\mu H)(\bar{d}\gamma^\mu d)$ \\

${\cal O}_{H\tilde{W}B}$ & $H^\dagger \tau^I H \tilde{W}^I_{\mu\nu} B^{\mu\nu}$ & ${\cal O}_{dB}$ & $(\bar{q}\sigma^{\mu\nu} d) H B^I_{\mu\nu}$ & ${\cal O}_{Hud}$ & $i(\tilde{H}^\dagger i D_\mu H)(\bar{u}\gamma^\mu d)$ \\


\hline 
\multicolumn{2}{|c|}{\color{blue} Type: $(\overline{L}L)(\overline{L}L)$} & \multicolumn{2}{c|}{\color{blue} Type: $(\overline{R}R)(\overline{R}R)$} & \multicolumn{2}{c|}{\color{blue} Type: $(\overline{L}L)(\overline{R}R)$}  \\

\hline 
${\cal O}_{\ell\ell}$ & $(\bar{\ell} \gamma^\mu \ell)(\bar{\ell} \gamma_\mu \ell)$ & ${\cal O}_{ee}$ & $(\bar{e} \gamma^\mu e)(\bar{e} \gamma_\mu e)$ & ${\cal O}_{\ell e}$ & $(\bar{\ell} \gamma^\mu \ell)(\bar{e} \gamma_\mu e)$ \\

${\cal O}_{qq}^{(1)}$ & $(\bar{q} \gamma^\mu q)(\bar{q} \gamma_\mu q)$ & ${\cal O}_{uu}$ & $(\bar{u} \gamma^\mu u)(\bar{u} \gamma_\mu u)$ & ${\cal O}_{\ell u}$ & $(\bar{\ell} \gamma^\mu \ell)(\bar{u} \gamma_\mu u)$ \\

${\cal O}_{qq}^{(3)}$ & $(\bar{q} \gamma^\mu \tau^I q)(\bar{q} \gamma_\mu \tau^I q)$ & ${\cal O}_{dd}$ & $(\bar{d} \gamma^\mu d)(\bar{d} \gamma_\mu d)$ & ${\cal O}_{\ell d}$ & $(\bar{\ell} \gamma^\mu \ell)(\bar{d} \gamma_\mu d)$ \\

${\cal O}_{\ell q}^{(1)}$ & $(\bar{\ell} \gamma^\mu \ell)(\bar{q} \gamma_\mu q)$ & ${\cal O}_{eu}$ & $(\bar{e} \gamma^\mu e)(\bar{u} \gamma_\mu u)$ & ${\cal O}_{qe}$ & $(\bar{q} \gamma^\mu q)(\bar{e} \gamma_\mu e)$ \\

${\cal O}_{\ell q}^{(3)}$ & $(\bar{\ell} \gamma^\mu \tau^I \ell)(\bar{q} \gamma_\mu \tau^I q)$ & ${\cal O}_{ed}$ & $(\bar{e} \gamma^\mu e)(\bar{d} \gamma_\mu d)$ & ${\cal O}_{qu}^{(1)}$ & $(\bar{q} \gamma^\mu q)(\bar{u} \gamma_\mu u)$ \\

& & ${\cal O}_{ud}^{(1)}$ & $(\bar{u} \gamma^\mu u)(\bar{d} \gamma_\mu d)$ & ${\cal O}_{qu}^{(8)}$ & $(\bar{q} \gamma^\mu T^A q)(\bar{u} \gamma_\mu T^A u)$ \\

\cline{1-2}
\multicolumn{2}{|c|}{\color{blue} Type: $(\overline{L}R)(\overline{R}L)$} & ${\cal O}_{ud}^{(8)}$ & $(\bar{u} \gamma^\mu T^A u)(\bar{d} \gamma_\mu T^A d)$ & ${\cal O}_{qd}^{(1)}$ & $(\bar{q} \gamma^\mu q)(\bar{d} \gamma_\mu d)$ \\

\cline{1-2}
${\cal O}_{\ell edq}$ & $(\bar{\ell} e)(\bar{d} q)$ & & & ${\cal O}_{qd}^{(8)}$ & $(\bar{q} \gamma^\mu T^A q)(\bar{d} \gamma_\mu T^A d)$ \\

\hline 
\multicolumn{2}{|c|}{\color{blue} Type: $(\overline{L}R)(\overline{L}R)$} & \multicolumn{4}{c|}{\color{blue} Type: $B$-violating}  \\

\hline 
${\cal O}_{quqd}^{(1)}$ & $(\bar{q}^j u)\epsilon_{jk}(\bar{q}^k d)$ & ${\cal O}_{duq}$ & \multicolumn{3}{c|}{ $\epsilon^{abc}\epsilon^{jk} (d^T_{a} C u_{b}) (q^T_{cj} C \ell_{k})$ } \\

${\cal O}_{quqd}^{(8)}$ & $(\bar{q}^j T^A u)\epsilon_{jk}(\bar{q}^k T^A d)$ & ${\cal O}_{qqu}$ & \multicolumn{3}{c|}{ $\epsilon^{abc}\epsilon_{jk} (q^T_{aj} C q_{bk}) (u^T_{c} C e)$ } \\

${\cal O}_{\ell equ}^{(1)}$ & $(\bar{\ell}^j e)\epsilon_{jk}(\bar{q}^k u)$ & ${\cal O}_{qqq}$ & \multicolumn{3}{c|}{ $\epsilon^{abc}\epsilon_{jn}\epsilon_{km} (q^T_{aj} C q_{bk}) (q^T_{cm} C \ell_{n})$ } \\

${\cal O}_{\ell equ}^{(3)}$ & $(\bar{\ell}^j \sigma_{\mu\nu} e)\epsilon_{jk}(\bar{q}^k \sigma^{\mu\nu} u)$ & ${\cal O}_{duu}$ & \multicolumn{3}{c|}{ $\epsilon^{abc} (d^T_{a} C u_{b}) (u^T_{c} C e)$ } \\







\hline
\end{tabular}
}\caption{The Warsaw basis~\cite{Grzadkowski:2010es}.}\label{tab:basis6}
\end{table}


\begin{table}[h]
  \centering
  \resizebox{1.0\textwidth}{!}{  
\begin{tabular}{|c|c|c|c|}
\hline 
\multicolumn{4}{|c|}{\color{blue} Only $L$-violating} \\

\hline
\multicolumn{2}{|c|}{\color{blue} Type: $\psi^2 H^4$} & \multicolumn{2}{c|}{\color{blue} Type: $\psi^2 H^3 D$} \\

\hline 
${\cal O}_{LH}$ & $\epsilon^{ik}\epsilon^{jl}(\ell^T_{i} C \ell_j) H_k H_l (H^\dagger H)$ & ${\cal O}_{LeHD}$ & $\epsilon^{ij}\epsilon^{kl} (\ell^T_{i} C \gamma^\mu e) H_j H_k (iD_\mu H_l) $ \\

\hline 
\multicolumn{2}{|c|}{\color{blue} Type: $\psi^2 H^2 D^2$} & \multicolumn{2}{c|}{\color{blue} Type: $\psi^2 H^2 X$} \\

\hline
${\cal O}_{LDH1}$ & $\epsilon^{ij}\epsilon^{kl} (\ell^T_{i} C D_{\mu} \ell_j) (H_k  D^\mu H_l)$ & ${\cal O}_{LHW}$ & $\epsilon^{ik}(\epsilon\tau^I)^{jl} (\ell^T_{i} C i\sigma^{\mu\nu} \ell_j) H_k H_l W^I_{\mu\nu} $\\

${\cal O}_{LDH2}$ & $\epsilon^{ik}\epsilon^{jl} (\ell^T_{i} C D_{\mu} \ell_j) (H_k  D^\mu H_l)$ & ${\cal O}_{LHB}$ & $\epsilon^{ik}\epsilon^{jl} (\ell^T_{i} C i\sigma^{\mu\nu} \ell_j) H_k H_l B_{\mu\nu} $ \\

{\color{gray}{${\cal R}_{LDH3}$}} & {\color{gray}{$\epsilon^{ik}\epsilon^{jl} (\ell^T_{i} C  \ell_j) H_k  H_l$}} & & \\

{\color{gray}{${\cal R}_{LDH4}$}} & {\color{gray}{$\epsilon^{ik}\epsilon^{jl} (\ell^T_{i} C \ell_j) (H_k D_{\mu}D^\nu H_l)$}} & & \\

{\color{gray}{${\cal R}_{LDH5}$}} & {\color{gray}{$\epsilon^{ik}\epsilon^{jl} (\ell^T_{i} C i\sigma^{\mu\nu} \ell_j) (D_{\mu}H_k) (D^\nu H_l)$}} & & \\

{\color{gray}{${\cal R}_{LDH6}$}} & {\color{gray}{$\epsilon^{ik}\epsilon^{jl} (\ell^T_{i} C i\sigma^{\mu\nu}D_\mu \ell_j) (H_k D_\nu H_l)$}} & & \\

\hline 
\multicolumn{2}{|c|}{\color{blue} Type: $\psi^4 D$} & \multicolumn{2}{c|}{\color{blue} Type: $\psi^4 H$} \\

\hline
${\cal O}_{duLLD}$ & $\epsilon^{ij} (\overline{d}^{a} \gamma^\mu u_{a}) (\ell^T_{i} C iD_{\mu}  \ell_j) $ & ${\cal O}_{eLLLH}$ & $\epsilon^{ij}\epsilon^{kl} (\overline{e}\ell_i) (\ell^T_{j} C \ell_k) H_l $ \\

{\color{gray}{${\cal R}_{duLLD2}$}} & {\color{gray}{$\epsilon^{ij} (\overline{d}^{a} iD_{\mu}  \ell_j ) (\ell^T_{i} C \gamma^\mu u_{a}) $}} & ${\cal O}_{dLQLH1}$ & $\epsilon^{ij}\epsilon^{kl} (\overline{d}^{a}\ell_{i}) (q^T_{aj} C \ell_k) H_l $ \\

{\color{gray}{${\cal R}_{duLLD3}$}} & {\color{gray}{$\epsilon^{ij} (\overline{d}^{a} \ell_i) (\ell^T_{j} C i\slashed{D} u_a) $}} & ${\cal O}_{dLQLH2}$ & $\epsilon^{ik}\epsilon^{jl} (\overline{d}^{a}\ell_{i}) (q^T_{aj} C \ell_k) H_l $ \\

& & ${\cal O}_{dLueH}$ & $\epsilon^{ij} (\overline{d}^{a} \ell_{i}) (u^T_{a} C e) H_j $ \\

& & ${\cal O}_{QuLLH}$ & $\epsilon^{ij} (\overline{q}^{ak} u_{a}) (\ell^T_{k} C \ell_i) H_j $ \\

\hline 
\multicolumn{4}{|c|}{\color{blue} $L$- and $B$-violating} \\

\hline 
\multicolumn{2}{|c|}{\color{blue} Type: $\psi^4 D$} & \multicolumn{2}{c|}{\color{blue} Type: $\psi^4 H$} \\

\hline
${\cal O}_{LQddD}$ & $\epsilon^{abc}(\overline{\ell}^{i} \gamma^\mu q_{ai}) (d^T_{b} C iD_{\mu} d_{c}) $ & ${\cal O}_{LdudH}$ & $\epsilon^{abc}\epsilon_{ij} (\overline{\ell}^i d_{a}) (u^T_{b} C d_{c}) H^{*j} $ \\

{\color{gray}{${\cal R}_{LQddD2}$}} & {\color{gray}{$\epsilon^{abc}(q_{ai}^T C \gamma_\mu d_b) (\bar{\ell}^i iD^\mu d_c) $}} & ${\cal O}_{LdddH}$ & $\epsilon^{abc}(\overline{\ell}^i d_{a}) (d^T_{b} C d_{c}) H_i $ \\

{\color{gray}{${\cal R}_{LQddD3}$}} & {\color{gray}{$\epsilon^{abc}(q_{ai}^T C i\overset{\leftarrow}{\slashed{D}} d_b) (\bar{\ell}^i d_c) $}} & ${\cal O}_{eQddH}$ & $-\epsilon^{abc}(\overline{e} Q_{ai}) (d^T_{b} C d_{c}) H^{*i} $ \\

${\cal O}_{edddD}$ & $\epsilon^{abc}(\overline{e} \gamma^\mu d_{a}) (d^T_{b} C iD_{\mu} d_{c}) $ & ${\cal O}_{LdQQH}$ & $-\epsilon^{abc}(\overline{\ell}^{k} d_{a}) (q^T_{bk} C q_{ci}) H^{*i} $ \\

\hline
\end{tabular}
}
\caption{The dimension-7 Green basis. The redundant operators ${\cal R}_i$ are marked in gray.}
\label{tab:basis7}
\end{table}

\newpage

\bibliographystyle{JHEP}
\bibliography{ref}

\end{document}